%% file: tese.tex
\documentclass[11pt,a4paper,twoside,openright]{book}
\usepackage{mythesis}
\usepackage{booktabs} 
\usepackage{textcomp} 
\usepackage{mathrsfs} 

\usepackage{amsmath, amsthm, amsfonts, amssymb, latexsym}
\usepackage{graphicx}
\usepackage{color}
\usepackage{appendix}
\usepackage{tabularx} 
\usepackage{ltablex}
\usepackage{cite} 

\usepackage{cases} 

\usepackage{afterpage}

\allowdisplaybreaks 

\newtheorem{thm}{Theorem} 

\usepackage[T1]{fontenc} 

\usepackage{caption}
\usepackage[perpage,symbol*]{footmisc}

\usepackage{paralist}

\usepackage{epigraph}

\usepackage{url}

\usepackage[pagebackref,unicode,hypertexnames=false]{hyperref}  

\hypersetup{
    unicode=true,          
    pdftitle={Quantum and classical aspects of scalar and vector fields around black holes},
    pdfauthor={Mengjie Wang},
    pdfsubject={PhD Thesis},
    pdfcreator={LaTeX},          
    pdfproducer={LaTeX},         
    pdfkeywords={General Relativity, Black Holes, Higher Dimensions, Scalar fields, Proca fields, Maxwell fields},
    colorlinks=false,            
    bookmarks=true,              
    linktoc=page,                
  }
\renewcommand*{\backref}[1]{}
\renewcommand*{\backrefalt}[4]{%
  \ifcase #1 { } \or (Cited on page~#2.) \else %
  (Cited on pages~#2.) \fi %
}

\let\origdoublepage\cleardoublepage
\newcommand{\clearemptydoublepage}{%
  \clearpage
  {\pagestyle{empty}\origdoublepage}%
}

\let\cleardoublepage\clearemptydoublepage

\title{Quantum and classical aspects of scalar and vector fields around black holes
}
\author{Mengjie Wang}
\date{Fevereiro de 2016}

\submitionplace{
Disserta\c{c}\~ao apresentada \`a Universidade de Aveiro no \^ambito do Programa Doutoral MAP-fis para
  cumprimento dos requisitos necess\'arios \`a obten\c{c}\~ao do grau de Doutor em F\'isica,
  realizada sob a orienta\c{c}\~ao cient\'ifica do Doutor Carlos Herdeiro, Professor Auxiliar com Agrega\c{c}\~ao, e co-orienta\c{c}\~ao do Doutor Marco O. P. Sampaio, do
  Departamento de F\'isica da Universidade de Aveiro.}

\begin{document}

\pagestyle{plain}

\coverp
\titlep
\dedication
\jury


\include{acks}
\cleardoublepage

\include{abs}
\cleardoublepage
\tableofcontents
\cleardoublepage
\listoftables
\cleardoublepage
\listoffigures
\cleardoublepage

\pagestyle{headings}
\numberwithin{equation}{section}

\include{intro}
\include{prelim}

\part{Scalar and Proca fields on asymptotically flat spacetimes} 
\label{main:p1} 
\include{GeneralInFlat}
\include{NeutralP}
\include{ChargedP}
\include{ChargedClouds}

\part{Scalar and Maxwell fields on asymptotically AdS spacetimes}
\label{main:p2}
\include{scalarHD}
\include{KerrAdS}

\include{conclusion}
\appendix
\include{app_neutralP}
\include{app_chargedP}
\include{app_ST1}
\include{app_angmomflux}

\include{pub-list}

\cleardoublepage
\phantomsection
\addcontentsline{toc}{chapter}{Bibliography}
\bibliographystyle{myutphys}
\bibliography{tese}

\end{document}

%% file: acks.tex
\chapter*{Acknowledgements}

Confucius said, ``\textit{lost time is never around the clock}''. Along my way to study physics during these years, I have received enormous encouragement and assistance from my family, my friends and my colleagues. I think it is the time to express my gratitude to all of them.

First of all, I am very thankful to my supervisor, Prof. Carlos Herdeiro, for his support and inspiration, for teaching me how to work on physics and for giving me the freedom to explore my own interests.

Special thanks are reserved for Marco Sampaio, for our enjoyable collaboration on various projects. As my co-supervisor, he has taught me a lot of physics and numerical techniques, while as a friend, his positiveness always reminds me to be optimistic when I face problems. I would also like to thank the other colleagues from the gravitation and high energy physics group at University of Aveiro, for all the great discussions about physics and others, they are Carolina Benone, Fl\'avio Coelho, Pedro Cunha, Juan Carlos Degollado, Jai Grover, Antonio Morais, Jo\~ao Rosa, Eugen Radu, and Helgi R\'unarsson. In particular, I am grateful to Juan Carlos Degollado and Jo\~ao Rosa for their interesting courses, and various discussions on the topic of superradiance.

I would like to acknowledge the financial support from Funda\c{c}\~ao para a Ci\^encia e a Tecnologia (FCT)--the International Doctorate Network in Particle Physics, Astrophysics and Cosmology (IDPASC) programme, with the grant SFRH/BD/51648/2011 during the completion of this thesis.

I owe my deepest gratitude to all my previous colleagues in China, especially Prof. Dezhi Huang, for encouraging me to work in a different field, and Prof. Jiliang Jing, for introducing me the black hole physics and supporting me to study aboard. I would also like to thank all my friends in China, who are too many to be listed individually, for their constant support. 

Finally, I would like to thank my parents, who teach me to be a virtuous man, and my sister, who is always on my side. Their endless love and support make me get better.

%% file: abs.tex
\begin{tabularx}{\textwidth}{lX}	
{\bf Keywords:} & Black holes, Proca fields, Hawking radiation, TeV gravity, scalar fields, Maxwell fields, asymptotically anti-de Sitter spacetimes, quasinormal modes, superradiance.\\
\\
{\bf Abstract:} & This thesis presents recent studies on test scalar and vector fields around black holes in the classical theory of General Relativity. It is separated in two parts according to the asymptotic properties of the spacetime under study. \\
& In the first part, we investigate scalar and Proca fields on an asymptotically flat background. For the Proca field, we obtain a complete set of equations of motion in higher dimensional spherically symmetric backgrounds. These equations are solved numerically, both to compute Hawking radiation spectra and quasi-bound states. In the former case, for the first time, we carry out a precise study of the longitudinal degrees of freedom induced by the mass of the field. This can be used to improve the modeling of evaporation of black holes coupled to massive vector fields, and black hole event generators currently used at the Large Hadron Collider to probe TeV gravity models with extra dimensions. Regarding quasi-bound states, we find arbitrarily long lived modes for a charged Proca field in a Reissner-Nordstr\"om black hole. As a comparison, we also find such long lived modes for a charged scalar field. \\
& The second part of this thesis presents research on superradiant instabilities of scalar and Maxwell fields on an asymptotically anti-de Sitter background. For the scalar case, we introduce a charge coupling between the field and the background, and show that superradiant instabilities do exist for all values of the total angular momentum, $\ell$, in higher dimensions. This result corrects a statement in the literature that such instabilities only appear in even dimensions. For the Maxwell case, we first propose a general prescription to impose boundary conditions on the Kerr-anti-de Sitter spacetime, and obtain two Robin boundary conditions which give two different quasinormal modes even in a simpler Schwarzschild-anti-de Sitter black hole. Then these
\end{tabularx}

\begin{tabularx}{\textwidth}{lX}	
{\color{white}{Abstract:}} & two boundary conditions are implemented to study superradiant unstable modes and vector clouds. In particular, we find that the new branch of quasinormal modes may be unstable in a larger parameter space. Furthermore, the existence of vector clouds indicates that one may find a vector hairy black hole solution for the Einstein-Maxwell-anti-de Sitter system at the nonlinear level, which implies, in such system, that the Kerr-Newman-anti-de Sitter black hole is not a unique solution.
\end{tabularx}

\cleardoublepage

\begin{tabularx}{\textwidth}{lX}	
{\bf Palavras-chave:} & Buracos negros, Campos de Proca, Radia\c{c}\~ao de Hawking, gravidade \`a escala do TeV, campos escalares, campos de Maxwell, asimt\'oticamente anti-de Sitter, modos quasi-normais, super-radi\^ancia.\\
\\
{\bf Resumo:} & Nesta tese apresentamos estudos recentes sobre campos escalares e vetoriais de teste, em torno de buracos negros na teoria cl\'assica da relatividade geral. A tese encontra-se dividida em duas partes, de acordo com as propriedades asimt\'oticas do espa\c{c}o-tempo em estudo. \\
& Na primeira parte, investigamos os campos escalar e de Proca num espa\c{c}o asimt\'oticamente plano. Para o campo de Proca, obtemos um conjunto completo de equa\c{c}\~oes do movimento em espa\c{c}os esfericamente sim\'etricos em dimens\~oes elevadas. Estas equa\c{c}\~oes s\~ao resolvidas numericamente, tanto para o c\'alculo de radia\c{c}\~ao de Hawking como para o c\'alculo de estados quasi-ligados. No primeiro c\'alculo, pela primeira vez, efetuamos um estudo preciso dos graus de liberdade longitudinais que s\~ao induzidos pelo termo de massa do campo. Este estudo pode ser usado para melhorar o modelo da evapora\c{c}\~ao de buracos negros acoplados a campos vetoriais massivos e geradores de eventos de buraco negro usados presentemente no Grande Colisor de H\'adrons para testar modelos de gravidade com dimens\~oes extra \`a escala do TeV. Relativamente aos estados quasi-ligados, encontramos estados com tempos de vida arbitrariamente longos para o campo de Proca carregado, no buraco negro de Reissner-Nordstr\"om. Como compara\c{c}\~ao, obtemos estados com tempos de vida arbitrariamente longos tamb\'em para o campo escalar.\\
& Na segunda parte da tese, apresentamos investiga\c{c}\~ao sobre instabilidades super-radiantes para os campos escalar e de Maxwell em espa\c{c}o asimt\'oticamente anti-de Sitter. No caso escalar introduzimos um acoplamento de carga entre o campo e o background e mostramos que instabilidades super-radiantes existem para todos os valores do momento angular total, $\ell$, em dimens\~oes mais elevadas. Este resultado
\end{tabularx}

\begin{tabularx}{\textwidth}{lX}	
{\color{white}{Abstract:}} & corrige a afirma\c{c}\~ao na literatura de que estas instabilidades aparecem apenas em dimens\~oes \'impares. Para o caso do campo de Maxwell, propomos primeiro uma prescri\c{c}\~ao para impor condi\c{c}\~oes fronteira no espa\c{c}o tempo de Kerr-anti-de Sitter obtendo duas condi\c{c}\~oes fronteira do tipo de Robin que originam dois tipos diferentes de modos quasi-normais, mesmo no caso mais simples do buraco negro de Schwarzschild-anti-de Sitter. Estas duas condi\c{c}\~oes fronteira s\~ao implementadas no estudo de modos super-radiantes inst\'aveis e nuvens vetoriais. Em particular, encontramos um novo ramo de modos quasi-normais que podem conter instabilidades mais fortes. Mostramos ainda que a exist\^encia de nuvens vetoriais indica a poss\'ivel exist\^encia de solu\c{c}\~oes de buraco negro com cabelo vetorial para o sistema Einstein-Maxwell-anti-de Sitter a n\'ivel n\~ao linear, o que implica, nesse sistema, que o buraco negro de Kerr-Newman-anti-de Sitter poder\'a n\~ao ser \'unico.
\end{tabularx}




%% file: intro.tex
\chapter{Introduction}
\label{ch:intro}



\section{Background and motivation}
\label{sc:BM}
General Relativity (GR), as one of the pillars of modern physics, was established one hundred years ago by Albert Einstein in 1915. It unifies space and time, and in particular gravity is described by the curvature of spacetime. Mathematically, the theory is formulated by the elegant Einstein field equations~\cite{Einstein:1916vd}, where the geometry is related to the distribution of matter and radiation. These equations were interpreted by John Wheeler in his famous statement that~\textit{matter tells spacetime how to curve, and spacetime tells matter how to move}. GR has been tested with high accuracy in the regime of weak gravity~\cite{Will:2014kxa}, while in the strong gravity regime the first direct observation on gravitational waves has been reported recently~\cite{Abbott:2016blz}. From an analysis of the waves, black holes are identified as the source for such an event~\cite{Abbott:2016blz}.


The concept of black hole (BH) can be dated back to the end of the eighteenth century. At that time, John Michell~\cite{Michell:1784xqa} and
Pierre-Simon Laplace~\cite{Israel:1987ae} put forward an idea that the largest bodies in the universe may be invisible since they are so massive
that even light could not escape, which were dubbed as \textit{dark stars}. This idea was revived by Robert Oppenheimer and his collaborators, more than one century later, in their studies of gravitational collapse, where they concluded that neutron stars above the Tolman-Oppenheimer-Volkoff limit (approximately 1.5 to 3 solar masses) would collapse~\cite{Oppenheimer:1939ne}. Such collapsed objects were called \textit{frozen stars}. In 1967, the term \textit{black hole} was introduced by John Wheeler~\cite{Wheeler:1998vs}, and since then it was quickly accepted for general use. 

From a modern viewpoint, BHs are the simplest macroscopic objects in nature, in the sense that they can be uniquely characterized by their mass, spin and charge. This is the well known~\textit{no hair conjecture}~\cite{Misner:1974qy}. This conjecture has been circumvented in different contexts, for example in the Einstein-Yang-Mills theory~\cite{Bizon:1990sr,Kuenzle:1990is,Volkov:1989fi,volkov1990black}, in the Horndeski theory~\cite{Sotiriou:2013qea,Sotiriou:2014pfa,Babichev:2013cya} and recently in the Einstein-Klein-Gordon system where a Kerr BH with scalar hair was found~\cite{Herdeiro:2014goa}. Among them, scalar hairy BH solutions are supported by the phenomenon of superradiance~\cite{Brito:2015oca}, which provides a mechanism to generate hairy BH solutions in general~\cite{Herdeiro:2014goa,Herdeiro:2014ima}. Another interesting phenomenon in BH physics is the Hawking radiation~\cite{Hawking:1974rv}. Hawking radiation has been attracting a lot of attention, not only because it relates gravity to quantum theory which may provide a connection to a \textit{quantum theory} of gravity, but also because it might be visible in high energy processes.

In this thesis, we are going to study BHs interacting with scalar and vector fields at the linear level, in the context of \textit{Hawking radiation}
and \textit{superradiance}. The motivation for these studies is as follows.

{\bf TeV gravity scenarios}\\
The study of gravitational theories in higher dimensions has been discussed for a century, at least since the works by Nordstr\"om~\cite{Nordstrom:1988fi}, Kaluza~\cite{Kaluza:1921tu} and Klein~\cite{Klein:1926tv}. During the last four decades, moreover, the naturalness of extra dimensions within supergravity and string theory made it a topic of intense research within high energy theoretical physics. At the end of the last century, this research led to models that, aiming at solving the hierarchy problem\footnote{The hierarchy problem refers to the relative weakness of gravity, around sixteen orders of magnitude, by comparing to the other fundamental interactions.}, predicted that the extra dimensions could be very large (or even infinite) in size, as compared to the traditional Planck scale.

Within such scenario, the true fundamental Planck scale could be as low as the TeV scale~\cite{Antoniadis:1990ew,Arkani-Hamed:1998rs,Antoniadis:1998ig,Arkani-Hamed:1998nn} so that the formation and evaporation of microscopic BHs could be visible in realistic man-made particle accelerators\footnote{See, for example, the latest reports from CMS~\cite{CMS:2015iwr} and ATLAS~\cite{Aad:2015mzg} in the search for microscopic BHs.}, such as the Large Hadron Collider (LHC). This motivates our study on Hawking radiation. In particular, the second run of the LHC is ongoing, with a center of mass energy of 13 TeV, where such scenarios will be properly tested. Therefore, any improvements on the phenomenology of these models are quite timely.

{\bf Asymptotically anti-de Sitter spacetimes}
\\
Anti-de Sitter (AdS) spacetime is the unique maximally symmetric solution of the vacuum Einstein equations with a negative cosmological constant. Asymptotically AdS spacetimes, referring to spacetimes which share the conformal boundary with AdS but may be different in the bulk, have attracted a lot of attention in theoretical physics. One reason is the AdS/CFT correspondence~\cite{Maldacena:1997re} which conjectures a duality between gravity in the $d$-dimensional AdS bulk and a quantum conformal field theory living in the $(d-1)$-dimensional conformal boundary. Another reason is the timelike property of the AdS boundary, which leads to interesting novel features, as compared to asymptotically flat spacetimes, such as the weak turbulent instability~\cite{Bizon:2011gg} and the superradiant instability for massless fields~\cite{Cardoso:2013pza,Uchikata:2009zz,Wang:2015fgp}.

With these motivations in mind, in the following we will briefly describe the physical phenomena of Hawking radiation and superradiance.

\section{Hawking radiation}
\label{sc:HR}
Hawking radiation~\cite{Hawking:1974rv} describes black body radiation that is predicted to be released by BHs, due to quantum effects
close to the event horizon. It is one of the most important features arising from quantum field theory in curved spacetime, discovered by Hawking \mbox{in 1974}. This effect was derived in a semiclassical framework, in the sense that the background geometry is classical (governed by classical gravitational theories) while the propagating fields are quantized. Since Hawking radiation connects classical gravity with quantum theory, it has inspired many works to re-derive Hawking
radiation through alternative methods, see for example~\cite{Hawking:1974sw,Hartle:1976tp,Damour:1976jd,Parikh:1999mf,Robinson:2005pd}, with an expectation to get a deeper understanding of gravity itself.

Physically, Hawking radiation can be understood through an intuitive picture by considering virtual particles generated from the vacuum\footnote{Note that this interpretation may lead to a flawed intuition on where does Hawking radiation originate~\cite{Giddings:2015uzr}.}. As it is well known since Dirac, the quantum vacuum is not completely empty but it contains fluctuations which produce particle-antiparticle pairs. Close to the event horizon of a BH, strong gravity effects may separate particle-antiparticle pairs, and if the antiparticle is attracted into the interior of the hole then the particle can escape to infinity thus generating Hawking radiation.

The phenomenon of Hawking radiation found recently an interesting application in TeV gravity models. In such models, scattering processes with center of mass
energy well above \mbox{the fundamental} Planck scale, should be dominated by classical gravitational interactions~\cite{'tHooft:1987rb}. Then, for sufficiently small impact parameter, miniature BHs should form in particle collisions, and in particular, Hawking radiation would be the main observable signature~\cite{Banks:1999gd,Dimopoulos:2001hw}. This motivation led to an intensive study of Hawking radiation from higher-dimensional BHs--
see~\cite{Kanti:2014vsa} for a recent review and reference therein.


If a microscopic BH is produced, it is expected that the decay process can be modeled by the following four phases~\cite{Giddings:2001bu}, namely
\begin{itemize}
\item \textit{Balding phase}: all original ``hair'' inherited from incoming particles (except mass and angular momentum) is lost through gravitational and Hawking
radiation and, at the end of this stage, the BH is axisymmetric and rotating. 
\item \textit{Spin-down phase}: then the BH emits Hawking radiation, losing mass and angular momentum evolving towards the end, into a spherically symmetric BH.
\item \textit{Schwarzschild phase}: the spherically symmetric BH continues to radiate losing its mass, until it reaches the Planck scale.
\item \textit{Planck phase}: the semiclassical approximation of Hawking radiation becomes invalid at this stage, and quantum gravity starts to play a significant role in the BH emission process.
\end{itemize}
It is believed that the spin-down and Schwarzschild phases will dominate the lifetime of the BH, therefore they are the most promising stages to generate
observational signatures of Hawking radiation. Indeed these phases have been modeled in BH event generators, such as \textsc{charybdis2}
~\cite{Frost:2009cf,sampaio2010production} and~\textsc{blackmax}~\cite{Dai:2007ki} currently in use at the LHC. These event generators, however, can still be improved.


One of the Hawking radiation channels that has not been properly addressed in the literature is that of massive vector bosons, both electrically neutral and electrically charged, to describe the emission of $Z$ and $W^{\pm}$ particles of the Standard Model. As our first goal in this thesis, we are going to study Hawing radiation
for both a neutral and a charged Proca field in higher dimensions, to bridge this gap.

\section{Superradiance}
\label{sc:SR}
Superradiance is a phenomenon which refers to a radiation enhancement process in several physical contexts. This term was coined by Dicke, to describe an effect in quantum optics that radiation of a group of emitters could be amplified in a coherent fashion~\cite{Dicke:1954zz}. In 1971, Zel'dovich~\cite{zeldovich1,zeldovich2} pointed out that scalar and electromagnetic radiation, impinging on a rotating cylinder with absorbing surfaces
can be amplified if the condition
\begin{equation}
\omega<m\Omega\;, \label{consup}
\end{equation}
holds. Here $\omega$ and $m$ are the waves' frequency and the azimuthal number, and $\Omega$ is the angular velocity of the cylinder.

Since the event horizon of a BH provides a natural absorbing surface at the classical level, Zel'dovich anticipated that a rotating BH with horizon
angular velocity $\Omega_H$ should display superradiant amplification within the regime given by Eq.~\eqref{consup}. This was indeed observed by Misner in
unpublished calculations, who found that certain modes of scalar fields are amplified by Kerr BHs, as a wave analogue of the Penrose process~\cite{Penrose:1969pc}. This work was then generalized and verified by Teukolsky and Press~\cite{Teukolsky:1974yv} who found that the amplification process also occurs for electromagnetic and gravitational waves on the same background. According to these observations, Bekenstein~\cite{Bekenstein:1973mi} realized that in order to satisfy Hawking's area theorem~\cite{Hawking:1971tu}, superradiant amplification is classically required when condition~\eqref{consup} holds. From the same reasoning he also derived that superradiance of charged bosonic waves by a charged BH exists when~\cite{Bekenstein:1973mi}
\begin{equation}
\omega<q\Phi_H\;,\label{consup2}
\end{equation}
is satisfied, where $q$ is the field charge and $\Phi_H$ is the electrostatic potential at the horizon.

This story changes dramatically for fermion fields. As first shown by Unruh~\cite{Unruh:1973bda}, superradiance is absent for neutral massless Dirac fields on the Kerr background. This conclusion was then generalized to massive Dirac fields~\cite{Iyer:1978du} (by correcting a previous work~\cite{Martellini:1977qf}) and on the Kerr-Newman background~\cite{Lee:1977ti}. Absence of superradiance for fermions originates from the fact that the net number current flowing down the horizon is positive definite, which implies that it is impossible for fermion fields to extract energy and angular momentum from the hole~\cite{Unruh:1973bda}. Moreover, the argument~\cite{Bekenstein:1973mi} based on the area theorem does not apply to a fermion field since its energy-momentum tensor does not obey the weak energy condition~\cite{Hawking:1974sw}. At the quantum level, it can also be understood as a consequence of the exclusion principle which does not allow for more than one particle in each outgoing wave and therefore the scattered wave can not be stronger than the incident wave~\cite{Hawking:1974sw}.

By placing a reflecting mirror around a rotating BH, the system composed with bosonic fields may become unstable. This was first addressed by Press and
Teukolsky~\cite{Press:1972zz}, and was dubbed as \textit{black hole bomb}. The role of the mirror is to feed back to the BH the amplified scattered wave,
as to recurrently extract rotational energy. Then, the wave bounces back and forth between the mirror and the BH until radiation pressure destroys the mirror.
In fact, the reflecting mirror is not necessarily artificial, and it has several realizations in nature. One realization is the field's mass. For a massive
bosonic field with mass $\mu$ satisfying the bound state condition $\omega < \mu$, the mass term can provide a confining mechanism similar to a
mirror~\cite{Damour:1976kh,Detweiler:1980uk,Zouros:1979iw,Dolan:2007mj,Rosa:2009ei,Hod:2009cp,Witek:2012tr,Rosa:2012uz,Pani:2012vp,Hod:2012zza}. Another realization is AdS asymptotics, which may also bind superradiant modes~\cite{Cardoso:2004hs,Cardoso:2006wa,Aliev:2008yk,Dias:2011at,Li:2012rx,Uchikata:2011zz,Cardoso:2013pza}.
Recently a considerable interest has been devoted to study superradiant scattering and instabilities using numerical methods at the non-linear level
\cite{Cardoso:2012qm,Yoshino:2012kn,Witek:2012tr,Yoshino:2013ofa,East:2013mfa,Okawa:2014nda,Dolan:2015dha,Sanchis-Gual:2015lje,Bosch:2016vcp}.

Since superradiance also exists for charged BHs under the condition in Eq.~\eqref{consup2}, it is natural to ask if charged superradiant modes can be confined by the field's mass and the AdS asymptotics as in the rotating case, and generate instabilities by extracting Coulomb energy and BH charge. As proved by Hod~\cite{Hod:2013nn,Hod:2013eea,Hod:2015hza} for the scalar case, the field's mass \textit{cannot} bound the superradiant modes since the superradiance
condition and the bound state condition for a charged massive scalar field in an asymptotically flat charged BH cannot be satisfied simultaneously. This statement has been generalized to charged Proca fields recently, showing that Reissner-Nordstr\"om BHs are also stable against those massive vector fields~\cite{Sampaio:2014swa}. There is still a possibility, however, to circumvent Hod's results by considering a non-minimal coupling between the field and the background since Hod's results only hold for minimally coupled scalar fields~\cite{Hod:2013eea,Hod:2015hza}. At this moment it might be safe to state that a charged BH bomb may be only achieved with AdS asymptotics~\cite{Uchikata:2011zz,Dias:2011tj,Wang:2014eha}\footnote{Alternatively, a charged BH bomb in an asymptotically flat spacetime can be also built artificially by imposing mirror-like boundary conditions~\cite{Herdeiro:2013pia,Hod:2013fvl,Degollado:2013bha,Sanchis-Gual:2015lje}.}.

The phenomenon of superradiance is interesting for various reasons\footnote{To find more applications of superradiance, we refer to a recent review on this
topic~\cite{Brito:2015oca}.}. Among others, we would like to mention that superradiance provides a systematic mechanism to construct \mbox{(quasi-) stationary} BH-field systems at the linear level, and generates hairy BH solutions at the nonlinear level~\cite{Herdeiro:2014goa,Herdeiro:2014ima}. As we mentioned before, in an asymptotically flat BH, the superradiant instability is absent for a charged massive bosonic field. But as shown in~\cite{Degollado:2013eqa,Sampaio:2014swa}, one can still find quasi-stationary states for such a BH-field system, when gravity is balanced with the electric field. Adding a mirror, as recently shown in~\cite{Dolan:2015dha}, a charged hairy BH solution was found. An even more interesting result was obtained on a rotating BH. As we explained above, the field mass can be used as a mirror and therefore the superradiant instability is present for a massive scalar field on a Kerr BH. At the linear level, stationary states can be found in such
a system~\cite{Hod:2012px,Benone:2014ssa,Herdeiro:2014pka}, which are dubbed as~\textit{clouds}. Considering the back reaction of a complex scalar field, scalar hairy Kerr solutions have been found in~\cite{Herdeiro:2014goa,Herdeiro:2015tia,Herdeiro:2015gia}.

Because the AdS asymptotics is another realization of the mirror, one may expect similar effects in an asymptotically AdS spacetime. Indeed the superradiant
instability for scalar and gravitational fields has been studied on a Kerr-anti-de Sitter (Kerr-AdS) BH~\cite{Uchikata:2009zz,Cardoso:2013pza}. In this thesis, we will complete this study by exploring the superradiant instability for the Maxwell field. Furthermore, since such an instability exists in AdS, one expects to find clouds for various fields on Kerr-AdS BHs. The only studied case before this thesis on a Kerr-AdS BH was the gravitational field~\cite{Cardoso:2013pza}. To fill this gap, we will present studies of scalar and vector clouds on the same background.


\section{Structure}
\label{sc:ST}
The structure of this thesis is as follows. In Chapter~\ref{ch:prelim}, we review two important perturbative methods, the Kodama-Ishibashi formalism and the
Newman-Penrose formalism, which set the foundations for the other chapters. In Chapter~\ref{ch:GeneralInFlat}, we employ the Kodama-Ishibashi formalism to derive
equations of motion for a Proca field (either neutral or charged), on a $D$-dimensional spherically symmetric background. Then these equations are used to study
a quantum semi-classical effect on BHs, i.e. Hawking radiation, for a neutral Proca field on a $D$-dimensional Schwarzschild BH in Chapter~\ref{ch:NeutralP}.
Following this line, in Chapter~\ref{ch:ChargedP} we explore Hawking radiation for a charged Proca field in a brane charged BH. We turn to classical properties in Chapter~\ref{ch:ChargedClouds}, to study quasi-bound states for a charged Proca field in a Reissner-Nordstr\"om BH. In Chapter~\ref{ch:scalarHD}, we study
superradiant instabilities for a charged scalar field in a $D$-dimensional Reissner-Nordstr\"om-AdS BH. This is then generalized to the Maxwell field on
a Kerr-AdS BH, leading to the study of superradiant instabilities and vector clouds, in Chapter~\ref{ch:KerrAdS}. Conclusions and outlook are drawn in
Chapter~\ref{ch:conclusion}.


%% file: prelim.tex
\chapter{Preliminaries}
\label{ch:prelim}


We start in this chapter by introducing the mathematical tools to deal with perturbations of test fields around BHs, as the foundation to perform the study for Proca and Maxwell fields. Two types of perturbation methods, the Kodama-Ishibashi formalism and the Newman-Penrose formalism, will be illustrated, respectively, in the following. 

Throughout this thesis we will use the signature $(-,+,...,+)$ and natural units $G=c=\hbar=1$, unless explicitly stated otherwise.

\section{Kodama-Ishibashi formalism}
\label{sc:KI}
The Kodama-Ishibashi (KI) formalism~\cite{Kodama:2000fa,Kodama:2003jz,Ishibashi:2003ap,Kodama:2003kk} (for a review see~\cite{Kodama:2007ph,Ishibashi:2011ws}), is the generalization of the Regge-Wheeler-Zerilli formalism~\cite{Regge:1957td,Zerilli:1970se} to higher dimensions. This method is applicable to any higher dimensional spacetime with maximal symmetry whose manifold structure can be locally written as a warped product between a Lorentzian manifold and an Einstein space. The basic idea of this method is to classify the perturbations into different types (scalar, vector and tensor types), based on their tensorial behavior in the Einstein space.

To be specific, let us consider the following gravitational background with the manifold structure $\mathcal{M}=\mathcal{N}\times\mathcal{K}$ in the form~\cite{Kodama:2000fa,Kodama:2007ph,Ishibashi:2011ws}
\begin{equation}
g_{MN}dx^Mdx^N=h_{ab}(y)dy^a dy^b+r(y)^2d\sigma_n^2 \;,\label{KI:metric}
\end{equation}
where $x^M=(y^a,z^j)$. Note that the Lorentzian manifold is denoted by $\mathcal{N}$ with metric $h_{ab}$, and the Einstein space is denoted by $\mathcal{K}$ with constant curvature $K (K=0,\pm1)$ and metric $\sigma_{ij}$,
\begin{equation}
d\sigma_n^2=\sigma_{ij}(z)dz^idz^j \; .
\end{equation}
Then the Riemann tensor and Ricci tensor on an Einstein space are given by
\begin{equation}
\hat{R}_{ijkl}=K (\sigma_{ik}\sigma_{jl}-\sigma_{il}\sigma_{jk})\;,\;\;\;\;\;\;\hat{R}_{ij}=(n-1)K \sigma_{ij}\;.
\end{equation}
We use indices $\{a,b,c,\ldots\}$ for the first set of coordinates, $\{y^a\}$, spanning on the $m$-dimensional space with metric $h_{ab}(y)$; and indices $\{i,j,k,\ldots\}$ for the second set of coordinates, $\{z^i \}$, spanning on the $n$-dimensional Einstein space. Then the spacetime dimension is $d=m+n$. We denote the covariant derivatives, the Christoffel connection coefficients and the Riemann tensors on the manifolds $\{\mathcal{M}$, $\mathcal{N}$, $\mathcal{K}\}$, by $\{\nabla_M$, $D_a$, $\hat{D}_i\}$, $\{\Gamma^M_{NL}, \bar{\Gamma}^a_{bc}, \hat{\Gamma}^i_{jk}\}$, and $\{R_{MNLS}, \bar{R}_{abcd}, \hat{R}_{ijkl}\}$, respectively. We also define the Laplace operator on the Einstein space as $\hat{\Delta}=\hat{D}_i\hat{D}^i$.
The metric form in Eq.~\eqref{KI:metric} covers several interesting cases such as $2+n$-dimensional spherically symmetric BHs or a singly rotating BH in $4+n$-dimensions.

The expressions of $\Gamma^M_{NL}$ and $R^M_{\;\;NLS}$ can be written in terms of the corresponding quantities on the manifold $\mathcal{N}$ with metric $h_{ab}(y)$ and on the Einstein space with metric $\sigma_{ij}$~\cite{Kodama:2000fa}, i.e.
\begin{equation}
\Gamma^a_{bc}=\bar{\Gamma}^a_{bc}\;,\;\;\;\Gamma^a_{ij}=-rD^ar\sigma_{ij}\;,\;\;\;\Gamma^i_{aj}=\dfrac{D_ar}{r}\delta^i_j\;,\;\;\;\Gamma^i_{jk}=\hat{\Gamma}^i_{jk}\;,
\end{equation}
where the other components of $\Gamma^M_{NL}$ vanish, and
\begin{align}
&R^a_{\;\;bcd}=\bar{R}^a_{\;\;bcd}\;,\;\;\;R^i_{\;\;ajb}=-\dfrac{D_aD_br}{r}\delta^i_j\;,\;\;\;R^a_{\;\;ibj}=-\dfrac{D^aD_br}{r}g_{ij}\;,\nonumber\\
&R^i_{\;\;jkl}=(K-D_arD^ar)(\delta^i_k\sigma_{jl}-\delta^i_l\sigma_{jk})\;.
\end{align}
Then the Ricci tensors and Einstein tensors can be derived directly~\cite{Kodama:2000fa}
\begin{equation}
R_{ab}=\bar{R}_{ab}-\dfrac{n}{r}D_aD_br\;,\;\;\;R_{ai}=0\;,\;\;\;R_{ij}=\left(-\dfrac{\bar{\Box} r}{r}+(n-1)\dfrac{K-D_arD^ar}{r^2}\right)g_{ij}\;,
\end{equation}
where we have defined $\bar{\Box}\equiv D^aD_a$, and
\begin{align}
&G_{ab}=\bar{G}_{ab}-\dfrac{n}{r}D_aD_br-\left(\dfrac{n(n-1)}{2r^2}(K-D_arD^ar)-\dfrac{n}{r}\bar{\Box} r\right)g_{ab}\;,\\
&G_{ij}=\left(-\dfrac{\bar{R}}{2}-\dfrac{(n-1)(n-2)}{2r^2}(K-D_arD^ar)+\dfrac{n-1}{r}\bar{\Box} r\right)g_{ij}\;,\\
&G_{ai}=0\;,
\end{align}
with the definition $G_{MN}=R_{MN}-\tfrac{1}{2}g_{MN}R$.

To write down the equations of motion, we shall decompose the perturbations in terms of their tensorial harmonics on $\mathcal{K}$~\cite{Kodama:2007ph,Ishibashi:2011ws}.\\
For a vector field $v_i$, it can be uniquely decomposed into a scalar field $v^{(s)}$ and a transverse vector field $v^{(t)}_i$ as
\begin{equation}
v_i=\hat{D}_iv^{(s)}+v^{(t)}_i\;,\;\;\;\;\;\;\hat{D}_iv^{(t)i}=0\;,\label{intro:decom}
\end{equation}
where $v^{(s)}$ and $v^{(t)}_i$ satisfy the corresponding scalar and vector eigenvalue equations
\begin{align}
&(\hat{\Delta}+\kappa_s^2)v^{(s)}=0\;,\label{intro:eigenscalar}\\
&(\hat{\Delta}+\kappa_v^2)v^{(t)}_i=0\;,\label{intro:eigenvector}
\end{align}
on an Einstein space with spherical topology, with $\kappa_s^2=\ell(\ell+n-1)$ and \mbox{$\kappa_v^2=\ell(\ell+n-1)-1$}. Note that the angular momentum quantum number, $\ell$, starts from zero in the scalar eigenvalue $\kappa_s$ and one in the vector eigenvalue $\kappa_v$, respectively. Then taking a derivative $\hat{D}_i$ on Eq.~\eqref{intro:decom}, we have the relation
\begin{equation}
\hat{\Delta}v^{(s)}=\hat{D}_iv^i\;,
\end{equation}
which determines $v^{(s)}$ from Eq.~\eqref{intro:eigenscalar}, and $v^{(t)}_i$ from Eq.~\eqref{intro:decom} after $v^{(s)}$ is obtained.

The scalar and vector harmonic functions on $n$-spheres, used in Eqs.~\eqref{intro:eigenscalar} and~\eqref{intro:eigenvector}, are defined as follows~\cite{Ishibashi:2011ws}. Let us denote the homogeneous cartesian coordinates on $n$-spheres by $\Omega^A$($A=1,\cdots,n+1$), and define the function $Y_{{\bf a}}$ by
\begin{equation}
Y_{{\bf a}}(\Omega)=a_{A_1\cdots A_\ell}\Omega^{A_1}\cdots \Omega^{A_\ell}\;,
\label{intro:ScalatrHarmonicFunction}
\end{equation}
in terms of a constant tensor ${\bf a}=(a_{A_1\cdots A_\ell})$ ($A_1,\cdots,A_\ell=1,\cdots,n+1$). Then $Y_{{\bf a}}$ is a scalar harmonic function with the eigenvalue $\kappa_s^2$ if and only if $\bf{a}$ satisfies the conditions
\begin{equation}
a_{A_1\cdots A_\ell}=a_{(A_1\cdots A_\ell)}\;,\;\;\;a_{A_1\cdots A_{\ell-2}}{}^B{}_B=0\quad (\ell\ge2)\;.
\end{equation}
Similarly, we can define the vector field $V_{{\bf b}}^i$ by
\begin{equation}
V_{{\bf b}}^i=b_{A_1\cdots A_\ell ;B}\Omega^{A_1}\cdots\Omega^{A_\ell }
                   \hat{D}^i\Omega^B \;,
\label{intro:HarmonicVector}
\end{equation}
in terms of a constant tensor ${\bf b}=(a_{A_1\cdots A_\ell ;B})$($A_1,\cdots,A_\ell,B=1,\cdots,n+1$). Then, $V_{{\bf b}}^i$ is a vector harmonic function on $n$-spheres with eigenvalue $\kappa_v^2$, if and only if the constant tensor ${\bf b}$ satisfies the conditions
\begin{equation}
b_{A_1\cdots A_\ell ;B}=b_{(A_1\cdots A_\ell );B}\;,\;\;\;b_{A_1\cdots A_{\ell -2}i}{}^i{}_{;B}=0\;,\;\;\;b_{(A_1\cdots A_\ell ;A_{\ell +1})}=0\;.
\end{equation}
For more details on these harmonic functions on $n$-spheres and the corresponding properties, we refer readers to~\cite{Ishibashi:2011ws}.

\section{Newman-Penrose formalism}
\label{sc:NP}
%
The Newman-Penrose formalism~\cite{Newman:1961qr}, as the name indicates, was developed by Newman and Penrose in 1962, as an alternative way to formulate field equations, such as the Einstein equations and the Maxwell equations. This formalism is extremely useful in various contexts in GR, for example to construct exact solutions of the Einstein equations~\cite{Stephani:2003tm}, to study perturbations of massless test fields on various BH backgrounds~\cite{Chandrasekhar:1985kt,Frolov:1998wf}, and to extract gravitational radiation in numerical relativity~\cite{baumgarte2010numerical}. For the problems we are interested in this thesis, we focus on the application of this formalism in the context of perturbation theory.

As first exhibited in the celebrated work of Teukolsky~\cite{Teukolsky:1972my}, linear perturbations of gravitational and electromagnetic fields on the Kerr background both separate and decouple, in terms of the Newman-Penrose variables. This was subsequently generalized to rotating BHs with a cosmological constant~\cite{Khanal:1983vb,Wu:2003qc,Yoshida:2010zzb,Dias:2012pp}. In this section, we review the Newman-Penrose formalism with application to the Maxwell field on Kerr-AdS BHs, and further present some new ingredients which have not been derived in the literature, in the presence of a cosmological constant, with details, including
\begin{itemize}
\item the derivation of Teukolsky-Starbinski identities, which was given in~\cite{Wu:2003qc} without proof,
\item the derivation for a complete set of solutions for the Maxwell field, in particular for $\Phi_1$, which is relevant for proving Appendix~\ref{app:angmomflux}.
\end{itemize}
\subsection{Basics}
%
%
In order to introduce the Newman-Penrose formalism, we first define a complex null tetrad $\{l^\mu,n^\mu,m^\mu,\bar{m}^\mu\}$, where the normalization conditions
\begin{equation}
l_\mu n^\mu=-1\;,\;\;\;\;\;\;m_\mu \bar{m}^\mu=1\;,\label{tetrad1}
\end{equation}
are satisfied, and all the other scalar products vanish. Note that here $\bar{m}^\mu$ is the complex conjugate of $m^\mu$. 
%
The tetrad is related with the metric\footnote{Note that this relation depends on the conventions.}
\begin{equation}
g_{\mu\nu}=-l_\mu n_\nu-l_\nu n_\mu+m_\mu\bar{m}_\nu+m_\nu\bar{m}_\mu\;.\label{teme1}
\end{equation}

Then spin coefficients are defined in terms of the tetrad
\begin{align}
-\kappa&=l_{\mu;\nu}m^\mu l^\nu\;,\;\;\;-\rho=l_{\mu;\nu}m^\mu \bar{m}^\nu\;,\;\;\;-\sigma=l_{\mu;\nu}m^\mu m^\nu\;,\;\;\;-\tau=l_{\mu;\nu}m^\mu n^\nu\;,\nonumber\\
\mu&=n_{\mu;\nu}\bar{m}^\mu m^\nu\;,\;\;\;\nu=n_{\mu;\nu}\bar{m}^\mu n^\nu\;,\;\;\;\;\;\;\lambda=n_{\mu;\nu}\bar{m}^\mu \bar{n}^\nu\;,\;\;\;\;\;\;\pi=n_{\mu;\nu}\bar{m}^\mu l^\mu\;,\nonumber\\
-\epsilon&=\frac{1}{2}(l_{\mu;\nu}n^\mu l^\nu-m_{\mu;\nu}\bar{m}^\mu l^\nu)\;,\;\;\;\;\;\;\;\;\;\;-\beta=\frac{1}{2}(l_{\mu;\nu}n^\mu m^\nu-m_{\mu;\nu}\bar{m}^\mu m^\nu)\;,\nonumber\\
-\gamma&=\frac{1}{2}(l_{\mu;\nu}n^\mu n^\nu-m_{\mu;\nu}\bar{m}^\mu n^\nu)\;,\;\;\;\;\;\;\;\;-\alpha=\frac{1}{2}(l_{\mu;\nu}n^\mu \bar{m}^\nu-m_{\mu;\nu}\bar{m}^\mu \bar{m}^\nu)\;.\label{spincoeffs}
\end{align}

Next we introduce the projection of the covariant derivatives in the null tetrad vectors, with the following notations
\begin{equation}
D=l^\mu\partial_\mu\;,\;\;\;\Delta=n^\mu\partial_\mu\;,\;\;\;\delta=m^\mu\partial_\mu\;,\;\;\;\bar{\delta}=\bar{m}^\mu\partial_\mu\;.\label{deriv}
\end{equation}

With the above spin coefficients in Eq.~\eqref{spincoeffs} and the directional derivatives in Eq.~\eqref{deriv} at hand, one may rewrite the Maxwell field equations. 
For the reader who is interested in the equations of motion for other spin fields in this formalism, such as the gravitational field and the Dirac field, a detailed account can be found in~\cite{Chandrasekhar:1985kt,Frolov:1998wf}.
%

In the Newman-Penrose formalism, the Maxwell tensor is decomposed into three complex scalars
\begin{equation}
\phi_0=F_{\mu\nu}l^{\mu}m^{\nu}\;,\;\;\;\;\;\;\phi_1=\frac{1}{2}F_{\mu\nu}(l^{\mu}n^{\nu}+\bar{m}^{\mu}m^{\nu})\;,\;\;\;\;\;\;\phi_2=F_{\mu\nu}\bar{m}^{\mu}n^{\nu}\;.\label{Maxwellscalars}
\end{equation}
Then the Maxwell equations become
\begin{align}
D\phi_1-\bar{\delta}\phi_0&=(\pi-2\alpha)\phi_0+2\rho\phi_1-\kappa\phi_2\;,\nonumber\\
D\phi_2-\bar{\delta}\phi_1&=-\lambda\phi_0+2\pi\phi_1+(\rho-2\epsilon)\phi_2\;,\nonumber\\
\delta\phi_1-\Delta\phi_0&=(\mu-2\gamma)\phi_0+2\tau\phi_1-\sigma\phi_2\;,\nonumber\\
\delta\phi_2-\Delta\phi_1&=-\nu\phi_0+2\mu\phi_1+(\tau-2\beta)\phi_2\;,\label{Maxwelleq1}
\end{align}
where the differential operators appearing on the left hand side are defined in Eq.~\eqref{deriv} and the spin coefficients on the right hand side are given by~Eq.\eqref{spincoeffs}.

To obtain the explicit form of the Maxwell equations, we shall specify a background geometry. For that purpose and for later application in the problems we are interested in, we first review the Kerr-AdS spacetimes in the next subsection. 
\subsection{Kerr-AdS black holes}
\label{subsec:KerrAdS}
In an asymptotically AdS background, the most general stationary and axisymmetric BH solution of the four dimensional Einstein-AdS system, is the Kerr-AdS BH. It was found by Carter~\cite{Carter:1968ks} firstly, a few years after the finding of the Kerr solution.

%
The line element for a Kerr-AdS BH, in Boyer-Lindquist coordinates, can be written as
\begin{equation}
ds^2=-\dfrac{\Delta_r}{\rho^2\Xi^2}\Big(dt-a\sin^2\theta d\varphi\Big)^2+\rho^2\left(\dfrac{dr^2}{\Delta_r}+\dfrac{d\theta^2}{\Delta_\theta}\right)+\dfrac{\Delta_\theta \sin^2\theta}{\rho^2\Xi^2}\Big(a\,dt-(r^2+a^2)d\varphi\Big)^2\;,\label{RKerrAdS}
\end{equation}
with metric functions
\begin{align}
&\rho^2\equiv\bar{\rho}\bar{\rho}^\ast=r^2+a^2\cos^2\theta\;,\;\;\Delta_r=\Big(r^2+a^2\Big)\left(1+\frac{r^2}{L^2}\right)-2Mr\;,\;\;\nonumber\\
&\Delta_\theta=1-\dfrac{a^2\cos^2\theta}{L^2}\;,\;\;\Xi=1-\dfrac{a^2}{L^2}\;,\label{RKerrAdSmetric}
\end{align}
where $\bar{\rho}^\ast$ is the complex conjugate of $\bar{\rho}$, and $\bar{\rho}=r+ia\cos\theta$.
The other parameters shown in Eq.~\eqref{RKerrAdSmetric} include, $L$, which is the AdS radius; $M$ and $a$, which are the mass and spin parameters and relate to the BH energy and angular momentum.

In this frame, the angular velocity of the event horizon and the Hawking temperature are given by
\begin{align}
&\Omega_H=\dfrac{a}{r_+^2+a^2}\;,\label{RHorizonV}\\
&T_H=\dfrac{1}{\Xi}\left[\dfrac{r_+}{2\pi}\left(1+\dfrac{r_+^2}{L^2}\right)\dfrac{1}{r_+^2+a^2}-\dfrac{1}{4\pi r_+}\left(1-\dfrac{r_+^2}{L^2}\right)\right]\;,\label{RTemp}
\end{align}
where the event horizon $r_+$ is determined as the largest root of $\Delta_r(r_+)=0$. For a given $r_+$, the mass parameter $M$ can be expressed as
\begin{equation}
M=\dfrac{(r_+^2+a^2)(L^2+r_+^2)}{2r_+L^2}\;.\nonumber
\end{equation}
To require the existence of BHs and to avoid singularities, one may impose the following constraints on the rotation parameter $a$
\begin{align}
&\dfrac{a}{L}\leq \dfrac{r_+}{L} \sqrt{\dfrac{3r_+^2+L^2}{L^2-r_+^2}}\;,\;\;\;\;\;\;\;\;\;\;\;\;\text{for}\;\;\;\;\;  \dfrac{r_+}{L}<\dfrac{1}{\sqrt{3}}\;,\nonumber\\
&\dfrac{a}{L}<1\;,\;\;\;\;\;\;\;\;\;\;\;\;\;\;\;\;\;\;\;\;\;\;\;\;\;\;\;\;\;\;\;\;\;\;\text{for}\;\;\;\;\;  \dfrac{r_+}{L}\geq\dfrac{1}{\sqrt{3}}\;,\label{intro_secKerrAdS_rotation}
\end{align}
where the equality sign in the first line corresponds to an extremal BH.

The Boyer-Lindquist coordinates are convenient to solve the perturbation equations for test fields. These coordinates, however, obscure the structure of the geometry at infinity. In fact, the metric in Eq.~\eqref{RKerrAdS} describes the Kerr-AdS spacetime in a rotating frame, which can be seen by calculating the angular velocity at infinity~\cite{Dias:2013sdc}
\begin{equation}
\Omega_\infty=\dfrac{a}{a^2-L^2}\;,
\end{equation}
which is apparently non-zero. In order to obtain a non-rotating frame, which is relevant to BH thermodynamics~\cite{Gibbons:2004ai}, we can make the following coordinate transformation~\cite{Henneaux:1985tv,Winstanley:2001nx}
\begin{equation}
\hat{t}=\dfrac{t}{\Xi}\;,\;\;\;\hat{\varphi}=\varphi+\dfrac{a}{L^2}\dfrac{t}{\Xi}\;,\;\;\;\hat{r}^2=\dfrac{r^2\Delta_\theta+a^2\sin^2\theta}{\Xi}\;,\;\;\;\hat{r}\cos\hat{\theta}=r\cos\theta\;,\label{RNTrans}
\end{equation}
so that in these new frame the Kerr-AdS geometry~\eqref{RKerrAdS} is simply AdS space in the usual spherical coordinates
\begin{equation}
ds^2=-\left(1+\dfrac{\hat{r}^2}{L^2}\right)d\hat{t}^2+\left(1+\dfrac{\hat{r}^2}{L^2}\right)^{-1}d\hat{r}^2+\hat{r}^2(d\hat{\theta}^2+\sin^2\hat\theta d\hat{\varphi}^2)\;,
\end{equation}
when $r\rightarrow\infty (\hat{r}\rightarrow\infty)$.
The angular velocity of the event horizon in these coordinates then becomes 
\begin{equation}
\hat{\Omega}_H=\Omega_H\Xi+\dfrac{a}{L^2}\;,\label{NRHAV}
\end{equation}
where $\Omega_H$, defined in Eq.~\eqref{RHorizonV}, is measured relative to a rotating observer at infinity.
Since the metric in coordinates $(\hat{t},\hat{r},\hat{\theta},\hat{\varphi})$ is complicated, and also because we have the coordinate transformations in Eq.~\eqref{RNTrans}, in practice we always work in Boyer-Lindquist coordinates. Our goal is to solve the Maxwell equations in the frequency domain, therefore it is useful to relate the frequencies in these two different coordinates. This is given by 
\begin{equation}
e^{-i\omega t}e^{im\varphi}=e^{-i\hat{\omega} \hat{t}}e^{im\hat{\varphi}}\;,\;\;\;\;\;\;\Rightarrow\;\;\;\;\;\;\hat{\omega}=\omega\Xi+m\dfrac{a}{L^2}\;.\label{nonrotatingfre}
\end{equation}
%

%
\subsection{Maxwell equations on Kerr-AdS}
By specializing the general Maxwell equations~\eqref{Maxwelleq1} to the Kerr-AdS background~\eqref{RKerrAdS}, and using a generalization of the Kinnersley tetrad~\cite{Khanal:1983vb}, i.e.
\begin{align}
l^\mu&=\left(\dfrac{(r^2+a^2)\Xi}{\Delta_r},1,0,\dfrac{a\Xi}{\Delta_r}\right)\;,\;\;\;\;\;\;
n^\mu=\dfrac{1}{2\rho^2}\Big((r^2+a^2)\Xi,-\Delta_r,0,a\Xi\Big)\;,\nonumber\\
m^\mu&=\dfrac{1}{\sqrt{2\Delta_\theta}\bar{\rho}}\left(ia\Xi\sin\theta,0,\Delta_\theta,\dfrac{i\Xi}{\sin\theta}\right)\;,\;\;\;\;\;\;\bar{m}^\mu=(m^\mu)^\ast\;,\label{pre:tetrad}
\end{align}
where $\bar{\rho}=r+ia\cos\theta$, one obtains
\begin{align}
\left(\mathscr{D}_0+\dfrac{1}{\bar{\rho}^\ast}\right)\Phi_1&=\sqrt{\Delta_\theta}\left(\mathscr{L}_1-\dfrac{ia\sin\theta}{\bar{\rho}^\ast}\right)\Phi_0\;,\label{Maxwelleq2_1}\\
\left(\mathscr{D}_0-\dfrac{1}{\bar{\rho}^\ast}\right)\Phi_2&=\sqrt{\Delta_\theta}\left(\mathscr{L}_0+\dfrac{ia\sin\theta}{\bar{\rho}^\ast}\right)\Phi_1\;,\label{Maxwelleq2_2}\\
-\Delta_r\left(\mathscr{D}_1^\dag-\dfrac{1}{\bar{\rho}^\ast}\right)\Phi_0&=\sqrt{\Delta_\theta}\left(\mathscr{L}_0^\dag+\dfrac{ia\sin\theta}{\bar{\rho}^\ast}\right)\Phi_1\;,\label{Maxwelleq2_3}\\
-\Delta_r\left(\mathscr{D}_0^\dag+\dfrac{1}{\bar{\rho}^\ast}\right)\Phi_1&=\sqrt{\Delta_\theta}\left(\mathscr{L}_1^\dag-\dfrac{ia\sin\theta}{\bar{\rho}^\ast}\right)\Phi_2\;,\label{Maxwelleq2_4}
\end{align}
where
\begin{align}
\mathscr{D}_n&=\dfrac{\partial}{\partial r}-\dfrac{i\Xi K}{\Delta_r}+\dfrac{n}{\Delta_r}\dfrac{d\Delta_r}{dr}\;,\nonumber\\
\mathscr{D}_n^\dag&=\dfrac{\partial}{\partial r}+\dfrac{i\Xi K}{\Delta_r}+\dfrac{n}{\Delta_r}\dfrac{d\Delta_r}{dr}\;,\nonumber\\
\mathscr{L}_n&=\dfrac{\partial}{\partial\theta}-\frac{\Xi H}{\Delta_\theta}+\dfrac{n}{\sqrt{\Delta_\theta}\sin\theta}\dfrac{d}{d\theta}\left(\sqrt{\Delta_\theta}\sin\theta\right)\;,\nonumber\\
\mathscr{L}_n^\dag&=\dfrac{\partial}{\partial\theta}+\frac{\Xi H}{\Delta_\theta}+\dfrac{n}{\sqrt{\Delta_\theta}\sin\theta}\dfrac{d}{d\theta}\left(\sqrt{\Delta_\theta}\sin\theta\right)\;,\label{Defoperator}
\end{align}
with
\begin{equation}
K=\omega(r^2+a^2)-am\;,\;\;\;H=a\omega\sin\theta-\dfrac{m}{\sin\theta}\;,\label{KHeq}
\end{equation}
and where the following transformations have been made
\begin{equation}
\phi_0=\Phi_0\;,\;\;\;\phi_1=\dfrac{1}{\sqrt{2}\bar{\rho}^\ast}\Phi_1\;,\;\;\;\phi_2=\dfrac{1}{2(\bar{\rho}^\ast)^2}\Phi_2\;.\label{phirelations}
\end{equation}
Note that the time and azimuthal dependence, $e^{-i\omega t+im\varphi}$, has been factored out. By acting with the operator $\sqrt{\Delta_\theta}(\mathscr{L}_0^\dag+\tfrac{ia\sin\theta}{\bar{\rho}^\ast})$ on Eq.~\eqref{Maxwelleq2_1}, and with the operator $(\mathscr{D}_0+\tfrac{1}{\bar{\rho}^\ast})$ on Eq.~\eqref{Maxwelleq2_3}, one obtains a decoupled equation for $\Phi_0$ by eliminating $\Phi_1$, i.e.
\begin{equation}
[\sqrt{\Delta_\theta}\mathscr{L}_0^\dag\sqrt{\Delta_\theta}\mathscr{L}_1+\Delta_r\mathscr{D}_1\mathscr{D}_1^\dag+2i\omega\Xi\bar{\rho}]\Phi_0=0\;,\label{Phi0eq}
\end{equation}
with aid of the identities 
%
\[
\begin{cases}
\left(\mathscr{D}_0+\dfrac{1}{\bar{\rho}^\ast}\right)\Delta_r\left(\mathscr{D}_1^\dag-\dfrac{1}{\bar{\rho}^\ast}\right)=\Delta_r\mathscr{D}_1\mathscr{D}_1^\dag+\dfrac{2i\Xi K}{\bar{\rho}^\ast}\;,\\
\sqrt{\Delta_\theta}\left(\mathscr{L}_0^\dag+\dfrac{ia\sin\theta}{\bar{\rho}^\ast}\right)\sqrt{\Delta_\theta}\left(\mathscr{L}_1-\dfrac{ia\sin\theta}{\bar{\rho}^\ast}\right)=\sqrt{\Delta_\theta}\mathscr{L}_0^\dag\sqrt{\Delta_\theta}\mathscr{L}_1
-\dfrac{2ia\sin\theta}{\bar{\rho}^\ast}\Xi H\;.
\end{cases}
\]

Following the same procedure, by acting with the operator $\sqrt{\Delta_\theta}(\mathscr{L}_0+\tfrac{ia\sin\theta}{\bar{\rho}^\ast})$ on Eq.~\eqref{Maxwelleq2_4} and with the operator $\Delta_r(\mathscr{D}_0^\dag+\tfrac{1}{\bar{\rho}^\ast})$ on Eq.~\eqref{Maxwelleq2_2}, $\Phi_1$ is again eliminated so that we obtain a decoupled equation for $\Phi_2$
\begin{equation}
[\sqrt{\Delta_\theta}\mathscr{L}_0\sqrt{\Delta_\theta}\mathscr{L}_1^\dag+\Delta_r\mathscr{D}_0^\dag\mathscr{D}_0-2i\omega\Xi\bar{\rho}]\Phi_2=0\;,\label{Phi2eq}
\end{equation}
with aid of the identities 
\[
\begin{cases}
\Delta_r\left(\mathscr{D}_0^\dag+\dfrac{1}{\bar{\rho}^\ast}\right)\left(\mathscr{D}_0-\dfrac{1}{\bar{\rho}^\ast}\right)=\Delta_r\mathscr{D}_0^\dag\mathscr{D}_0-\dfrac{2i\Xi K}{\bar{\rho}^\ast}\;,\\
\sqrt{\Delta_\theta}\left(\mathscr{L}_0+\dfrac{ia\sin\theta}{\bar{\rho}^\ast}\right)\sqrt{\Delta_\theta}\left(\mathscr{L}_1^\dag-\dfrac{ia\sin\theta}{\bar{\rho}^\ast}\right)=\sqrt{\Delta_\theta}\mathscr{L}_0\sqrt{\Delta_\theta}\mathscr{L}_1^\dag
+\dfrac{2ia\sin\theta}{\bar{\rho}^\ast}\Xi H\;.
\end{cases}
\]
%
Now taking 
\begin{equation}
\Phi_0=R_{+1}(r)S_{+1}(\theta)\;\;\;\text{and}\;\;\;\Phi_2=R_{-1}(r)S_{-1}(\theta)\;,\label{decompfields}
\end{equation}
we finally obtain the Maxwell equations with separated variables
\begin{align}
&\left(\Delta_r\mathscr{D}_1\mathscr{D}_1^\dag+2i\omega\Xi r\right)R_{+1}=\lambda R_{+1}\;,\label{Rpluseq}\\
&\left(\sqrt{\Delta_\theta}\mathscr{L}_0^\dag\sqrt{\Delta_\theta}\mathscr{L}_1-2a\omega\Xi\cos\theta\right)S_{+1}=-\lambda S_{+1}\;,\label{Spluseq}
\end{align}
and
\begin{align}
&\left(\Delta_r\mathscr{D}_0^\dag\mathscr{D}_0-2i\omega\Xi r\right)R_{-1}=\lambda R_{-1}\;,\label{Rminuseq}\\
&\left(\sqrt{\Delta_\theta}\mathscr{L}_0\sqrt{\Delta_\theta}\mathscr{L}_1^\dag+2a\omega\Xi\cos\theta\right)S_{-1}=-\lambda S_{-1}\;,\label{Sminuseq}
\end{align}
from Eqs.~\eqref{Phi0eq} and~\eqref{Phi2eq}. Note that $\lambda$ refers to the separation constant from now on, and should not be confused with the spin coefficient.
Using the commutative property $\Delta_r\mathscr{D}_{n+1}=\mathscr{D}_{n}\Delta_r$, Eq.~\eqref{Rpluseq} can be rewritten as
\begin{equation}
\left(\Delta_r\mathscr{D}_0\mathscr{D}_0^\dag+2i\omega\Xi r\right)\left(\Delta_rR_{+1}\right)=\lambda \left(\Delta_rR_{+1}\right)\;,\label{Ppluseq}
\end{equation}
which shows that $\Delta_rR_{+1}$ and $R_{-1}$ satisfy complex conjugate equations, by comparing with Eq.~\eqref{Rminuseq}.

The decoupled equations~\eqref{Maxwelleq2_1}-~\eqref{Maxwelleq2_4} provide solutions for $\Phi_0$, $\Phi_1$ and $\Phi_2$. In particular, $\Phi_0$ and $\Phi_2$ can be separated into radial parts which satisfy Eqs.~\eqref{Rpluseq} and~\eqref{Rminuseq}, and angular parts which satisfy Eqs.~\eqref{Spluseq} and~\eqref{Sminuseq}. The solution of $\Phi_1$ is ignored for now and we will be back to this problem later.
There is an additional issue to discuss, which is related to the solutions of $\Phi_0$ and $\Phi_2$.
%
%

As we have already shown in Eqs~\eqref{Rminuseq} and~\eqref{Ppluseq}, $\Phi_0$ and $\Phi_2$ satisfy equations which are complex conjugate of each other, but the relative normalization between these two solutions still remains to be determined. The answer to this problem is given by the famous Starobinsky-Teukolsky identities. In the following we will address this issue by proving the Starobinsky-Teukolsky identities for the Maxwell field on the Kerr-AdS background.

\underline{\bf Starobinsky-Teukolsky identities}
\begin{thm}\label{thm1}
$\bar{\mathscr{L}_0}\bar{\mathscr{L}_1}S_{+1}$ is a constant multiple of $S_{-1}$,\;i.e. $\bar{\mathscr{L}_0}\bar{\mathscr{L}_1}S_{+1}=BS_{-1}$, \\
{\color{white}{and}}$\;\;\;\;\;\;\;\;\;\;\;\;\;\;\;\;\;\bar{\mathscr{L}_0}^\dag\bar{\mathscr{L}_1}^\dag S_{-1}$ is a constant multiple of $S_{+1}$, i.e. $\bar{\mathscr{L}_0}^\dag\bar{\mathscr{L}_1}^\dag S_{-1}=BS_{+1}$.
\end{thm}
Note that the new angular operators are defined as
\begin{equation}
\bar{\mathscr{L}_n}=\sqrt{\Delta_\theta}\mathscr{L}_n\;,\;\;\;\;\;\;\bar{\mathscr{L}_n}^\dag=\sqrt{\Delta_\theta}\mathscr{L}_n^\dag\;,\label{newangoperator}
\end{equation}
so that this theorem can be written in the same form as its counterpart for the Kerr background~\cite{Chandrasekhar:1985kt}. Also note that the two proportionality constants in this theorem have been set to the same (denoted by $B$), which is guaranteed if we normalize both $S_{+1}$ and $S_{-1}$ to unity, i.e.
\begin{equation}
\int_0^\pi S_{+1}^2\sin\theta d\theta=\int_0^\pi S_{-1}^2\sin\theta d\theta=1\;.\label{angnorm}
\end{equation}
Theorem~\ref{thm1} is proved in Appendix~\ref{app:ST1} with details, by taking $\bar{\mathscr{L}_0}\bar{\mathscr{L}_1}S_{+1}=BS_{-1}$, as an example.

To evaluate $B$, we start by applying the operator $\bar{\mathscr{L}_0}^\dag\bar{\mathscr{L}_1}^\dag$ to the first expression in Theorem~\ref{thm1} and with aid of the second expression in this theorem, then we have 
\begin{align}
B^2S_{+1}&=\bar{\mathscr{L}_0}^\dag\bar{\mathscr{L}_1}^\dag\bar{\mathscr{L}_0}\bar{\mathscr{L}_1}S_{+1}=\bar{\mathscr{L}_0}^\dag(\bar{\mathscr{L}_1}+2\mathcal{Q}\sqrt{\Delta_\theta})(\bar{\mathscr{L}_0}^\dag-2\mathcal{Q}\sqrt{\Delta_\theta})\bar{\mathscr{L}_1}S_{+1}\nonumber\\
&=\bar{\mathscr{L}_0}^\dag\left(\bar{\mathscr{L}_1}\bar{\mathscr{L}_0}^\dag-\dfrac{2}{\sin\theta}\dfrac{d}{d\theta}\left(\sin\theta H\right)\right)\bar{\mathscr{L}_1}S_{+1}\nonumber\\
&=\bar{\mathscr{L}_0}^\dag\bar{\mathscr{L}_1}(2a\omega\Xi\cos\theta-\lambda)S_{+1}-4a\omega\Xi\bar{\mathscr{L}_0}^\dag\cos\theta\bar{\mathscr{L}_1}S_{+1}\nonumber\\
&=2a\omega\Xi\bar{\mathscr{L}_0}^\dag(\cos\theta\bar{\mathscr{L}_1}-\sin\theta\sqrt{\Delta_\theta})S_{+1}-\lambda\bar{\mathscr{L}_0}^\dag\bar{\mathscr{L}_1}S_{+1}-4a\omega\Xi\bar{\mathscr{L}_0}^\dag\cos\theta\bar{\mathscr{L}_1}S_{+1}\nonumber\\
&=-2a\omega\Xi(\cos\theta\bar{\mathscr{L}_0}^\dag-\sin\theta\sqrt{\Delta_\theta})\bar{\mathscr{L}_1}S_{+1}-2a\omega\Xi\sin\theta\sqrt{\Delta_\theta}\bar{\mathscr{L}_1}^\dag S_{+1}-\lambda\bar{\mathscr{L}_0}^\dag\bar{\mathscr{L}_1}S_{+1}\nonumber\\
&=-(2a\omega\Xi\cos\theta+\lambda)\bar{\mathscr{L}_0}^\dag\bar{\mathscr{L}_1}S_{+1}+2a\omega\Xi\sin\theta\sqrt{\Delta_\theta}(\bar{\mathscr{L}_1} -\bar{\mathscr{L}_1}^\dag)S_{+1}\nonumber\\
&=-(2a\omega\Xi\cos\theta+\lambda)(2a\omega\Xi\cos\theta-\lambda)S_{+1}-4a\omega\sin\theta\Xi^2HS_{+1}\nonumber\\
&=\left(\lambda^2-4\omega^2\Xi^2(a^2-\dfrac{am}{\omega})\right)S_{+1}\;,\label{derivaB}
\end{align}
where the angular equation~\eqref{Spluseq} was used in the above derivations, and $\mathcal{Q}$ is defined as
\begin{equation}
\mathcal{Q}=\dfrac{\Xi H}{\Delta_\theta}\;. \label{Qdef}
\end{equation}
Eq.~\eqref{derivaB} finally gives the value of $B$, i.e.
\begin{equation}
B^2=\lambda^2-4\omega\Xi^2(\omega a^2-ma)\;.\label{Bvalue}
\end{equation}
The sign of $B$ can be fixed by comparing with the spherical case $(a=0)$ when the angular functions reduce to the spin-weighted spherical harmonics~\cite{goldberg1967spin}. This comparison requires us to choose the positive square root in Eq.~\eqref{Bvalue}.
\begin{thm}\label{thm2}
$\Delta_r\mathscr{D}_0\mathscr{D}_0R_{-1}$ is a constant multiple of $\Delta_rR_{+1}$,\;i.e. $\Delta_r\mathscr{D}_0\mathscr{D}_0R_{-1}=B\Delta_rR_{+1}$, \\
{\color{white}{and}}$\;\;\;\;\;\;\;\;\;\;\;\;\;\;\;\;\;\Delta_r\mathscr{D}_0^\dag\mathscr{D}_0^\dag\Delta_rR_{+1}$ is a constant multiple of $R_{-1}$, i.e. $\Delta_r\mathscr{D}_0^\dag\mathscr{D}_0^\dag\Delta_rR_{+1}=BR_{-1}$.
\end{thm}
The proportionality stated in this theorem can be proved as follows, by taking the first expression as an example. By applying the operator $\mathscr{D}_0\mathscr{D}_0$ to Eq.~\eqref{Rminuseq}, we have
\begin{align}
\lambda\mathscr{D}_0\mathscr{D}_0R_{-1}&=\mathscr{D}_0\mathscr{D}_0(\Delta_r\mathscr{D}_0^\dag\mathscr{D}_0-2i\omega\Xi r)R_{-1}\nonumber\\
&=\mathscr{D}_0\mathscr{D}_0\Delta_r\mathscr{D}_0\mathscr{D}_0R_{-1}+2i\Xi\mathscr{D}_0\mathscr{D}_0K\mathscr{D}_0R_{-1}-2i\omega\Xi\mathscr{D}_0\mathscr{D}_0(rR_{-1})
\nonumber\\
&=\mathscr{D}_0\Delta_r\mathscr{D}_1\mathscr{D}_0\mathscr{D}_0R_{-1}+2i\Xi\mathscr{D}_0(K\mathscr{D}_0+2\omega r)\mathscr{D}_0R_{-1}
-2i\omega\Xi\mathscr{D}_0\mathscr{D}_0(rR_{-1})\nonumber\\
&=\mathscr{D}_0(\Delta_r\mathscr{D}_1+2i\Xi K)\mathscr{D}_0\mathscr{D}_0R_{-1}+2i\omega\Xi(2\mathscr{D}_0r\mathscr{D}_0R_{-1}-r\mathscr{D}_0\mathscr{D}_0R_{-1}
+2\mathscr{D}_0R_{-1})\nonumber\\
&=\mathscr{D}_0\Delta_r\mathscr{D}_1^\dag\mathscr{D}_0\mathscr{D}_0R_{-1}+2i\omega\Xi r\mathscr{D}_0\mathscr{D}_0R_{-1}\nonumber\\
&=(\Delta_r\mathscr{D}_1\mathscr{D}_1^\dag+2i\omega\Xi r)(\mathscr{D}_0\mathscr{D}_0R_{-1})\;.
\end{align}
Therefore, $\mathscr{D}_0\mathscr{D}_0R_{-1}$ satisfies the same equation as $R_{+1}$, by comparing with Eq.~\eqref{Rpluseq}. The second part in this theorem can be proved following the same logic. Notice that the constants of proportionality in this theorem have been fixed to $B$, the same constant used in Theorem~\ref{thm1}. This fact, indeed, is a consequence of Theorem~\ref{thm1}, and can be understood as follows. The Maxwell scalars $\Phi_0$ and $\Phi_2$ are governed by their equations \eqref{Phi0eq} and \eqref{Phi2eq}, but their relative normalization is still undetermined. To obtain this relative normalization constant, by applying the operator $\sqrt{\Delta_\theta}(\mathscr{L}_0+ia\sin\theta/\bar{\rho}^\ast)$ to Eq.~\eqref{Maxwelleq2_1} and $(\mathscr{D}_0+1/\bar{\rho}^\ast)$ to Eq.~\eqref{Maxwelleq2_2}, and eliminating $\Phi_1$, we obtain
\begin{equation}
\left(\mathscr{D}_0+\dfrac{1}{\bar{\rho}^\ast}\right)\left(\mathscr{D}_0-\dfrac{1}{\bar{\rho}^\ast}\right)\Phi_2
=\sqrt{\Delta_\theta}\left(\mathscr{L}_0+\dfrac{ia\sin\theta}{\bar{\rho}^\ast}\right)
\sqrt{\Delta_\theta}\left(\mathscr{L}_1-\dfrac{ia\sin\theta}{\bar{\rho}^\ast}\right)\Phi_0
\end{equation}
which can be further simplified
\begin{equation}
\mathscr{D}_0\mathscr{D}_0\Phi_2=\bar{\mathscr{L}_0}\bar{\mathscr{L}_1}\Phi_1\;,\label{complete1}
\end{equation}
with aid of the commutative property of the angular operator
\begin{equation}
\sqrt{\Delta_\theta}\sin\theta\mathscr{L}_{n+1}=\mathscr{L}_n\sqrt{\Delta_\theta}\sin\theta\;.
\end{equation}
With the field decompositions~\eqref{decompfields} and the identity $\bar{\mathscr{L}_0}\bar{\mathscr{L}_1}S_{+1}=BS_{-1}$ in Theorem~\ref{thm1}, Eq.~\eqref{complete1} becomes
\begin{equation}
\mathscr{D}_0\mathscr{D}_0R_{-1}=BR_{+1}\;.
\end{equation}
This gives the first identity in Theorem~\ref{thm2}, by multiplying $\Delta_r$ on both sides.

\underline{\bf The solution for $\Phi_1$}

To complete the solutions, now we step back to look for the solution for $\Phi_1$. 
We start from Eq.~\eqref{Maxwelleq2_1}, by multiplying $\bar{\rho}^\ast$ on both sides, then we have
\begin{equation}
\Big(\bar{\rho}^\ast\mathscr{D}_0+1\Big)\Phi_1=\left(\bar{\rho}^\ast\bar{\mathscr{L}_1}-ia\sin\theta\sqrt{\Delta_\theta}\right)\Phi_0\;,
\end{equation}
which, from the definition of $\mathscr{D}_0$, may be rewritten as
\begin{equation}
\mathscr{D}_0\Big(\bar{\rho}^\ast\Phi_1\Big)=\left(\bar{\rho}^\ast\bar{\mathscr{L}_1}-ia\sin\theta\sqrt{\Delta_\theta}\right)\Phi_0\;.\label{Phi1sol1}
\end{equation}
Then multiplying by $\Delta_r$ and expanding $\Phi_0$ as in Eq.~\eqref{decompfields}, Eq.~\eqref{Phi1sol1} becomes
\begin{align}
\Delta_r\mathscr{D}_0\Big(\bar{\rho}^\ast\Phi_1\Big)&=\left(\bar{\rho}^\ast\bar{\mathscr{L}_1}-ia\sin\theta\sqrt{\Delta_\theta}\right)\Big(\Delta_r\Phi_0\Big)\nonumber\\
&=r\Big(\Delta_rR_{+1}\Big)\bar{\mathscr{L}_1}S_{+1}-ia\Big(\Delta_rR_{+1}\Big)\left(\cos\theta\bar{\mathscr{L}_1}S_{+1}+\sin\theta\sqrt{\Delta_\theta}S_{+1}\right)\nonumber\\
&=\Delta_r\Big(\mathscr{D}_0g_{+1}\bar{\mathscr{L}_1}S_{+1}-iaf_{-1}\mathscr{D}_0\mathscr{D}_0P_{-1}\Big)\;,
\end{align}
which gives
\begin{equation}
\mathscr{D}_0\Big(\bar{\rho}^\ast\Phi_1\Big)=\mathscr{D}_0\Big(g_{+1}\bar{\mathscr{L}_1}S_{+1}-iaf_{-1}\mathscr{D}_0P_{-1}\Big)\;,\label{Phi11eq}
\end{equation}
where we defined
\begin{align}
&g_{+1}=\dfrac{1}{B}\Big(r\mathscr{D}_0P_{-1}-P_{-1}\Big)\label{gpluseq}\;,\\
&f_{-1}=\dfrac{1}{B}\left(\cos\theta\bar{\mathscr{L}_1}S_{+1}+\sin\theta\sqrt{\Delta_\theta}S_{+1}\right)\;,\label{fminuseq}
\end{align}
and
\begin{equation}
\Delta_r\Phi_0\equiv P_{+1}S_{+1}\;,\;\;\;\Phi_2\equiv P_{-1}S_{-1}\;,\;\;\;\;
\end{equation}
and where the Starobinski-Teukolsky identity in Theorem~\ref{thm2}, i.e.
\begin{equation}
\Delta_r\mathscr{D}_0\mathscr{D}_0P_{-1}=BP_{+1}\;,
\end{equation}
has been used.\\
Applying a similar procedure to Eq.~\eqref{Maxwelleq2_2}, we obtain
\begin{equation}
\bar{\mathscr{L}_0}\Big(\bar{\rho}^\ast\Phi_1\Big)=\bar{\mathscr{L}_0}\Big(g_{+1}\bar{\mathscr{L}_1}S_{+1}-iaf_{-1}\mathscr{D}_0P_{-1}\Big)\;.\label{Phi12eq}
\end{equation}
By comparing Eqs.~\eqref{Phi11eq} and~\eqref{Phi12eq}, and considering that $\mathscr{D}_0$ is the differential operator only for the radial part while $\bar{\mathscr{L}_0}$ is the differential operator only for the angular part, we conclude that\footnote{Strictly, we could add an extra function, say $F$, satisfying homogenous equations $\mathscr{D}_0(F)=\bar{\mathscr{L}_0}(F)=0$, to the solution of $\Phi_1$ in Eq.~\eqref{Phi1eq}. The solution of $F$, however, is singular at $\theta=0$ and $\theta=\pi/2$, similar to the Kerr case~\cite{Chandrasekhar:1985kt}. Therefore, we have not included $F$ in the solution for $\Phi_1$.}
\begin{equation}
\bar{\rho}^\ast\Phi_1=g_{+1}\bar{\mathscr{L}_1}S_{+1}-iaf_{-1}\mathscr{D}_0P_{-1}\;,\label{Phi1eq}
\end{equation}
which determines $\Phi_1$ uniquely. This equation is relevant to derive the angular momentum flux for the Maxwell field on Kerr-AdS background, see Appendix~\ref{app:angmomflux} for details.

Although the scalar $\Phi_1$ is obtained in Eq.~\eqref{Phi1eq}, $\bar{\mathscr{L}_1}S_{+1}$ is still yet unknown. Indeed $\bar{\mathscr{L}_1}S_{+1}$ can be expressed in terms of $S_{+1}$ and $S_{-1}$, by
%
\begin{numcases}{}
\bar{\mathscr{L}_1}S_{+1}=\dfrac{(2a\omega\Xi\cos\theta-\lambda)S_{+1}-BS_{-1}}{2\sqrt{\Delta_\theta}\mathcal{Q}}\;,\label{L1Sp1}\\
\bar{\mathscr{L}_1}^\dag S_{-1}=\dfrac{(2a\omega\Xi\cos\theta+\lambda)S_{-1}+BS_{+1}}{2\sqrt{\Delta_\theta}\mathcal{Q}}\;,\label{L1Sm1}
\end{numcases}
where $B$ and $\mathcal{Q}$ are given in Eqs.~\eqref{Bvalue} and~\eqref{Qdef}.

To prove Eq.~\eqref{L1Sp1}, we start from Eq.~\eqref{Spluseq},
\begin{align}
\left(\bar{\mathscr{L}_0}^\dag\bar{\mathscr{L}_1}-2a\omega\Xi\cos\theta+\lambda\right)S_{+1}=0\Rightarrow
\left(\bar{\mathscr{L}_0}+2\mathcal{Q}\sqrt{\Delta_\theta}\right)\bar{\mathscr{L}_1}S_{+1}+\left(\lambda-2a\omega\Xi\cos\theta\right)S_{+1}=0\;,
\end{align}
and considering $\bar{\mathscr{L}_0}\bar{\mathscr{L}_1}S_{+1}=BS_{-1}$ in Theorem~\ref{thm2}, we obtain
\begin{equation}
BS_{-1}+2\mathcal{Q}\sqrt{\Delta_\theta}\bar{\mathscr{L}_1}S_{+1}+\left(\lambda-2a\omega\Xi\cos\theta\right)S_{+1}=0\;,
\end{equation}
which finally gives Eq.~\eqref{L1Sp1}.
Following a similar procedure, Eq.~\eqref{L1Sm1} can be proved.

%% file: GeneralInFlat.tex
\chapter{Proca field equations}
\label{ch:GeneralInFlat}
%
%
In the first part of this thesis, we are going to study Hawking radiation, for neutral and charged Proca fields; and quasi-bound states, for charged scalar and Proca fields. To explore these problems on a particular background, the first step is to obtain the wave equations for those fields. Two key properties that will allow the study of these wave equations are: separation of variables and decoupling of degrees of freedom. For scalar fields, which are governed by the Klein-Gordon equation, there is only one degree of freedom, and variables in general can be separated on Kerr-like spacetimes~\cite{Carter:1968rr}. This is in part the reason why scalar fields have been explored so intensively in the literature\footnote{It is now known that at least one fundamental scalar field exists in Nature with the discovery of the Higgs boson~\cite{Chatrchyan:2012xdj,Aad:2012tfa}.}. The situation becomes complicated for massive bosonic fields, with spin.
In the standard model of particle physics (SM), massive spin-1 (Proca) fields describe the $Z$ and $W$ particles, where the former is neutral and the latter is charged. Perturbations of a neutral Proca field were first studied in~\cite{Pawl:2004bx,Konoplya:2005hr} on spherically symmetric backgrounds. In these studies only the $\ell=0$ mode was considered since the corresponding wave equation is directly decoupled and separated. Later on by using the Kodama-Ishibashi formalism, see Section~\ref{sc:KI} for details, we were able to obtain a set of equations on $D$-dimensional spherically symmetric backgrounds, to cover all the possible modes for both neutral and charged Proca fields. This was done in the context of Hawking radiation and quasi-bound states~\cite{Herdeiro:2011uu,Wang:2012tk,Sampaio:2014swa}. Similar equations for neutral Proca fields on Schwarzschild BHs were shortly after obtained using different perturbation variables, and these equations were applied to study quasinormal modes and quasi-bound states~\cite{Rosa:2011my}. In the Kerr background, separation of variables for Proca fields can be only achieved in the slow rotation limit, and the corresponding coupled wave equations were obtained in~\cite{paniPRL,paniPRD}, to study superradiant instabilities. A fully numerical study of the BH-Proca system was recently implemented~\cite{Zilhao:2015tya}, in nonrotating spacetimes.

In this chapter, we present a complete set of Proca equations on spherically symmetric backgrounds, by using the Kodama-Ishibashi formalism introduced in Section~\ref{sc:KI}.


We start by describing scalar and Proca fields, which may be complex and charged under a $U(1)$ electromagnetic field, with the Lagrangian
\begin{equation}
\mathcal{L}=-(\mathcal{D}_\mu\Psi)^\ast\mathcal{D}^\mu\Psi-\mu_s^2\Psi^\ast\Psi -\dfrac{1}{2}W^\dagger_{\mu\nu}W^{\mu\nu}-\mu_p^2W_\mu^\dagger W^\mu-iqW_\mu^\dagger W_\nu F^{\mu\nu} \;,\label{Lagrangian}
\end{equation}
where $W_{\mu\nu}=\mathcal{D}_\mu W_\nu-\mathcal{D}_\nu W_\mu$, $\mathcal{D}_\mu\equiv \partial_\mu-i q A_\mu$ and the field charge is $q$\footnote{In general, scalar and Proca fields may have different charge. Since they will be studied separately, for simplicity here we denote both charges by the same symbol $q$.}. Scalar and Proca fields are denoted by $\Psi$ and $W_\mu$, with mass $\mu_s$ and $\mu_p$, respectively. As one observes from the above Lagrangian, both scalar and Proca fields are coupled to the electromagnetic potential $A_\mu$ through the gauge covariant derivative, while the latter one also couples to the electromagnetic field strength tensor $F_{\mu\nu}=\partial_\mu A_\nu-\partial_\nu A_\mu$, as determined by gauge invariance in the SM.



The equations of motion for Proca fields, when all the background fields are fixed, are
\begin{equation}
\dfrac{1}{\sqrt{-g}}\mathcal{D}_\nu\left(\sqrt{-g}W^{\mu\nu}\right)+\mu_p^2W^\mu+iqW_\nu F^{\mu\nu}=0 \;.\label{Procaeq}
\end{equation}
For the gravitational background with an Einstein symmetric space shown in Eq.~\eqref{KI:metric}, the Kodama-Ishibashi formalism~\cite{Kodama:2000fa} can be applied.
%
%
For a Proca field $W_\mu=\{W_a,W_i\}$, $W_a$ are $m$-scalars, with respect to the Einstein space $\sigma_n$, so they must obey the scalar eigenvalue equations~\eqref{intro:eigenscalar}. $W_i$ is a covector field, so it can be decomposed into a scalar $\Phi$ obeying the scalar eigenvalue equation~\eqref{intro:eigenscalar}, and a transverse covector $W^{T}_i$ obeying the vector eigenvalue equation~\eqref{intro:eigenvector}.

This decomposition allows for an expansion of the various degrees of freedom $\{W_a,\Phi,W^T_i\}$ in a basis of harmonics of the Einstein space.
Furthermore, this decomposition allows for a decoupling of the field equations into an independent vector mode $W^T_i$ and $m+1$ coupled scalar fields for each set of quantum numbers labeling the basis of harmonic functions. Observe, however, that not all these modes correspond to physically independent degrees of freedom, as we will show. 
\\
In the following we shall consider, separately, two cases according to the scalar eigenvalue, i.e. $\kappa_s$.

\section{Modes with $\kappa_s\neq 0$}
Expanding the field equations~\eqref{Procaeq}, with the decomposition~\eqref{intro:decom} for $W_i$, and using conditions~\eqref{intro:eigenscalar} for $\{W_a,\Phi\}$ and~\eqref{intro:eigenvector} for $W^T_i$, we obtain
\begin{align}
&\dfrac{\kappa_s^2}{r^2}B_a+\dfrac{h_{af}}{r^n}\mathcal{D}_b\left[h^{db}h^{cf}r^n\left(\mathcal{D}_cB_d-\mathcal{D}_dB_c\right)\right]-\mu_p^2B_a-\mathcal{D}_a\left[\dfrac{1}{r^{n-2}}\mathcal{D}_b\left(r^{n-2}h^{bc}B_c\right)\right]+iqB_bF_a^{\;\;b}\nonumber\\&=0 \ , \hspace{12mm}\label{Ba}\\
&\dfrac{1}{r^{n-2}}\mathcal{D}_a\left(r^{n-2}h^{ab}B_b\right)-\mu_p^2\Phi=0 \ , \label{Ge:PHIeq}\\
&\left[\dfrac{1}{r^2}\left(\kappa_v^2+\dfrac{\hat{R}}{n}\right)+\mu_p^2+\dfrac{1}{r^{n-2}}\mathcal{D}_a\left(r^{n-2}h^{ab}\mathcal{D}_b\right)\right]\hat
W^{Tj}=0 \label{WT}\; ,
\end{align}
with definition\footnote{Here we use the variables $B_a$, since $\Phi$ can be determined by $B_a$ and becomes a non-dynamical degree of freedom, as shown in Eq.~\eqref{Ge:PHIeq}.} $B_a\equiv W_a-\mathcal{D}_a\Phi$, and $\hat{W}^{Ti}=r(y)^2W^{Ti}$.
We consider the spherically symmetric case, by specifying the metric in Eq.~\eqref{KI:metric} with $\{y^a\}=\{t,r\}$,
$|h|=1$, $h_{ab}$ is diagonal,
$h_{tt}=-1/h_{rr}\equiv -V$. Since $\Phi$ is given by the second equation in terms of the other fields, it is a non-dynamical degree of freedom. In four dimensions, this agrees with the fact that a spin-1 massive field has three possible physical polarizations which in this case will be the two dynamical scalars and the transverse vector. In higher dimensions we will see that the transverse vector on the $n$-sphere contains more (degenerate) polarizations.

We can factor out the spherical harmonics through the decomposition
\begin{align}
B_a&=\beta_a^\Lambda(y)\mathcal{Y}_\Lambda(x)\ , \nonumber \\
\hat{W}^{T i}&=q^\Lambda(y)\mathcal{Y}^i_{\Lambda}(x) \; ,\label{Lexpansion}
\end{align}
where $\mathcal{Y}_\Lambda$ is the scalar harmonic function with the definition~\eqref{intro:ScalatrHarmonicFunction}, $\mathcal{Y}^i_\Lambda$ is the vector harmonic function with the definition~\eqref{intro:HarmonicVector}, and $\Lambda$ denotes the mode eigenvalues for the corresponding harmonic functions. Furthermore, making the ansatz
\[
\beta^{\Lambda}_t=e^{-i\omega t}\psi(r) \ , \qquad \beta^{\Lambda}_r=e^{-i\omega t}\dfrac{\chi(r)}{V} \ , \qquad q^{\Lambda}=e^{-i\omega t}\Upsilon(r)\ ,
\]
 and using Eqs.~\eqref{Ba} and~\eqref{WT}, we obtain\footnote{Note that there is a symmetry for the coupled system, i.e. for real $\omega$, if $(\psi,\chi)$ is a solution to the equations, $(-\psi^\ast,\chi^\ast)$ is also a solution. For complex $\omega$, this statement still can be made as: if $(\psi,\chi)e^{-i\omega t}$ is a solution to the equations, then $(-\psi^\ast,\chi^\ast)e^{-i\omega^\ast t}$ is also a solution.}
\begin{align}
&\left[V^2\dfrac{d}{dr}\left(\dfrac{1}{r^{n-2}}\dfrac{d}{dr}r^{n-2}\right)+(\omega+qA_t)^2-\left(\dfrac{\kappa_s^2}{r^2}+\mu_p^2\right)V \right]\chi-i\left((\omega+qA_t) V^{\prime}+2qA_t\dfrac{V}{r}\right)\psi\nonumber\\& =0\ ,\label{originalsys1}\\
&\left[\dfrac{V^2}{r^n}\dfrac{d}{dr}\left(r^n\dfrac{d}{dr}\right)+(\omega+qA_t)^2-\left(\dfrac{\kappa_s^2}{r^2}+\mu_p^2\right)V\right]\psi+i\left(\dfrac{2\omega V}{r}-(\omega+qA_t)V^{\prime}\right)\chi \nonumber\\&=0 \ ,\label{originalsys2} \\
&\left[\dfrac{V}{r^{n-2}}\dfrac{d}{dr}\left(r^{n-2}V\dfrac{d}{dr}\right)+(\omega+qA_t)^2-\left(\dfrac{\kappa_v^2+\frac{\hat{R}}{n}}{r^2}+\mu_p^2\right)V\right] \Upsilon=0 \label{transverse}
\; ,
\end{align}
where $A_t$ is the only nonvanishing component of electromagnetic potential due to the spherically symmetric background. Thus we obtain two second order coupled radial equations for $\left\{\psi,\chi\right\}$ and a decoupled equation for $\Upsilon$. Note that $\kappa_s^2=\ell(\ell+n-1)$ and $\kappa_v^2=\ell(\ell+n-1)-1$ with $\ell$ starting at zero and one respectively. The third combination is $\kappa_v^2+\frac{\hat{R}}{n}=\ell(\ell+n-1)+n-2$.

The manipulations leading to the two coupled equations above are only valid for non-zero $\mu_p$. In the exactly massless theory, a similar calculation leads to a single decoupled equation for one of the scalar modes which is
\begin{equation}
\left[V\dfrac{d}{dr}\left(\dfrac{V}{r^{n-2}}\dfrac{d}{dr}r^{n-2}\right)-\dfrac{2qd_rA_tV^2}{(\omega+qA_t)r^{n-2}}\dfrac{d}{dr}r^{n-2}
+\Big(\omega+qA_t\Big)^2-\dfrac{\kappa_s^2}{r^2}V\right] \chi=0 \; , \label{Max}
\end{equation}
whereas the other mode $\psi=i V d_r(r^{n-2}\chi)/((\omega+qA_t) r^{n-2})$ becomes non-dynamical\footnote{Note that the coupled equations have only one dynamical degree of freedom in the massless limit. One can use either $\psi$ or $\chi$ to describe the dynamical mode, then the other one becomes non-dynamical and is determined in terms of the dynamical mode.}. Here $d_r\equiv d/dr$. The transverse mode -- described by equation (\ref{transverse}) -- remains the same for any $\mu_p$; in particular, for $\mu_p=0$, and (only) $n=2$ it becomes equivalent to (\ref{Max}). This will be manifest in the numerical results.
\section{Modes with $\kappa_s= 0$}
For the exceptional modes with $\kappa_s=0$, $\Phi$ does not enter the wave equation so it is a free non-dynamical field. The corresponding equation for $W^{(0)}_a$ is (the superscript denotes it is the exceptional mode)
\begin{equation}
\dfrac{h_{af}}{r^n\sqrt{|h|}}\mathcal{D}_b\left[h^{db}h^{cf}r^n\sqrt{|h|}\left(\mathcal{D}_cW^{(0)}_d-\mathcal{D}_dW^{(0)}_c\right)\right]+\mu_p^2W^{(0)}_a+iqW_b^{(0)}F_a^{\;\;b}=0 \; .
\end{equation}
When $\mu_p^2\neq 0$ one uses an ansatz similar to the previous section to obtain a radial equation for a dynamical degree of freedom
\begin{align}
&\left[\dfrac{\mu_p^2}{r^n}\dfrac{d}{dr}\left(\dfrac{r^nV}{(\omega+qA_t)^2-\mu_p^2V}\dfrac{d}{dr}\right)+\dfrac{\mu_p^2}{V}+\dfrac{\mu_p^2qd_rA_t}
{((\omega+qA_t)^2-\mu_p^2V)^2}\Big(2qVd_rA_t-(\omega+qA_t)V'\Big)\right.\nonumber\\
&\left.-\dfrac{\omega+qA_t}{r^n((\omega+qA_t)^2-\mu_p^2V)}
\dfrac{d}{dr}\Big(qr^nd_rA_t\Big)\right]\psi^{(0)}=0 \; ,\label{k0M}
\end{align}
and a non-dynamical one,
\begin{equation}
\chi^{(0)}=\dfrac{i V}{(\omega+qA_t)^2-\mu_p^2V}\left((\omega+qA_t)d_r\psi^{(0)}-q\psi^{(0)}d_rA_t\right)\;,\nonumber
\end{equation}
where $d_r\equiv d/dr$.
Otherwise, for $\mu_p^2=0$, we recover the well known result that all the exceptional modes are non-dynamical (see e.g. \cite{Konoplya:2005hr}).

Now that we have covered all possibilities, several comments are in order. First there is a discrete difference between the small mass limit and the exactly massless theory since we have different sets of equations for each case. This should not be surprising since there is an extra longitudinal mode for massive vector bosons. Second, the equations for the Maxwell theory case are all decoupled, in agreement with previous work \cite{Page:1976df}.
%

With all of the Proca equations at hand, we are going to apply them to study Hawking radiation in Chapter~\ref{ch:NeutralP} for a neutral Proca field, and in Chapter~\ref{ch:ChargedP} for a charged Proca field. In Chapter~\ref{ch:ChargedClouds}, we will then apply them to the computation of quasi--bound states in the Reissner-Nordstr\"om BH. 

%% file: NeutralP.tex
\chapter{Hawking radiation for a Proca field: neutral case}
\label{ch:NeutralP}



\section{Introduction}

Hawking radiation, as we described in Section~\ref{sc:HR}, is an interesting quantum phenomenon. It has been widely studied on different backgrounds, for various types of emitting particles, and in diverse gravitational theories. In particular, the study of Hawking radiation from higher-dimensional BHs has gained a lot attention recently, motivated by the possibility of producing microscopic BHs at the LHC. This is a prediction of TeV gravity scenarios in which the fundamental Planck scale could be as low as the TeV scale. This motivation led to an intensive study of Hawking radiation from higher-dimensional BHs~\cite{Ida:2002ez,Harris:2003eg,Harris:2005jx,Ida:2005ax,Duffy:2005ns,Casals:2005sa,Cardoso:2005vb,Cardoso:2005mh,Ida:2006tf,Casals:2006xp,Casals:2008pq,Sampaio:2009ra,Sampaio:2009tp,Herdeiro:2011uu,Wang:2012tk}.

One of the Hawking radiation channels that has not been properly addressed in the literature is that of massive vector bosons, both neutral and charged, to describe the emission of $Z$ and $W^{\pm}$ particles of the SM. The basic difficulty is that the Proca equations do not decouple even in a spherically symmetric BH background\footnote{In a rotating background, it appears that the Proca field can neither be decoupled nor separated.}. To bridge this gap, in this chapter, we study Hawking radiation for a neutral Proca field, by solving the Proca equations derived in Chapter~\ref{ch:GeneralInFlat}, in the background of a $D$-dimensional spherically symmetric Schwarzschild BH~\cite{Tangherlini:1963bw}.

In order to perform such a study, we have designed a numerical strategy to solve the coupled equations\footnote{The decoupled equations have been solved simultaneously.} without decoupling and showed that the coupled system may be treated by an {\bf S}-matrix type formalism which allows decoupling in the asymptotic regions. This {\bf S}-matrix is used to define a transmission matrix which gives the transmission factors as its eigenvalues. We have computed transmission factors for various modes, masses and spacetime dimensions. The mass term lifts the degeneracy between transverse modes, in $D=4$, and excites the longitudinal modes, in particular the $s$-wave. Moreover, it increases the contribution of waves with larger $\ell$, which can be dominant at intermediate energies. The transmission factors are then used to obtain the Hawking fluxes in this channel.

This chapter is organized as follows. In Section~\ref{sec:NPnearhorizon} we study the near horizon and far region asymptotic behaviors of the coupled Proca equations, which can be used to extract relevant information to study Hawking radiation. In Section~\ref{sec:scatteringM} we discuss how the scattering matrix is used to compute the transmission factor and the Hawking spectrum. In Section~\ref{sec:resultsNP} we discuss the numerical method and results, and we summarize our results in Section~\ref{sec:sumProneu}. Some technical relations are left to Appendix~\ref{app:neutralP}.

\section{Boundary conditions and radial system}
\label{sec:NPnearhorizon}
In this section, we study the wave equation for a neutral Proca field by setting the field charge $q=0$ in all the field equations derived in Chapter~\ref{ch:GeneralInFlat}, i.e. Eqs.~\eqref{originalsys1},~\eqref{originalsys2},~\eqref{Max} and~\eqref{k0M}, in the Schwarzschild-Tangherlini background with the metric function $V=1-\frac{M}{r^{n-1}}$. In the numerics, we choose units such that the horizon radius is $r_H=1$, so then $M=r_H^{n-1}$. Since decoupled radial equations have been extensively studied in the literature we will not present the details of our analysis of such modes and refer to the method in~\cite{Sampaio:2009ra,Sampaio:2009tp}. Thus in the following we will focus on the solutions of the coupled system for the massive theory, which will be used in conjunction with the decoupled modes to obtain the full Hawking spectrum in Section~\ref{sec:resultsNP}.

We start by finding a series expansion of the solution near the horizon for the coupled system $\left\{\psi,\chi\right\}$. This will be used to initialize the corresponding fields for the radial integration at $r=1.001$. If we define $y=r-1$, Eqs.~\eqref{originalsys1} and
~\eqref{originalsys2} become
\begin{eqnarray}
\left[M(r)\dfrac{d^2}{dy^2}+N(r)\dfrac{d}{dy}+P(r)\right]\psi+Q(r)\chi=0\label{NPsysterm1}\ ,\\
\left[\tilde M(r)\dfrac{d^2}{dy^2}+\tilde
N(r)\dfrac{d}{dy}+\tilde P(r)\right]\chi+\tilde Q(r)\psi=0\label{NPsysterm2}\ ,
\end{eqnarray}
where the polynomials are defined in Appendix~\ref{app:neutralP}. Making use of Frobenius' method to expand $\psi$ and $\chi$ as
\begin{equation}
\psi=y^\rho\sum^{\infty}_{j=0}{\mu_jy^j}\label{defpsi} \; , \qquad \chi=y^\rho\sum^{\infty}_{j=0}{\nu_jy^j} \ ,
\end{equation}
and inserting the above two equations into Eqs.~\eqref{NPsysterm1} and
~\eqref{NPsysterm2}, we obtain
\begin{equation}
\rho=\pm\frac{i\omega}{n-1} \; \; \mathrm{or} \; \; \rho = 1\pm\frac{i\omega}{n-1} \; .\label{rhoexp}
\end{equation}
We want to impose an ingoing boundary condition at the horizon, so we must choose the minus sign. Furthermore, after this sign choice, the right hand side solution produces a series expansion which is a special case of the left hand side (where the first coefficient is set to zero), so without loss of generality we choose $\rho = -i\omega/(n-1)$. One then writes down the recurrence relations and concludes that a general solution, close to the horizon, can be parameterized by the two coefficients $\nu_0$ and $\nu_1$. The other coefficients are generated by the recurrence relations~\eqref{NPrecurone}.

To understand the asymptotic behavior of the waves at infinity we now study a large $r$ asymptotic expansion in the form
\begin{equation}
\psi=e^{\beta r}r^{p}\sum_{j=0}\dfrac{a_j}{r^j}\label{psifar} \; , \qquad \chi=e^{\beta r}r^{p}\sum_{j=0}\dfrac{b_j}{r^j} \ .
\end{equation}
Inserting this into Eqs.~\eqref{originalsys1} and~\eqref{originalsys2} we obtain, at leading order,
\begin{equation}
\beta = \pm ik\;,\;\;\;\;\;\;\;k=\sqrt{\omega^2-\mu_p^2}\;,\qquad p = 1-\frac{n}{2}\pm i\varphi \; \; {\rm or} \; \; p =-\frac{n}{2}\pm i\varphi \; ,
\end{equation}
where $\varphi=\delta_{n,2}(\omega^2+k^2)/(2k)$. Thus one can show that  asymptotically\footnote{We have used, without loss of generality, the leading power behavior for $p$ and discarded the second option which produces the same solution, similarly to the near horizon expansion.}
\begin{equation}
\psi \rightarrow \dfrac{1}{r^{\frac{n}{2}-1}}\left[\left(a_0^++\dfrac{a_1^+}{r}+\ldots\right)e^{i\Phi}+\left(a_0^-+\dfrac{a_1^-}{r}+\ldots\right)e^{-i\Phi}\right] \ , \nonumber
\end{equation}
\begin{equation}
\chi \rightarrow \dfrac{1}{r^{\frac{n}{2}-1}}\left[\left(\left(-\frac{k}{\omega}+\dfrac{c^+}{r}\right)a_0^++\ldots\right)e^{i\Phi}+\left(\left(\frac{k}{\omega}+\dfrac{c^-}{r}\right)a_0^-+\ldots\right)e^{-i\Phi}\right] \; ,
\end{equation}
where $\Phi\equiv kr+\varphi \log r$ and $c^\pm$ is defined in the Appendix, Eq.~\eqref{eq:NPcpm}.
So as expected, each field is a combination of ingoing and outgoing waves at infinity. This asymptotic expansion also shows that, for a generic wave at infinity, we can choose four independent constants $\left\{a_0^\pm,a_1^\pm\right\}$ to characterize the solution. This is expected, since we have two coupled scalar fields and for each scalar degree of freedom we must have an associated ingoing wave and outgoing wave. Thus we can define four new fields $\left\{\chi^\pm,\psi^\pm\right\}$ (which will asymptote respectively to $\left\{a_0^\pm,a_1^\pm\right\}$), by truncating the expansion for the fields and for their first derivatives at infinity. Such a transformation can be written in matrix form by defining the 4-vector $\mathbf{\Psi}^T=(\psi_{+},\psi_{-},\chi_{+},\chi_{-})$ for the new fields, and another 4-vector $\mathbf{V}^T=(\psi,d_r\psi,\chi,d_r\chi)$ for the original fields and derivatives. Then the transformation is given in terms of an $r$-dependent matrix $\mathbf{T}$ defined through
\begin{equation}
\mathbf{V}= \mathbf{T} \mathbf{\Psi} \; ,
\end{equation}
which we provide in the Appendix, Eq.~\eqref{eq:NPT}. In order to obtain a first order system of ordinary differential equations (ODEs) for the new fields, we first define a matrix $\mathbf{X}$ through
\begin{equation}
\dfrac{d\mathbf{V}}{dr}=\mathbf{X}\mathbf{V} \; ,
\end{equation}
which is read out from the original system in Eqs.~\eqref{originalsys1} and~\eqref{originalsys2}. Its explicit form is given in the Appendix, Eq.~\eqref{eq:X}.
Then we obtain
\begin{equation}
\dfrac{d\mathbf{\Psi}}{dr}=\mathbf{T}^{-1}\left(\mathbf{X}\mathbf{T}-\dfrac{d\mathbf{T}}{dr}\right) \mathbf{\Psi} \ .
\end{equation} We can write other equivalent systems using different $\mathbf{T}$ matrices. In particular, we have also integrated a first order system using the fields $\psi_{s}=k \psi-isd_r\psi$ and $\chi_{s}=k \chi-isd_r\chi$ with $s=\pm$, which produced numerically equivalent results. The only difference is that for such fields we need to extract $\mathcal{O}(r^{-1})$ coefficients to obtain $a_1^s$.

\section{The Hawking spectrum}
\label{sec:scatteringM}
The boundary conditions we have chosen in Section~\ref{sec:NPnearhorizon} are suitable for the computation of the Hawking spectrum of radiated quanta from the BH. The Hawking spectrum is generically given by a sum over a complete set of modes with labels $\zeta$, of the transmission factor $\mathbb{T}_\zeta$ times a thermal average number of quanta produced at the horizon $\left<n_\zeta\right>$. This is defined for a basis of decoupled modes. In our problem, we have a sub-set of modes, the transverse vector mode, and the $\ell=0$ ($\kappa_s=0$) mode, which are decoupled. But we also have a tower of modes which are coupled two by two for each $\ell>0$, the two scalars $\psi$ and $\chi$. It is not obvious how to decouple them for all $r$ through an explicit transformation. Instead, let us try to understand how to extract the relevant information in the asymptotic regions.

Let us denote the two coupled fields by a 2-vector $\mathbf{U^T}=(\psi,\chi)$ and represent the coupled system of radial equations through a (linear) second order matrix differential operator $\mathcal{D}^{(2)}$ acting on $\mathbf{U}$, i.e. $\mathcal{D}^{(2)}\mathbf{U}=0$. The system is coupled because of the off diagonal elements of the $\mathcal{D}^{(2)}$ operator. To decouple the system we would have to find a transformation of the fields $\mathbf{U}=\mathcal{A} \mathbf{\bar U}$, such that the new differential operator $\bar{\mathcal{D}}^{(2)}=\mathcal{D}^{(2)}$\textopenbullet$\mathcal{A}$ is diagonal, i.e.
\begin{equation}
\bar{\mathcal{D}}^{(2)}=\left(\begin{array}{cc}
\bar{\mathcal{D}}^{(2)}_1 & 0 \\ 0 & \bar{\mathcal{D}}^{(2)}_2
\end{array}\right) \ .
\end{equation}
Even without finding such a transformation explicitly, one can draw some conclusions by assuming its existence\footnote{In fact, for example, if we consider $\mathcal{A}$ to be a general $r$-dependent matrix, we can write down two conditions for the four arbitrary functions of such a matrix. Thus, in principle, there is enough freedom.}. In particular we may establish a map between our general solution of the coupled system and the actual decoupled solution, for each of the asymptotic regions (horizon and far field). To find such a map let us first summarize the information we have on the general solution of the coupled system.

In Section~\ref{sec:NPnearhorizon} we have found that a general solution is parameterized by four independent coefficients in one of the asymptotic regions; either at the horizon or at infinity. Once we have chosen one set of coefficients, say at the horizon, due to the linearity of the equations, the four independent wave components at infinity are a linear combination of the four coefficients at the horizon. Let us formally denote the ingoing and outgoing wave coefficients at the horizon ($+/-$ respectively) by
 \[\vec{\mathbf{h}}=({\mathbf h}^+,{\mathbf h}^-)=(h^+_{i},h^-_{i}) \ , \] where $i=1,2$ since we have two fields.
Similarly, the coefficients at infinity are defined as the large $r$ limit of the $\mathbf{\Psi}$ field components (up to linear transformation which we will define next), i.e.
  \[\vec{\mathbf{y}}=({\mathbf y}^+,{\mathbf y}^-)=(y^+_{i},y^-_{i}) \ , \]
with $i=1,2$ for $\psi$ and $\chi$ respectively. Due to linearity, we can define a scattering matrix
\begin{eqnarray}
\vec{\mathbf{y}}=\mathbf{S} \vec{\mathbf{h}} &&\ \ \Leftrightarrow \ \ \left(\begin{array}{c} {\mathbf y}^+ \\ {\mathbf y}^- \end{array} \right)=\left(\begin{array}{c|c} {\mathbf S}^{++} & {\mathbf S}^{+-} \\ \hline {\mathbf S}^{-+}  & {\mathbf S}^{--} \end{array} \right)\left(\begin{array}{c} {\mathbf h}^+ \\ {\mathbf h}^- \end{array} \right) \;,\nonumber\\  &&\ \ \Leftrightarrow \ \ \left(\begin{array}{c} {y}^+_i \\ { y}^-_i \end{array} \right)=\sum_{j}\left(\begin{array}{c|c} S_{i j}^{++} & S_{i j}^{+-} \\ \hline S_{i j}^{-+}  & S_{i j}^{--} \end{array} \right)\left(\begin{array}{c} h_{j}^+ \\ h_{j}^- \end{array} \right) \; ,
\end{eqnarray}
which is a set of numbers (depending on energy, angular momentum, etc\ldots) containing all the information on the scattering process. It can be fully determined by considering specific modes at the horizon and integrating them outwards.
In our problem, we have imposed an ingoing boundary condition at the horizon which is simply ${\mathbf h}^{+}=0$. Then
\begin{equation}\label{NPscattering_ingoing}
{\mathbf y}^s={\mathbf S}^{s-}{\mathbf h}^{-} \; .
\end{equation}
Taking the $s=-$ component, and denoting the inverse matrix of ${\mathbf S}^{--}$ by $({\mathbf S}^{--})^{-1}$, we invert~\eqref{NPscattering_ingoing} to obtain the wave at the horizon given the ingoing wave at infinity
\begin{equation}
{\mathbf h}^{-} =\left({\mathbf S}^{--}\right)^{-1}{\mathbf y}^{-}\; .
\end{equation}
Inserting this relation back in the $s=+$ component of~\eqref{NPscattering_ingoing}, we obtain the outgoing wave in terms of the ingoing wave, at infinity
\begin{equation}\label{Reflection}
{\mathbf y}^+={\mathbf S}^{+-}({\mathbf S}^{--})^{-1}{\mathbf y}^-\equiv{\mathbf R} \, {\mathbf y}^-\; ,
\end{equation}
where in the last line we have defined the reflection matrix $\mathbf{R}$. Before proceeding, we note that there is still some freedom in the definition of the asymptotic coefficients since any (non-singular) linear combination is equally good from the point of view of satisfying the boundary condition. This freedom can be written in terms of 3 matrices $\mathbf{M}^s$, $\mathbf{M}_H^-$ relating some new fields (hatted) to the old fields
\begin{equation}
{\mathbf y}^s=\mathbf{M}^s\hat{\mathbf y}^s \; ,\; \; \; \; \; \; \; \; {\mathbf h}^{-}=\mathbf{M}_H^-\hat{\mathbf h}^{-} \; .
\end{equation}
Since this represents the most general parametrization of the solution in the asymptotic regions, there must be a choice which decouples the fields in those regions. To find the correct transformation we need a physical prescription.

To obtain the transmission factor for the  decoupled components, it is instructive to remind ourselves of the calculation of the transmission factor for a single decoupled field. It is defined as the fraction of the incident wave which is transmitted to the horizon. If we look at a wave with energy $\omega$ (for an observer at infinity), with ingoing/outgoing amplitudes $Y^{(\infty)}_{\mp}$, then~\cite{Kanti:2004nr}
\begin{equation}
\mathbb{T}=\dfrac{|Y^{(\infty)}_{-}|^2-|Y^{(\infty)}_{+}|^2}{|Y^{(\infty)}_{-}|^2}=\dfrac{\omega\left(|Y^{(\infty)}_{-}|^2-|Y^{(\infty)}_{+}|^2\right)}{\omega |Y^{(\infty)}_{-}|^2}=\dfrac{\mathcal{F}^{in}_H}{\mathcal{F}^{in}_\infty} \ ,\label{deftrans}
\end{equation}
where in the last step we note that $\mathbb{T}$ can be re-expressed as a ratio between the total incident energy flux $\mathcal{F}^{in}_H$ (which is the difference between the energy carried by the ingoing wave and the energy of the outgoing wave) and the incident energy flux associated with the ingoing wave at infinity ($\mathcal{F}^{in}_\infty$). The former is the flux of energy transmitted down to the horizon.

We now compute the energy fluxes through a sphere at radius $r$ using the energy-momentum tensor. This will allow us to identify the decoupled fields at infinity and at the horizon, and in particular, the ingoing and outgoing decoupled waves at infinity. Such a flux is shown to be conserved in our background, by using the conservation law for the energy-momentum tensor, combined with the fact that the spatial integral of $T_t^{\phantom{t} t}$ for each energy eigen-mode is constant. It is defined, evaluated at $r$, as
\begin{equation}\label{eq:FluxR}
\mathcal{F}|_r=-\int_{S^n} d\Sigma\, T_t^{\; r}
\end{equation}
where $d\Sigma$ is the volume element on a $t,r={\rm constant}$ hyper-surface. The energy-momentum tensor for the complex neutral Proca field is
\begin{equation}
T^{\mu\nu}=-\dfrac{1}{2}\left(W^{\dagger \mu \alpha}W^{\nu}_{\; \alpha}-\mu_p^2W^{\dagger \mu}W^{\nu}+c.c.\right)-\dfrac{g^{\mu \nu}}{2}\mathcal{L} \; ,
\end{equation}
up to an irrelevant normalization. If we insert this in~\eqref{eq:FluxR}, assume a field configuration with a well defined energy/frequency $\omega$, and make use of the equations of motion, then, for the non-trivial case of $\mu_p^2\neq 0 \neq \kappa_s^2$, we obtain
\begin{equation}\label{eq:FluxR2}
\mathcal{F}|_r=\sum_\Lambda\dfrac{i\omega V\Upsilon^{\dagger}_\Lambda}{2r^2}\dfrac{d\Upsilon_\Lambda}{dr}+\sum_\Lambda\left\{\dfrac{\kappa_s^2}{2r^2}\psi^\dagger \chi-\dfrac{1}{2\mu_p^2}\left[\dfrac{V}{r^n}\dfrac{d(r^n\xi^\dagger)}{dr}-\dfrac{\kappa_s^2\psi^\dagger}{r^2}\right]\left[i\omega\xi+\dfrac{\kappa_s^2\chi}{r^2}\right]\right\}+c.c
\end{equation}
where $\Lambda$ denotes the mode eigenvalues for the corresponding harmonic functions, and for convenience we define $\xi=d_r\psi+i\omega\chi$. Modes with different angular momentum eigenvalues are clearly decoupled, as are the transverse vector mode contributions in the first sum. The terms in the second sum couple two fields for fixed $\Lambda$. We can compute the flux at infinity and close to the horizon and express it in terms of the asymptotic coefficients in the corresponding region. Focusing on a specific mode and in the coupled part of the flux (second sum in~\eqref{eq:FluxR2})
\begin{equation}\label{eq:FluxInf}
\mathcal{F}^{\mathrm{coupled}}_\infty=|y_0^-|^2-|y_0^+|^2+|y_1^-|^2-|y_1^+|^2\equiv(\mathbf{y}^-)^\dagger \mathbf{y}^--(\mathbf{y}^+)^\dagger \mathbf{y}^+ \; ,
\end{equation}
where $y_i^s$ are linear combinations of the asymptotic coefficients $a_i^s$ given in the Appendix, Eq.~\eqref{eq:NPyplus}. This choice of $y_i^s$ is already in a form close to decoupled, since we have separated the modulus square of the incident contribution from the reflected contribution, without interference terms. This form is invariant under separate unitary transformation of $\mathbf{y}^\pm$. Using the reflection matrix we obtain
\begin{equation}\label{eq:FluxInfT}
\mathcal{F}^{\mathrm{coupled}}_\infty=(\mathbf{y}^-)^\dagger\left(\mathbf{1}-\mathbf{R}^\dagger\mathbf{R}\right) \mathbf{y}^-\equiv (\mathbf{y}^-)^\dagger\mathbf{T}\, \mathbf{y}^-  \; ,
\end{equation}
where we have defined a (hermitian) transmission matrix $\mathbf{T}$. Note that this matrix is composed of the transmission matrix given in Eq.~\eqref{Reflection}, and it is a generalization of the transmission factor, defined in Eq.~\eqref{deftrans}, to the coupled system. The transmission matrix can be diagonalized through a unitary transformation which is the remaining freedom we have for $\mathbf{y}^-$. In fact we can do even better, and diagonalize the reflection matrix $\mathbf{R}$ with a bi-unitary transformation using the arbitrary unitary $\mathbf{M}^{\pm}$ transformations. Then the fields are manifestly decoupled at infinity, both at the level of the reflection matrix and the transmission matrix. As a consequence, in the decoupled basis, an incident wave is reflected back in the same decoupled mode without interference with the other mode. Finally, the transmission factors are simply the eigenvalues of $\mathbf{T}$, since they are each associated with a decoupled component.

Furthermore, one can use the conservation law for the flux, to find an alternative expression for the transmission matrix, at the horizon (this will be useful to control numerical errors). The total flux at the horizon is
\begin{equation}\label{eq:FluxH}
\mathcal{F}^{\mathrm{coupled}}_H=\left(\mathbf{h^-}\right)^\dagger\mathbf{h^-} \; ,
\end{equation}
where the $h^-_i$ coefficients are linear combinations of the two independent $\nu_i$ coefficients ($i=0,1$), given in the Appendix, Eqs.~\eqref{eq:NPhminus}. Eq.~\eqref{eq:FluxH} establishes the important point that the flux is positive definite, so the transmission factors must be positive definite (as expected since there is no superradiance in Schwarzschild spacetime). Finally, using the relation between $\mathbf{y}^-$ and $\mathbf{h}^-$ through\footnote{Note that the relation between $\mathbf{h}^-$ and $(\mathbf{M}^-)^{-1}\mathbf{y}^-$ can be made diagonal using $\mathbf{M}^{-}_H$, so the problem is also decoupled at the horizon.} $\mathbf{S}^{--}$, we find
\begin{equation}\label{eq:FluxH2}
\mathbf{T}=(\mathbf{S}^{--}\mathbf{S}^{\dagger--})^{-1} \; .
\end{equation}
Once we have obtained the transmission factors, the number and energy fluxes are given by the standard result
\begin{equation}\label{eq:HawkFlux}
\dfrac{d\left\{N,E\right\}}{dt d\omega}=\dfrac{1}{2\pi}\sum_\ell\sum_\zeta \dfrac{\left\{1,\omega \right\}}{\exp(\omega/T_H)-1}d_\zeta \mathbb{T}_{\zeta} \ ,
\end{equation}
where $\zeta$ is a label running over the final set of decoupled scalar modes and the transverse mode, and $d_\zeta$ are the degeneracies of the corresponding spherical harmonics. Labeling the scalar and vector harmonic degeneracies by $d_S$ and $d_V$ respectively we have \cite{Ishibashi:2011ws}
\begin{eqnarray}\label{eq:degen}
d_S&=& \dfrac{(n+2\ell-1)(n+\ell-2)!}{(n-1)!\ell!}\ , \\
d_V&=& \dfrac{(n+2\ell-1)(n+\ell-1)(n+\ell-3)!}{(\ell+1)(\ell-1)!(n-2)!} \; .
\end{eqnarray}
The Hawking temperature in horizon radius units is
\begin{equation}
T_H=\dfrac{n-1}{4\pi} \; .
\end{equation}

\section{Results}
\label{sec:resultsNP}
In this section we present a selection of numerical results to illustrate the behavior of the transmission factors and the corresponding Hawking fluxes. To integrate the coupled and decoupled radial equations, we first wrote test codes in \textsc{mathematica} and then a code in the \textsc{c++} language, using the numerical integration routines of the Gnu Standard Library (GSL). Besides using different programming frameworks we have also tested different integration strategies which all agreed within relative numerical errors smaller than 0.1~\%. In fact, most of our numerical points have a precision which is one order of magnitude better. To check numerical errors we have integrated the radial equations up to a large radius of typically $r=10^4r_H$ and varied this up to a factor of 3 to check the precision. Furthermore we have used the two expressions for the transmission factor from Eqs.~\eqref{eq:FluxInfT} and~\eqref{eq:FluxH2} which agree within the quoted precision for almost all energies. The exception is for small energy, where the first definition converges poorly. This can be explained by a simple analysis of propagation of errors combined with the fact that the $\mathbf{y}^{\pm}$ coefficients grow very fast as we decrease energy, thus requiring a very large precision for some fine cancellations to occur. The second expression is thus more natural in that limit since it does not need such cancellations and does not require such large precision.

We have generated several samples of transmission factors, some of which are displayed in Fig.~\ref{fig:Tfacs}. Hereafter, we shall denominate the partial waves associated to the different modes of the Proca field by $\ell_1,\ell_2,\ell_T$ and $\ell=0$, where $\ell_1,\ell_2$ correspond to the two coupled modes described by Eqs.~\eqref{originalsys1} and~\eqref{originalsys2}, $\ell_T$ to the decoupled mode described by Eq.~\eqref{transverse} and $\ell=0$ to the $\kappa_s=0$ mode, described by Eq.~\eqref{k0M}. Moreover, partial waves associated to the Maxwell field shall be denoted by $\ell_E$, and are described by Eq.~\eqref{Max}.
%
\afterpage{
\begin{figure}[t]
\includegraphics[scale=0.68,clip=true,trim= 0 0 0 0]{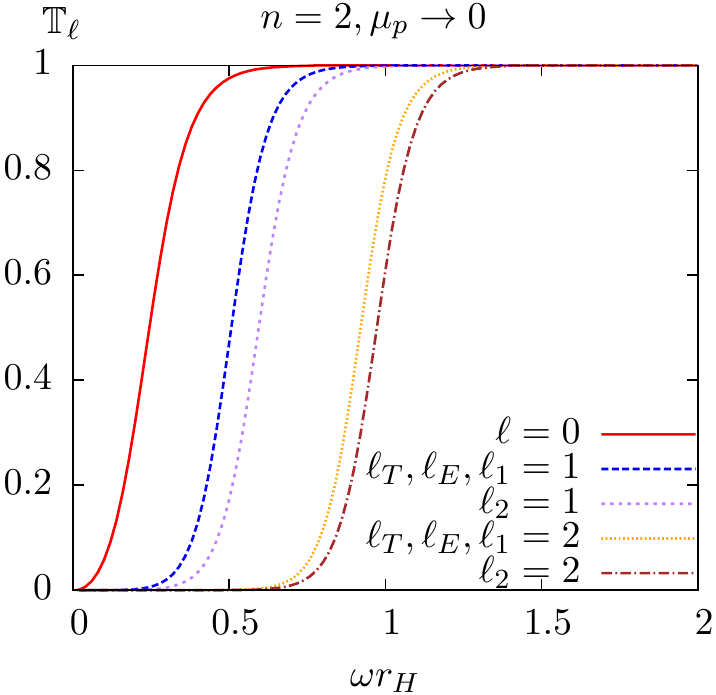}
\hspace{-1.5mm}
\includegraphics[scale=0.68,clip=true,trim= 0 0 0 0]{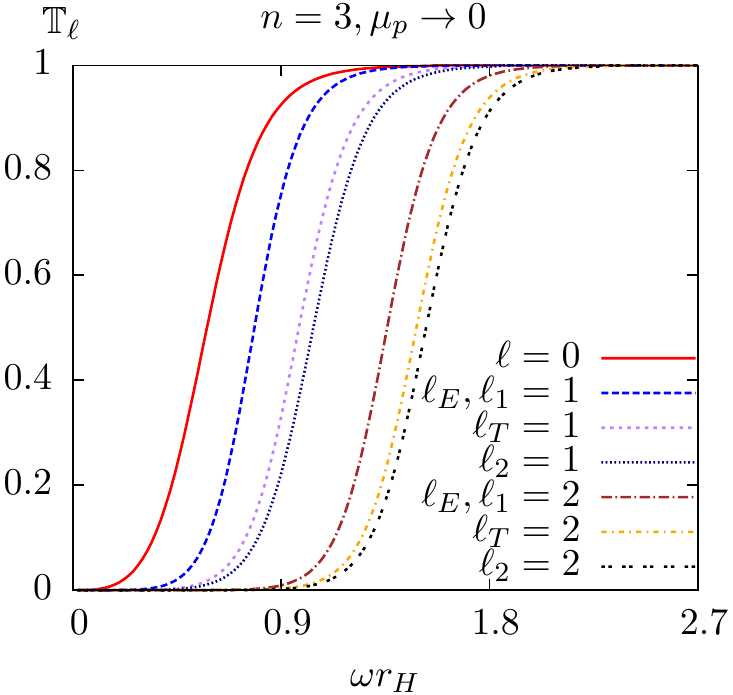}
\hspace{-2.5mm}
\includegraphics[scale=0.68,clip=true,trim= 0 0 0 0]{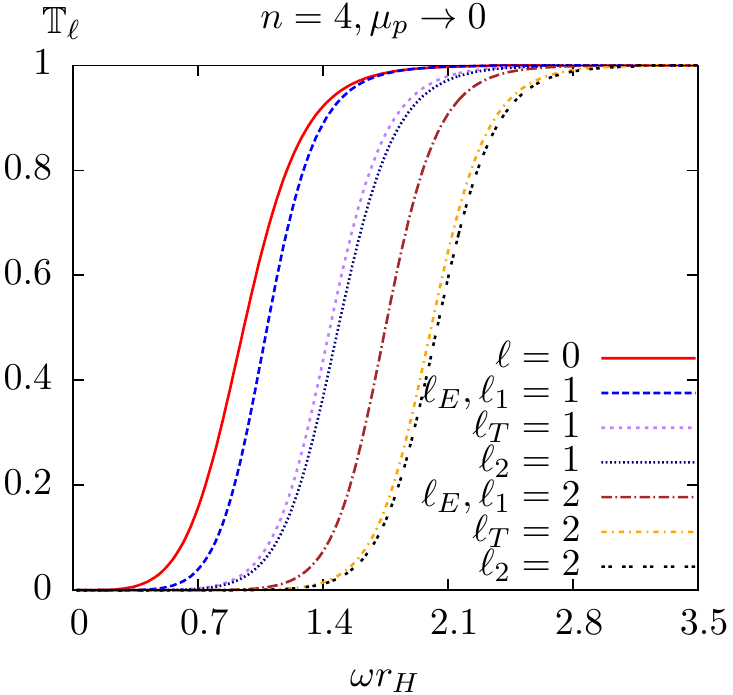} \vspace{3mm}\\
\includegraphics[scale=0.68,clip=true,trim= 0 0 0 0]{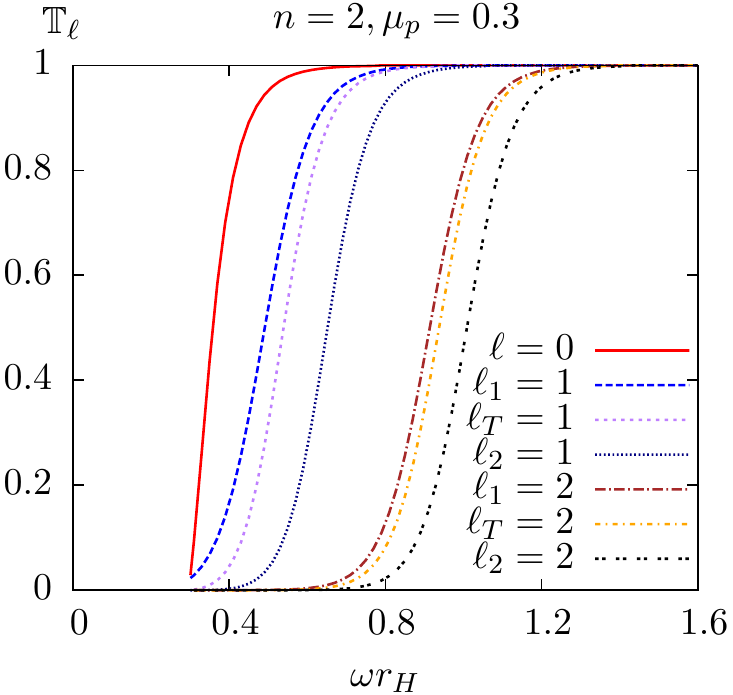} \hspace{-2.5mm} \includegraphics[scale=0.68,clip=true,trim= 0 0 0 0]{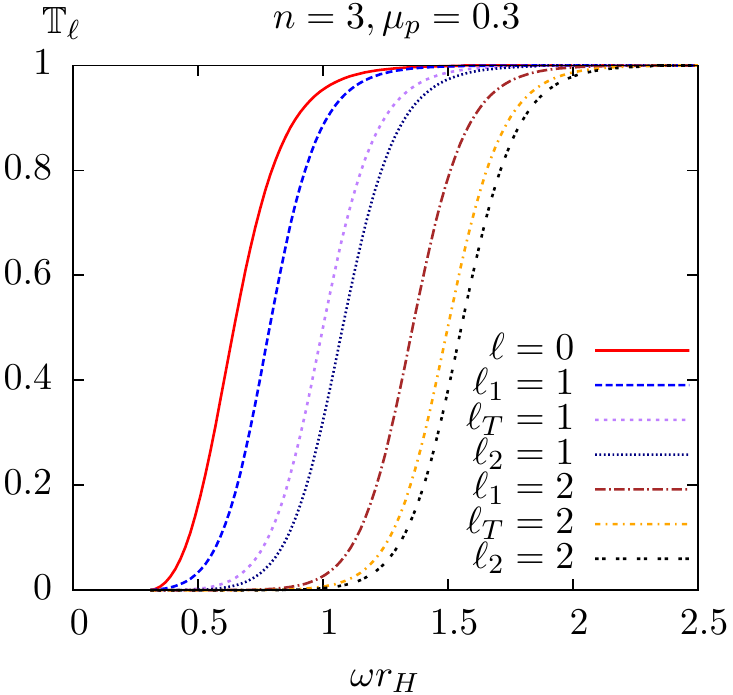} \hspace{-2.5mm} \includegraphics[scale=0.68,clip=true,trim= 0 0 0 0]{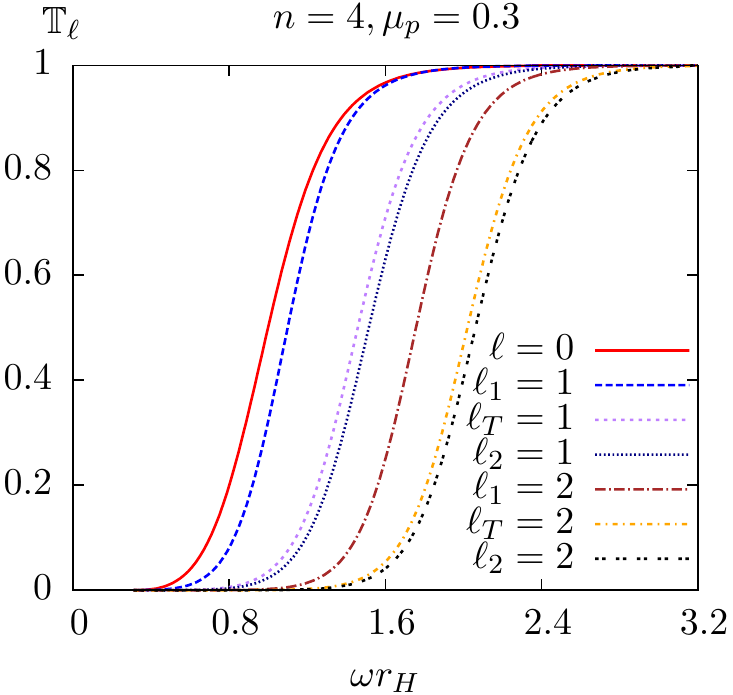} \vspace{3mm} \\
\includegraphics[scale=0.68,clip=true,trim= 0 0 0 0]{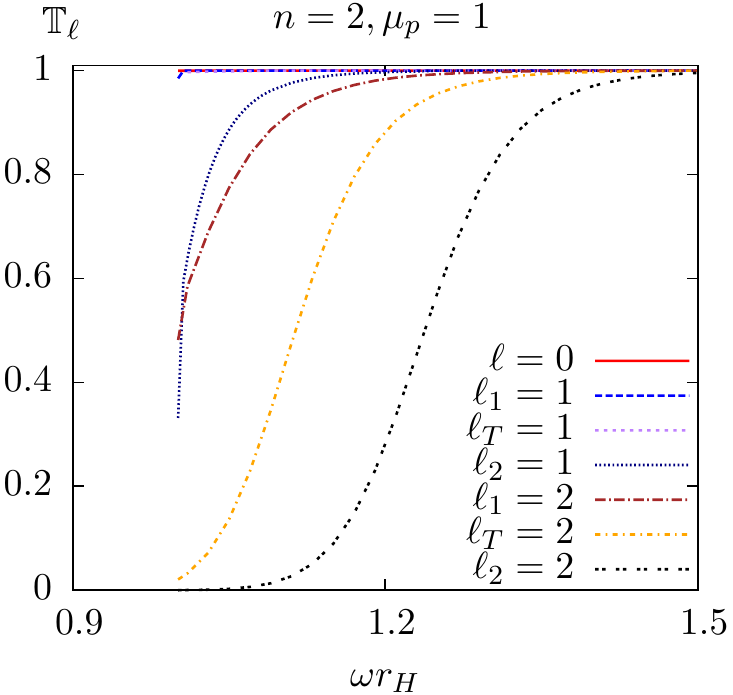} \hspace{-2.5mm} \includegraphics[scale=0.68,clip=true,trim= 0 0 0 0]{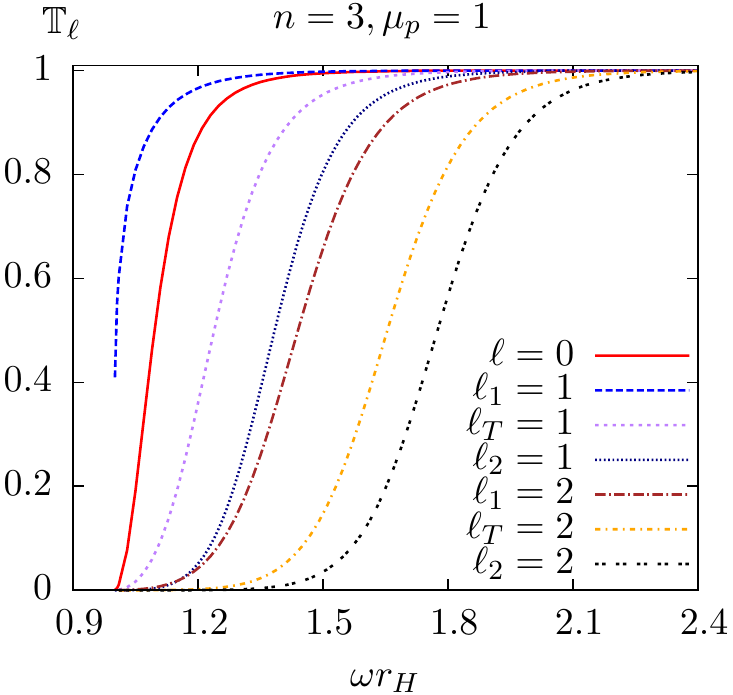} \hspace{-2.5mm} \includegraphics[scale=0.68,clip=true,trim= 0 0 0 0]{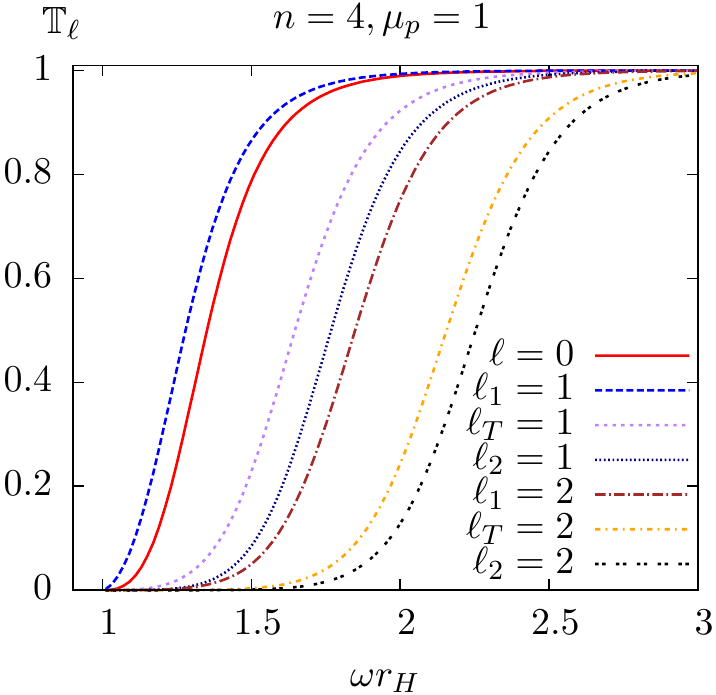}
\caption{\label{fig:Tfacs} {\em Transmission factors:} The three rows of panels, show the first few partial waves contributing to the Hawking spectrum. Each row corresponds to a fixed mass and each column to a fixed dimension. In particular, the first row shows the small mass limit of the Proca theory in order to compare it with Maxwell's theory.
}
\end{figure}
\clearpage
}

In the top row panels of Fig.~\ref{fig:Tfacs}, we show the partial wave contributions for $n=2,3,4$ in the zero mass limit. Some general properties are as follows. The $\mathbb{T}_\ell$ curve becomes shifted towards higher frequencies both as $\ell$ is increased, for $n$ fixed, and as $n$ is increased, for $\ell$ fixed. The former can be understood from standard geometrical optics arguments. Moreover, for this choice, there is always a numerical coincidence between one of the partial waves ($\ell_1$) obtained from the two coupled fields and the electromagnetic partial wave $\ell_E$. The $\ell=0$ and $\ell_2$ modes are always absent in the Maxwell theory, so they can be associated with the longitudinal polarization of the massive vector field. Similarly, the $\ell_T$ and $\ell_1$ partial waves are associated with the transverse polarizations of the field. A qualitative dependence on dimension is that for $n=2$, $\ell_T$ and $\ell_1$ (or $\ell_E$) modes are all equal. Curiously, this is in agreement with the fact that they describe the same number of transverse degrees of freedom as can be seen from the degeneracies~\eqref{eq:degen} specialized for $n=2$. This degeneracy is lifted for $n>2$.

For non-zero mass (middle and bottom row panels of Fig.~\ref{fig:Tfacs}), the degeneracy observed for $n=2$ in the massless limit is lifted. Also, we observe, for all $n$, that modes with higher $\ell$ partial waves  (especially $\ell_1$ modes) become a more dominant contribution at lower energies, as compared to lower $\ell$ partial waves of other modes. In particular for $\mu_p=1$, the transmission factor for $\ell_1$ becomes the largest for small energy. This effect of excitation of sub-dominant partial waves is well known to exist for example as we increase $n$ (and we can also observe such effect in our plots) as well as with the introduction of BH rotation \cite{sampaio2010production}. If this effect persists cumulatively on a rotating background, then we may have enhanced angular correlations for massive Proca fields emitted from the BH, since higher $\ell$ partial waves are less uniform.

Another outstanding point is that for large mass, when $n=2, 3$, it can be seen that the transmission factor starts from a constant non-zero value at the threshold $\omega=\mu_p$ ($k=0$), at least for small $\ell$. We have checked that this does not happen for $n\geq 4$ for masses as large as $\mu_p=10\sim 15$, where the curves always asymptote smoothly to zero at $k=0$. Note that the parameter in the radial equations is $\mu_p^2$ so these are very large masses. A possible explanation for this phenomenon can be motivated from considerations about the range of the gravitational field in Rutherford scattering. In $n=2$, the total cross-section for Rutherford scattering diverges, so the Newtonian gravitational potential is long ranged. This means that the effective size of the gravitational potential is infinite. The same happens in $n=3$ but only at zero momentum $k=0$. This indicates that a possible reason is that an incident wave at infinity with a very small momentum will still be sufficiently attracted by the gravitational field so that a constant non-zero fraction is still absorbed by the potential. In particular we note that some of the radial equations are similar in form to those obeyed by massive scalar and massive fermion fields, so the same effect exists for such fields. To our knowledge, this feature has not been noted or discussed in the literature. The only exception is the paper by Nakamura and Sato~\cite{Nakamura:1976nc} in four dimensions, where it is claimed that the reflection factor for a scalar field always goes to $1$ at $\omega=\mu_s$ (and thus the transmission factor goes to zero). Their result seems, however, inconsistent with Figs.~1,~2 and~3 of the paper by Page~\cite{Page:1977um} (also in four dimensions), where the Hawking fluxes for massive fermions become constant at the $k=0$ threshold (in agreement with our result).

Once we obtain the transmission factors, the computation of the Hawking fluxes~\eqref{eq:HawkFlux} follows straightforwardly by summing up partial waves with the appropriate degeneracy factors~\eqref{eq:degen}. We have chosen to show the flux of number of particles. The flux of energy has similar features and is simply related by multiplying each point in the plots by $\omega$.

\begin{figure}[t]
\includegraphics[scale=0.65,clip=true,trim= 0 0 0 0]{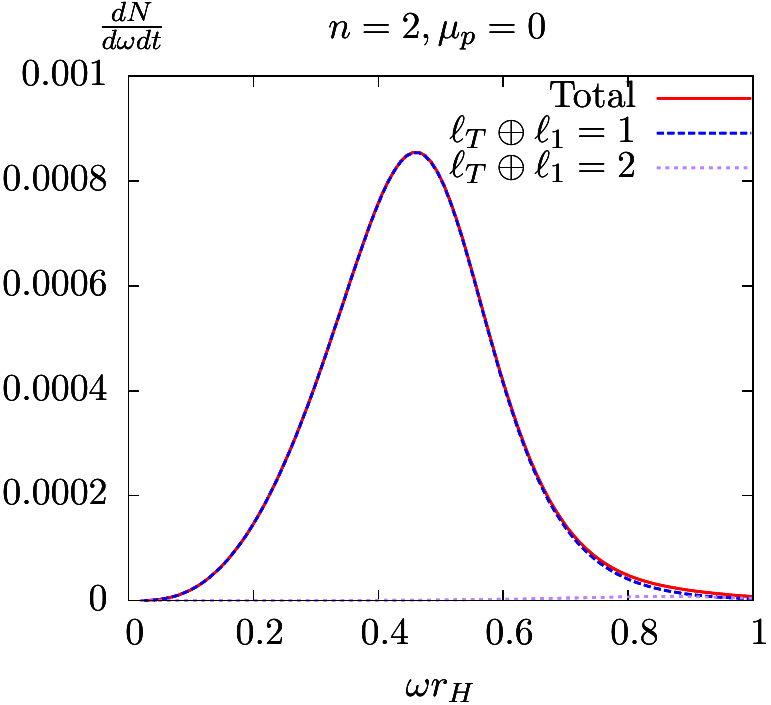}\hspace{0mm}  \includegraphics[scale=0.65,clip=true,trim= 0 0 0 0]{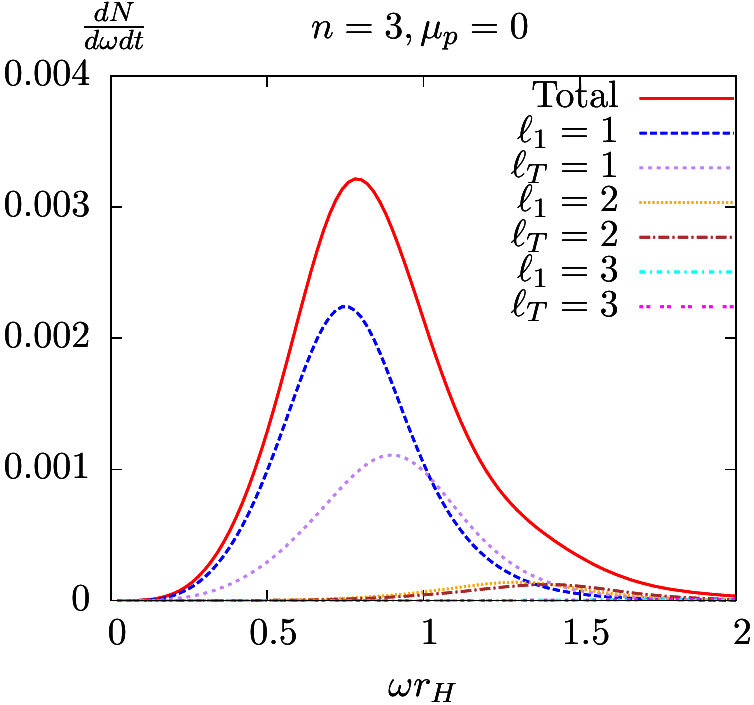} \hspace{-2.3mm} \includegraphics[scale=0.64,clip=true,trim= 0 0 0 0]{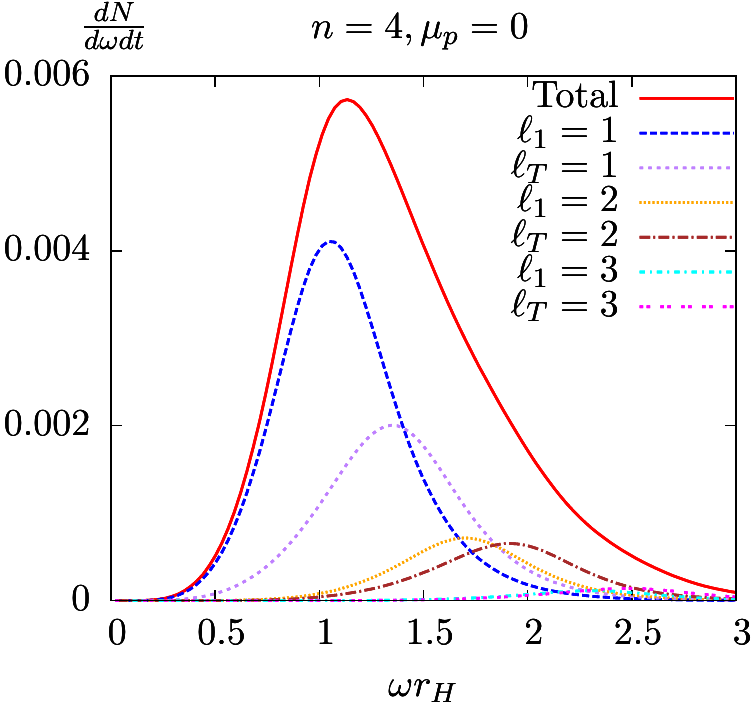} \vspace{0mm}\\
\includegraphics[scale=0.653,clip=true,trim= 0 0 0 0]{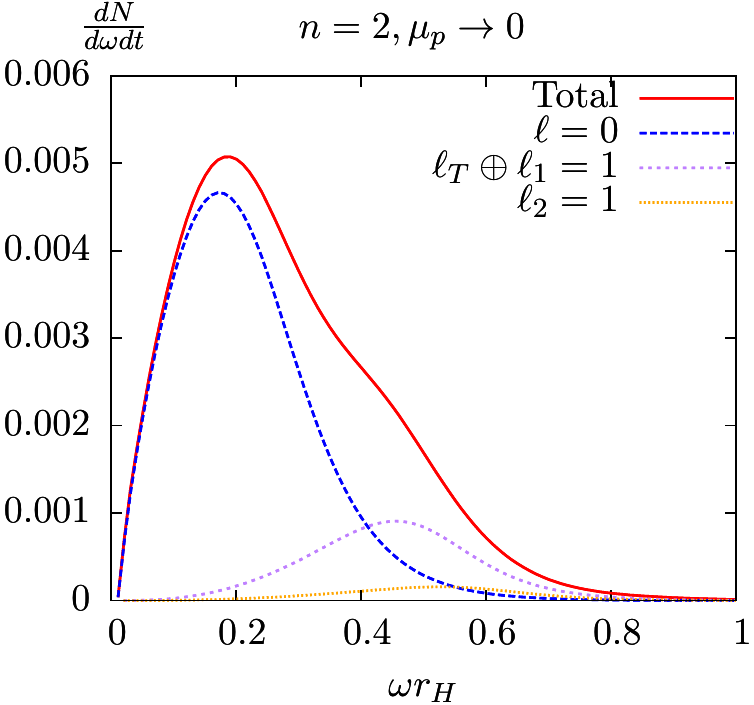} \hspace{-2.5mm} \includegraphics[scale=0.653,clip=true,trim= 0 0 0 0]{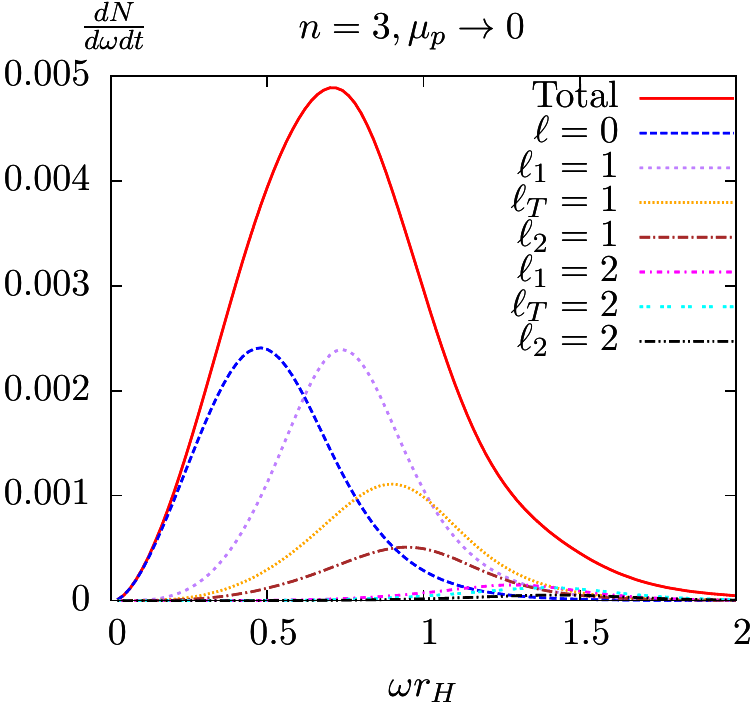} \hspace{-2.5mm} \includegraphics[scale=0.653,clip=true,trim= 0 0 0 0]{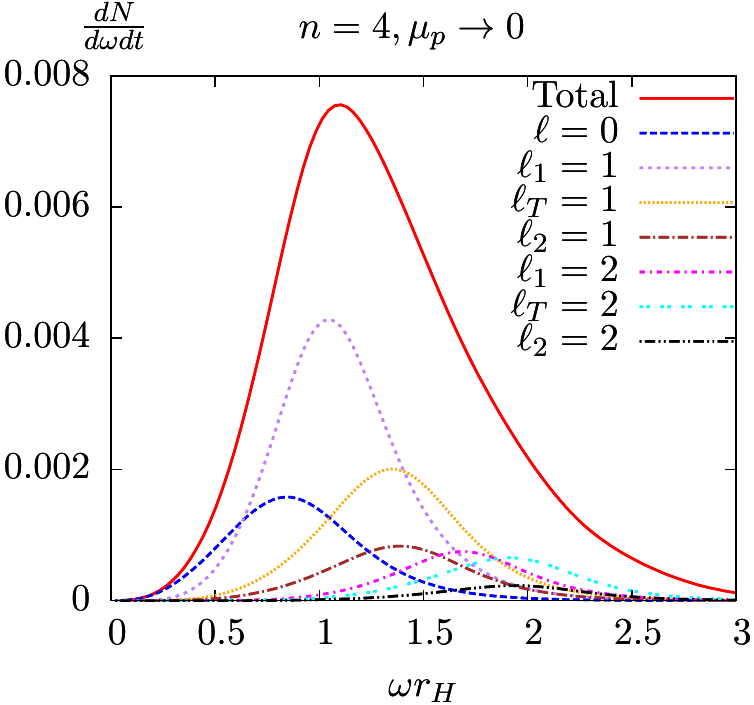} \vspace{0mm}
\caption{\label{fig:NfluxM0} {\em Number fluxes for $\mu_p=0$ (top panels) and $\mu_p\rightarrow 0$ (bottom panels):} The red solid curve of the top panels shows the Hawking flux of particles summed over the dominant partial waves for the Maxwell theory. The different partial waves are multiplied by the corresponding degeneracies. In the bottom panels the small (but non zero) $\mu_p$ limit of the Proca theory is shown for comparison. The $\oplus$ symbol denotes the addition of modes which are numerically equal.}
\end{figure}

In Fig.~\ref{fig:NfluxM0} we compare the Hawking fluxes of the Maxwell theory with the small mass limit of the Proca theory. For the particular case of $n=2$ we have reproduced the results by Page~\cite{Page:1976ki} for the electromagnetic field and found very good agreement.  All panels show a red solid curve corresponding to the total Hawking flux summed up over partial waves. The partial waves included in the sum are also represented, scaled up by the appropriate degeneracy factor. As claimed in the discussion of the transmission factors, as we increase $n$, partial waves with larger $\ell$ become more important for both Maxwell and Proca fields. One can clearly see that there is a large contribution to the total flux from the longitudinal degrees of freedom, since the vertical scales are larger for the Proca field. In particular the $\ell=0$ mode enhances the spectrum greatly at small energies. Note that these extra contributions associated with the longitudinal degrees of freedom cannot in general (for arbitrary mass) be described by a scalar field, since there is always a contribution from the coupled modes $\ell_1,\ell_2$. That is, however, the approximation done so far in BH event generators, where the $W$ and $Z$ fields Hawking spectra in use are those of the electromagnetic field (for transverse polarizations) and a scalar field (for the longitudinal polarization). Thus, our methods can be readily applied to improve this phenomenological modeling.

\begin{figure}[t]
\includegraphics[scale=0.652,clip=true,trim= 0 0 0 0]{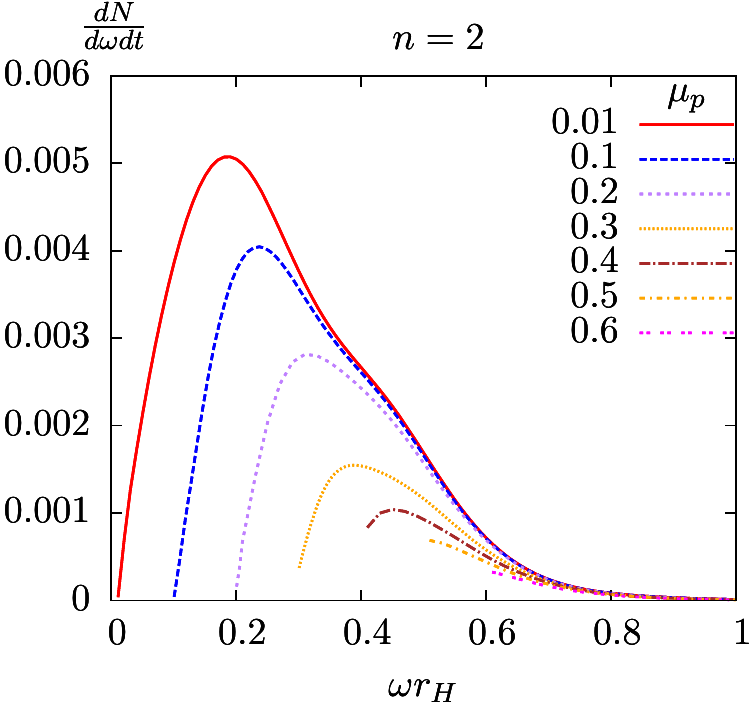} \hspace{-2.5mm} \includegraphics[scale=0.652,clip=true,trim= 0 0 0 0]{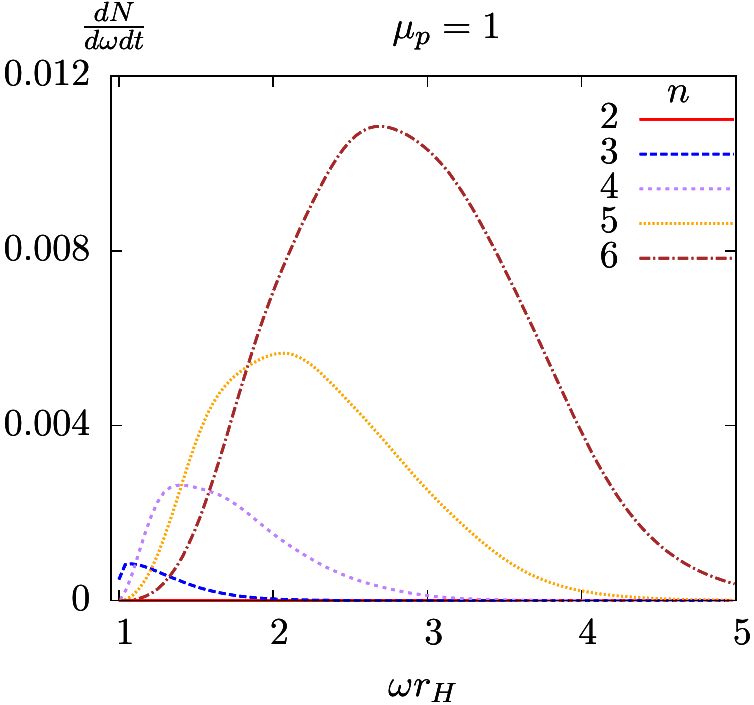} \hspace{-2.5mm} \includegraphics[scale=0.652,clip=true,trim= 0 0 0 0]{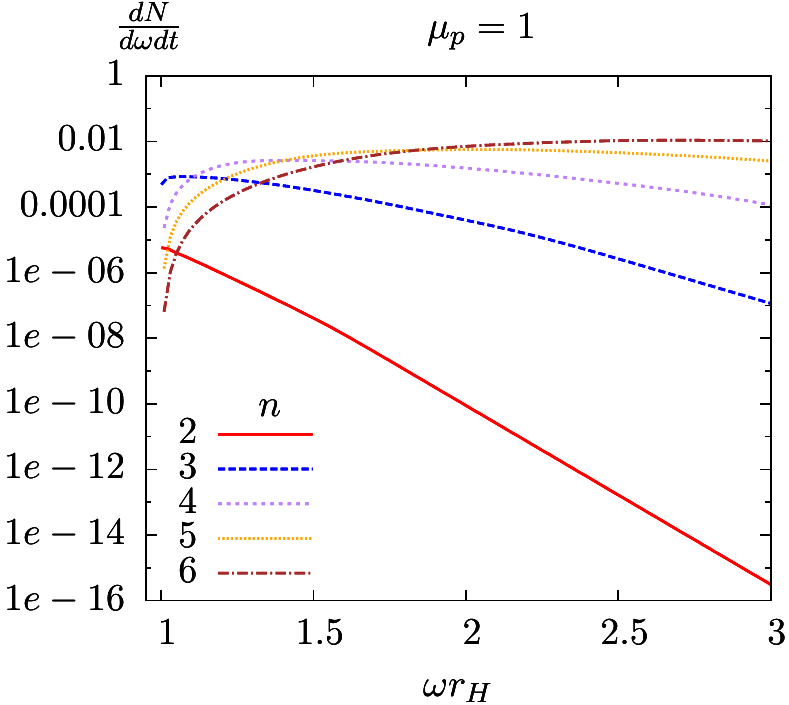} \vspace{0mm}
\caption{\label{fig:NfluxComparison} {\em Number fluxes for various $\mu_p$ and $n$:} (Left panel) Variation of the flux of particles for fixed $n=2$ and variable mass. (Middle and right panels) Variation of the flux with $n$ in a linear and logarithmic scale respectively. The logarithmic scale shows more clearly that the limiting flux at $k=0$ is finite for $n=2,3$.}
\end{figure}

In Fig.~\ref{fig:NfluxComparison} we show the variation of the total number flux with $n$ and $\mu_p$. The left panel shows the expected variation with $\mu_p$: that the flux not only gets cutoff at the energy threshold $\omega=\mu_p$, but it is also suppressed with $\mu_p$ (the same holds for $n>2$). This is the same behavior as found in~\cite{Sampaio:2009tp,Sampaio:2009ra}. As pointed out already, in event generators massive vector particles are modeled using the Hawking fluxes for the Maxwell field and a massless scalar, with a cutoff at the mass threshold. In \cite{Sampaio:2009ra,Sampaio:2009tp} it was shown that simply imposing a sharp cut-off on the fluxes of massless scalars and fermions over-shoots the real amount of Hawking radiation emitted in the massive scalar and fermion channel. Qualitative inspection of our results suggests a similar effect for the $W$ and $Z$ channels in the evaporation. A quantitative comparison, however, requires a consideration of a Proca field confined to a thin brane, which will be studied in Chapter~\ref{ch:ChargedP}. The middle and right panels show variation with $n$. In addition to the well known large scaling of the  area under the curve and the shift of the spectrum to larger energies, we can also see that more partial waves start contributing to the shape of the curve which becomes more wavy. This is particularly true because the degeneracy factors for fixed $\ell$ increase rapidly with $n$, which is a consequence of the larger number of polarizations available for a vector boson in higher dimensions. Finally, regarding $n=2,3$ we confirm the feature that the flux becomes a constant at $k=0$. This can be seen more clearly in the right panel in a logarithmic scale where the lines for $n\geq 4$ curve down very sharply around that point, whereas for $n=2,3$ they tend to a constant.
%

\section{Summary}
\label{sec:sumProneu}
%

In this chapter, we have studied Hawking radiation for a neutral Proca field, by solving the coupled wave equations as well as decoupled equations numerically, on a $D$-dimensional Schwarzschild BH. Our results exhibit distinctive features as we introduce the mass term, such as the lifting of the degeneracy of the two transverse modes in four dimensions, the appearance of longitudinal mode contributions (absent for Maxwell's theory) and in particular the $s$-wave. As we have shown, there is a large contribution from the longitudinal modes, to the Hawking fluxes. 

One feature that appears not to have been discussed in the literature is that in four and five spacetime dimensions, the transmission factor has a non-vanishing value in the limit of zero spatial momentum. We also find the expected suppression with mass of the Proca field, but perhaps the most relevant feature is to notice the increasing importance of the longitudinal modes and larger $\ell$ partial waves.

Our results could be applied to improve the model used in the \textsc{charybdis2} Monte Carlo event generator \cite{Frost:2009cf}. This simulates the production and decay of higher dimensional BHs in parton-parton collisions, a scenario which is being constrained at the second run of the LHC. It is therefore quite timely to improve the phenomenology of these models. Indeed our knowledge of Hawking evaporation process can still be improved greatly through the numerical study of various wave equations in BH backgrounds, which approximate the ones that could be produced at the LHC. This is illustrated by our results in this study, which alert for the importance of modelling the longitudinal modes correctly, instead of treating them as decoupled scalars as in current BH event generators.



%% file: ChargedP.tex
\chapter{Hawking radiation for a Proca field: charged case}
\label{ch:ChargedP}


\section{Introduction}
In this chapter, we are going to study Hawking radiation for a charged Proca field, by solving the charged Proca equations derived in Chapter~\ref{ch:GeneralInFlat} numerically. This is a generalization of the study presented in Chapter~\ref{ch:NeutralP}, by adding charge both to the field and the BH, on the 3+1 dimensional SM brane. This charged brane background is motivated by TeV gravity scenarios, in which the SM particles are confined on a 4-dimensional brane, while gravity propagates in extra dimensions, see Section~\ref{sc:BM} for more details. As shown in~\cite{Sampaio:2009ra}, the Schwinger emission alone does not suffice to discharge the BH, which makes the study of Hawking radiation for charged BHs essential. The effects of charge on the Hawking evaporation process, for scalar and fermion fields, have been performed in~\cite{Sampaio:2009ra,Sampaio:2009tp}. As a first motivation for this study, we are going to complete this picture by exploring the charged Proca fields.

Compared to the study for neutral Proca fields in Chapter~\ref{ch:NeutralP}, a new feature due to the charge of the background and of the field is the existence of superradiant modes. These modes are amplified through the extraction of Coulomb energy, as well as charge, from the charged BH. Furthermore, in a rotating background, the Proca field equation variables are not known to separate, which presents an extra difficulty added to the non decoupling of the modes, making it difficult to study exactly the superradiance phenomenon -- see~\cite{paniPRL,paniPRD} for a recent study in the slow rotation approximation. Thus, as a second motivation for this study, the charged BH background with spherical symmetry, yields a setup where superradiance of a massive spin 1 field can be explored without any approximation, albeit numerically. Such analysis will be performed herein.

The structure of this chapter is organized as follows. In Section~\ref{sec:nearhorizon} we introduce the background geometry and study the near horizon and asymptotic behaviors of the coupled charged Proca equations. In Section~\ref{fluxes} we discuss how to construct the scattering matrix from the electric current. The numerical results for the transmission factor and the associated Hawking fluxes are presented in Section \ref{chargedhresults} and we summarize our results in the last section. To keep the main part of this chapter compact and clear, some technical relations are left to Appendix~\ref{app:chargedP}.

\section{Boundary conditions and first order system}\label{sec:nearhorizon}
Before starting to deal with the Proca equations, we first present the following brane BH geometry
\begin{equation}
ds^2=-V(r)dt^2+\dfrac{1}{V(r)}dr^2+r^2(d\theta^2+\sin^2\theta d\varphi^2) \; ,\label{chargebrane}
\end{equation}
with metric function
\begin{equation}
V(r)=1-\dfrac{M}{r^{n-1}}+\dfrac{Q^2}{r^2}\;,\label{metricchbrane}
\end{equation}
where $M$ and $Q$ are the parameters related with BH bulk mass and brane charge. For numerical convenience, we choose units such that the outer horizon radius is $r_H=1$, i.e. $M=1+Q^2$ at the outer horizon.

It is easy to map the line element in Eq.~\eqref{chargebrane}, to the general background geometry with the Einstein space given in Eq.~\eqref{KI:metric}. Then the charged Proca field equations we are interested herein, can be obtained by setting $n=2$ in Eqs.~\eqref{originalsys1},~\eqref{originalsys2},~\eqref{Max} and~\eqref{k0M}\footnote{Note that the explicit $n$ appearing in these wave equations only depends on the dimension of the Einstein space. Therefore, we set $n=2$ for the background in Eq.~\eqref{chargebrane}. But we keep $n$ in general in the metric function~\eqref{metricchbrane} since gravity may propagate in the extra dimensions.}.

The procedure to rewrite the second order wave equation for the coupled system into a first order form, to impose an ingoing boundary condition at the horizon, and to extract the asymptotic expansion coefficients, is similar to what was done in Section~\ref{sec:NPnearhorizon}. For clarity, we summarize the main steps in the reminder of this section.

To determine the transmission factors, we need to integrate the radial equations from the horizon to the far away region with ingoing boundary conditions. The standard procedure is to find a series expansion of the solution near the horizon, which can be used to initialize the solution (we do so at $r=1.001$). Focusing on the coupled system $\left\{\psi,\chi\right\}$, if we define $y=r-1$, Eqs.~\eqref{originalsys1} and
~\eqref{originalsys2} become
\begin{eqnarray}
\left[A(r)\dfrac{d^2}{dy^2}+B(r)\dfrac{d}{dy}+C(r)\right]\psi+ E(r)\chi&=&0\label{systerm1}\ ,\\
\left[\tilde A(r)\dfrac{d^2}{dy^2}+\tilde
B(r)\dfrac{d}{dy}+\tilde C(r)\right]\chi+\tilde E(r)\psi&=&0\label{systerm2}\ ,
\end{eqnarray}
where the polynomials $A,B,C,E,\tilde{A},\tilde{B},\tilde{C},\tilde{E}$ are defined in Appendix~\ref{app:chargedP}. Making use of Frobenius' method to expand $\psi$ and $\chi$, we insert the following expansions into Eqs.~\eqref{systerm1} and~\eqref{systerm2}
\begin{equation}
\psi=y^\rho\sum^{\infty}_{j=0}{\mu_jy^j}\label{defpsi} \; , \;\;\; \chi=y^\rho\sum^{\infty}_{j=0}{\nu_jy^j} \; , \;\;\; \rho=\dfrac{-i(\omega-qQ)}{(n-1)+(n-3)Q^2}\;,
\end{equation}
where the sign of $\rho$ was chosen to impose an ingoing boundary condition. We then obtain the recurrence relations~\eqref{CPrecurone} for the coefficients $\mu_j$ and $\nu_j$ found in the Appendix. A general solution close to the horizon can be parameterized by two free coefficients $\nu_0$ and $\nu_1$.

Similarly, to understand the asymptotic behavior of the waves at infinity we now expand $\psi$ and $\chi$ as
\begin{equation}
\psi=e^{\beta r}r^{p}\sum_{j=0}\dfrac{a_j}{r^j}\label{psifar} \; , \qquad \chi=e^{\beta r}r^{p}\sum_{j=0}\dfrac{b_j}{r^j} \ ,
\end{equation}
which, after insertion into Eqs.~\eqref{originalsys1} and~\eqref{originalsys2}, yield
\begin{equation}
\beta = \pm ik\; , \;\;\;\;\;\;\;k=\sqrt{\omega^2-\mu_p^2}\;,\qquad p = \pm i\varphi \; ,
\end{equation}
where $\varphi=\delta_{n,2}(\omega^2+k^2)(1+Q^2)/(2k)-qQ\omega/k$. Thus one can show that  asymptotically
\begin{equation}
\psi \rightarrow \left(a_0^++\dfrac{a_1^+}{r}+\ldots\right)e^{i\Phi}+\left(a_0^-+\dfrac{a_1^-}{r}+\ldots\right)e^{-i\Phi} \; , \label{asymptoticpsi}
\end{equation}
\begin{eqnarray}
\chi \rightarrow
\left[\left(-\frac{k}{\omega}+\dfrac{c^+}{r}\right)a_0^++\ldots\right]e^{i\Phi}
+\left[\left(\frac{k}{\omega}+\dfrac{c^-}{r}\right)a_0^-+\ldots\right]e^{-i\Phi} \; ,\label{asymptoticpchi}
\end{eqnarray}
where $\Phi\equiv kr+\varphi \log r$ and $c^\pm$ is defined in the Appendix, Eq.~\eqref{CPcpm}.
Thus, as expected, each field is a combination of ingoing and outgoing waves at infinity.
Asymptotically, the solution is parameterized by four independent coefficients $\left\{a_0^\pm,a_1^\pm\right\}$, two for each independent mode in the coupled system. In the same way as in Section~\ref{sec:NPnearhorizon}, one can define a first order system of ODEs containing four radial functions $\left\{\chi^\pm,\psi^\pm\right\}$ which coincide with such coefficients at infinity, allowing for an easy extraction of the wave amplitudes. Our target system, which will be solved numerically in the remainder, is then
\begin{equation}\label{eq:ODEcoupled}
\dfrac{d\mathbf{\Psi}}{dr}=\mathbf{T}^{-1}\left(\mathbf{X}\mathbf{T}-\dfrac{d\mathbf{T}}{dr}\right) \mathbf{\Psi} \ ,
\end{equation}
with $\mathbf{\Psi}^T=(\psi_{+},\psi_{-},\chi_{+},\chi_{-})$. The definition of the matrices $\mathbf{X}$ and $\mathbf{T}$, and how they relate with~\eqref{systerm1} and~\eqref{systerm2} can be found in Appendix~\ref{app:chargedP}.

\section{Hawking fluxes}
\label{fluxes}
We shall now calculate the transmission factor for the coupled system as well as the Hawking fluxes generated from all the modes. In contrast to constructing a conserved flux from the energy-momentum tensor in Chapter~\ref{ch:NeutralP} which was simple enough to extract the transmission factors, in the present case, we use the conserved electric current which is naturally defined for this charged field. One can show that such a current is given by
\begin{equation}
\mathcal{J}^{\alpha}=W^{\dagger\alpha\mu}W_{\mu}+\dfrac{1}{\sqrt{-g}}\partial_{\beta}\left(\sqrt{-g}W^{\dagger\beta}W^{\alpha}\right)-c.c. \;\label{current}
\end{equation}
The radial flux at $r$ fixed is obtained by integrating the $\alpha=r$ component on the sphere. We note that only the first term in Eq.~\eqref{current} (denoted from now on $\mathcal{J}^{\alpha}_{\uppercase\expandafter{\romannumeral1}}$) contributes, since the second term becomes a total derivative on the sphere.

The contribution for the flux of the coupled fields at infinity, is then found by simplifying the radial component of Eq.~\eqref{current} using the equations of motion, and inserting the far away expansion at infinity
\begin{equation}\label{eq:currentInf}
\mathcal{J}^{r, \infty}_{\uppercase\expandafter{\romannumeral1}-couple} =|y_0^-|^2-|y_0^+|^2+|y_1^-|^2-|y_1^+|^2
\equiv(\mathbf{y}^-)^\dagger \mathbf{y}^--(\mathbf{y}^+)^\dagger \mathbf{y}^+ \; ,
\end{equation}
where $y_i^s(s=\pm; i=0,1)$ are linear combinations of the asymptotic coefficients $a_i^s$ given in the Appendix, Eq.~\eqref{eq:CPyplus}.
Using the reflection matrix $(\mathbf{R})$ defined in Chapter~\ref{ch:NeutralP}, we obtain
\begin{equation}\label{eq:TF1}
\mathcal{J}^{r, \infty}_{\uppercase\expandafter{\romannumeral1}-couple}=(\mathbf{y}^-)^\dagger\left(\mathbf{1}-\mathbf{R}^\dagger\mathbf{R}\right) \mathbf{y}^-\equiv (\mathbf{y}^-)^\dagger\mathbf{T}\, \mathbf{y}^-  \; ,
\end{equation}
where we have defined a (hermitian) transmission matrix $\mathbf{T}$, which can be diagonalised to find the decoupled asymptotic fields.

Following the same procedure, one can also calculate the electric current at the horizon
\begin{equation}\label{eq:currentH}
\mathcal{J}^{r, H}_{\uppercase\expandafter{\romannumeral1}-couple}=\dfrac{1}{\omega-qQ}\left(\mathbf{h^-}\right)^\dagger\mathbf{h^-} \; ,
\end{equation}
where the $h^-_i$ coefficients are linear combinations of the two independent $\nu_i$ coefficients \mbox{($i=0,1$)}, given in the Appendix, Eq.~\eqref{eq:hminus}. It shows an important point in Eq.~\eqref{eq:currentH} that the current can be positive or negative. This is expected, because for a bosonic field, the electric coupling can trigger superradiance.

Furthermore, from the conservation law of the electric current, one can find an alternative expression for the transmission matrix. Using the scattering matrix $(\mathbf{S}^{--})$ defined in Chapter~\ref{ch:NeutralP}, we find
\begin{equation}\label{eq:TF2}
\mathbf{T}=\dfrac{1}{\omega-qQ}(\mathbf{S}^{--}\mathbf{S}^{\dagger--})^{-1} \; .
\end{equation}

Once we have obtained the transmission factors, the number and energy fluxes are given by
\begin{equation}\label{eq:HawkFlux}
\dfrac{d\left\{N,E\right\}}{dt d\omega}=\dfrac{1}{2\pi}\sum_{\ell,\zeta} \dfrac{(2\ell+1)\left\{1,\omega \right\}}{\exp((\omega-qQ)/T_H)-1} \mathbb{T}_{\ell,\zeta} \;,
\end{equation}
where $\zeta$ labels the mode and  $T_H$ is the Hawking temperature which, in our units, is
\begin{eqnarray}
T_H=\dfrac{(n-1)+(n-3)Q^2}{4\pi}\;.\label{hawkingtemperature}
\end{eqnarray}

\section{Numerical Results}
\label{chargedhresults}
We now present a selection of numerical results for the transmission factor and the corresponding Hawking fluxes. In order to integrate the decoupled and coupled radial equations, we wrote independent codes in \textsc{mathematica} and in \textsc{c++}, finding agreement between the two codes. Using them we have generated a set of figures that we now describe.

\begin{figure*}
\begin{center}
\begin{tabular}{ccc}
\includegraphics[clip=true,width=0.31\textwidth]{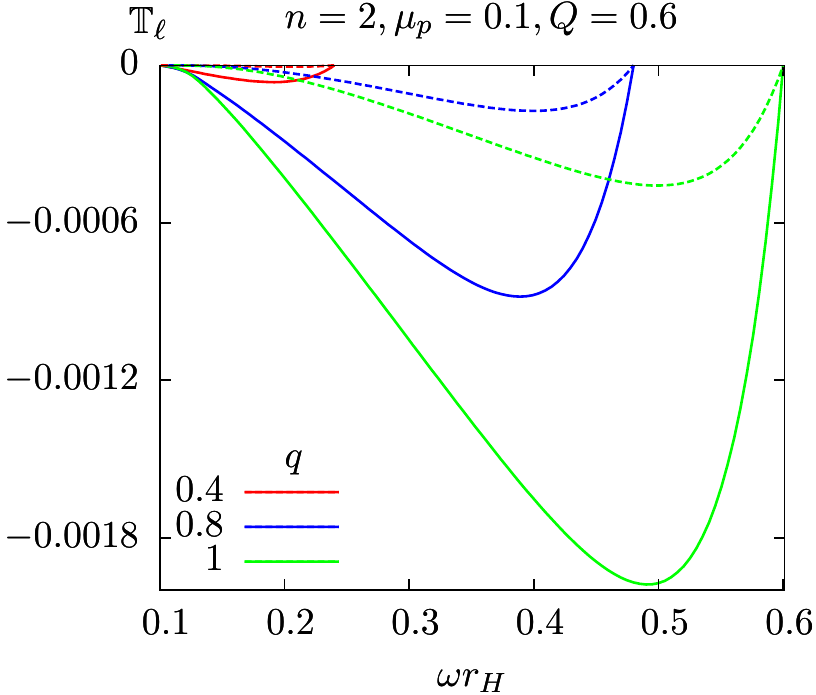}
\includegraphics[clip=true,width=0.292\textwidth]{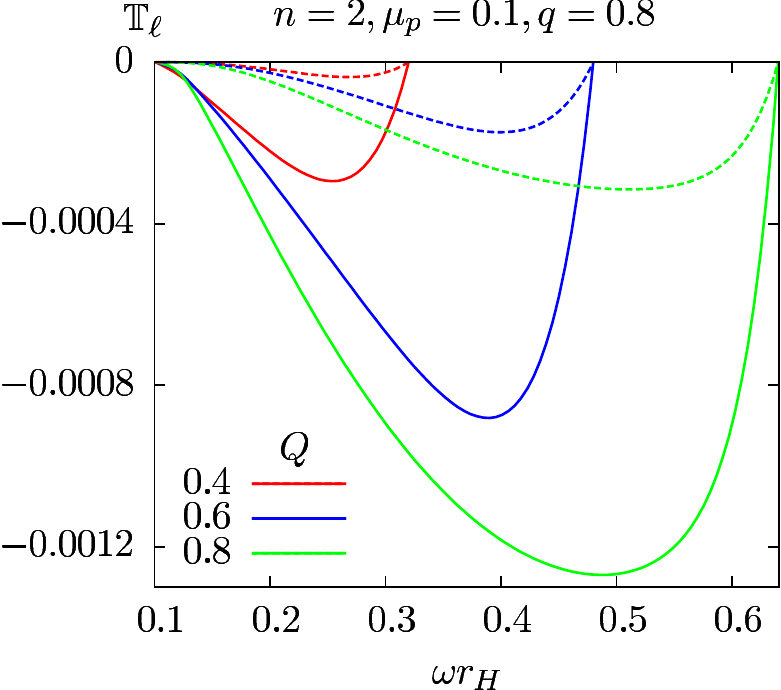} \hspace{2mm}
\includegraphics[clip=true,width=0.31\textwidth]{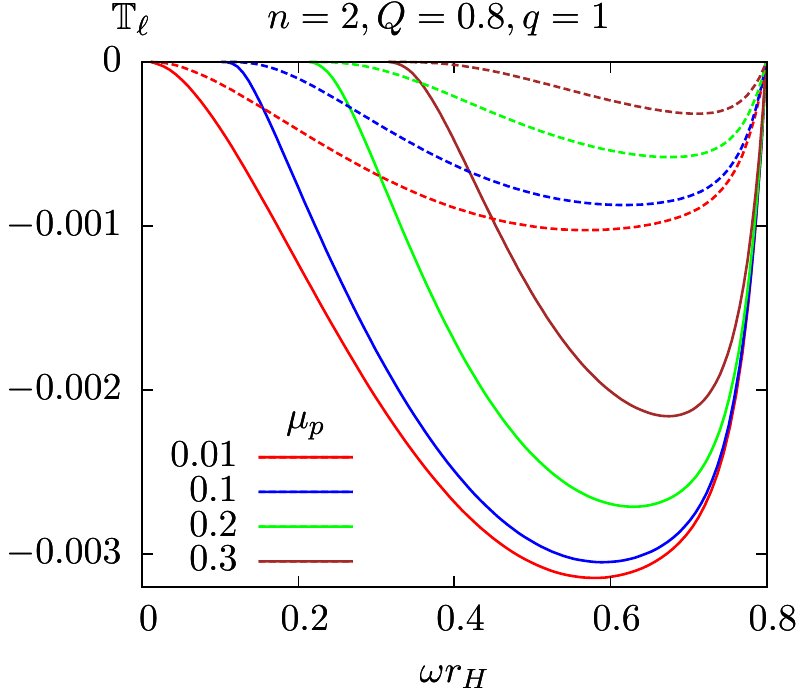}\\
\includegraphics[clip=true,width=0.31\textwidth]{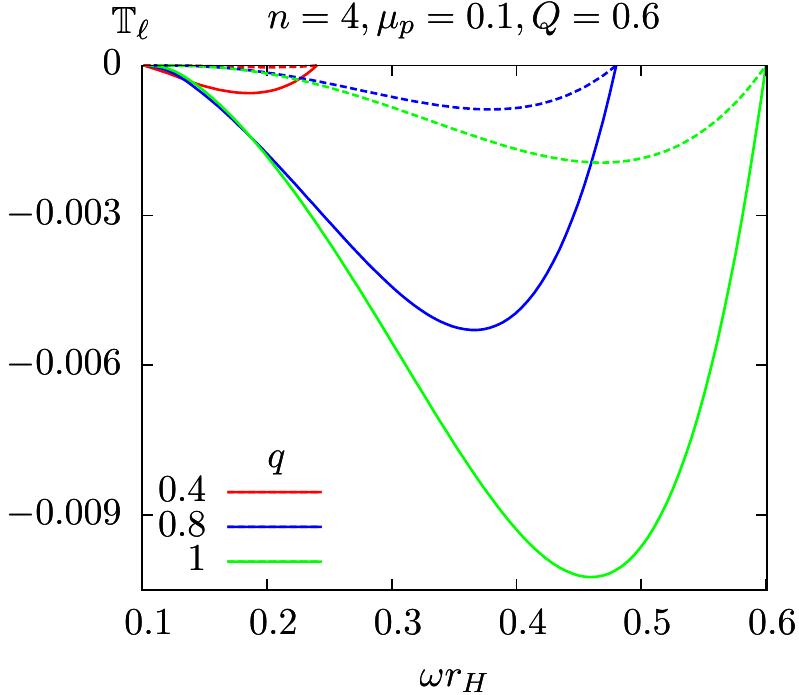}
\includegraphics[clip=true,width=0.292\textwidth]{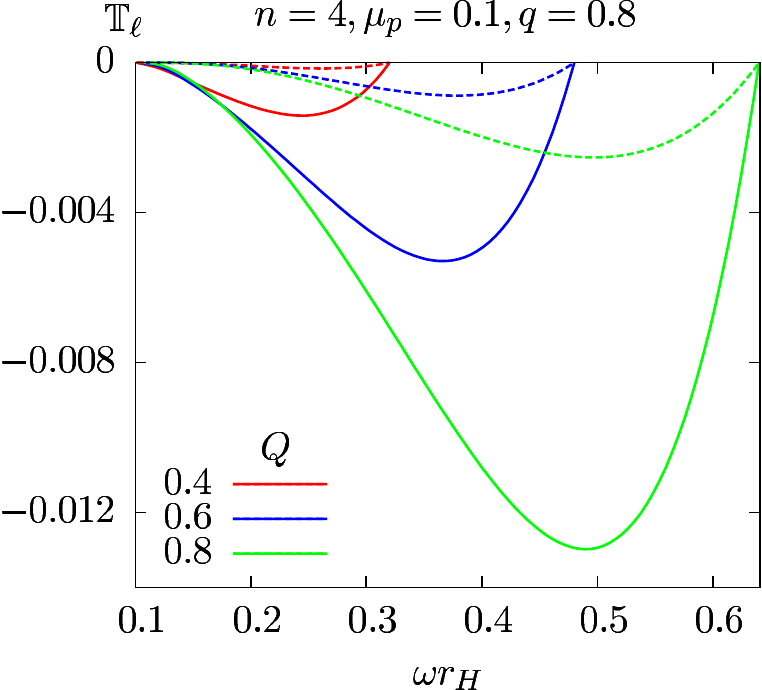}\hspace{2mm}
\includegraphics[clip=true,width=0.31\textwidth]{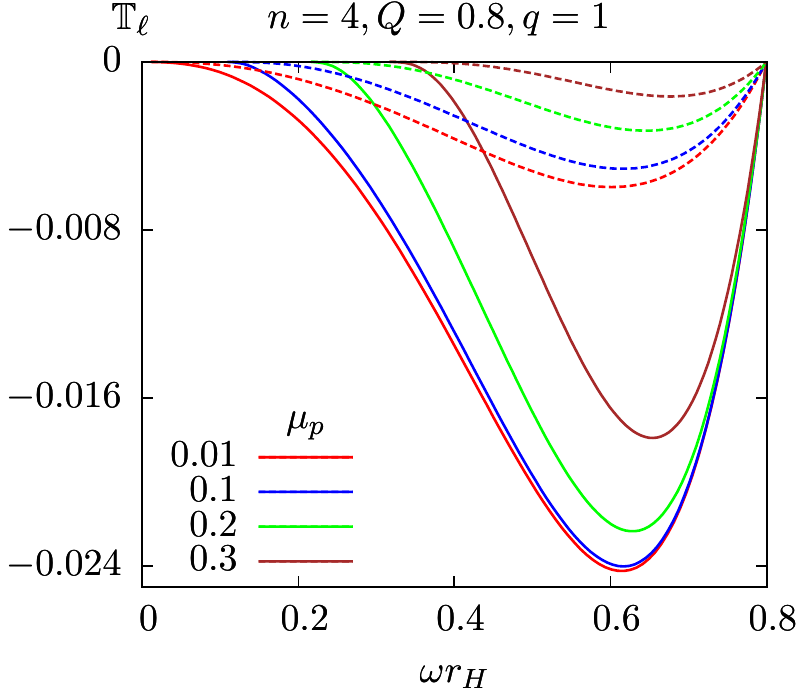}
\end{tabular}
\end{center}
\caption{\label{Superradiance01} Negative transmission factors that show the superradiant amplification of the coupled modes with $\ell= 1$ for different parameters. The two coupled modes are represented with the same color (solid and dashed lines). The top and bottom rows differ in the spacetime dimension. (Left panel)  Variation with the field charge $q$; (Middle panel) variation with the background charge $Q$; (Right panel) variation with the field mass.}
\end{figure*}

In Fig.~\ref{Superradiance01} transmission factors for different masses, spacetime dimensions and charges are shown to exhibit the superradiance phenomenon. As explained in Chapter~\ref{ch:intro}, superradiant amplification of a bosonic field in a charged and/or rotating BH occurs since there is Coulomb and/or rotational energy that can be extracted without decreasing the BH area. The general condition of superradiance is $\omega<m\Omega_H+q\Phi_H$; when this condition is verified, the transmission factor becomes negative and the scattered mode is amplified. In order to make the results clearer and more readable, here we just show the two coupled modes with $\ell=1$, as an example. Other cases are qualitatively similar. In the left panel of Fig.~\ref{Superradiance01}, we show the superradiance dependence on the field charge for $n=2$ (top row) and $n=4$ (bottom row). It is clear that superradiant amplification is enhanced with growing field charge except in the small energy regime; in this regime one observes, at least for the cases plotted, that the amplification decreases with increasing field charge (this effect is more noticeable for higher dimensions, see bottom row). The middle panel of Fig.~\ref{Superradiance01} shows the dependence on the background charge -- the trend is the same as when the field charge is varied, i.e. a generic superradiance enhancement with the charge except for small energies. In the right panel of Fig~\ref{Superradiance01}, we show the superradiance  dependence on the field mass; a suppression effect with increasing mass is observed, which is a generic behavior for the transmission factor (independently of the spin of the field or of being in the superradiant regime)~\cite{Sampaio:2009ra,Sampaio:2009tp}.
%
\begin{figure*}
\begin{center}
\begin{tabular}{cc}
\includegraphics[clip=true,width=0.338\textwidth]{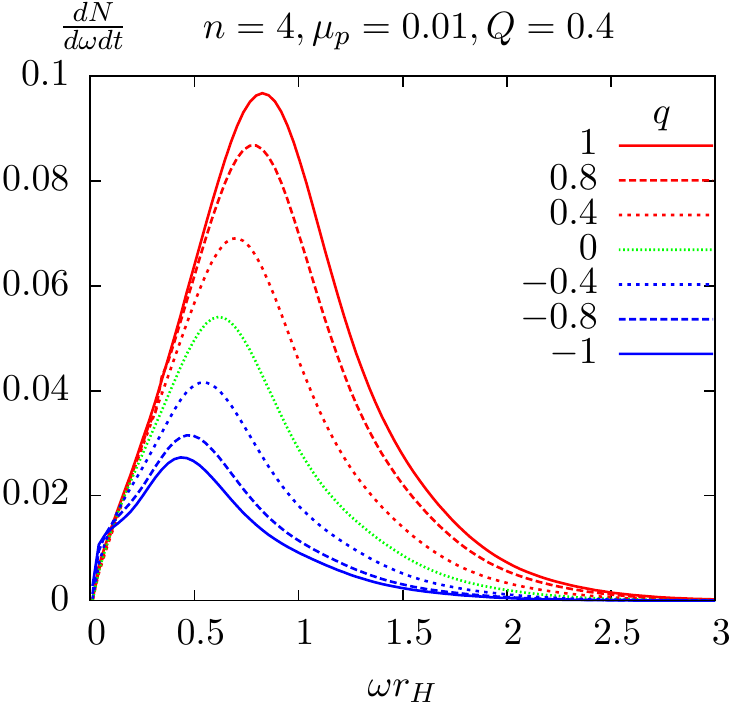}
\includegraphics[clip=true,width=0.45\textwidth]{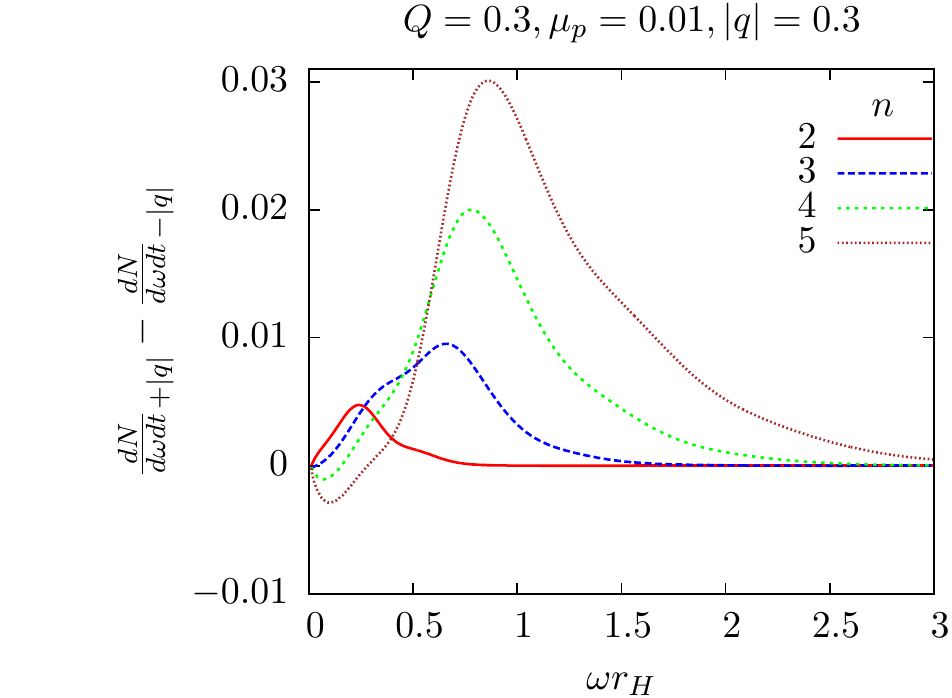}
\end{tabular}
\end{center}
\caption{\label{asymmetryeffect} (Left panel) Variation of the number fluxes with field charge for fixed $n$ and BH parameters in the small field mass limit. (Right panel) Variation of the difference between positive charge and negative charge flux with $n$.}
\end{figure*}

After we obtain the transmission factors, the Hawking fluxes (\ref{eq:HawkFlux}) can be calculated directly. In the remainder we have chosen to show the flux of the number of particles. The flux of energy has similar features and can be simply calculated by multiplying each point with $\omega$.

In Fig~\ref{asymmetryeffect} we illustrate the number fluxes dependence on the field charge. On the left plots, we have kept the background charge parameter fixed (and positive), with $Q=0.4$ and $n=4$, varying the field charge $q$ from $-1$ to $1$. The plots show that there is a region at low energy, where the negative field charge flux is larger, whereas in the remaining part of the spectrum, positive charge emission is favored. It is also clear that if we integrate over the curves the emission of positive charge is always favored. This low energy behavior where negative charges are favored, only occurs for $n>3$ as can be seen on the right panel, where the difference between the positive charge and negative charge flux spectrum is presented for various $n$. This inverted charge splitting effect was also observed for scalars and fermions~\cite{Sampaio:2009tp}, and it results from the interplay between the thermal factor and the transmission factor appearing in the expression for the number fluxes. Whereas the thermal factor always favors same charge emission, the transmission factor favors opposite charge emission and these factors dominate different parts of the spectrum. We have considered the same parameters as in~\cite{Sampaio:2009tp}, $Q=|q|=0.3$, wherein scalars and fermions have been studied, to allow for an easy comparison.

\begin{figure*}
\begin{center}
\begin{tabular}{ccc}
\hspace{-4.5mm}\includegraphics[clip=true,width=0.32\textwidth]{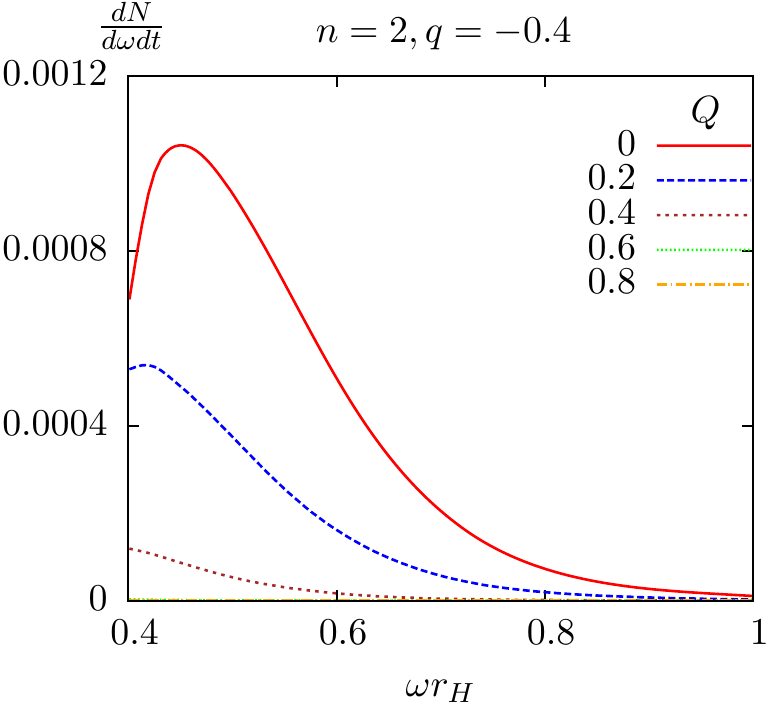} & \hspace{-4.5mm}
\includegraphics[clip=true,width=0.32\textwidth]{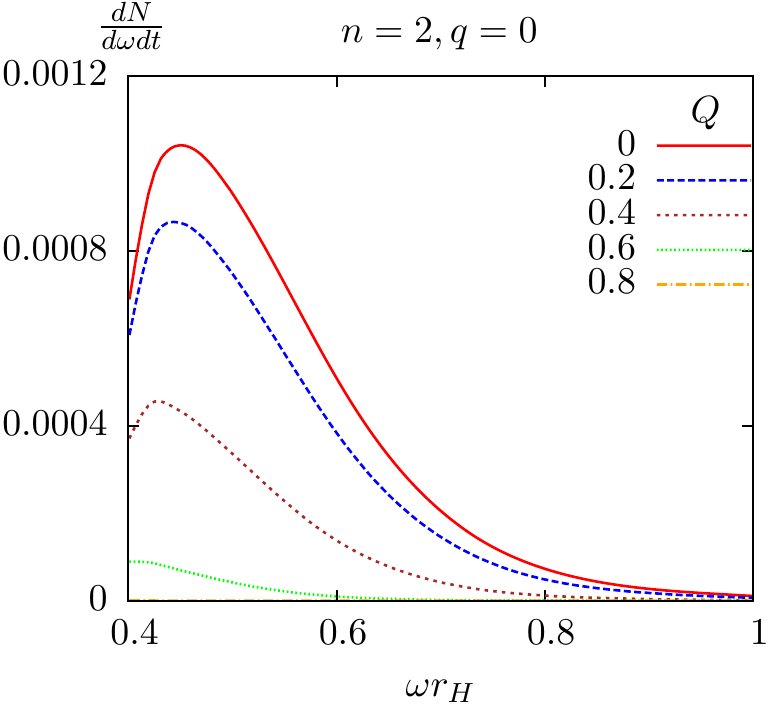} & \hspace{-5mm}
\includegraphics[clip=true,width=0.32\textwidth]{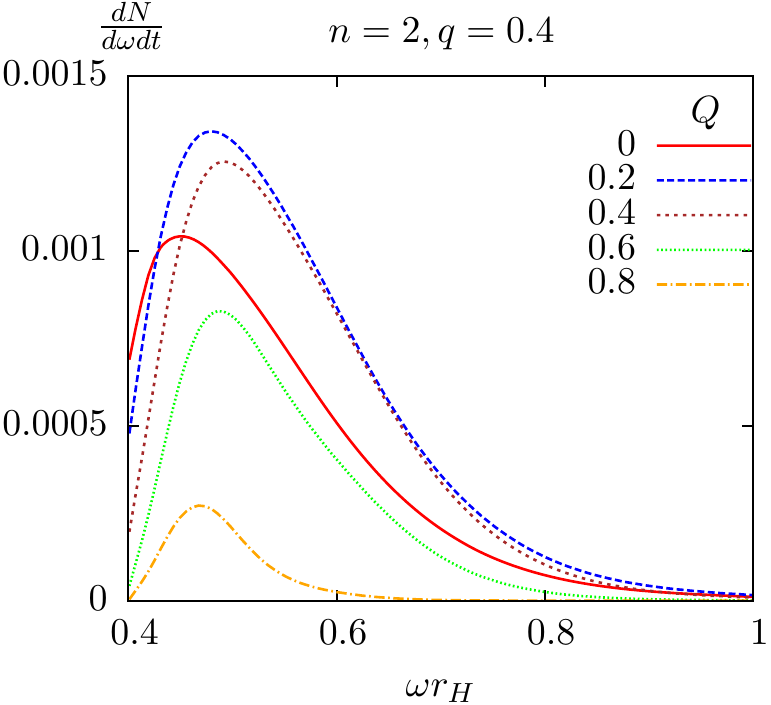}
\\
\hspace{-2mm}\includegraphics[clip=true,width=0.323\textwidth]{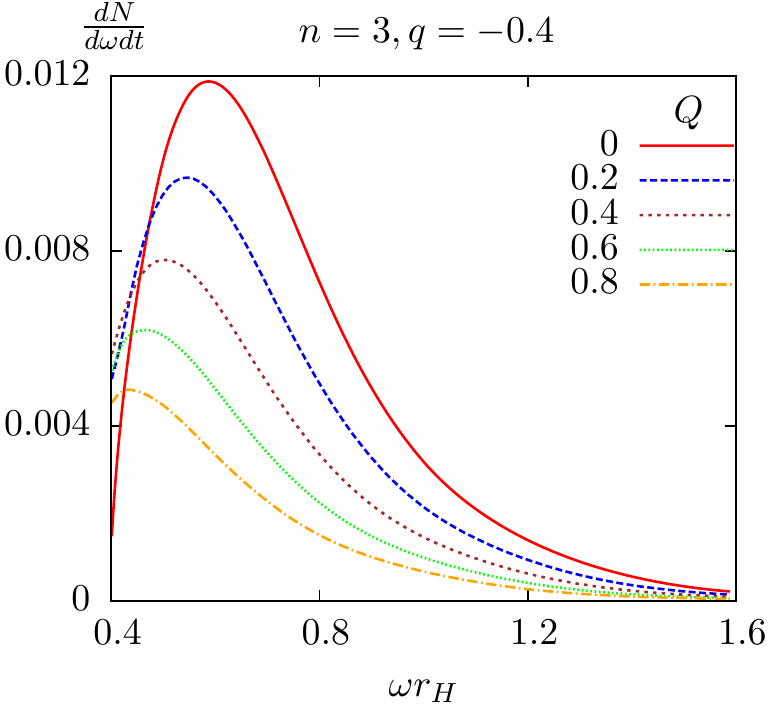} &
\hspace{-1mm}\includegraphics[clip=true,width=0.324\textwidth]{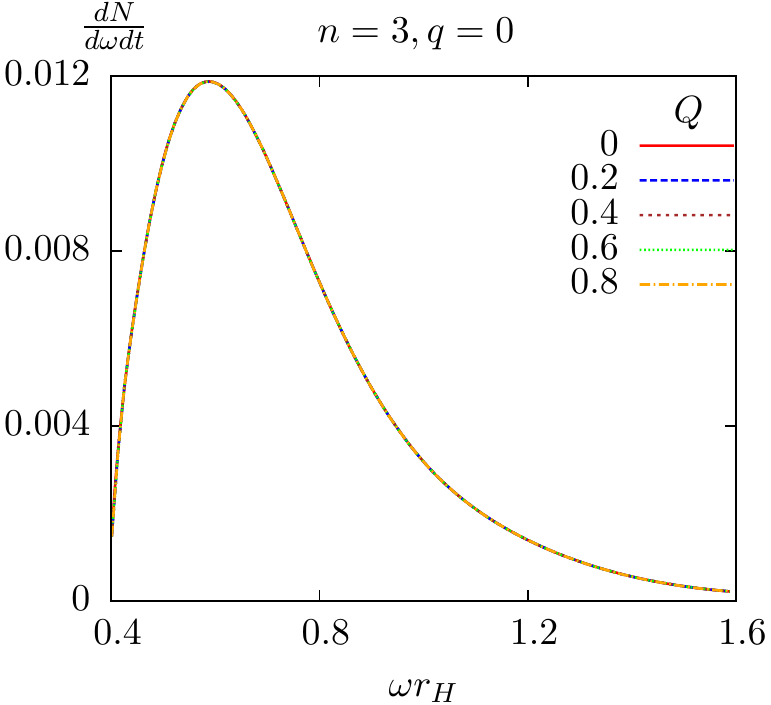} &
\hspace{-3mm}\includegraphics[clip=true,width=0.318\textwidth]{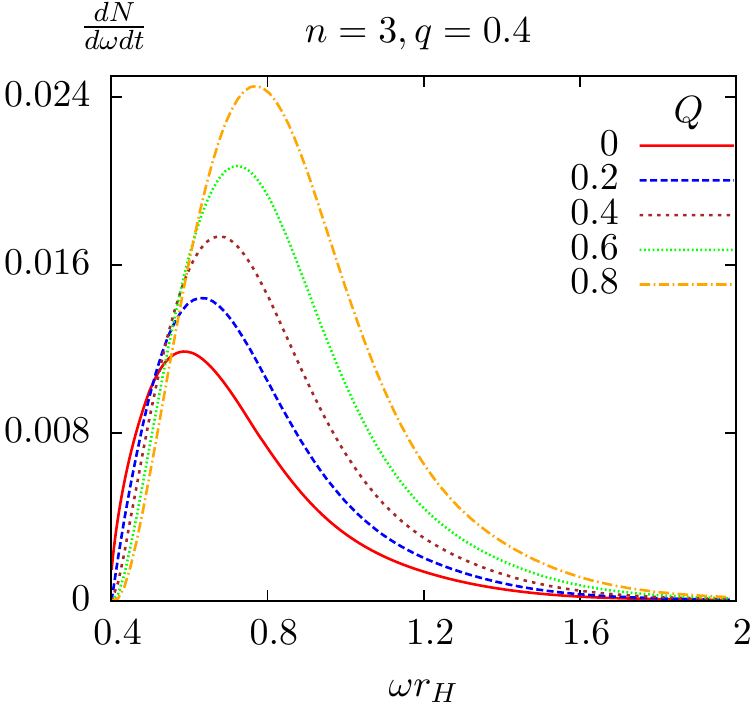}
\\
\includegraphics[clip=true,width=0.32\textwidth]{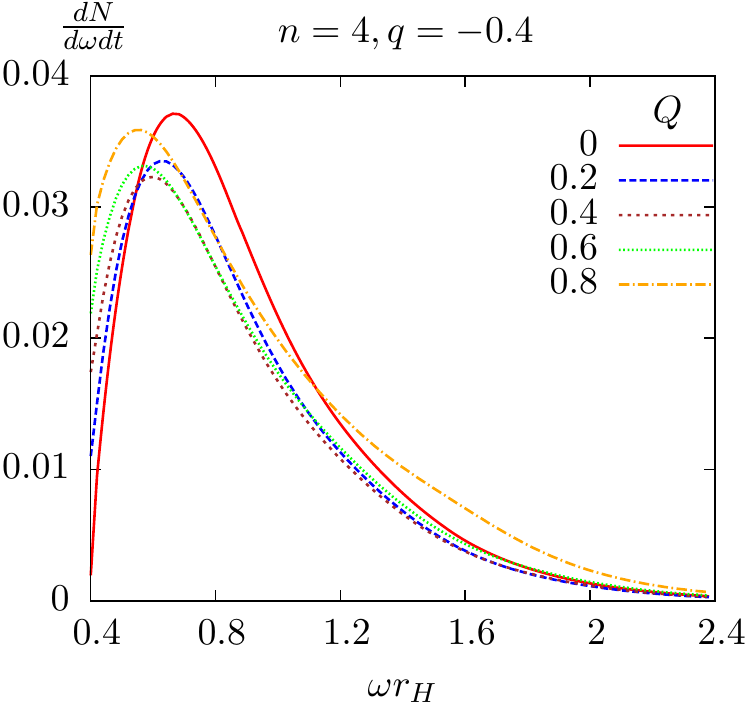} &
\includegraphics[clip=false,width=0.32\textwidth]{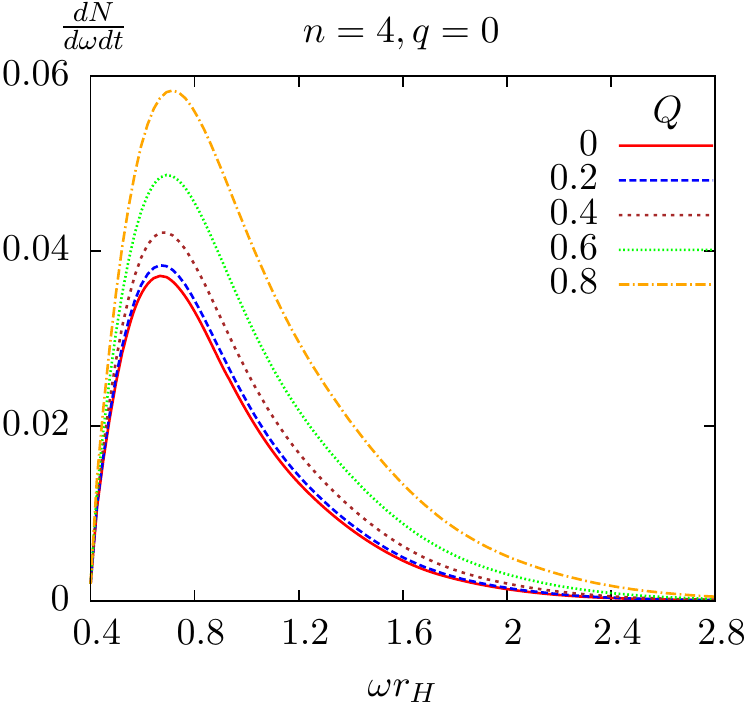} &
\includegraphics[clip=true,width=0.32\textwidth]{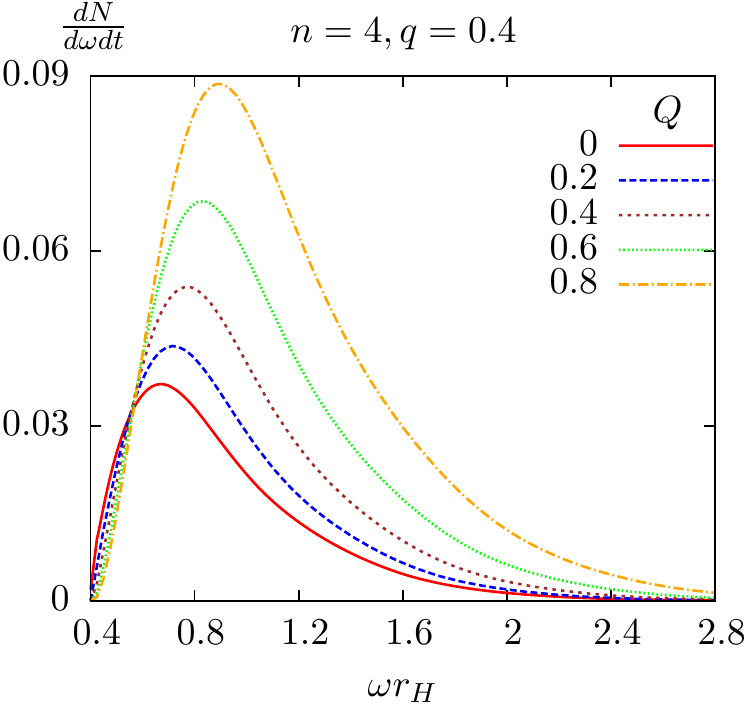}
\end{tabular}
\end{center}
\caption{\label{FluxVarQ} Number fluxes dependence on the background charge for different spacetime dimensions and field charge, with fixed field mass $\mu_p=0.4$. We vary the field charge in the same row and vary the spacetime dimension in the same column.
}
\end{figure*}

In Fig.~\ref{FluxVarQ}, we present the number flux dependence on the background charge for different field charge, in the same row, and different spacetime dimensions, in the same column. Consider first the $q=0$ case (middle column); it shows that the fluxes are suppressed/maintained/enhanced with the increase of background charge for $n=2$/$n=3$/$n=4$. To understand this behavior observe, from the definition of Hawking temperature, Eq.~\eqref{hawkingtemperature},  that the Hawking temperature is decreased/maintained/increased with increasing background charge for $n=2$/$n=3$/$n=4$, in these units. Since we are using horizon radius units, as we vary the charge parameter $Q$, we are actually varying the mass of the BH as well as the charge while keeping $r_H=1$. Nevertheless, it is easy to see that, up to a stretching of the horizontal axis, if we fix the BH mass and vary the dimensionful charge, these conclusions for the variation of the height of the curves do not change since the number flux is dimensionless~\footnote{The integrated flux however will not be the same for all background charges as expected, scaling as $r_H^{-1}$ for fixed BH mass and varying charge.}. For higher temperature, one expects a larger flux of particles, which is indeed the behavior shown in the second column of Fig.~\ref{FluxVarQ}. Turning on the field charge we observe a more involved behavior. For $n=3$, in which the Hawking temperature does not vary with $Q$, we see in the first/third column and for sufficiently large energies a monotonic suppression/enhancement of the Hawking flux when the BH has the opposite/same charge as the field. This is in agreement with the discussion of the left panel of Figure~\ref{asymmetryeffect}. For $n=2,4$, varying $Q$ one also varies the Hawking temperature and more complex patterns are observed.  Another trend is that the number fluxes increase as the spacetime dimension increases, which may be understood from the existence of more modes that contribute to the transmission factor.
%

\subsection{Mass effect and bulk/brane emission}
\begin{figure*}
\begin{center}
\begin{tabular}{ccc}
\includegraphics[clip=true,width=0.33\textwidth]{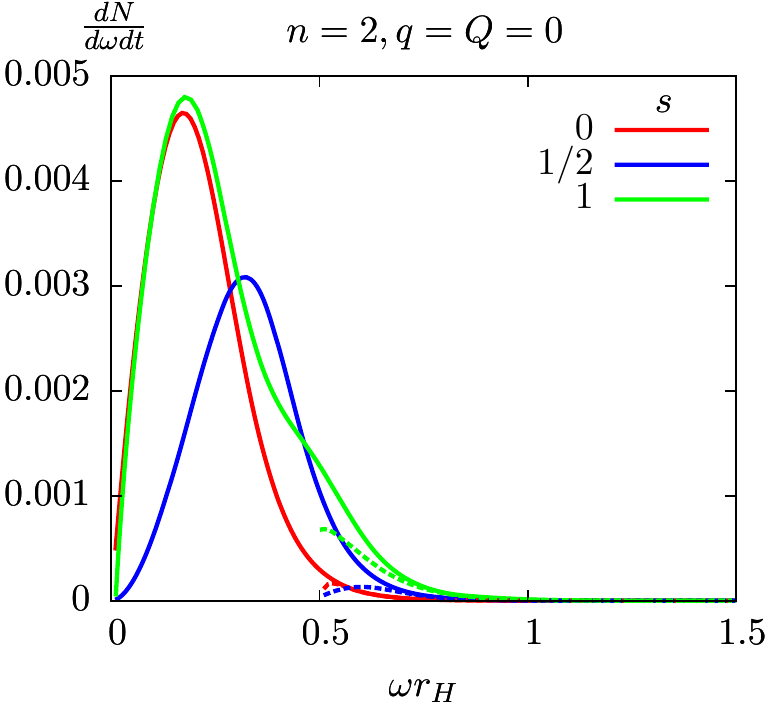}
\includegraphics[clip=true,width=0.331\textwidth]{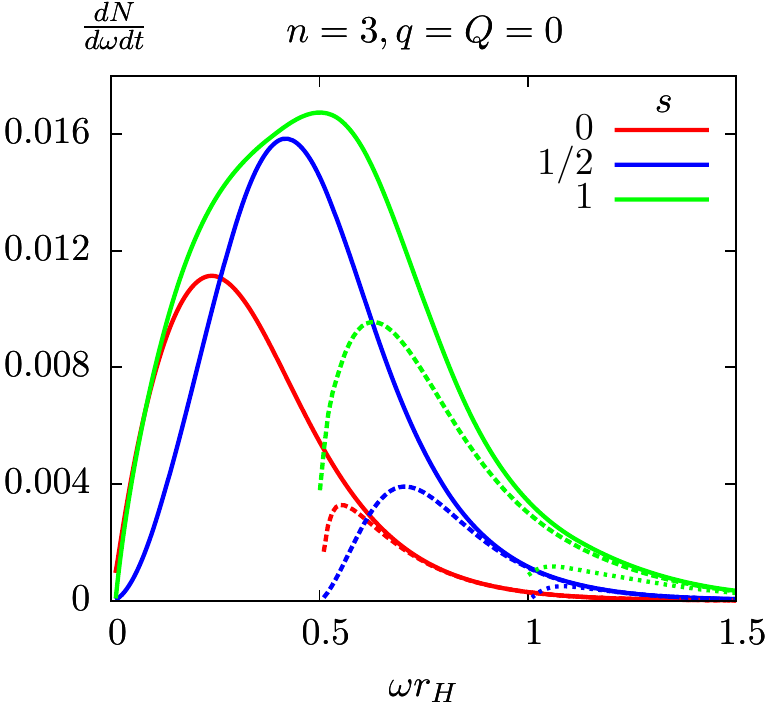}
\includegraphics[clip=true,width=0.324\textwidth]{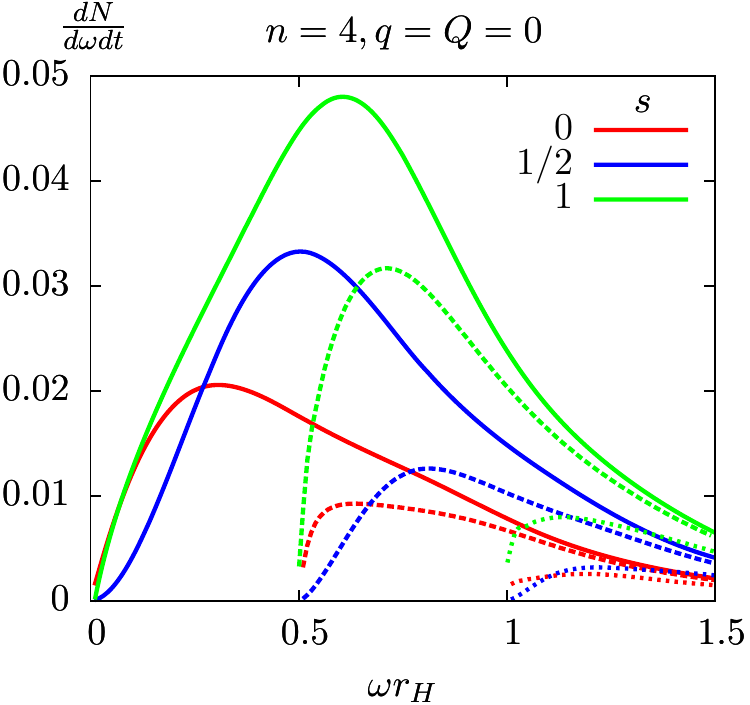}
\end{tabular}
\end{center}
\caption{\label{Mass_spin} Variation of the number fluxes for neutral particles with different spins (scalar, fermion and Proca) on the brane. For each type of particles we consider three different masses ($\mu=0$, $0.5$ and $1$), which can be identified by the starting point of the curve. Note we have included the two helicities for the fermion field and all the three modes for the Proca field (we have used a small mass $\mu_p=0.01$ for the latter, instead of $\mu_p=0$).}
\end{figure*}

In Figure~\ref{Mass_spin}, we perform a comparison of the effect of introducing a mass term for the various spins which are relevant for the Standard Model brane degrees of freedom. We have used the data of~\cite{Sampaio:2009tp} for scalars and fermions. Note that for fermions we have multiplied the data by a factor of two to take into account the two helicities of the Dirac field, since here we are also considering all the three modes for the Proca field. The dashed curves are for increasingly larger field mass (we have used the cases $\mu\footnote{In this subsection, the parameter $\mu$, if not specified, referes to the mass for different spin fields, include scalar, fermion and Proca.}=0$, $0.5$ and $1$, as can be seen from the threshold points where the curves start). For $D=n+2=4$ (left panel), we observe a striking similarity between the Proca ($s=1$) flux with the scalar ($s=0$) flux for $\mu=0$ (except at the high energy tail), which is due to the dominance of the $\ell=0$ mode. We can see this feature is always true at small energies for larger $n$ (middle and right panels);  as we increase $n$, however, higher modes of the Proca field enhance the flux as compared to scalars. The extra modes of the Proca field contributing at higher energy, also explain the fact that the mass suppression is not as large as for Dirac fermions or scalars, as we see from the dashed curves corresponding to $\mu=0.5$ for example. The high energy behavior contrasts with the low energy behavior, where scalar and Proca fields are dominated by the $s$-wave, whereas Dirac fermions are suppressed since they do not allow an $s$-wave. At high energy, we observe tails which are in the ratio $1:2:3$ following the number of degrees of freedom for the scalar, fermion and Proca fields respectively. This is in agreement with the fact that all transmission factors tend to one at high energy.

In the remainder, we combine our results for the Proca field on the brane with those in Chapter~\ref{ch:NeutralP} for a Proca field in the bulk, to analyze the relative bulk-to-brane emissivity in the neutral case.

\begin{table}[h] 
\begin{center}
\small
\begin{tabular}{|c|c|c|c|c|c|c|}
\hline
 & $n=2$ & $n=3$ & $n=4$ & $n=5$ & $n=6$ & $n=7$
\\
\hline
$\mu_p=0.1$ & 1 & 0.46 & 0.38 & 0.41 & 0.53 & 0.83
\\
\hline
$\mu_p=0.3$ & 1 & 0.49 & 0.40 & 0.42 & 0.55 & 0.86
\\
\hline
$\mu_p=0.6$ & 1 & 0.59 & 0.47 & 0.49 & 0.62 & 0.95
\\
\hline
\end{tabular}
\end{center}
\caption{Bulk-to-Brane relative energy emission rates for massive neutral vector fields for different mass $\mu_p$ in terms of the spacetime dimension $D=2+n$.}
\label{bulkbranecomparison}
\end{table}

The total energy rate (or number rate) emitted in Hawking radiation for a given field, is obtained by integrating the fluxes of Eq.~\eqref{eq:HawkFlux} (or their counterparts in the bulk) over $\omega$. The bulk-to-brane energy emissivity ratio, for massive neutral Proca fields of different mass is shown in Table~\ref{bulkbranecomparison} as a function of $n$. We have used the data in Chapter~\ref{ch:NeutralP} to obtain the total emissivity for bulk fields, as well as the data presented here, for the brane emissivity. The entries of Table~\ref{bulkbranecomparison}, show clearly that the emission of energy into brane-localized Proca particles is dominant, for $n$ larger than two, which is consistent with the argument that BHs radiate mainly on the brane~\cite{EHMprl2000}. For fixed mass, as $n$ increases, the bulk-to-brane energy emission ratio initially decreases, reaching a minimum value for an intermediate $n$. However, if one increases $n$ further, the bulk-to-brane energy emission ratio increases again. Furthermore, we have also observed that the bulk-to-brane energy emission ratio increases with the field mass. A similar behavior  was observed for scalar fields~\cite{Harris:2003eg,Kanti:2010mk}. Finally, we found that the bulk-to-brane energy emission ratio for the Proca field is larger than that for the scalar field for all $n$, being more noticeable for large $n$, say $n=6$ and $n=7$ (note the different notation for $D=2+n$ in this chapter as compared with~\cite{Harris:2003eg,Kanti:2010mk}).

\section{Summary}
\label{discussion}
In this chapter we have completed the analysis started in Chapter~\ref{ch:NeutralP}, by computing the transmission factor for a charged Proca field propagating in the background of a charged BH on a brane. Furthermore, this work completes the study of the effect of mass and charge for particles evaporating on the brane, for all spins relevant for the SM in brane world scenarios~\cite{Sampaio:2009ra,Sampaio:2009tp}.

One of the main, and novel features arising from considering a charged Proca field on a charged background is the existence of negative transmission factors, a signature of superradiant scattering, which we have presented and described. Our results are consistent with the condition of superradiance as allowed by the area theorem, and we found that generically increasing the field or background charge amplifies the effect (though some inversions are possible at small energies). As we explained in Chapter~\ref{ch:intro}, it is worth commenting that although superradiant scattering is observed for both rotating and charged BHs, rotating BHs exhibit superradiant instabilities against massive scalar fields, and in particular against the Proca field~\cite{paniPRL,paniPRD}, whereas charged (non-rotating) BHs do not exhibit analogous instabilities against charged, massive bosonic fields. This will be explored for the Proca field in Chapter~\ref{ch:ChargedClouds}, by computing the frequencies of quasi-bound states, a study that can also teach us how long lived Proca hair around charged BHs can be.

An effect observed herein, is that similarly to scalars and fermions, there is an inverted charge splitting effect at small energies for more than one extra spacetime dimension, where particles with charge opposite to the BH are emitted dominantly. Nevertheless, for larger energies, the normal splitting order, favoring the emission of same charge particles as to discharge the BH is restored and overall (by integrating out the flux), this channel tends to discharge the BH. This effect has been suggested to be another signature that could be found in the energy spectrum of charged fermions \cite{Sampaio:2009tp}. Since fermions in the final state are also produced indirectly from the decay of vector bosons such as $W^{\pm}$ or the $Z$ particles one may ask whether the effect survives. Here we have verified that the effect is present for $W^{\pm}$, so at least for the case when the final state decay products are a charged lepton $\ell^\pm$ and a neutrino, these contributions will certainly enhance the effect.

In the neutral case, we have performed two comparisons. First we have compared the effect of the mass and spin on the Hawking fluxes, for all spins in the SM. Our main findings are that the Proca field spectrum departs from being similar to a scalar field in four dimensions and becomes increasingly dominant for larger number of dimensions, peaking and extending towards larger energies. This also means that the mass suppression effect is smaller for the Proca field than it is for scalars and fermions. Second, we have compared the bulk-to-brane emission ratio for Proca fields, confirming brane dominance in general, and the suppression of brane dominance with increasing mass.

These results can be used to improve the modelling of BH evaporation in TeV gravity scenarios, in the BH event generators~\cite{Frost:2009cf,Dai:2007ki} that are in use at the ATLAS and CMS experiments to put bounds on extra dimensions in this channel~\cite{CMS:2012yf,ATLAS-CONF-2011-065,ATLAS-CONF-2011-068,Gingrich:2012vs,Aad:2015mzg}.

%% file: ChargedClouds.tex
\chapter{Marginal charged clouds around charged black holes}
\label{ch:ChargedClouds}

\section{Introduction}
Scattering processes for the Proca field were studied in Chapters~\ref{ch:NeutralP} and~\ref{ch:ChargedP}. An interesting generalization is to study this field in the context of quasinormal modes and quasi-bound states. In fact, such studies have been performed for a neutral Proca field on a Schwarzschild BH~\cite{Rosa:2011my}, and on a slowly rotating Kerr BH~\cite{Pani:2012vp}. In this chapter, we are going to study quasi-bound states for a charged Proca field (as well as a charged scalar field) on a Reissner-N\"ordstrom (RN) BH. This study is particularly interesting since we would like to know if the charge couplings between the background and the Proca field could balance the gravitational attraction, similarly to what occurs to scalar clouds.
%

Scalar clouds~\cite{Hod:2012px,Hod:2013zza,Herdeiro:2014goa} are equilibrium configurations of a complex, massive scalar field in the background of a Kerr BH. They are stationary bound states, with a real frequency. These configurations are possible due to the existence of two qualitatively different types of quasi-bound states -- i.e. bound field configurations with a \textit{complex} frequency:
\begin{itemize}
\item[i)] Time decaying quasi-bound states; this is the generic behavior expected for matter around a BH, due to the purely ingoing boundary condition at the horizon. Indeed, this is the only kind of quasi-bound state that can be found around Schwarzschild BHs.
\item[ii)] Time growing quasi-bound states; this occurs for Kerr BHs in the superradiant regime, 
    i.e. when the real part of the frequency, $\omega$, of the quasi-bound state obeys $\omega<m\Omega_H$, where $m$ is the azimuthal harmonic index of the field mode and $\Omega_H$ is the horizon angular velocity of the Kerr BH. This yields an instability of Kerr BHs in the presence of any field for which quasi-bound states can be found in the superradiant regime. 
\end{itemize}
Scalar clouds exist at the boundary between these two regimes, i.e. when $\omega=m\Omega_H$, which is compatible with the bound state condition $\omega<\mu$ for scalar fields with mass $\mu$ on the Kerr BH. Renewed interest concerning these clouds has been triggered by the recent observation that there are Kerr BHs with scalar hair~\cite{Herdeiro:2014goa,Herdeiro:2014ima,Herdeiro:2014jaa}, found as exact solutions of the Einstein-Klein-Gordon system, and which correspond to the nonlinear realization of these clouds.

It has long been known that charged, i.e. RN BHs can amplify charged scalar fields through superradiant scattering~\cite{Bekenstein:1973mi}. This is a process that has qualitative similarities with the superradiant scattering of (neutral) scalar fields by Kerr BHs. For the charged case, as explained in Section~\ref{sc:SR}, there is no analogue of a charged superradiant instability for asymptotically flat charged BHs~\cite{Furuhashi:2004jk,Hod:2013nn,Hod:2013eea}. As such, one concludes that there will be no charged scalar clouds around RN BHs analogue to the ones discussed above for Kerr BH, i.e. as true bound states. An open question is whether the same holds for higher spin massive fields, where an effective potential analysis (which indicates if there is a potential well to support quasi-bound states) is not always straightforward. Here we study the charged Proca field around a RN BH, and show that similar statements to those for a scalar field hold.

Although no stationary bound states exist for charged scalar or Proca fields around RN BHs, we show in this chapter that stationary \textit{marginally bound states} do exist. The first observation of such states for scalar case was reported in~\cite{Degollado:2013eqa} in the double extremal limit wherein the BH charge tends to its mass $|Q|\rightarrow M$ and the field charge tends to its mass $|q|\rightarrow \mu$. It was noticed that, in this limit, the imaginary part of the quasi-bound states vanishes and the real part tends to the mass of the field. Moreover, in this limit, the scalar field does not trivialize and it can be interpreted as a collection of scalar particles at rest, outside the horizon, in a no-force configuration with the BH, due to a balance of electromagnetic and gravitational forces. These particles are only marginally bound to the BH and we shall therefore dub the corresponding field configuration as \textit{marginal (charged) clouds}.

In fact we show that one needs not take a double extremal limit (as in~\cite{Degollado:2013eqa}) to get marginal (charged) scalar clouds around RN BHs. It is enough that the \textit{threshold condition}
\begin{equation}
qQ=\mu M \ ,
\label{tc}
\end{equation}
is obeyed, clearly indicating that the equilibrium is due to a force balance. Moreover, we show that a completely analogous behavior is found for a charged Proca field around RN BHs. It is worth emphasizing that although these clouds are only marginally bound, arbitrarily long lived quasi-bound states exist close to the limit where these clouds appear, which, for many practical purposes, may be faced as eternal clouds.

The structure of this chapter is organized as follows. In Section~\ref{sec:qProca} we review the background geometry as well as the field equations for the charged Proca and scalar fields. In particular, in Section~\ref{sec:eff_potential}, we analyze the effective potential for a decoupled mode of the Proca field, to illustrate some expected features of the result. In Section~\ref{sec:num_strategy} we describe the numerical strategy to calculate the quasi-bound state frequencies for coupled and decoupled modes, summarize some known results in the small frequency limit and combine them to predict an approximation for the (hydrogen-like) real part of spectrum for the charged Proca field. In Section~\ref{results} we present numerical results illustrating the variation of the quasi-bound state frequencies with particle charge and mass as to identify the condition for long lived states. In Section~\ref{sec_analytic} we demonstrate that there are indeed solutions when the threshold condition holds, which means that the imaginary part of the frequency vanishes. In a double extremal limit we even find closed form solutions for both the scalar (already discussed in~\cite{Degollado:2013eqa}) and also the Proca case. Finally we conclude with a summary of our results in Section~\ref{Discussion}, in particular we discuss the possible existence of nonlinear realizations of these marginal clouds.



\section{Charged Proca and scalar fields on a charged black hole}
\label{sec:qProca}
\subsection{The Proca field}

We consider first a Proca field $W^{\mu}$ with mass $\mu$\footnote{In this chapter, we use $\mu$ in general to stand for the mass both for the Proca fields and scalar fields, which were denoted by $\mu_p$ and $\mu_s$, respectively, in previous chapters.}, which is charged under a $U(1)$ gauge symmetry associated with the gauge field $A_\mu$.  The field $W^{\mu}$ can describe $W$ particles in the SM coupled to gravity and its Lagrangian density is given in Eq.~\eqref{Lagrangian}.
The background geometry for a $U(1)$ charged BH is given by the RN solution with line element
\begin{equation}
ds^2=-U(r)dt^2+\dfrac{1}{U(r)}dr^2+r^2(d\theta^2+\sin^2\theta d\varphi^2) \ ,
\end{equation}
where
\begin{equation}
U(r)=1-\dfrac{2M}{r}+\dfrac{Q^2}{r^2}\;.\nonumber
\end{equation}
$M$ and $Q$ are the BH mass and charge parameters, respectively. For numerical convenience, we change to units such that the outer horizon radius is at $r_H=1$, i.e.
\begin{equation}
2M=1+Q^2 \ .
\end{equation}
The field equations for a Proca field on an Einstein-symmetric (and in particular spherically symmetric) background have been derived in Eqs.~\eqref{originalsys1},~\eqref{originalsys2},~\eqref{Max} and~\eqref{k0M}. In the following we set $n=2$ in those equations, and identify the metric function as $U(r)$. Then we obtain a system of two coupled modes $(\psi,\chi)$ obeying
\begin{align}
&\left[U^2\dfrac{d}{dr}\left(r^2\dfrac{d}{dr}\right)+\left(\omega r -qQ\right)^2-U\left(\ell(\ell+1)+\mu^2r^2\right)\right]\psi+\left[2iU\omega r-ir\left(\omega r-qQ\right)\dfrac{dU}{dr}\right]\chi=0 \;, \nonumber \\
&\left[U^2r^2\dfrac{d^2}{dr^2}+\left(\omega r -qQ\right)^2-U\left(\ell(\ell+1)+\mu^2r^2\right)\right]\chi  +\left[2iqQU-i r\left(\omega r-qQ\right)\dfrac{dU}{dr}\right]\psi=0 \; \;, \; \; \; \; \label{coupledequations}
\end{align}
and a decoupled transverse mode $\Upsilon$
\begin{equation}
\left[r^2U\dfrac{d}{dr}\left(U\dfrac{d}{dr}\right)+\left(\omega r  -qQ\right)^2-\left(\ell(\ell+1)+\mu^2r^2\right)U\right] \Upsilon =0  \label{TmodeEq}\ ,
\end{equation}
where $\ell=1,2,\ldots$. There is also an exceptional mode for $\ell=0$ which is described by a decoupled radial equation
\begin{equation}
\left[\dfrac{d}{dr}\left(\dfrac{Ur^4}{(\omega r-qQ)^2-Ur^2\mu^2}\dfrac{d}{dr}\right)+\dfrac{r^2}{U}-\dfrac{qQr^2\left((\omega r-qQ) r U'-2qQU\right)}{\left[(\omega r-qQ)^2-Ur^2\mu^2\right]^2}\right]\psi^{(0)}=0 \ .
\end{equation}
The asymptotics of the solutions of such systems has been detailed in Chapter~\ref{ch:NeutralP} and Chapter~\ref{ch:ChargedP}. Let us denote a generic field mode (coupled or not to others) by $\psi_i$. Then the near horizon asymptotics for all modes takes the form
\begin{equation}\label{Eq:NH_expansion}
\psi_i= (r-1)^{\mp i\frac{\omega-qQ}{1-Q^2}}\sum_{n=0}^{+\infty} a_i^{(n)} (r-1)^n\ ,
\end{equation}
where the minus sign corresponds to an ingoing boundary condition at the horizon. On the other hand, in the asymptotic region $r\rightarrow +\infty$ the expansion is
\begin{equation}
\psi_i= e^{ i \Phi}\sum_{n=0}^{+\infty}\dfrac{c_{i,+}^{(n)}}{r^n}+ e^{-i \Phi} \sum_{n=0}^{+\infty}\dfrac{c_{i,-}^{(n)}}{r^n}\label{eq:FFexpansion}
\end{equation}
with
\begin{align}
 \Phi &\equiv k r+\varphi \log r\;,\nonumber \\
 \varphi&\equiv \dfrac{(\omega^2+k^2)(1+Q^2)-2qQ\omega}{2k} \;,\\
 k&\equiv \sqrt{\omega^2-\mu^2} \; .
\end{align}
All the necessary expansion coefficients were provided in Chapters~\ref{ch:NeutralP} and~\ref{ch:ChargedP}. As shown originally by Press and Teukolsky~\cite{Press:1973zz}, if we consider initial data given on a compact support, the late time dynamics of generic linear perturbation is governed by a superposition of a discrete set of solutions in Fourier space. After imposing an ingoing boundary condition at the horizon, there are basically two types of solutions depending on the boundary conditions when $r\rightarrow +\infty$ (considering a mode oscillating as $e^{i\Phi}$, i.e. $\Re(k)>0$ in the square root and assuming, without loss of generality, $\Re(\omega)>0$):
\begin{itemize}
\item $\Im(k)<0$: these are the quasinormal modes which describe time decaying oscillations. These modes are free to escape the BH potential and asymptotically grow exponentially~\cite{Nollert:1999ji} (for large $r$).
\item $\Im(k)>0$, these are quasi-bound states, i.e. they describe field configurations which are confined in the outside region of the BH and decay exponentially when $r\rightarrow+\infty$. They are possible if there is a confining potential well where the field can accumulate. The boundary condition for these states can thus be recast as
\begin{equation}
\lim_{r\rightarrow+\infty}\psi_i=0\;.
\end{equation}

\end{itemize}
Quasi-bound states are very interesting in the presence of superradiance, since they provide the possibility of an instability as discussed in Chapter~\ref{ch:intro}. If they exist within the superradiant regime, the field is able to extract energy from the BH which accumulates in the confining potential. This would be signaled by an exponential growth of the wave amplitude. The condition for the instability to appear is $\omega_I\equiv\Im{(\omega)}>0$, i.e. the time-dependence is
\begin{equation}
\psi_i\sim e^{-i\omega t}=e^{-i\omega_Rt+\omega_It} \; .
\end{equation}
In the absence of instabilities (i.e. $\omega_I<0$), the wave amplitude decays exponentially with a lifetime
\begin{equation}
\tau\equiv |\omega_I|^{-1}\; .
\end{equation}
If $\omega_I\rightarrow 0$ is possible, then the state can be truly bound, or marginally bound if $\omega_R\rightarrow \mu \Leftrightarrow k_R\rightarrow 0$. Observe that in this exact limit the wavelength diverges so the state becomes de-localized. Actually, $\mu-\omega_R$ is basically the binding energy of the state; since this goes to zero, in the limit, the state becomes marginally bound. Moreover, one can have arbitrarily long lived quasi-bound states near this limit which have a large amplitude in a compact domain (similarly to true bound states -- see the results in Section~\ref{results}). If $\tau$ is very large, then the state may be effectively considered bound (rather than quasi-bound) for many practical purposes.

\subsection{Effective potential for the transverse mode}
\label{sec:eff_potential}
\begin{figure}
\begin{center}
\includegraphics[clip=true,trim = 40 450 50 50, width=0.44\textwidth]{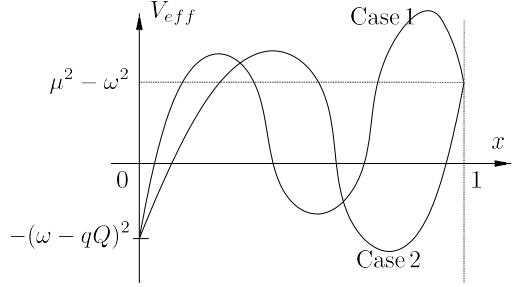}
\end{center}
\caption{\label{EffectiveVschematics} Schematic representation of the effective potential (in the compactified coordinate $x$) for the two possible cases where a well may form.
}
\end{figure}

A natural strategy to investigate the possibility of quasi-bound states to appear is to recast the radial equations for the fields in a Schr\"odinger like form with an effective potential. In the case of the Proca field, this is not so straightforward for the coupled system. However, the transverse mode Eq.~\eqref{TmodeEq}, can be easily recast in a Schr\"odinger like form ($dr_\star \equiv dr/U $)
\begin{eqnarray}
&&\left[-\dfrac{d^2}{dr_\star^2}+V_{\rm eff}\right] \Upsilon=0\;,\\
&&V_{\rm eff}=\left(\tfrac{\ell(\ell+1)}{r^2}+\mu^2\right)U-\left(\omega -\tfrac{qQ}{r}\right)^2 \; .
\end{eqnarray}
To classify the effective potential and investigate when a well forms, it is convenient to define a compactified coordinate $x\equiv 1-1/r\in [0,1]$ such that
\begin{equation}\label{eq:Veff}
V_{\rm eff}(x) = \sum_{k=0}^4 b_k x^k \ ,
\end{equation}
where the coefficients $b_k$ are
\begin{align}
b_0&=-(\omega-qQ)^2 \;,\nonumber\\
b_1&= (1-Q^2)(\mu^2+\ell(\ell+1))-2qQ(\omega-qQ)\;,\nonumber\\
b_2&= -(2-3Q^2)\ell(\ell+1)+Q^2(\mu^2-q^2)\;,\label{eq:bcoeffs}\\
b_3&= \ell(\ell+1)(1-3Q^2)\;,\nonumber\\
b_4&= \ell(\ell+1)Q^2\; . \nonumber
\end{align}
Thus we are dealing with a quartic polynomial. Taking into account its values at the end points a well can only form for two possible configurations with three roots (shown schematically in Fig.~\ref{EffectiveVschematics}).
In case~1 the derivative $V'_{\rm eff}(x)$ must have three positive roots $x_1,x_2,x_3$. Then its form is $V'_{\rm eff}(x)=A(x-x_1)(x-x_2)(x-x_3)$. But we also know that
\begin{eqnarray}
  V'_{\rm eff}(x)&=&\sum_{k=0}^3 (k+1) b_{k+1} x^k\ .
\end{eqnarray}
Equating the two forms we conclude that $x_1x_2x_3=-b_1/(4b_4)$. We are interested in finding a well in the superradiant regime, $\omega<qQ$; but using Eqs.~\eqref{eq:bcoeffs} we conclude that $-b_1/b_4<0$, which is inconsistent with the positivity of the roots of the derivative. Thus case~1 is not possible in the superradiant regime.

Case~2 on the other hand is defined by\footnote{In this case we could not extract any other condition as for case~1.}
\begin{equation}
V'_{\rm eff}(1)>0 \Leftrightarrow \mu^2 >\dfrac{2qQ\omega}{1+Q^2} \Leftrightarrow \mu M >qQ\dfrac{\omega}{\mu} \Leftrightarrow \dfrac{\omega}{\mu}<\sqrt{\dfrac{M}{r_H}}\;,\label{chargedclouds:case2relation}
\end{equation}
where in the last step we have used the superradiance condition $\omega r_H<qQ$. Except in the near-extremal case, Eq.~\eqref{chargedclouds:case2relation} is not consistent with hydrogen-like spectrum. We have also made a numerical scan of the parameter space and found none of parameters satisfying the condition in Eq.~\eqref{chargedclouds:case2relation}.

This analysis indicates that no potential well is possible in the superradiant regime for the transverse modes. For the coupled system it is not simple to make a similar analysis, so we proceed in the following sections investigating the quasi-bound state frequencies numerically and attempting to flow the parameters into the superradiant regime.

We should note at this point that Furuhashi and Nambu~\cite{Furuhashi:2004jk} found a condition for bound states to exist for a charged scalar field, working in a small frequency and charge approximation, and using an analytic matching technique. They found the condition that $\mu M\gtrsim qQ$, which, from the applicability of their analysis, is not necessarily true for large parameters. We will find, however, a perfect agreement with this threshold condition for large parameters as well, therefore away from the applicability of the analytic matching approximation in~\cite{Furuhashi:2004jk}.

\subsection{The scalar field}
\label{sec:scalar}
In this study we have also solved the problem for a charged massive scalar field. In addition to being interesting on its own this also serves as a comparison to check some of the features we have found for the Proca field.

Discussions of the charged massive scalar field can be found, for example, in~\cite{Sampaio:2009ra,Sampaio:2009tp,sampaio2010production,Degollado:2013eqa} and in~\cite{Furuhashi:2004jk} where the rotating case was also analyzed. Here we only present the potential since the structure is very similar to the $\Upsilon$ mode of the Proca field. After decomposing the scalar field $\varphi$ in spherical harmonics $\varphi=e^{-i\omega t}Y_\ell^m(\phi,\theta)R(r)/r$, the radial equation takes exactly the same Schr\"odinger like form as for $\Upsilon$ with a modified effective potential
\begin{equation}
V_{\rm eff}^{\mathrm{(scalar)}}=\left(\tfrac{\ell(\ell+1)}{r^2}+\mu^2+\dfrac{1}{r}\dfrac{dU}{dr}\right)U-\left(\omega -\tfrac{qQ}{r}\right)^2 \; .
\end{equation}
Then all the asymptotic analysis and boundary conditions follow our previous discussion, Eqs.~\eqref{Eq:NH_expansion} and~\eqref{eq:FFexpansion}, for the Proca field modes.

\section{Numerical strategy}
\label{sec:num_strategy}
To find the quasi-bound state frequencies, we scan over $\omega$ on the complex plane. We first define a function of $\omega$ which returns the value of the solutions $\psi_i$ at a large $r_{far} \sim (50\sim100)/k_R$ ($k_R=\Re(k)$), i.e. at a very large distance in multiples of the typical wavelength of the wave. For each value of $\omega$ this is done by integrating the radial equations outwards from the horizon, with initial conditions given by the power series expansions Eq.~\eqref{Eq:NH_expansion} evaluated at $r=1.001$, with typically twenty terms.  This function is expected to be exponentially large for generic values of $\omega$ except in the vicinity of a quasi-bound state where it should be very small. Numerically, the quasi-bound state frequency is never attained exactly since there is always some numerical contamination of the exponentially growing solution. For example for a decoupled mode
\begin{equation}\label{eq:illustratedecay}
\psi_i\sim c_{i,+}^{(0)}e^{ik_Rr-|k_I|r}+\ldots+c_{i,-}^{(0)}e^{-ik_Rr+|k_I|r} \; .
\end{equation}
In the quasi-bound state limit, $c_{i,-}^{(0)}\rightarrow 0$. Numerically this corresponds to frequencies that minimize $|\psi_i|_{r_{far}}$.
\begin{figure*}
\begin{center}
\hspace{-3mm}\includegraphics[clip=true,trim = 90 400 50 30, width=0.51\textwidth]{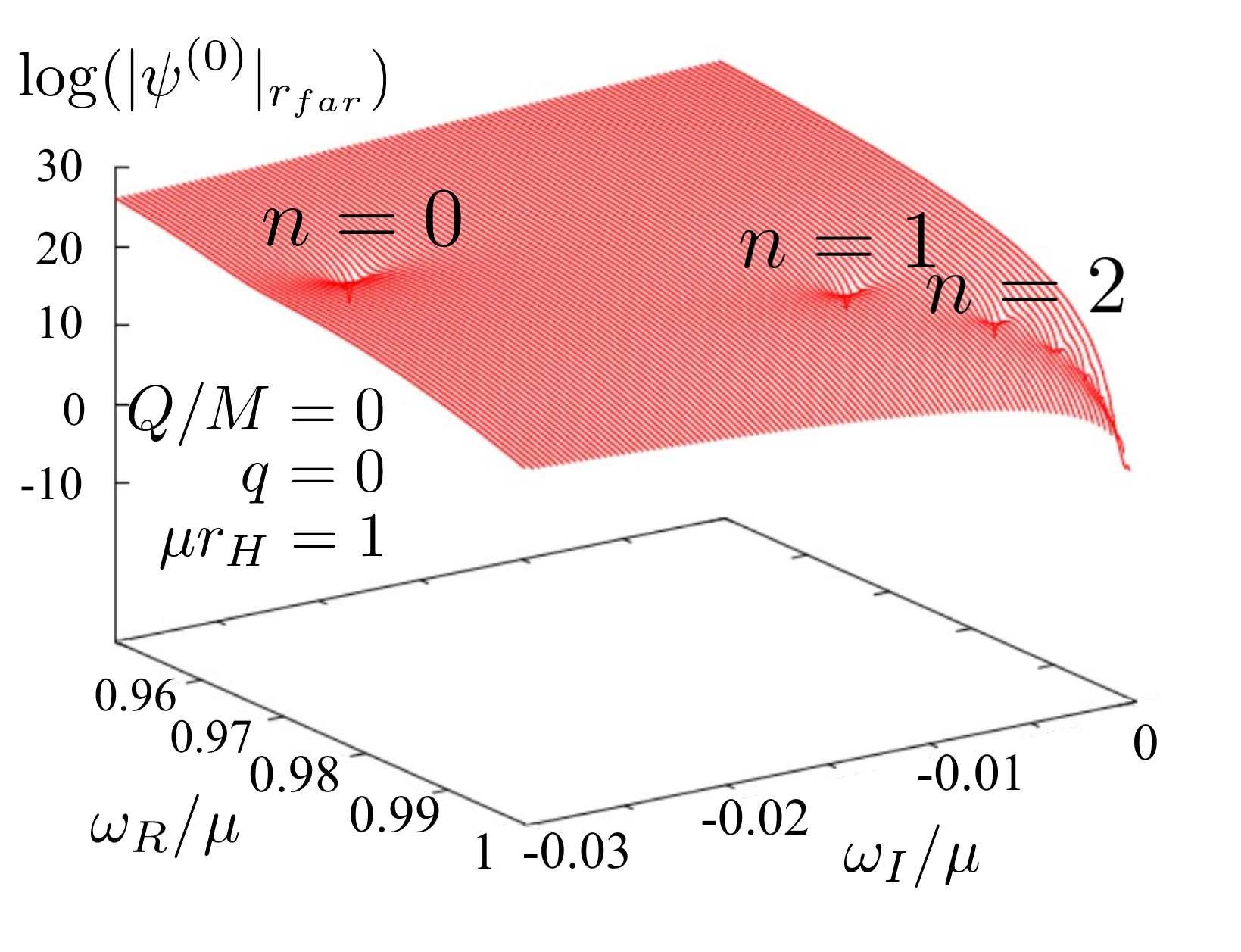}\includegraphics[clip=true,trim = 90 30 50 380, width=0.51\textwidth]{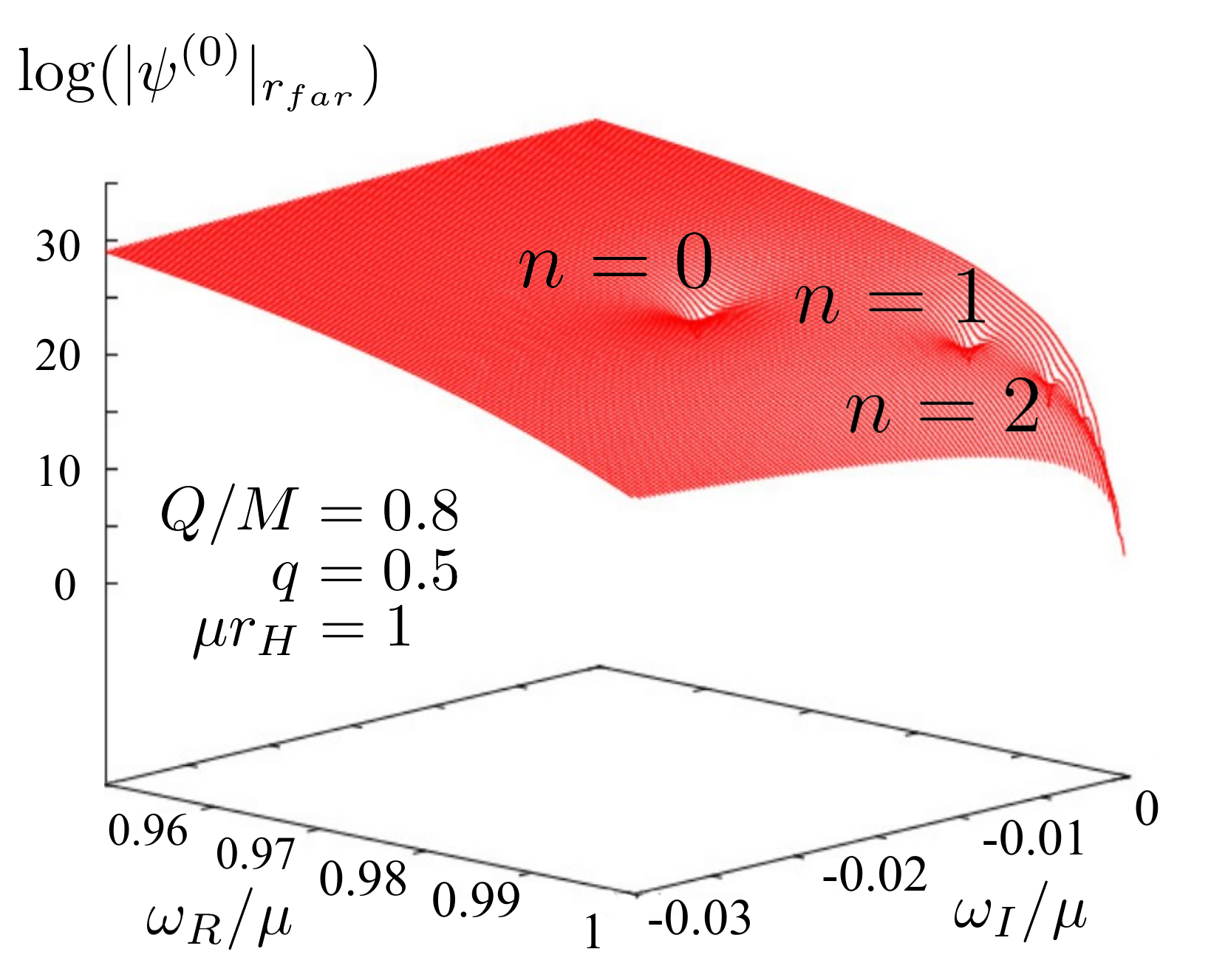}
\end{center}
\caption{\label{SurfacesPsi0} Magnitude of the $\psi^{(0)}$ mode at $r_{far}$ (i.e. in the asymptotic far region), as a function of the complex frequency $\omega$ in the neutral limit (left) and for non-zero charges (right). The quasi-bound state frequencies occur at sharp dips indicated by the level numbers $n$.
}
\end{figure*}
In Fig.~\ref{SurfacesPsi0} we show this quantity for the $\ell=0$ decoupled mode in the neutral case (left) and in the charged case (right). The quasi-bound state frequencies are clearly seen as sharp dips in the surface plots (note that in the vertical axis the logarithm of the scalar field value is represented). We observe that turning on the charge, with $qQ>0$, the real part of the frequency, $\omega_R$, moves to values closer to the field mass, $\mu$, and the (negative) imaginary part $\omega_I$ moves to values closer to zero.

For the coupled sector, the procedure is analogous if one recalls the scattering matrices defined in Chapter~\ref{ch:NeutralP}. First one notes that the general solution of a system of $n$ coupled fields $\psi_i$, with second order linear equations, can be represented by $2n$ integration constants. Those constants can be defined either at the event horizon or at infinity. The linearity of the system implies that a linear transformation relates the integration constants at the horizon, with those at infinity. We denote the ingoing and outgoing wave coefficients at the horizon ($+/-$ respectively)
 \[\vec{\mathbf{h}}=({\mathbf h}^+,{\mathbf h}^-)=(h^+_{i},h^-_{i}) \ , \] where $i=1,2$ for the current coupled system, and, similarly, the coefficients at infinity are defined
  \[\vec{\mathbf{y}}=({\mathbf y}^+,{\mathbf y}^-)=(y^+_{i},y^-_{i}) \ . \]
The linear transformation is then represented as
\begin{equation}
\vec{\mathbf{y}}=\mathbf{S} \vec{\mathbf{h}} \ \ \Leftrightarrow \ \ \left(\begin{array}{c} {\mathbf y}^+ \\ {\mathbf y}^- \end{array} \right)=\left(\begin{array}{c|c} {\mathbf S}^{++} & {\mathbf S}^{+-} \\ \hline {\mathbf S}^{-+}  & {\mathbf S}^{--} \end{array} \right)\left(\begin{array}{c} {\mathbf h}^+ \\ {\mathbf h}^- \end{array} \right)  ,
\end{equation}
where the scattering matrix $\mathbf{S}$ depends on $\omega$, $\ell$, field couplings and the background. It encodes all the information on the scattering process and it can be constructed from specific combinations of modes with boundary conditions set at the horizon.

\begin{figure*}
\begin{center}
\hspace{-3mm}\includegraphics[clip=true,trim = 90 400 50 30, width=0.505\textwidth]{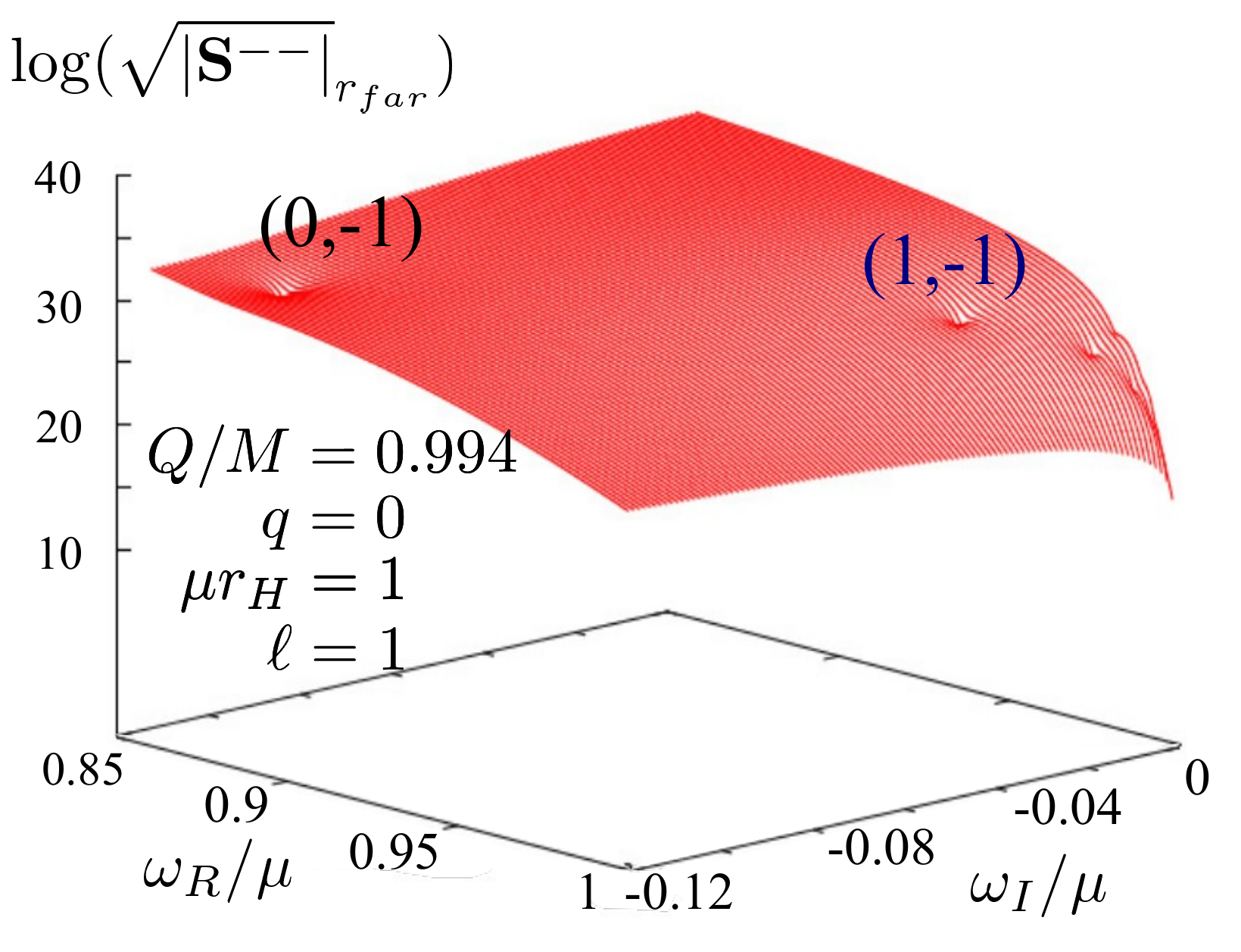}\includegraphics[clip=true,trim = 90 30 50 380, width=0.505\textwidth]{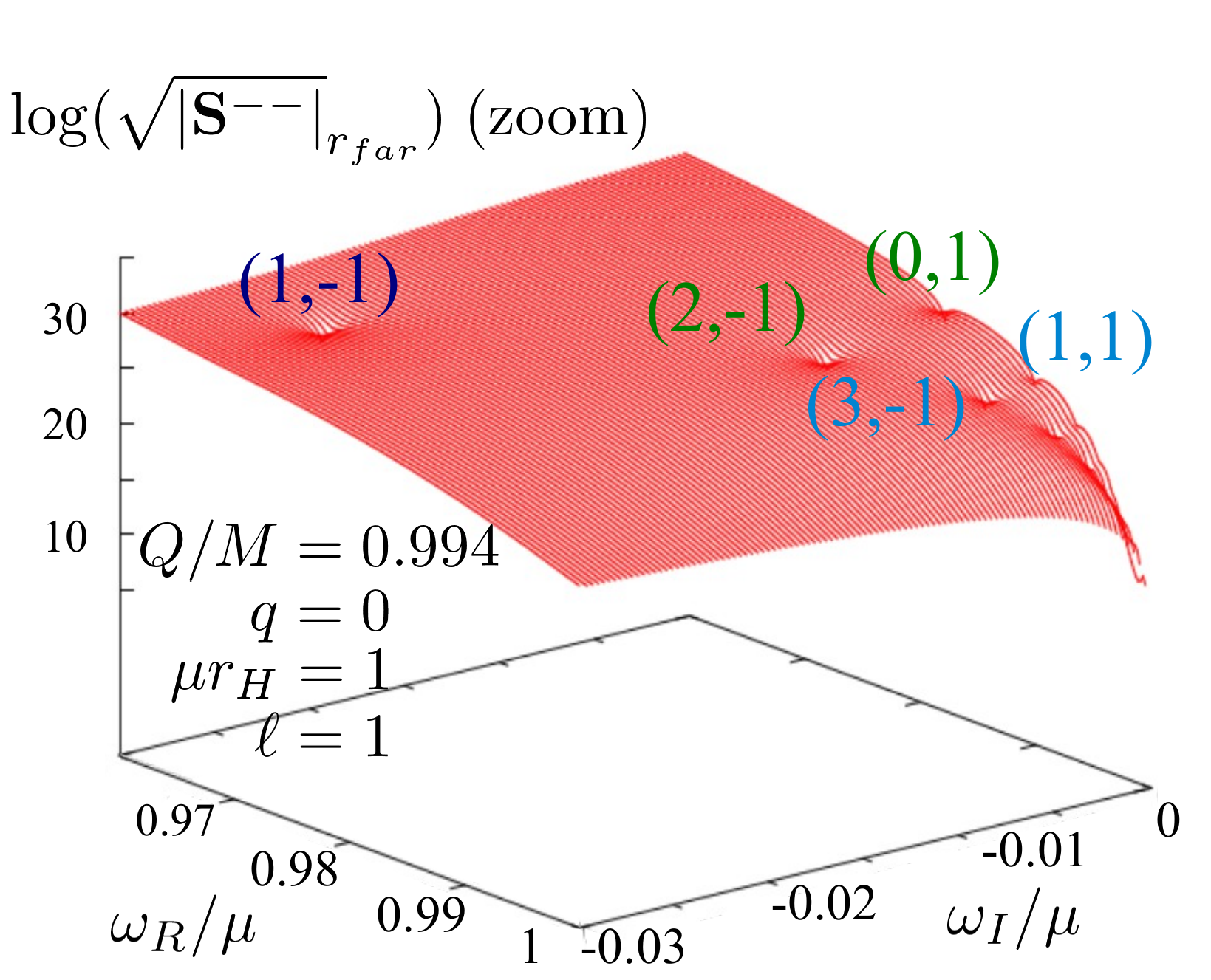}
\end{center}
\caption{\label{SurfacesCoupled} Magnitude of the determinant of ${\mathbf S}^{--}$ for the coupled system at $r_{far}$ (i.e. in the asymptotic far region), as a function of the complex frequency $\omega$ in the neutral limit for $\ell=1$. The various levels are indexed by $(n,S)$ as discussed in the text. The right panel shows a zoom to better display the positions of a few higher levels.
}
\end{figure*}

Similarly to the decoupled modes, we want to impose an ingoing boundary condition at the horizon i.e. ${\mathbf h}^{+}=0$. Then the solution coefficients far away are
\begin{equation}\label{scattering_ingoing}
{\mathbf y}^s={\mathbf S}^{s-}{\mathbf h}^{-} \; .
\end{equation}
We then must impose that the coefficients of the solution which grows exponentially fast vanish (in analogy to Eq.~\eqref{eq:illustratedecay}), i.e. we need that there is a particular initial condition $\hat{\mathbf h}$ at the horizon such that $\mathbf{y}^{-}=0$, i.e.
\begin{equation}
0={\mathbf S}^{--}\hat{\mathbf h}^{-} \; .
\end{equation}
Thus we need to choose the eigenvector with zero eigenvalue of ${\mathbf S}^{--}$. The condition for this solution to be possible is then
\begin{equation}
\det{\mathbf S}^{--}=0 \; ,
\end{equation}
which will occur at the quasi-bound state frequencies. Furthermore, one can check that (up to a normalization constant) the radial profile of the quasi-bound state solution is obtained using the following initial condition at the horizon
\begin{equation}
\left(\begin{array}{c}\hat{h}^{-}_1 \\ \hat{h}^{-}_2\end{array}\right)\propto\left(\begin{array}{c}-S^{--}_{12} \\ S^{--}_{11}\end{array}\right) \; .
\end{equation}
In practice, there is numerical contamination of exponentially growing eigenmodes, which means we must again employ a minimization condition (similarly to the decoupled modes)
\begin{equation}
  \min_{\omega}\left[|{\mathbf S}^{--}|_{r_{far}}\right] \; .
\end{equation}
In Fig.~\ref{SurfacesCoupled} we show an example of this quantity for $\ell=1$ coupled modes with a non-zero background charge and a zoom around the corner where the higher frequency levels pile up (right). We discuss in more detail the effect of the field charge in Sections~\ref{results} and~\ref{Discussion} so here we only highlight the differences compared with the decoupled cases.
Once again the procedure is to find the local minimum for each valley found in the figure.

Since the search for the quasi-bound state frequencies reduces to a (local) minimization problem, one needs in general a good starting guess which is close enough to the valley. Several analytic estimates have been developed in the literature for small parameters. The general conclusion is that $\omega_R$ follows a hydrogen like spectrum. Of particular relevance to our problem is the charged scalar spectrum approximation found in~\cite{Furuhashi:2004jk}:
\begin{equation}\label{AnalyticFN}
\omega_R\simeq \mu\left[1-\frac{1}{2}\frac{(\mu M-qQ)^2}{N^2}\right]\;,\;N=\ell+n+1\;,
\end{equation}
with $n\footnote{In the remaining part of this chapter, $n$ refers to the overtone number.}\in \mathbb{N}_0$.
This approximation suggests that, as the charge is turned on to positive values, there is a critical value at\footnote{In this limit $\omega_R=\mu$ and in fact $\omega_I=0$ (see~\cite{Furuhashi:2004jk}).} $\mu M =qQ$, after which we expect an exit from the quasi-bound state regime (since the exponentially decaying tail disappears at this threshold).

Also in the low energy limit, but in the neutral case ($q=Q=0$), a low energy approximation was found for the Proca field in~\cite{Rosa:2011my}. Their analytic approximations can in principle be generalized to our case by introducing a charge dependence. Even though in this study we have not developed such analytic matching calculation, we found an excellent agreement (for small parameters) with the following ansatz
\begin{equation}\label{eq:ProcaOmegaR}
\omega_R\simeq \mu\left[1-\frac{1}{2}\frac{(\mu M-qQ)^2}{N^2}\right]\; \;, \;  N=\ell+S+n+1
\end{equation}
which results from shifting $\mu M\rightarrow \mu M-qQ$ in the formula of~\cite{Rosa:2011my} as suggested by Eq.~\eqref{AnalyticFN}. Here $S=0$ for the transverse modes and $S=1,-1$ for the coupled modes (for the exceptional mode $\ell=0$ there is only $S=1$).

In Fig.~\ref{SurfacesCoupled} we indicate the labels $(n,S)$ corresponding to this approximation, for the first few pairs of levels which arise in the coupled system. As expected, due to the degeneracy in $N$, i.e the same $N$ can be obtained by adding different combinations of $(n,S)$ for the same $\ell$, the number of degrees of freedom is doubled. This can be observed in the existence of two distinct lines of valleys in the right panel. By continuity, this classification must hold also for larger parameters, so a natural strategy to obtain new frequencies is to perform a flow of the parameters in small steps from some reference quasi-bound state frequencies. This can be pictured as a flow of the valleys in the plots of Figs.~\ref{SurfacesPsi0} and~\ref{SurfacesCoupled}.

The two parameters we will want to vary continuously are the field mass $\mu$ and charge $q$. Taking the mass $\mu$ as an example, let us assume we have obtained a high precision quasi-bound state frequency $\omega$ by minimization near a valley (with given mode labels $\{\ell,n,S\}$ and fixed $\mu,q,Q,M$). Then by continuity the frequency for this mode at a nearby mass $\mu\rightarrow \mu+\delta \mu$ will shift by a small amount $\omega\rightarrow\omega+\delta \omega$ (all other parameters are fixed). If the step is small enough, we can use the previous $\omega$ as a first guess and refine it by minimization to obtain the new frequency. Thus, by using small steps, we can iterate this procedure to flow the frequencies as a function of a continuous parameter. The same procedure can be applied to flow the frequency with charge $q$.

Even though analytic approximations give a very useful guide to the dependence of the frequencies, and help understanding the labels of the various levels, in practice (especially for larger parameters) we determined various reference initial estimates for the frequencies of each state graphically (using plots such as Fig.~\ref{SurfacesPsi0}) and then refined them through minimization before using as seeds to the flows. In Table~\ref{TabSeeds}, we provide a set of seed frequencies that were used as starting points to produce various of the plots in the results of Section~\ref{results} for the Proca field.  Some scalar frequencies  that were used to obtain scalar profiles to compare with the Proca field in the region of parameters where long lived states were found (see Sections~\ref{results} and~\ref{Discussion}) are shown in Table~\ref{TabSeedsScalar}.
\begin{table}
\begin{minipage}{\textwidth}
\begin{center}
$Qr_H=0$\footnote{Note that the results shown in this table agree with the data used in~\cite{Rosa:2011my}, to all of the significant figures quoted.}\vspace{1mm}\\
\begin{tabular}{||c|c|c||}
\hline
 $N$ & $(\ell,n,S)$ &$\mu^{-1}(\omega_R,\omega_I)$ \\
\hline
$1$ & $(1,0,-1)$ &$\phantom{..}(0.95972157,-0.0037679893)\phantom{.}$\\
\hline
    & $(1,1,-1)$ & $(0.99032863,-0.00062350043)$\\
$2$ & $(0,0,1)$  & $(0.99031646,-0.00051223870)$\\
    & $(1,0,0)$  & $(0.99131023,-0.00001346974)$\\
\hline
\end{tabular}\vspace{2mm}\\
$Qr_H=0.3 \; (Q/M\simeq 0.55\;,\;\mu M \simeq 0.27)$ \vspace{1mm}\\
\begin{tabular}{||c|c|c||}
\hline
 $N$ & $(\ell,n,S)$ &$\mu^{-1}(\omega_R,\omega_I)$ \\
\hline
$1$ & $(1,0,-1)$ & $(0.95270408,-0.0046419907)$\\
\hline
     & $(1,1,-1)$ & $(0.98838598,-0.00084694923)$\\
 $2$ & $(0,0,1)$  & $(0.98812624,-0.00095739793)$\\
     & $(1,0,0)$  & $(0.98953577,-0.00002375643)$\\
\hline
\end{tabular}\vspace{2mm}\\
$Qr_H=0.9\;(Q/M\simeq 0.994\;,\;\mu M \simeq 0.45)$ \vspace{1mm}\\
\begin{tabular}{||c|c|c||}
\hline
 $N$ & $(\ell,n,S)$ &$\mu^{-1}(\omega_R,\omega_I)$ \\
\hline
$1$ & $(1,0,-1)$ & $\phantom{..}(0.87274751,-0.031112471)\phantom{...}$\\
\hline
    & $(1,1,-1)$ & $(0.96469871,-0.011098631)$\\
$2$ & $(0,0,1)$  & $(0.96824273,-0.016591447)$\\
    & $(1,0,0)$  & $(0.96354486,-0.001705819)$\\
\hline
\end{tabular}
\end{center}
\caption{\label{TabSeeds} Some reference frequencies (found through minimization) used in the flow for the Proca field, with $\mu r_H=0.5$ and $q=0$. We indicate the background charge both in inverse horizon radius units, and also in BH mass units for comparison.}
\begin{center}
\begin{tabular}{||c|c|c||}
\hline
 $N$ & $(\ell,n)$ &$\mu^{-1}(\omega_R,\omega_I)$ \\
\hline
$1$ & $(0,0)$ & $\phantom{..}(0.93713433,-0.083684629)\phantom{...}$\\
\hline
    & $(0,1)$ & $(0.97694630,-0.016936300)$\\
$2$ & $(1,0)$ & $(0.96780454,-0.000797486)$\\
\hline
\end{tabular}
\end{center}
\caption{\label{TabSeedsScalar} Some reference frequencies (found through minimization) used in the flow for the scalar field, with $\mu r_H=0.5$ and $q=0$. We have focused on $Q/M=0.994$ to compare with the Proca field.}
\end{minipage}
\end{table}
All frequencies in these tables were obtained by refining initial estimates through minimization. The initial estimates were obtained graphically by zooming surfaces such as Figs.~\ref{SurfacesPsi0} and~\ref{SurfacesCoupled} around each valley. Once an initial estimate is found which is inside the valley, the minimization routines refine the result to the required precision. These high precision frequencies can be used to flow either the mass $\mu$ or the charge $q$ in small steps, to other values as shown in the plots of Section~\ref{results}.

\section{Results}
\label{results}
In this section, we present a selection of numerical results, focusing mostly on the positive charge coupling case. First, we note that we did not find any quasi-bound states in the superradiant region $\omega<qQ$.
\begin{figure*}
\begin{center}
\includegraphics[clip=true, width=0.33\textwidth]{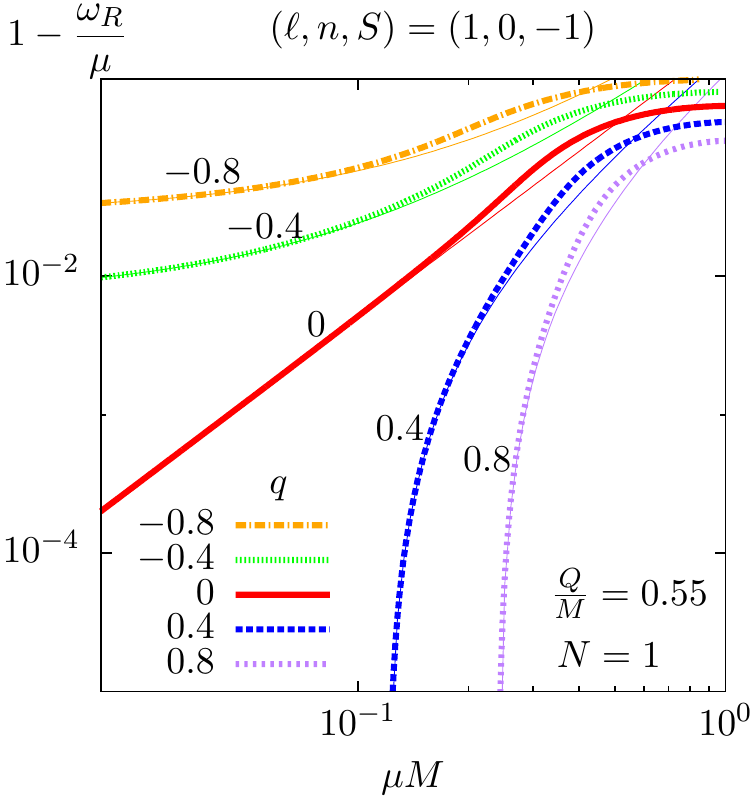}\includegraphics[clip=true, width=0.33\textwidth]{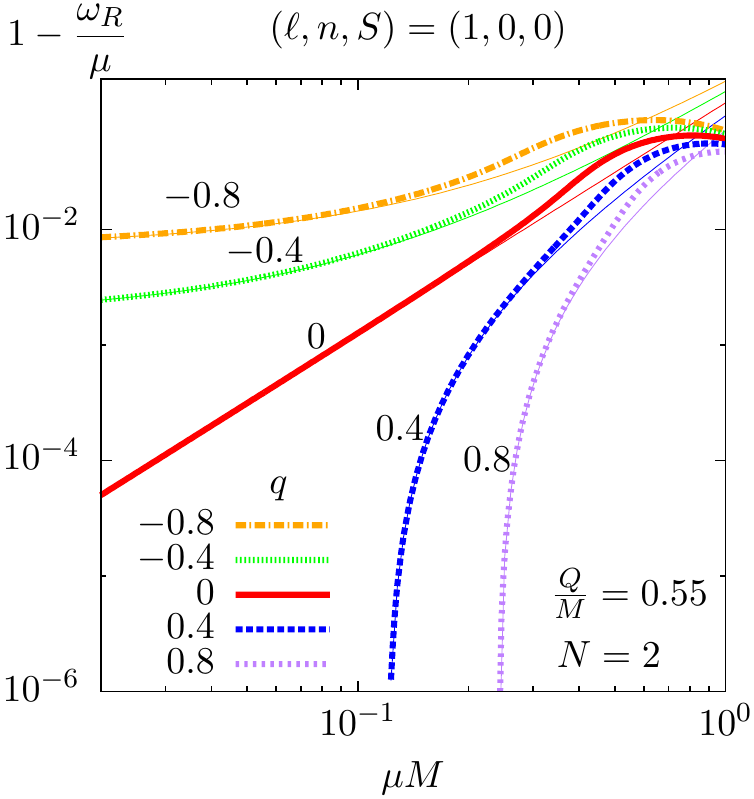}\includegraphics[clip=true, width=0.33\textwidth]{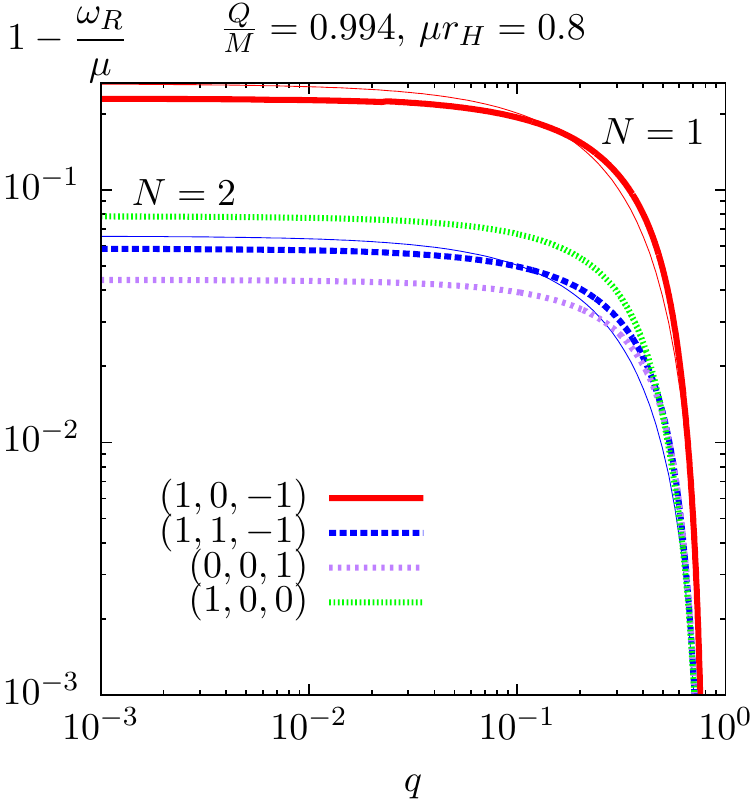} \vspace{2mm}\\
\includegraphics[clip=true, width=0.33\textwidth]{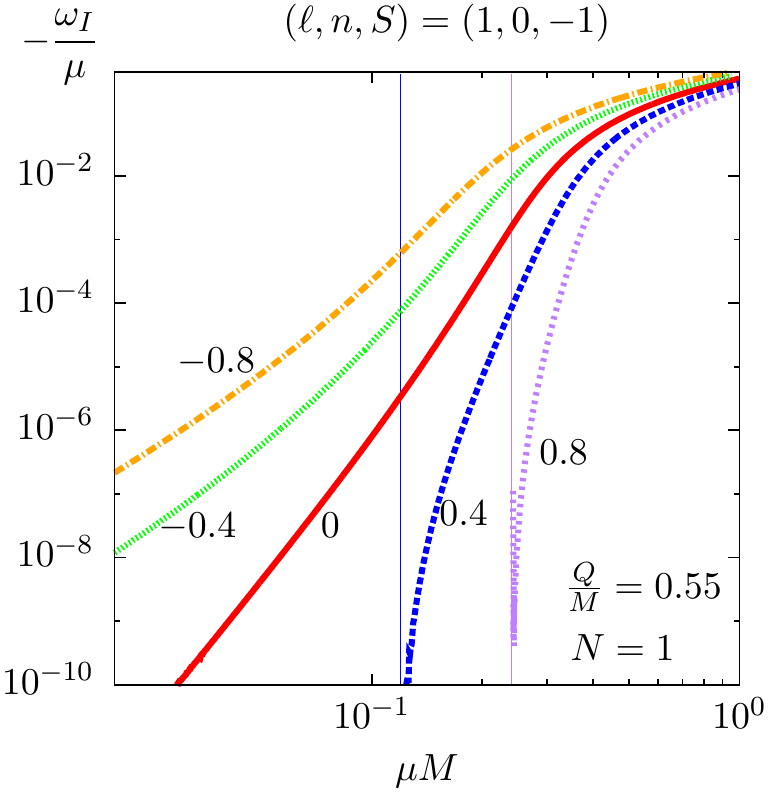}\includegraphics[clip=true, width=0.33\textwidth]{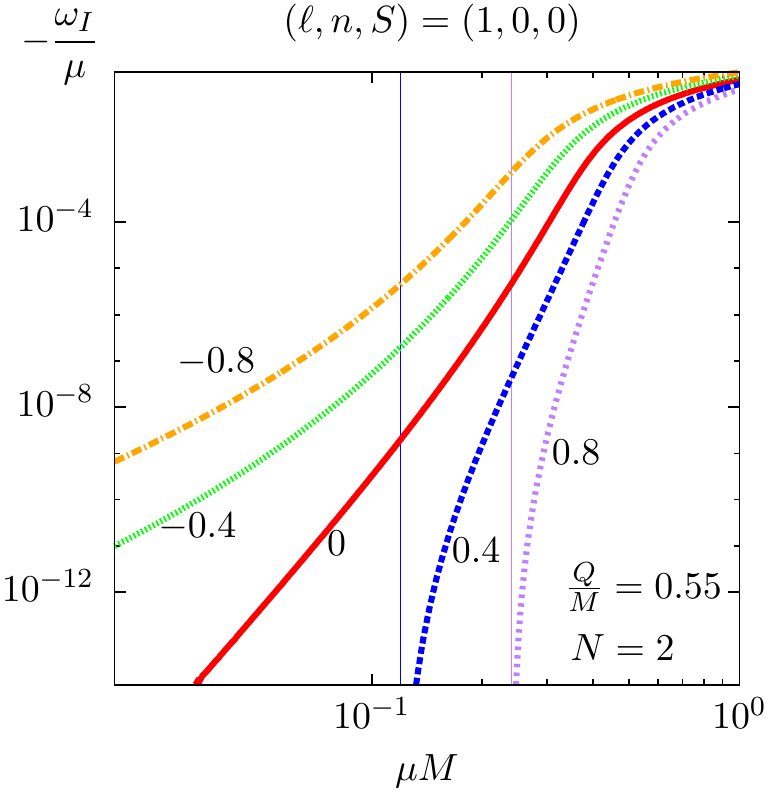}\includegraphics[clip=true, width=0.325\textwidth]{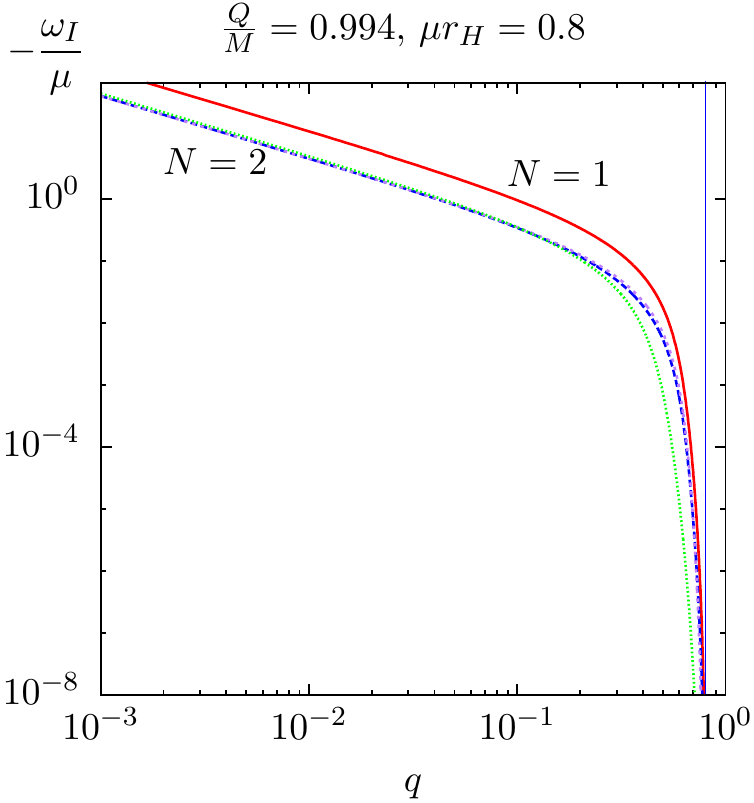}
\end{center}
\caption{\label{Fundamental} Variation of the quasi-bound state frequencies with field mass $\mu$ and charge $q$ for the Proca field. In the top panels we represent curves for the real part $\omega_R$ whereas the corresponding curves for the imaginary part are in the bottom panels. The thick curves labeled in the key to the figures are obtained by numerical integration. The thin solid lines correspond to the analytic formula Eq.~\eqref{eq:ProcaOmegaR}. The vertical lines give, for each case $\mu M=qQ$.
}
\end{figure*}
To understand why this is the case we analyze first Fig.~\ref{Fundamental} where we represent the variation of the frequencies with charge and mass couplings on a fixed background. In the top row of panels we represent the real part of the frequency (or more precisely $1-\omega_R/\mu$). The thick curves correspond to the numerical results that we have generated through direct integration. We have requested a minimum relative precision of six digits (i.e. $10^{-6}$ of relative error). The thin solid lines (matching the color of the solid curves) correspond to the analytic guess, Eq.~\eqref{eq:ProcaOmegaR}. In the bottom panels we represent the imaginary part which is always negative. We also indicate with vertical lines the value where the  threshold condition  $\mu M=qQ$ is attained.

In the left panels we represent the fundamental mode frequency of the Proca system ($N=1$) with various curves for different positive and negative field charge as a function of mass coupling $\mu M$. The middle panels are similar except that we represent a first excited state ($N=2$). The main observations on these are:
\begin{itemize}
\item Generically, $\omega_I$ is negative and its modulus decreases for smaller mass.
\item For negative field charge\footnote{Throughout we assume the convention that the BH charge is positive.} $q$, as it goes to more negative values, $|\omega_I|$ grows for fixed mass, which means that $\tau$ decreases so the scalar mode should be quickly absorbed by the BH.
\item For positive $q$, however, as we flow to smaller mass, there is a sharp threshold in $\omega_I$ precisely at $\mu M= qQ$, i.e. when the mass coupling balances out exactly the electromagnetic coupling. At this threshold $\tau\rightarrow +\infty$, which means that we can have arbitrarily long lived Proca quasi-bound states.
\item Furthermore, at the threshold $\mu M= qQ$, the real part of the frequency also tends sharply to $\mu$. This means that as we try to flow into the superradiant regime ($\omega<qQ$), we leave the quasi-bound state regime into the scattering regime, before reaching it. This explains the fact that we did not find any quasi-bound states in that regime.
\item Finally, one can see a very good agreement between our guess, Eq.~\eqref{eq:ProcaOmegaR}, (thin solid lines) not only for small mass, but also near the long lived state threshold even for larger mass and charge couplings.
\end{itemize}
These features are confirmed in the right panels of Fig.~\ref{Fundamental}, where we have performed a flow of the charge $q$, with $Q$ close to the extremal BH background limit, for all the $N=1,2$ modes of the Proca system. Here it can be seen more clearly that as the charge $q$ is increased to larger positive values, there is a maximum value allowed which is given by $q_{max}=\mu M/Q$ (in the extremal limit this becomes $q=\mu$).

\begin{figure*}
\begin{center}
\includegraphics[clip=true, trim = 0 0 13 0, width=0.32\textwidth]{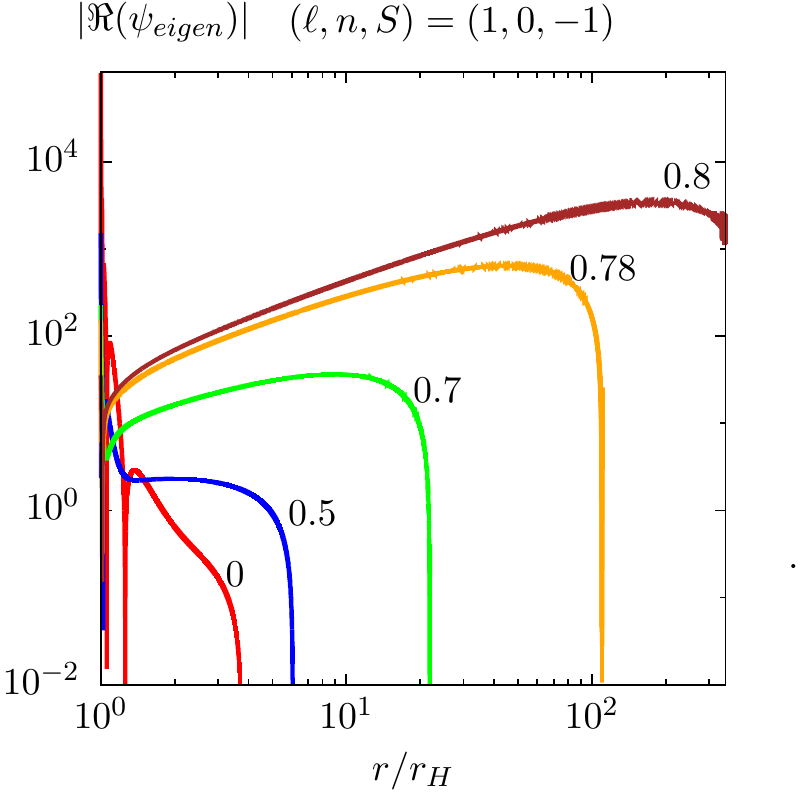}\hspace{1mm}\includegraphics[clip=true, width=0.32\textwidth]{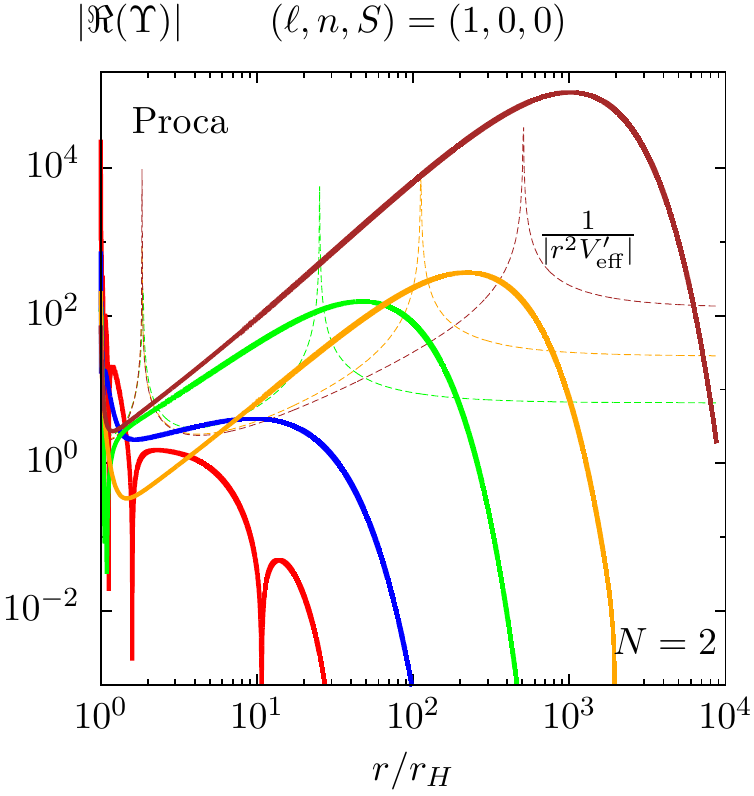}\hspace{1mm}\includegraphics[clip=true, width=0.32\textwidth]{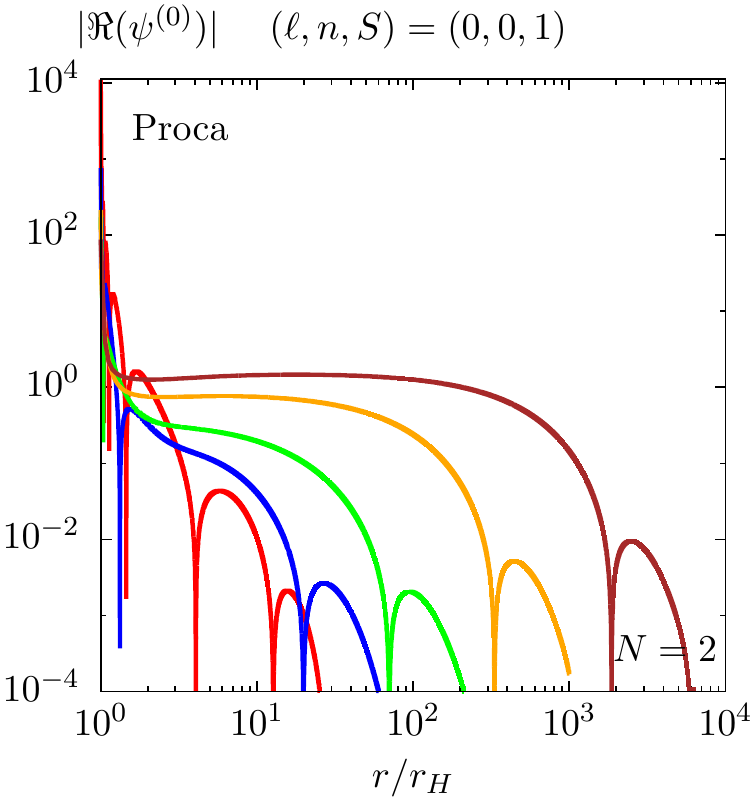} \vspace{2mm}\\
\includegraphics[clip=true, trim = 0 0 13 0, width=0.436\textwidth]{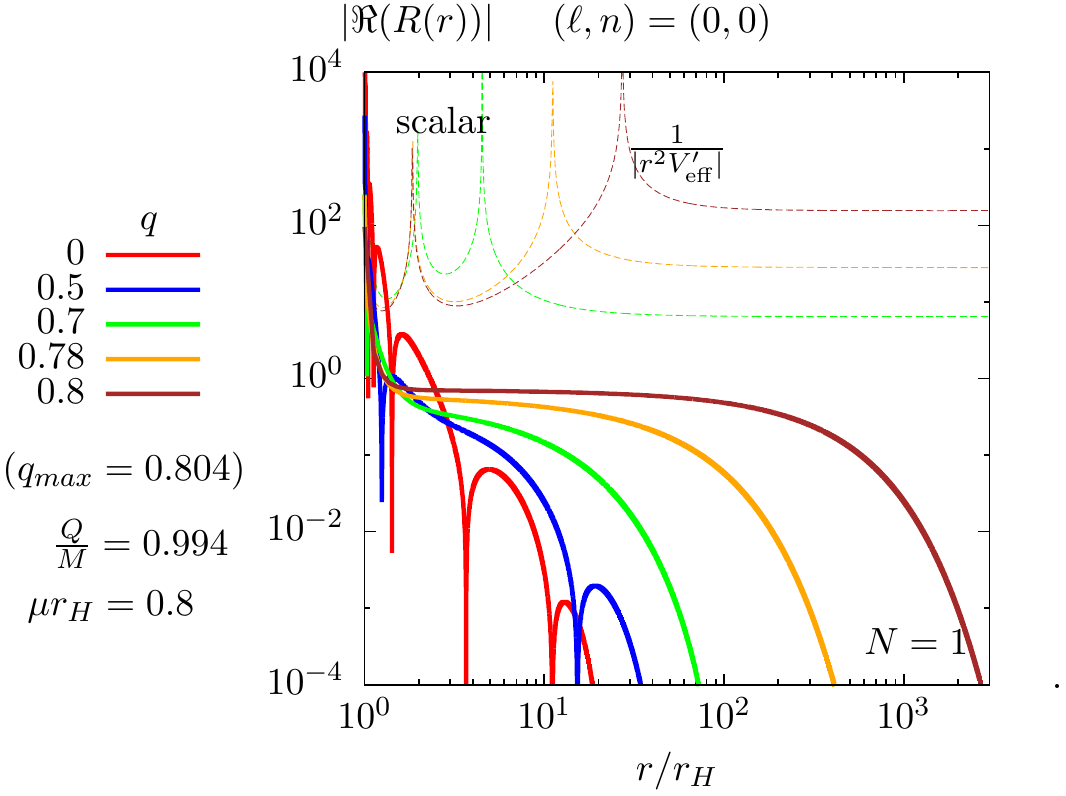}\hspace{3mm}\includegraphics[clip=true, trim = 0 0 13 0, width=0.35\textwidth]{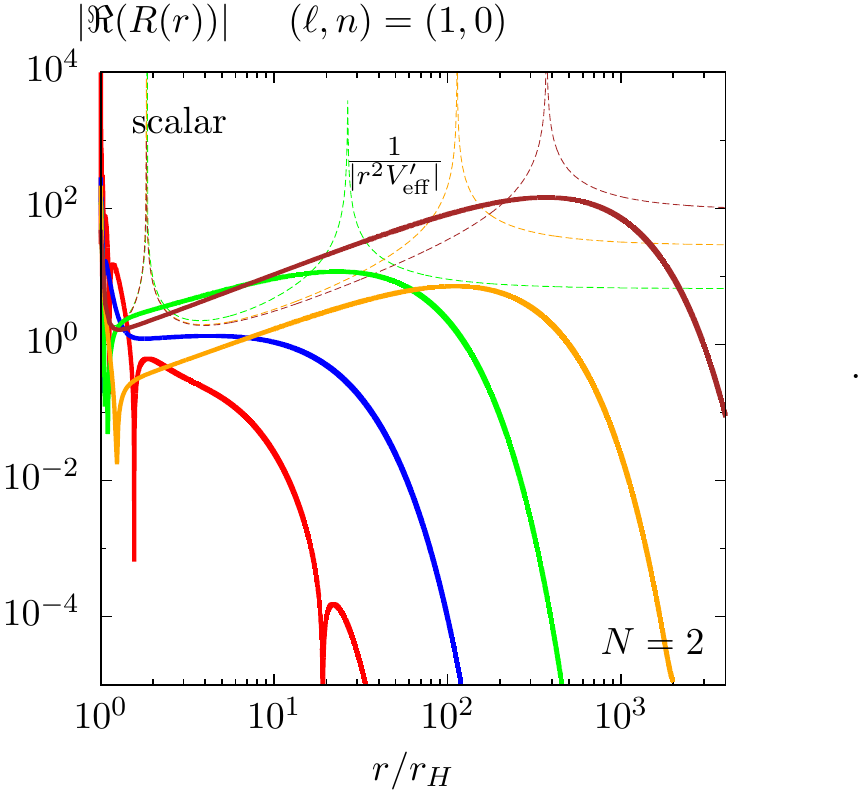}
\end{center}
\caption{\label{WFuncs} Radial profiles for some quasi-bound state modes for the Proca field (top row) and scalar field (bottom row) for comparison (key in the bottom row applies to all plots). We also indicate for some cases $|r^2dV_{\rm eff}/dr|^{-1}$ whose poles indicate the local maximum and minimum of the effective potential (left and right peaks in each panel where it is represented in thin dashed lines). In the top left panel we also indicate the charge $q$ associated with each curve (the same ordering applies to all other panels).
}
\end{figure*}

In Fig.~\ref{WFuncs} we display some examples of quasi-bound state radial profiles. In all panels we vary the field charge from zero up to a value close to the maximum allowed by the threshold condition, to observe how the field binds around the BH close to that limit. The top row is for modes of the Proca field and we have also included, for comparison, some modes for the scalar field in the bottom row. Whenever an effective potential is known, we display the quantity $|r^2dV_{\rm eff}/dr|^{-1}$ (i.e. the inverse derivative). The latter has poles, one at a local maximum of the effective potential near the horizon and another at a minimum (further away), which define the location of the potential well.

In the top left panel we display the fundamental mode of the Proca field. This mode is in the coupled system and it is the eigenvector with zero eigenvalue of the $\mathbf{S}^{--}$ matrix. It results from the linear combination of two exponentially growing solutions, so for large $r$ the fine cancellation that must occur in the linear combination becomes numerically more difficult. This can be seen in the brown ($q=0.8$) curve which starts oscillating due to numerical errors on the right tail region (in $r$). It is clear from the curves that as we approach the threshold charge the field spreads out to a larger radius and the oscillations close to the horizon (which are associated with the field's absorption by the BH) become less extended. This shows that the field configuration becomes more and more bound state like (rather than quasi-bound) since the field oscillations associated with absorption near the horizon become smaller in amplitude.

In the top middle panel we show an $N=2$ level obtained from the transverse $S=0$ equation. The features are similar to the left one except that the radial profile peaks more strongly. The peak is clearly correlated with the depth of the potential well where the field is more tightly bound, as seen in the dashed lines indicating the maximum and minimum of the potential.

The top right panel, shows the first $\ell=0$ level. The main difference is that the curves in the bound region tend to be much more flat. This is due to the absence of orbital angular momentum for these states. In fact, in the bottom left panel we show the corresponding mode in the scalar case (for comparison) where exactly the same happens for $\ell=0$.

In the scalar case (bottom row), similar features apply. In particular one can verify very clearly that the field profile in the long lived limit, extends inside the (increasingly wide) potential well.

\section{Analytic solutions}
\label{sec_analytic}
We will now show that in the limit in which the imaginary part of the frequency goes to zero, there are indeed non-trivial solutions, which are the limiting behavior of the modes found numerically in the previous section. This follows closely the analysis in~\cite{Degollado:2013eqa} but extends it, not only because we also consider the Proca field but also because we do not restrict ourselves to the double extremal limit considered therein.

From the results of Section~\ref{results}, in the limit of the threshold condition
 \begin{equation}
\mu M=qQ\Leftrightarrow qQ=\dfrac{(1+Q^2)}{2}\mu\;,
\label{forcebalance}
\end{equation}
the frequency becomes real and equal to the field mass
\begin{equation}
\omega = \mu \; .
\end{equation}
We replace these two conditions in the original equations, for both the Proca and the scalar field, and define $\rho\equiv r-1$ and
\begin{equation}
\epsilon\equiv 1-Q^2 \ .
\end{equation}
If $\epsilon=0$, then $Q^2=1$; going back to natural units, this means $Q=M$ and $\mu=q$. So this corresponds to a \textit{double extremal limit}, i.e. an extremal limit for both the BH and the field. Thus $\epsilon$ parameterizes the deviation from the double extremal limit, but keeping the force balance condition given by Eq.~\eqref{forcebalance}. We will show that there are exact solutions for the Proca field when $\epsilon=0$, an analogous situation to that observed in~\cite{Degollado:2013eqa} for the scalar field case. But even for $\epsilon\neq 0$, as we will show later, non-trivial solutions of the Proca fields are existed.

With the definitions above, the equations for the Proca field simplify as follows. The transverse mode equation becomes
\begin{equation}
\left[\dfrac{d^2}{d\rho^2}+\left(\frac{1}{\rho} - \frac{2}{1+\rho} + \frac{1}{\epsilon+\rho}\right)\dfrac{d}{d\rho}-\dfrac{\ell(\ell+1)}{\rho(\epsilon+\rho)}+\left(\dfrac{\epsilon\mu(1+\rho)}{2\rho(\epsilon+\rho)}\right)^2\right] \Upsilon=0 \;.\label{TmodeEq2}
\end{equation}
The $\ell=0$ mode equation is in fact very similar
\begin{equation}
\left[\dfrac{d^2}{d\rho^2}+\left(\frac{1}{\rho} + \frac{2}{1+\rho} + \frac{1}{\epsilon+\rho}\right)\dfrac{d}{d\rho}-\dfrac{2-\epsilon}{\rho(1+\rho)(\epsilon+\rho)}+\left(\dfrac{\epsilon\mu(1+\rho)}{2\rho(\epsilon+\rho)}\right)^2\right] \psi^{(0)}=0 \;.\label{Psi0_2}
\end{equation}
The coupled system is now
\begin{align}
&\left[\dfrac{d^2}{d\rho^2}+ \frac{2}{1+\rho}\dfrac{d}{d\rho}-\dfrac{\ell(\ell+1)}{\rho(\epsilon+\rho)}+\left(\dfrac{\epsilon\mu(1+\rho)}{2\rho(\epsilon+\rho)}\right)^2\right]\psi+i\mu\dfrac{4\rho^3+6\epsilon \rho^2+\epsilon^2(\rho-1)}{2\rho^2(\epsilon+\rho)^2}\chi=0 \;, \nonumber \\
&\left[\dfrac{d^2}{d\rho^2}-\dfrac{\ell(\ell+1)}{\rho(\epsilon+\rho)}+\left(\dfrac{\epsilon\mu(1+\rho)}{2\rho(\epsilon+\rho)}\right)^2\right]\chi-\dfrac{i\epsilon^2\mu(1+\rho)}{2\rho^2(\epsilon+\rho)^2}\psi=0.
\label{coupledequations2}
\end{align}
Finally, for the scalar field we have
\begin{equation}
\left[\dfrac{d^2}{d\rho^2}+\left(\frac{1}{\rho} - \frac{2}{1+\rho} + \frac{1}{\epsilon+\rho}\right)\dfrac{d}{d\rho}-\dfrac{\ell(\ell+1)}{\rho(\epsilon+\rho)}+\left(\dfrac{\epsilon\mu(1+\rho)}{2\rho(\epsilon+\rho)}\right)^2+\dfrac{\epsilon(\rho-1)-2\rho}{\rho(1+\rho)^2(\epsilon+\rho)}\right] R=0  \;.\label{RscalarEq2}
\end{equation}

We first consider the scalar field case with $\epsilon=0$, and obtain
\begin{equation}
R_\ell=(\rho+1)\left(A^{(R)}\rho^{\ell}+B^{(R)}\rho^{-\ell-1}\right)\; .\label{scalarrecover}
\end{equation}
Then the scalar field can be written as a linear combination of harmonic functions\footnote{One may observe that the harmonic modes with $e^{-i\mu t}$ can be transformed to be static by making a gauge transformation, $\varphi\rightarrow\varphi e^{i\xi}$ with $\xi=\mu t$, and $A_\mu\rightarrow A_\mu+q^{-1}\partial_\mu\xi$. This is also true for the Proca case, following the same transformation but $W_\mu\rightarrow W_\mu e^{i\xi}$. We would like to emphasize, however, that even without the above transformations, the energy momentum tensors for marginal scalar and Proca clouds are time independent, which imply that they are static.}
\begin{equation}
\varphi=e^{-i\mu t}\sum_{\ell,m}Y_{\ell m}(\theta,\phi)\left(A^{(R)}\rho^{\ell}+B^{(R)}\rho^{-\ell-1}\right)\;.\label{scalarsolnew}
\end{equation}
To interpret the meaning of the modes~\eqref{scalarsolnew}, it is useful to rewrite the extremal RN BH, in isotropic coordinates
\begin{equation}
ds^2=-H^{-2}dt^2+H^2\delta_{ij}dx^idx^j\; , \; A=(H^{-1}-1)dt \ ,\label{extremalRNnew}
\end{equation}
where $H(x^i)$ is a harmonic function on Euclidean 3-space, i.e. $\Delta_{\mathbb{E}^3}H=0$. To describe a single extremal RN BH, $H$ is taken to have a single pole localized at the origin, $H=1+M/|{\bf x}|$. But multi-centered solutions, of Majumdar-Papapetrou type~\cite{Majumdar:1947eu,Papapetrou1945}, are also possible if one chooses a harmonic function with multiple poles. Then one can solve the scalar field equation, by taking $\varphi(t,x^i)=e^{-i\mu t}U(x^i)$, with $\mu=q$ on the background in Eq.~\eqref{extremalRNnew}. The field equation thus becomes 
\begin{equation}
\Delta U(\mathbf{x})=-4\pi\mu\delta^3(\mathbf{x}-\mathbf{x^\prime})\;,
\end{equation}
with the solution $U(\mathbf{x})=\mu/|\mathbf{x}-\mathbf{x^\prime}|$, and where the $\delta$ function is introduced to describe that the field is localized at $\mathbf{x^\prime}$ (not coincide with the BH horizon). In terms of spherical coordinates, $(\rho,\theta,\phi)$, we have\footnote{This corrects a statement made in~\cite{Sampaio:2014swa} and~\cite{Degollado:2013eqa} that regular solutions at the horizon can be constructed from divergent partial waves. We would like to thank Sam Dolan for pointing out that only inhomogeneous solutions can be made regular at both boundaries.}
\begin{equation}
U(\mathbf{x})=\dfrac{\mu}{|\mathbf{x}-\mathbf{x^\prime}|}=\dfrac{4\pi\mu}{\rho^\prime}\sum_{\ell,m}\dfrac{Y^\ast_{\ell m}(\theta^\prime,\phi^\prime)}{2\ell+1}F_\ell(\rho)Y_{\ell m}(\theta,\phi)\;,\label{Usol}
\end{equation}
where the function $F_\ell$ is constructed from two homogeneous solutions of Eq.~\eqref{scalarsolnew}
\begin{equation}
F_l(\rho) \equiv \begin{cases} (\rho / \rho^\prime)^\ell , & \rho \le \rho^\prime , \\ (\rho^\prime / \rho)^{\ell+1}, & \rho \ge \rho^\prime . \end{cases}
\end{equation}
It is obvious that the solution in Eq.~\eqref{Usol} is regular both at the horizon and at infinity, and describe scalar particles at some spatial location in equilibrium with the BH (due to a force cancellation).

Considering the Proca field equations and $\epsilon=0$, we find exact general solutions which are qualitatively similar to the scalar solution:
\begin{align}
&\Upsilon=((\rho+1)\ell+1)A^{(\Upsilon)}\rho^{\ell}+((\rho+1)\ell+\rho)B^{(\Upsilon)}\rho^{-\ell-1}\ , \nonumber\\
&\psi^{(0)}=\dfrac{1}{1+\rho}\left(A^{(0)}\rho+B^{(0)}\rho^{-2}\right)\ , \nonumber\\
&\chi= A^{(\chi)}\rho^{\ell+1}+B^{(\chi)}\rho^{-\ell}\ , \nonumber\\
&\psi= \dfrac{1}{1+\rho}\left[A^{(\psi)}\rho^{\ell+1}+\dfrac{B^{(\psi)}}{\rho^{\ell}}-i\mu\left(A^{(\chi)}\dfrac{\rho^{\ell+2}(3+\rho+\ell(2+\rho))}{(3+5\ell+2\ell^2)}-B^{(\chi)}\dfrac{\rho^{1-\ell}(\ell(2+\rho)-1)}{\ell(2\ell-1)}\right)\right] .\label{extremalsol}
\end{align}
%

We now apply a scalar-vector decomposition, using invariant tensors on the Euclidean 3-space, of the form
\begin{equation}
W_0=H^{-1}\Psi\;,\;\;\;\;\;\;
\vec{W}=\nabla\Phi +\vec{V}\;, \;\;\;\;\;\;\nabla\cdot \vec{V}=0 \ ,
\end{equation}
where $W_\mu=(W_0,\vec{W})$. All vectors and differential operators are now in Euclidean 3-space. Then, $\Phi$ becomes the non-dynamical mode which disappears due to the (gauged) Lorentz condition; this condition is a consequence of the Proca equations. As such, we have (as before) a longitudinal degree of freedom, $\Psi$, and two transverse degrees of freedom, $\vec{V}$. From the time and space components of the Proca equation, these obey a Laplace and a Maxwell-Faraday like equation, respectively:
\begin{equation}
\Delta \Psi=0\;,\;\;\;\;\;\;
\nabla\times \vec{\beta}=-i\mu \nabla\Psi\;,
\end{equation}
where we have defined
\begin{equation}
\vec{\beta}=\dfrac{1}{H^2}\nabla \times \vec{V} \; .
\end{equation}
Clearly, the longitudinal mode equation can be solved by a point like source centered at any point
\begin{equation}
\Psi=\dfrac{A}{|\mathbf{x}-\mathbf{x}^\prime|}\; ,
\label{pc}
\end{equation}
and it serves as a source to the transverse modes. Furthermore, due to the linear structure, we can superpose any linear combination of solutions.

The bottom line we wish to emphasize is that the marginal (charged) Proca clouds exist in the double extremal limit and, as the scalar clouds, can be made regular at both boundaries. This can be achieved, as shown for scalar case, by constructing an inhomogeneous solution of the form \eqref{pc} with ${\bf x^\prime}$ not coinciding with the BH horizon, from two homogeneous solutions of each equation in~\eqref{extremalsol}.

When $\epsilon \neq 0$ we cannot find a closed form solution. Nevertheless, after asymptotically expanding the equations in the limit $\rho \gg 1$ we find
\begin{align}
\Upsilon&\rightarrow \rho\left(A^{(\Upsilon)}\rho^{J_{(0)}}+B^{(\Upsilon)}\rho^{-1-J_{(0)}}\right)\;, \nonumber\\
\psi^{(0)}&\rightarrow \dfrac{1}{\rho}\left(A^{(0)}\rho^{J_{(+)}}+B^{(0)}\rho^{-J_{(+)}-1}\right)\;, \\
\chi&\rightarrow A_1\rho^{J_{(-)}}+B_1\rho^{-1-J_{(-)}}+A_2\rho^{J_{(+)}}+B_2\rho^{-1-J_{(+)}}\;,\nonumber \\
\rho\,\psi&\rightarrow \hat{A}_1\rho^{2+J_{(-)}}+\hat{B}_1\rho^{1-J_{(-)}}+\hat{A}_2\rho^{2+J_{(+)}}+\hat{B}_2\rho^{1-J_{(+)}}\;,\nonumber
\end{align}
where the hatted constants $\hat{A}_i,\;\hat{B}_i$ are proportional to $A_i,\;B_i$ (the constants are unimportant) and we have defined the effective total angular momentum
\begin{equation}
J_{(S)}\equiv \sqrt{\left(\frac{1}{2}+\ell+S\right)^2-\left(\frac{\epsilon \mu}{2}\right)^2}-\frac{1}{2} \; .
\end{equation}
For the scalar field
\begin{equation}
R\rightarrow \rho\left(A^{(R)}\rho^{J_{(0)}}+B^{(R)}\rho^{-1-J_{(0)}}\right)\; .
\end{equation}
This result, justifies the existence of non-trivial solutions when the threshold condition $\mu M=qQ$ is obeyed,  even away from double extremality.

\section{Summary}
\label{Discussion}

In this chapter we have studied quasi-bound state for test field configurations of charged Proca and scalar fields in the background of the RN BH. We have found some interesting properties which are present for both types of fields and are likely to be generic for massive charged bosonic fields.

First, despite the existence of superradiant scattering of charged Proca fields by RN BHs as found in Chapter~\ref{ch:ChargedP}, we did not find any quasi-bound states in the superradiant regime, hence no superradiant instabilities due to charged Proca fields. This agrees with what occurs for charged scalar fields~\cite{Furuhashi:2004jk,Hod:2013nn,Hod:2013eea}.

Second, we have found, considering $\mu M> q Q >0$, that either decreasing the mass $\mu$  or increasing the charge $q$ as to achieve the threshold condition  $\mu M=qQ$, one gets arbitrarily long lived states, since the imaginary part of the quasi-bound state frequency tends to zero. When taking this limit the frequency becomes equal to the field mass $\omega\rightarrow \mu$, and so the states become marginally bound. Since they do not trivialize, we obtain configurations we have dubbed as marginal (charged) scalar and Proca clouds around RN BHs. But observe that these clouds are qualitatively different from those found in the Kerr case~\cite{Hod:2012px,Hod:2013zza,Herdeiro:2014goa}, and recently extended to the Kerr-Newman case~\cite{Hod:2014baa}, which are real bound states.

By analyzing the behavior of the field profiles when approaching the threshold condition, the region where the field has a large amplitude agrees well with the region where a potential well is present. The width of the well typically increases when approaching the marginally bound limit meaning that the field is large on a wider region away from the BH horizon. Since the quasi-bound state is localized away from the horizon, the gravitational and electromagnetic field interactions which are responsible for supporting the bound state should be dominated by their asymptotic Newtonian and Coulombian limit respectively. Thus the threshold  condition $\mu M=qQ$ should in fact correspond to a force balance condition between the Newtonian force and the electrostatic force.

Scalar clouds around Kerr BHs can be promoted to nonlinear hair. Indeed Kerr BHs with scalar hair exist that connect precisely to the Kerr solutions that allow the existence of the corresponding clouds~\cite{Herdeiro:2014goa,Herdeiro:2014ima,Herdeiro:2014jaa}. One may therefore ask if the existence of these marginal (charged) scalar and Proca clouds hints at the existence of nonlinear solutions of RN BHs with scalar or Proca hair. One possibility is along the lines of the discussion in~\cite{Degollado:2013eqa}, for the scalar case. It was observed therein that the marginal scalar clouds (albeit this terminology was not used therein) can be seen as the partial waves of a distribution of charged scalar particles in a no-force balance with the RN BH. This suggests the existence of multi-object nonlinear solutions -- of Majumdar-Papapetrou type~\cite{Majumdar:1947eu,Papapetrou1945}, but not necessarily multiple BHs. One other possibility, however, is that since the clouds that are regular at the horizon are partial waves with $\ell\neq 0$, the corresponding nonlinear configurations will have non-zero angular momentum and thus will be a Kerr-Newman BH with charged scalar (or Proca) hair, a possibility already anticipated in~\cite{Herdeiro:2014goa,Hod:2014baa}. Of course, the existence (or nonexistence) of any of these possible solutions, can only be decided by a fully nonlinear analysis of the Einstein-(charged)-Klein-Gordon or Einstein-(charged)-Proca systems.

%% file: scalarHD.tex
\chapter{Superradiant instabilities in higher dimensions}
\label{ch:scalarHD}


\section{Introduction}
\label{intro}
Superradiant instabilities are interesting phenomena with various applications from high energy physics to astrophysics, see a recent review~\cite{Brito:2015oca} and references therein. Most of these studies focus on four dimensional spacetimes\footnote{Exceptionally a few works were carried out in five dimensional spacetimes, see~\cite{Aliev:2008yk,Dias:2011tj}, for example.}, but a generic analysis for superradiant instabilities in higher dimensions is still missing. In order to perform such a study, we choose asymptotically AdS BHs, one of the confining mechanisms discussed in Section~\ref{sc:SR}, in which bound states of perturbation fields can be supported.

The goal of this chapter is to present a complete study of superradiant instabilities triggered by a charged scalar field interacting with a $D$-dimensional Reissner-Nordstr\"om-anti-de Sitter (RN-AdS) BH. As a spin off we shall clarify a claim, made in~\cite{Aliev:2008yk}, that, in $D=5$,  a subset of field modes -- those with odd angular momentum quantum number, $\ell$ -- do not develop superradiant instabilities. We shall show otherwise: that in fact \textit{all} $\ell$ modes in \textit{all} dimensions can develop superradiant instabilities.

%
In addition to the study of superradiant instabilities for a minimally coupled charged scalar field, other studies have been considered for charged AdS BHs. It was first observed in \cite{Gubser:2000ec,Gubser:2000mm} that for BHs obtained within $\mathcal{N}=8$ supergravity\footnote{This is a theory containing four $U(1)$ gauge fields and three real scalar fields \textit{non-minimally} coupled to the Maxwell fields.} in $D=4$, \textit{large} RN-AdS BHs are dynamically  unstable due to the existence of a tachyonic mode in the scalar field perturbations. On the other hand, in purely Einstein-Maxwell theory in $D$ dimensions, RN-AdS BHs have been argued to be stable against gravitational perturbations~\cite{Konoplya:2008rq}. Thus the existence of scalar fields with some coupling to Maxwell fields is central to the instability of~\cite{Gubser:2000ec,Gubser:2000mm}.

A qualitatively different instability of RN-AdS BHs has been discussed in the context of holographic superconductors. It occurs in the presence of a massive scalar field that may or may not be charged and leads to the formation of scalar hair around the BH \cite{Hartnoll:2008kx}. Unlike the superradiant instability, this other instability occurs even if the scalar field is not charged. In that case the, say, $D=4$ RN-AdS BH should be nearly extremal, which means its near horizon geometry has a two dimensional AdS factor; moreover, the scalar field should have a tachyonic mass above the Breitenlohner-Freedman  bound \cite{Breitenlohner:1982bm,Breitenlohner:1982jf} of the four dimensional AdS space, but below the Breitenlohner-Freedman bound of the two dimensional AdS factor in the near horizon geometry of the BH. This is why the scalar field provides an instability to the BH geometry, but not to four dimensional AdS space.

In order to investigate the superradiant instability of a charged, massive scalar field in a $D$-dimensional RN-AdS background we shall employ both an analytical and a numerical method. First, we solve the charged Klein-Gordon equation using a matching method, and obtain a solution near the BH region and another one far away from the BH. This is done for small BHs, i.e. BHs obeying $r_+\ll L$, where $r_+$ and $L$ stand for the BH event horizon and the AdS radius, respectively. The analytical quasinormal frequency for small BHs is then obtained by matching the near and far solutions in an intermediate region. We find that the relation between $\ell$ and $D$ plays a central role in determining the analytical quasinormal frequency formula. When $\ell=(p+\frac{1}{2})(D-3)$, where $p$ is a non-negative integer, the matching method fails. The reason is that the near region solution and the far region solution have different functional dependence in terms of the radial coordinate, which makes our matching impossible. Such difficulty in employing the matching method also occurs for extremal BHs as discussed in~\cite{Rosa:2009ei}, where an alternative point matching method was used. For all other values of $\ell$ and $D$, the matching method works and it may be used to show that a superradiance instability exists for all $\ell$ modes, in a region of the parameters space.

After the (approximate) analytical analysis, then we solve the charged Klein-Gordon equation numerically both to check the analytical results and to explore the special case where the analytical method fails. We find good agreement between the two methods in the regime where both are valid. For the special case $\ell=(p+\frac{1}{2})(D-3)$, the numerical results show that the superradiant instability does exist.

The structure of this chapter is organized as follows. In Section~\ref{seceq} we introduce the background geometry and the scalar field equation. In Section~\ref{secmatching} we solve the scalar field equation analytically using the matching method and obtain an analytical quasinormal frequency formula. We analyze this formula for different relations between $\ell$ and $D$, and show the reason why the matching method fails for the special case $\ell=(p+\frac{1}{2})(D-3)$ in Section~\ref{result analysis}. To confirm our analytical results and to be able to investigate if there is a superradiant instability for that special case, we resort to a numerical method to solve the Klein-Gordon equation in Section~\ref{numerical}.
Conclusions are presented in the last section.
%
\section{Background and field equation}
\label{seceq}
We consider a $D$-dimensional RN-AdS BH with the line element
\begin{equation}
ds^2=-f(r)dt^2+\dfrac{1}{f(r)}dr^2+r^2d\Omega^2_n \;,\label{metric}
\end{equation}
where $d\Omega^2_n$ is the metric on the unit $n$-sphere. In the following it will be convenient to use $n=D-2$ rather than $D$ to parameterize the spacetime dimension. The metric function $f(r)$ takes the form
\begin{equation}
f(r)=1-\dfrac{M}{r^{n-1}}+\dfrac{Q^2}{r^{2(n-1)}}+\dfrac{r^2}{L^2} \;,\label{metricfunc}
\end{equation}
where the parameters $M, Q$ and $L$ are related to the BH mass $\mathcal{M}$, charge $\mathbb{Q}$ and cosmological constant $\Lambda$ through
\begin{equation}
M=\dfrac{16\pi G\mathcal{M}}{nS_n}\;\;,Q^2=\dfrac{8\pi G\mathbb{Q}^2}{n(n-1)}\;\;,L^2=-\dfrac{n(n+1)}{2\Lambda}\;,\nonumber
\end{equation}
and the area of a unit $n$-sphere is $S_n=\frac{2\pi^{\frac{n+1}{2}}}{\Gamma(\frac{n+1}{2})}$.
The Hawking temperature is given by
\begin{equation}
T=\dfrac{1}{4\pi}\left[\frac{(n-1)M}{r_+^n}-\frac{2(n-1)Q^2}{r_+^{2n-1}}+\frac{2r_+}{L^2}\right]\;,\label{hawkingtemp}
\end{equation}
where the event horizon $r_+$ is determined as the largest root of $f(r_+)=0$. For non-extremal BHs, we have $Q<Q_c$, where the critical charge $Q_c$ corresponds to the maximal charge (i.e the charge of an  extremal BH) and is given by
\begin{equation}
Q_c\equiv r_+^{n-1}\sqrt{1+\dfrac{n+1}{n-1}\left(\dfrac{r_+}{L}\right)^2}\;.\label{critialq}
\end{equation}
The electromagnetic potential of the charged BH is
\begin{equation}
A=\left(-\sqrt{\dfrac{n}{2(n-1)}}\dfrac{Q}{r^{n-1}}+C\right)dt\;,\nonumber
\end{equation}
where the choice of the constant $C$ is a gauge choice. For instance, we should fix $C$ as
\begin{equation}
C=\sqrt{\dfrac{n}{2(n-1)}}\dfrac{Q}{r_+^{n-1}}\;,\nonumber
\end{equation}
in order to have a vanishing electromagnetic potential at the event horizon, a choice used in the context of the AdS/CFT correspondence~\cite{Horowitz:2010gk}. As argued in~\cite{Uchikata:2011zz}, however, this constant just shifts the real part of quasinormal frequency as Re$(\omega)\rightarrow$ Re$(\omega)+qC$ ($q$ is the field charge) without affecting the imaginary part of it. Therefore, as we are mostly interested in the superradiant instability, which is determined by the imaginary part of the quasinormal frequency, we take in the following $C=0$.

For a charged massive scalar field, the corresponding Klein-Gordon (K-G) equation can be written as
\begin{equation}
\dfrac{1}{\sqrt{-g}}D_\mu\left[\sqrt{-g}g^{\mu\nu}D_\nu\right]\phi-\mu^2\phi=0 \;,\label{sacalarEq}
\end{equation}
where $D_\mu=\partial_\mu -iqA_\mu$, $q$ and $\mu$ are the field charge and mass, respectively. The scalar field $\phi$ can be decomposed in terms of spherical scalar harmonics due to the spherical symmetry of the background
\begin{equation}
\phi={\rm e}^{-i\omega t}\rm{R}(r)\rm{Y}(\theta_i) \;,\label{phicecompose}
\end{equation}
where $\rm{Y}(\theta_i)$ is a scalar spherical harmonic on the $n$-sphere. Substituting the metric in Eq.~\eqref{metric} and the field decomposition in Eq.~\eqref{phicecompose} into the K-G equation~\eqref{sacalarEq}, we have
\begin{equation}
\dfrac{\Delta}{r^{n-2}}\dfrac{d}{dr}\left(\dfrac{\Delta}{r^{n-2}}\dfrac{d{\rm R}}{dr}\right)-\Big(\lambda+\mu^2r^2\Big)\Delta{\rm R}
+r^{2n}\Big(\omega+qA_t\Big)^2 {\rm R}=0 \;,\label{RadialEq}
\end{equation}
with
\begin{equation}
\Delta\equiv r^{2(n-1)}-M r^{n-1}+Q^2+\dfrac{r^{2n}}{L^2}\;,\nonumber
\end{equation}
and the $n-$dimensional spherical harmonic eigenvalue $\lambda$ is given by
\begin{equation}
\lambda\equiv \ell(\ell+n-1)\;,\nonumber
\end{equation}
where $\ell$ is the angular momentum quantum number.

For numerical convenience, we may rewrite Eq.~\eqref{RadialEq} in terms of a new function $\rm{X}$
\begin{align}
&f(r)^2\dfrac{d^2\rm X}{dr^2} + f(r)f'(r) \dfrac{d \rm X}{dr} + \left[(\omega+qA_t)^2 -f(r) \left(\dfrac{\lambda}{r^2}+\mu^2+\dfrac{nf'(r)}{2r}+\dfrac{n(n-2)f(r)}{4r^2}\right)\right] \rm X\nonumber\\&=0\;,\label{RadialEq2}
\end{align}
where ${\rm X}\equiv r^{n/2} \rm R$.

In order to determine the quasinormal frequency, by solving Eq.~\eqref{RadialEq2}, one has to impose boundary conditions. Near the event horizon, we impose an ingoing boundary condition
\begin{equation}
{\rm X} \sim {\rm e}^{-i\left(\omega-\omega_0\right)r_\ast}\;,\label{boundaryingoing}
\end{equation}
where $\omega_0\equiv -qA_t(r_+)$ and the tortoise coordinate $r_\ast$ is defined by
\begin{equation}
\dfrac{dr_\ast}{dr}=\dfrac{1}{f(r)}\;.\nonumber
\end{equation}
At infinity, we impose a decaying boundary condition
\begin{equation}
{\rm X} \sim r^{-\frac{1}{2}(1+\sqrt{4\mu^2L^2+(n+1)^2})}\; ,\label{boundarydecaying}
\end{equation}
then
\begin{equation}
{\rm R} \sim r^{-\frac{1}{2}(n+1+\sqrt{4\mu^2L^2+(n+1)^2})}\;.\label{boundarydecaying2}
\end{equation}
For a massless field in $D$ dimensions, the radial function goes, asymptotically as ${\rm R}\sim 1/r^{D-1}$, as expected for a normalizable massless scalar perturbation in $D$-dimensional AdS space.

\section{Analytic calculations}
\label{secmatching}
In this section, we present the analytic calculations of quasinormal frequencies for a charged massive scalar field in a higher dimensional RN-AdS BH. Using a standard procedure, we shall divide the space outside the event horizon into two regions: the \textit{near region}, defined by the condition $r-r_+\ll1/\omega$, and the \textit{far region}, defined by the condition $r_+\ll r-r_+$. Then, in order to perform a matching of the two solutions we consider a low frequency condition, i.e. $r_+\ll1/\omega$, and match the two solutions in an  \textit{intermediate region} defined by $r_+\ll r-r_+\ll1/\omega$. In the following analysis we focus on small AdS BH $(r_+\ll L)$. This allows us to treat the frequencies for the RN-AdS BH as a perturbation of the AdS normal frequencies.

\subsection{Near region solution}
For the near region analysis, we rewrite Eq.~\eqref{RadialEq} as
\begin{equation}
\left(n-1\right)^2\Delta\dfrac{d}{dx}\left(\Delta\dfrac{d {\rm R}}{dx}\right)-\Big(\lambda+\mu^2r^2\Big)\Delta {\rm R}
+ r^{2n}\Big(\omega+qA_t\Big)^2 {\rm R}=0 \;,\label{NearEq1}
\end{equation}
where $x\equiv r^{n-1}$. In the following we shall neglect the mass term in the first line of Eq.~\eqref{NearEq1}. This amounts to say that  $r\ll\frac{\ell}{\mu}$, which is obeyed if the Compton wave length of the scalar particles is much larger than the BH horizon size and indeed becomes the near region condition if, moreover, the scalar particle Compton wave length is much smaller than the AdS radius. Observe that the condition  $r\ll\frac{\ell}{\mu}$ fails for $\ell=0$ modes, but it turns out that even in that case the analytical results we shall obtain are in good agreement with the numerical results for sufficiently small mass (cf. Section~\ref{numerical}). Under this approximation Eq.~\eqref{NearEq1} becomes
\begin{equation}
(n-1)^2\Delta\dfrac{d}{dx}\left(\Delta\dfrac{d{\rm R}}{dx}\right)-\lambda\Delta {\rm R} + r_+^{2n}\Big(\omega+qA_t\Big)^2 \rm R=0 \;.\label{NearEq2}
\end{equation}
For convenience in the analytic calculations, one can define a new dimensionless variable
\begin{equation}
z\equiv \dfrac{x-x_+}{x-x_-}\;,\nonumber
\end{equation}
with $x_+=r_+^{n-1}$ and $x_-=r_-^{n-1}$,
where $r_+$ and $r_-$ refer to the event horizon and the Cauchy horizon, respectively.
Then Eq.~\eqref{NearEq2} can be transformed into
\begin{equation}
z\dfrac{d}{dz}\left(z\dfrac{d{\rm R}}{dz}\right)+\left[\bar{\omega}^2-\dfrac{\lambda}{(n-1)^2}\dfrac{z}{(1-z)^2}\right] \rm R=0\;,\label{NearEq}
\end{equation}
with
\begin{equation}
\bar{\omega}\equiv \dfrac{x_+^{\frac{n}{n-1}}}{(n-1)(x_+ - x_-)}\left(\omega-\sqrt{\dfrac{n}{2(n-1)}}\dfrac{qQ}{x_+}\right) \;.\label{superradiance}
\end{equation}
Observe that $\bar{\omega}<0$ for $\omega<\sqrt{\frac{n}{2(n-1)}}\frac{qQ}{x_+}$. This will be shown below to correspond to the superradiant regime.

One may now obtain a solution for Eq.~\eqref{NearEq} with ingoing boundary condition in terms of a hypergeometric function:
\begin{equation}
{\rm R} \sim z^{-i\bar{\omega}}(1-z)^\alpha F(\alpha,\alpha-2i\bar{\omega},1-2i\bar{\omega};z) \;,\label{NearSol}
\end{equation}
where
\begin{equation}
\alpha\equiv 1+\dfrac{\ell}{n-1}\;.\label{alpha}
\end{equation}
In order to match the far region solution, one must expand the near region solution, \mbox{Eq.~\eqref{NearSol}}, at large $r$. To achieve this we take the $z\rightarrow1$ limit and use the properties of the hypergeometric function~\cite{abramowitz+stegun}, then obtain
\begin{equation}
{\rm R} \sim \Gamma(1-2i\bar{\omega})\left[\dfrac{{\rm R}^{\rm near}_{1/r}}{r^{n-1+\ell}} +{\rm R}^{\rm near}_r r^{\ell}\right]
\;,\label{NearsolFar}
\end{equation}
where
\begin{align}
{\rm R}^{\rm near}_{1/r} & \equiv  \dfrac{\Gamma(1-2\alpha) (r_+^{n-1} - r_-^{n-1})^\alpha}{\Gamma(1-\alpha)\Gamma(1-\alpha-2i\bar{\omega})} \ ,\nonumber \\
{\rm R}^{\rm near}_r &\equiv \dfrac{\Gamma(2\alpha-1)(r_+^{n-1}-r_-^{n-1})^{1-\alpha}}{\Gamma(\alpha)\Gamma(\alpha-2i\bar{\omega})} \ .
\end{align}
Since the Gamma function has poles at negative integers, one observes that special care must be taken with the factor $\Gamma(1-2\alpha)/\Gamma(1-\alpha)$. Its analysis will play a role below.

\subsection{Far region solution}
In the far region, $r-r_+\gg r_+$, the BH effects can be neglected ($M\rightarrow0,\;Q\rightarrow0$) so that
\begin{equation}
\Delta\simeq r^{2n} \left(\dfrac{1}{r^2}+\dfrac{1}{L^2}\right)\; .\nonumber
\end{equation}
Then Eq.~\eqref{RadialEq} becomes
\begin{equation}
\left(1+\dfrac{r^2}{L^2}\right)\dfrac{d^2{\rm R}}{dr^2}+\left(\dfrac{n}{r}+\dfrac{(n+2)r}{L^2}\right)\dfrac{d {\rm R}}{dr}
+\left[\omega^2\left(1+\dfrac{r^2}{L^2}\right)^{-1}-\left(\dfrac{\lambda}{r^2}+\mu^2\right)\right]\rm R=0\;.\label{fareq1}
\end{equation}
Defining a new variable
\begin{equation}
y\equiv 1+\dfrac{r^2}{L^2}\;,\nonumber
\end{equation}
Eq.~\eqref{fareq1} becomes
\begin{equation}
y\left(1-y\right)\dfrac{d^2{\rm R}}{dy^2}+\left(1-\dfrac{n+3}{2}y\right)\dfrac{d {\rm R}}{dy}-\left[\dfrac{\omega^2L^2}{4y}-\dfrac{\mu^2L^2}{4}+\dfrac{\lambda}{4(1-y)}\right]\rm R=0\;.\label{fareq2}
\end{equation}
The above equation has a hypergeometric equation structure, which can be shown explicitly through the transformation
\begin{equation}
{\rm R}=y^{\frac{\omega L}{2}}(1-y)^{\frac{\ell}{2}}F(a,b;c;y) \;,\nonumber
\end{equation}
with parameters
\begin{align}
&a\equiv \dfrac{n+1}{4}+\dfrac{\omega L}{2}+\dfrac{\ell}{2}+\dfrac{1}{2}\sqrt{\mu^2L^2+\left(\frac{n+1}{2}\right)^2}\;,\nonumber\\
&b\equiv \dfrac{n+1}{4}+\dfrac{\omega L}{2}+\dfrac{\ell}{2}-\dfrac{1}{2}\sqrt{\mu^2L^2+\left(\frac{n+1}{2}\right)^2}\;,\nonumber\\
&c\equiv 1+\omega L\;.\label{hyperparameter}
\end{align}
Then considering a decaying boundary condition at infinity given in Eq.~\eqref{boundarydecaying2}, one finds a solution for Eq.~\eqref{fareq2} in the form
\begin{equation}
{\rm R}\thicksim (1-y)^{\frac{\ell}{2}}y^{\frac{\omega L}{2}-a} F(a,1+a-c;1+a-b;\frac{1}{y})\;.\label{farsolution}
\end{equation}
To achieve the small $r$ behavior of Eq.~\eqref{farsolution}, making the transformation $\frac{1}{y}\rightarrow1-y$ and using properties of the hypergeometric function~\cite{abramowitz+stegun}, we obtain
\begin{equation}
{\rm R} \sim  \Gamma(1+a-b)\left[\dfrac{{\rm R}^{\rm far}_{1/r}}{r^{n-1+\ell}} +{\rm R}^{\rm far}_r r^{\ell}\right]   \;,\label{farsolution@near}
\end{equation}
where
\begin{equation}
{\rm R}^{\rm far}_{1/r} \equiv  \dfrac{\Gamma(\ell+\frac{n-1}{2})L^{n-1+\ell}}{\Gamma(a)\Gamma(1+a-c)} \;,\;\;\;\;\;\;
{\rm R}^{\rm far}_r \equiv \dfrac{\Gamma(-\ell-\frac{n-1}{2})L^{-\ell}}{\Gamma(1-b)\Gamma(c-b)}\ .
\label{far_branches}
\end{equation}
The solution in Eq.~\eqref{farsolution@near} is for pure AdS. Regularity of the above solution at the origin $(r=0)$ of AdS requires\footnote{We remark that, alternatively, one can also demand $\Gamma(a)=0$, which gives the negative AdS spectrum. Without loss of generality, we only consider the positive spectrum here.}
\begin{equation}
\Gamma(1+a-c)=\infty\  \Rightarrow  1+a-c=-N ,\nonumber
\end{equation}
which gives the discrete spectrum
\begin{equation}
\omega_{N} L=2N+\frac{n+1}{2}+\ell+\sqrt{\mu^2L^2+\left(\frac{n+1}{2}\right)^2}\;,\label{AdSfrequency}
\end{equation}
where $N$ is a non-negative integer.
Observe that the AdS frequencies remain real even for tachyonic modes when $0>\mu^2\ge -(n+1)^2/(4L^2)$. This is the well known Breitenlohner-Freedman bound already discussed in the introduction. In particular one may see that the bound is more negative for higher dimensional spaces. This is the reason why one may violate the bound for two dimensional AdS but obey it for four dimensional AdS. 

When the BH effects are taken into account, a correction to the frequency (which may be complex) will be generated
\begin{equation}
\omega=\omega_N+i\delta\;,
\end{equation}
where the real part of $\delta$ is used to describe the damping of the quasinormal modes.
Then, for small BHs, using the approximation $1/\Gamma(-N+\epsilon)\simeq (-1)^NN!\epsilon$ for small $\epsilon$, the first term inside the bracket of Eq.~\eqref{farsolution@near} becomes
\begin{equation}
{\rm R}^{\rm far}_{1/r} = (-1)^{N+1} i\delta N! \dfrac{\Gamma(\ell+\frac{n-1}{2})L^{n+\ell}}{2\Gamma(a)} \;.\nonumber
\end{equation}
Finally, observe that there appears to be extra poles in ${\rm R}_r^{\rm far}$, Eq.~\eqref{far_branches}, due to the Gamma function $\Gamma(-\ell-\frac{n-1}{2})$ for odd $n$. In the ${\rm R}_r^{\rm far}$ expression, however, due to Eq.~\eqref{AdSfrequency}, $\Gamma(1-b)=\Gamma(-\ell-\frac{n-1}{2}-N)$, which cancels the former poles.

\subsection{Overlap region}
To match the near region solution~(\ref{NearsolFar}) and the far region solution~(\ref{farsolution@near}) in the intermediate region, we impose the matching condition ${\rm R}^{\rm near}_r{\rm R}^{\rm far}_{1/r}={\rm R}^{\rm far}_r{\rm R}^{\rm near}_{1/r}$. Then $\delta$ can be obtained perturbatively
\begin{align}
\delta=&(-1)^{N}2i\dfrac{(r_+^{n-1}-r_-^{n-1})^{2\alpha-1}}{N!L^{2\ell+n}} \dfrac{\Gamma(1-2\alpha)\Gamma(\alpha)}{\Gamma(2\alpha-1)\Gamma(1-\alpha)} \dfrac{\Gamma(a)}{\Gamma(1-b)\Gamma(c-b)} \dfrac{\Gamma(-\ell-\frac{n-1}{2})}{\Gamma(\ell+\frac{n-1}{2})} \nonumber\\ &\times \dfrac{\Gamma(\alpha-2i\bar{\omega})}{\Gamma(1-\alpha-2i\bar{\omega})}\;.\label{imaginarypart}
\end{align}
%

\section{Analytical result analysis}
\label{result analysis}
To analyze Eq.~\eqref{imaginarypart}, we shall simplify the Gamma functions therein. Firstly, the following combination, which is independent of the relation between $\ell$ and $n$, can be simplified as
\begin{align}
\dfrac{\Gamma(a)}{\Gamma(1-b)\Gamma(c-b)}\dfrac{\Gamma\left(-\ell-\frac{n-1}{2}\right)}{\Gamma\left(\ell+\frac{n-1}{2}\right)}
=&\dfrac{(-1)^N}{\Gamma\left(\ell+\frac{n-1}{2}\right)} \dfrac{\Gamma\left(N+\frac{n+1}{2}+\ell+\sqrt{\mu^2L^2+(\frac{n+1}{2})^2}\right)}{\Gamma\left(N+1+\sqrt{\mu^2L^2+(\frac{n+1}{2})^2}\right)}\nonumber\\
&\times\prod_{k=1}^{N}\left(\ell+\frac{n-1}{2}+k\right)\;.\nonumber
\end{align}
Then one has to consider the following cases separately, because the simplification for the other Gamma functions in Eq.~\eqref{imaginarypart} depends on the relation between $\ell$ and $n$.

\subsection{$\ell$ is an integer multiple of $(n-1)$}
For this case we can write $\ell=p(n-1)$, where $p$ is a non-negative integer.
Then, the corresponding Gamma functions in Eq.~\eqref{imaginarypart} can be simplified to
\begin{equation}
\dfrac{\Gamma(1-2\alpha)\Gamma(\alpha)}{\Gamma(2\alpha-1)\Gamma(1-\alpha)}=\dfrac{(-1)^{p+1}}{2}\dfrac{(p!)^2}{(2p)!(2p+1)!}\;,\nonumber
\end{equation}
\begin{equation}
\dfrac{\Gamma(\alpha-2i\bar{\omega})}{\Gamma(1-\alpha-2i\bar{\omega})}=(-1)^{p+1} 2i\bar{\omega} \prod_{k^{\prime}=1}^p(k^{\prime 2}+4\bar{\omega}^2)\;.\nonumber
\end{equation}
Therefore, Eq.~\eqref{imaginarypart} becomes
\begin{align}
\delta=&-2\bar{\omega}\dfrac{(r_+^{n-1}-r_-^{n-1})^{1+\frac{2\ell}{n-1}}}{N! L^{2\ell+n}\Gamma\left(\ell+\frac{n-1}{2}\right)}\dfrac{(p!)^2}{(2p)!(2p+1)!}
\dfrac{\Gamma\left(N+\frac{n+1}{2}+\ell+\sqrt{\mu^2L^2+(\frac{n+1}{2})^2}\right)}{\Gamma\left(N+1+\sqrt{\mu^2L^2+(\frac{n+1}{2})^2}\right)}\nonumber\\ &\times\prod_{k=1}^N\left(\ell+\frac{n-1}{2}+k\right) \prod_{k^{\prime}=1}^p (k^{\prime 2}+4\bar{\omega}^2)\;.\label{case1}
\end{align}
This equation shares a similar structure to the corresponding result in $D=4$. From the definition of $\bar{\omega}$ in Eq.~\eqref{superradiance}, we find that in the superradiant regime, $\bar{\omega}<0$ which implies $\delta>0$. In this superradiant regime the wave function of the scalar field will grow with time which means the BH is unstable. Moreover, from Eq.~\eqref{superradiance}, one may get a condition for the onset of the superradiant instability, i.e.
\begin{equation}
\dfrac{Q}{Q_c} > \sqrt{\frac{2(n-1)}{n}} \dfrac{\omega_N}{q}\;,\label{onset1}
\end{equation}
where $\omega_N$ is given in Eq.~\eqref{AdSfrequency}. For a massless field, Eq.~\eqref{onset1} simplifies to
\begin{equation}
\dfrac{Q}{Q_c} > \sqrt{\frac{2(n-1)}{n}} \dfrac{2N+n+1+\ell}{q L}\;.\label{onset2}
\end{equation}

\subsection{$\ell$ is not an integer multiple of $(n-1)$}
For this case, the corresponding Gamma function in Eq.~\eqref{imaginarypart} can be simplified as
\begin{equation}
\dfrac{\Gamma(1-2\alpha)\Gamma(\alpha)}{\Gamma(2\alpha-1)\Gamma(1-\alpha)}=-\dfrac{1}{2\cos\frac{\pi \ell}{n-1}}\dfrac{\Gamma^2(1+\frac{\ell}{n-1})}{\Gamma(1+\frac{2\ell}{n-1})\Gamma(2+\frac{2\ell}{n-1})}\;.\nonumber
\end{equation}
If $\dfrac{\ell}{n-1}\neq p+\frac{1}{2}$, then cos$\frac{\pi\ell}{n-1}\neq0$, and the parameter $\delta$ becomes complex (not simply real as in the previous case).
In this case the real part of $\delta$ reflects the instability, which is given by
\begin{align}
\mbox{Re} \delta=&-2\bar{\omega}\dfrac{(r_+^{n-1}-r_-^{n-1})^{1+\frac{2\ell}{n-1}}}{N! L^{2\ell+n}} \dfrac{\Gamma^4(1+\frac{\ell}{n-1})}{\Gamma(1+\frac{2\ell}{n-1})\Gamma(2+\frac{2\ell}{n-1})\Gamma(\ell+\frac{n-1}{2})}\prod_{k=1}^N(\ell+\frac{n-1}{2}+k) \nonumber\\
&\times\dfrac{\Gamma(N+\frac{n+1}{2}+\ell+\sqrt{\mu^2L^2+(\frac{n+1}{2})^2})}{\Gamma(N+1+\sqrt{\mu^2L^2+(\frac{n+1}{2})^2})}\;,\label{case2}
\end{align}
where we have expanded the terms $\Gamma(x-2i\bar{\omega})$ around small $\bar{\omega}$ to clearly distinguish the superradiant regime. Thus, when $\bar{\omega}<0$, we obtain Re$\delta$ $>0$ which implies that the BH is also unstable, and the corresponding onset of such instability is governed by Eq.~\eqref{onset1} for a massive field and Eq.~\eqref{onset2} for a massless field.

If $\frac{\ell}{n-1}=p+\frac{1}{2}$, the matching method fails; a similar situation occurs for extremal Kerr BHs~\cite{Rosa:2009ei}. In order to make this point clear, we can do the following analysis. First, from the definition of $\alpha$ in Eq.~\eqref{alpha} and the condition $\frac{\ell}{n-1}=p+\frac{1}{2}$, one observes that the first expansion term inside the brackets of  Eq.~\eqref{NearsolFar} is divergent, which means that we cannot expand Eq.~\eqref{NearSol} into Eq.~\eqref{NearsolFar} anymore when $\frac{\ell}{n-1}=p+\frac{1}{2}$. Alternatively, using a property of the hypergeometric function~\cite{abramowitz+stegun}, we shall expand Eq.~\eqref{NearSol} as
\begin{equation}
{\rm R} \sim \Gamma(1-2i\bar{\omega})\left[-\dfrac{(r_+^{n-1} - r_-^{n-1})^\alpha \zeta}{\Gamma(1-\alpha)\Gamma(1-\alpha-2i\bar{\omega})\Gamma(2\alpha)}\dfrac{1}{r^{n-1+\ell}}+\dfrac{\Gamma(2\alpha-1)(r_+^{n-1}-r_-^{n-1})^{1-\alpha}}{\Gamma(\alpha)\Gamma(\alpha-2i\bar{\omega})}r^{\ell}\right]\;,
\label{NearsolFar2}
\end{equation}
with
\begin{equation}
\zeta=\log\left(\dfrac{r_+^{n-1}-r_-^{n-1}}{r^{n-1}}\right)+\gamma+\psi(\alpha)-\psi(2\alpha)+\psi(\alpha-2i\bar{\omega})\;,\nonumber
\end{equation}
where $\gamma$ is the Euler constant and $\psi(x)$ denotes the digamma function.
Because the $\log r$ term is associated with distinct powers of $r$, it is impossible to match Eqs.~\eqref{farsolution@near} and \eqref{NearsolFar2}. For this case, we have to resort to a numerical solution, which is discussed in the next section.

\section{Numerical results}
\label{numerical}
In order to confirm the above analytical results and to calculate the quasinormal frequencies for the special cases $\frac{\ell}{n-1}=p+\frac{1}{2}$ where the analytical method fails, we shall solve, in this section, Eq.~\eqref{RadialEq2} numerically. We use a direct numerical integration method to obtain the quasinormal frequency of the BH. To do so, taking the boundary conditions near the horizon in Eq.~\eqref{boundaryingoing} and at infinity in Eq.~\eqref{boundarydecaying}, we expand the radial function near the horizon as
\begin{equation}
{\rm X} \sim e^{-i(\omega-\omega_0)r_\ast} \sum_{j=0}^{\infty}\alpha_j(r-r_+)^j\;,\label{nearHExp}
\end{equation}
and at infinity as
\begin{equation}
{\rm X} \sim r^{-\frac{1}{2}(1+\sqrt{4\mu^2L^2+(n+1)^2})} \sum_{j=0}^{\infty}\frac{\beta_j}{r^j}\;.\label{infExp}
\end{equation}
The series expansion coefficients can be derived directly after inserting these expansions into Eq.~(\ref{RadialEq2}).
We use the series expansion near the horizon Eq.~\eqref{nearHExp} to initialize the radial system Eq.~\eqref{RadialEq2} from a point $r_s$ which is close to $r_+$ through the relation $r_s=(1+0.01)r_+$, and integrate the radial system outwards up to a radial value $r_m$. Similarly we can also use Eq.~\eqref{infExp} as initial condition to integrate the radial system inward from $r_l=1000r_+$ down to $r_m$. Then we have two solutions at an intermediate radius $r_m$, and these two solutions are linearly dependent if their Wronskian vanishes at $r_m$. Using a secant method one can solve $W(\omega,r_m)=0$ iteratively to look for the quasinormal frequency of the BH. We also varied $r_s$, $r_m$ and $r_l$ to check the numerical accuracy.

We list some numerical results in Tables~\ref{3DRNAdS}-\ref{compmass2}. Note that all physical quantities are normalized by the AdS radius $L$ and we set $L=1$. In the first three tables, we focus on the fundamental modes of massless fields because they are typically the most unstable modes. To check the mass effect on the validity of the analytical formulas, we also consider $\mu=0.5$ and $\mu=3.0$ in the last two tables. As a check of our numerical method, we have calculated the quasinormal frequencies for small Schwarzschild-AdS BHs and we obtained results which are in good agreement with those reported in~\cite{Konoplya:2002zu}.

In order to address the special case for which the analytical method fails, i.e when $\ell=(p+\frac{1}{2})(n-1)$, we chose $n=3$ (five dimensional spacetime) and $\ell=1$, corresponding to $p=0$ in our condition. The results are shown in Table~\ref{3DRNAdS}, with $r_+=0.1$, field mass $\mu=0$ and field charge $q=8$. It shows clearly that a superradiant instability appears when $Q/Q_c$ satisfies the condition in Eq.~\eqref{onset2}. Moreover, we also list numerical results for the $\ell=0$ mode in Table~\ref{3DRNAdS}. It shows that the frequencies of the odd modes ($\ell=1$) and even modes ($\ell=0$) have a similar behavior; in other words, there is nothing special for odd modes.
\begin{table}
\caption{\label{3DRNAdS} Frequencies of the fundamental modes with different $\ell$ for a BH with $r_+=0.1$, $q=8$, $\mu=0$ in $D=5$.}
\begin{center}
\begin{tabular*}{\textwidth}{@{\extracolsep{\fill}} l l l }
\hline
\hline
$Q/Q_c$ & $\ell$=0 & $\ell$=1 \\
\hline
0.1 & 3.958 - 1.335$\times 10^{-2}$ i & 4.978 - 2.689$\times 10^{-4}$ i\\
0.3 & 3.997 - 6.435$\times 10^{-3}$ i & 4.998 - 1.367$\times 10^{-4}$ i\\
0.5 & 4.030 - 1.522$\times 10^{-3}$ i & 5.014 - 5.053$\times 10^{-5}$ i\\
0.7 & 4.058 + 1.996$\times 10^{-3}$ i & 5.028 - 2.596$\times 10^{-6}$ i\\
0.8 & 4.070 + 3.198$\times 10^{-3}$ i & 5.034 + 7.524$\times 10^{-6}$ i\\
0.9 & 4.081 + 3.954$\times 10^{-3}$ i & 5.040 + 9.597$\times 10^{-6}$ i\\
\hline
\hline
\end{tabular*}
\end{center}
\end{table}

To confirm the validity of the analytical quasinormal frequency formulas in Eqs.~\eqref{case1} and \eqref{case2}, we also compare some analytical results with numerical data in Tables~\ref{comp1}-\ref{compmass2}. In Table~\ref{comp1}, we present analytical results obtained from Eq.~\eqref{case1} and numerical results with $r_+=0.01$, $q=6$ for the $\ell=0$ massless fundamental mode in five dimensional spacetimes. They show good agreement; the difference is smaller than 1\%. In Table~\ref{comp2}, we present analytical results obtained from Eq.~\eqref{case2} and numerical data for the $\ell=1$ fundamental mode with $r_+=0.01$, $q=10$, $\mu=0$ in $D=6$, and they show good agreement as well. From these two tables, we confirm the validity of the analytical matching method for \mbox{$\mu=0$}. Results for non-zero mass are reported in Tables~\ref{compmass1} and~\ref{compmass2}. Two conclusions may be drawn from these tables. First, as the mass increases the agreement between the analytical and numerical method becomes worse. This is expected in view of the discussion of the approximation employed in Section~\ref{secmatching}. Second, as the mass increases, the mode with $Q/Q_c=0.9$ becomes stable. This is in agreement with Eq.~\eqref{onset1} since, for the parameters in Table~\ref{compmass2}, superradiance is only expected for $Q/Q_c \gtrsim 1.1$.
\begin{table}[h]
\caption{\label{comp1} Comparison of the frequencies for the $\ell=0$ fundamental modes of a BH with $r_+=0.01$, $q=6$, $\mu=0$ in $D=5$.}
\begin{tabular*}{\textwidth}{@{\extracolsep{\fill}} l l l }
\hline
\hline
$Q/Q_c$ & Im$(\omega)$ (numerical) & Im$(\omega)$ (analytical) \\
\hline
0.1 & -1.053$\times 10^{-5}$  & -1.0441$\times 10^{-5}$ \\
0.3 & -7.369$\times 10^{-6}$  & -7.3230$\times 10^{-6}$ \\
0.5 & -4.222$\times 10^{-6}$  & -4.2050$\times 10^{-6}$ \\
0.7 & -1.088$\times 10^{-6}$  & -1.0870$\times 10^{-6}$ \\
0.9 & 2.023$\times 10^{-6}$  & 2.0310$\times 10^{-6}$ \\
\hline
\hline
\end{tabular*}
\end{table}
%
%
\begin{table}[h]
\caption{\label{comp2} Comparison of the frequencies for the $\ell=1$ fundamental modes of a BH with $r_+=0.01$, $q=10$, $\mu=0$ in $D=6$.}
\begin{tabular*}{\textwidth}{@{\extracolsep{\fill}} l l l }
\hline
\hline
$Q/Q_c$ & Im$(\omega)$ (numerical) & Im$(\omega)$ (analytical) \\
\hline
0.1 & -4.377$\times 10^{-11}$  & -4.3678$\times 10^{-11}$ \\
0.3 & -2.830$\times 10^{-11}$  & -2.8274$\times 10^{-11}$ \\
0.5 & -1.342$\times 10^{-11}$  & -1.3418$\times 10^{-11}$ \\
0.7 & -1.538$\times 10^{-12}$  & -1.5371$\times 10^{-12}$ \\
0.8 & 2.283$\times 10^{-12}$  & 2.2846$\times 10^{-12}$ \\
0.9 & 3.778$\times 10^{-12}$  & 3.7844$\times 10^{-12}$ \\
\hline
\hline
\end{tabular*}
\end{table}
%
%
\begin{table}[h]
\caption{\label{compmass1} Comparison of the frequencies for the $\ell=0$ fundamental modes of a BH with $r_+=0.01$, $q=6$, $\mu=0.5$ in $D=5$.}
\begin{tabular*}{\textwidth}{@{\extracolsep{\fill}} l l l }
\hline
\hline
$Q/Q_c$ & Im$(\omega)$ (numerical) & Im$(\omega)$ (analytical) \\
\hline
0.1 & -1.093$\times 10^{-5}$  & -1.0844$\times 10^{-5}$ \\
0.3 & -7.711$\times 10^{-6}$  & -7.6617$\times 10^{-6}$ \\
0.5 & -4.498$\times 10^{-6}$  & -4.4797$\times 10^{-6}$ \\
0.7 & -1.300$\times 10^{-6}$  & -1.2977$\times 10^{-6}$ \\
0.9 &  1.878$\times 10^{-6}$  & 1.8842$\times 10^{-6}$ \\
\hline
\hline
\end{tabular*}
\end{table}
%
%
\begin{table}
\caption{\label{compmass2} Comparison of the frequencies for the $\ell=0$ fundamental modes of a BH with $r_+=0.01$, $q=6$, $\mu=3.0$ in $D=5$.}
\begin{tabular*}{\textwidth}{@{\extracolsep{\fill}} l l l }
\hline
\hline
$Q/Q_c$ & Im$(\omega)$ (numerical) & Im$(\omega)$ (analytical) \\
\hline
0.1 & -2.379$\times 10^{-5}$  & -2.3423$\times 10^{-5}$ \\
0.3 & -1.886$\times 10^{-5}$  & -1.8637$\times 10^{-5}$ \\
0.5 & -1.399$\times 10^{-5}$  & -1.3850$\times 10^{-5}$ \\
0.7 & -9.155$\times 10^{-6}$  & -9.0632$\times 10^{-6}$ \\
0.9 & -4.321$\times 10^{-6}$  & -4.2765$\times 10^{-6}$ \\
\hline
\hline
\end{tabular*}
\end{table}

\section{Summary}
\label{discussion}
We have studied  the superradiant instability of small charged AdS BHs in $D$ dimensions, in the presence of a charged scalar field. Very recently our result has been generalized to the $D$-dimensional singly rotating Myers-Perry BHs, where the same conclusion was drawn, i.e. the superradiant instability exist for all $\ell$ modes in \textit{all} dimensions~\cite{Aliev:2015wla,Delice:2015zga}.

First, we solved the Klein-Gordon equation for a charged scalar field in charged AdS BHs with a standard matching method. We found that the relation between the angular momentum quantum number $\ell$ and the spacetime dimension $D$ plays an important role in determining the analytical quasinormal frequency formula. When $\ell=p(D-3)$, for a non-negative integer $p$, we found that the quasinormal frequencies of the small RN-AdS BHs have only an imaginary correction to the AdS normal frequencies. This is the case for all modes (i.e. all $\ell$) in $D=4$, even $\ell$ in $D=5$, $\ell=0,3,6,9,\dots$ in $D=6$, $\ell=0,4,8,12,\dots$ in $D=7$ and so on.

A more subtle case occurs when $\ell=(p+\frac{1}{2})(D-3)$. For this case the matching method fails because a $\log r$ term appears in the near region solution --~Eq.~\eqref{NearsolFar2} -- which cannot be matched to Eq.~\eqref{farsolution@near}. Failure to observe this limitation has led to a claim that odd $\ell$ modes in $D=5$ did not exhibit superradiance~\cite{Aliev:2008yk}. Here we have shown otherwise that the superradiant instability indeed exists using a numerical method which is mandatory for analyzing this case, in view of the invalidity of the matching method. A similar conclusion applies to all cases defined by  $\ell=(p+\frac{1}{2})(D-3)$, i.e, odd $\ell$ in $D=5$, $\ell=2,6,10,14,\dots$ in $D=7$ and so on. Observe that this case can only occur in odd dimensions.

Finally, all other cases have a complex correction to the AdS normal frequencies, i.e. the real part of the frequency is also shifted.

Our analytic results show good agreement with the numerical results in Section~\ref{numerical}. In particular a central conclusion is that all $\ell$ modes in all dimensions, for sufficiently large field charge $q$ display superradiance. Moreover, in $D=4$, the dependence of the instability on the various parameters seems to be in qualitative agreement with the study of cavity BHs in $D=4$~\cite{Herdeiro:2013pia,Hod:2013fvl,Degollado:2013bha}, and it would be interesting to make a more detailed comparison between the two cases.

Let us close this study with two questions. First, is there a simple pattern for the behaviour of the frequencies as $D\rightarrow \infty$? A preliminary analysis could not unveil a simple formula. Finding such behavior would be relevant in view of the recent interest on General Relativity in the large $D$ limit \cite{Emparan:2013moa}. Second, can one follow this instability numerically into the non-linear regime? It has been recently shown that in four dimensions the end-point of the instability is a hairy charged AdS BH~\cite{Bosch:2016vcp}. It would be interesting to generalize such study to higher dimensions, to display how the properties of hairy BHs depend on the spacetime dimension.

%% file: KerrAdS.tex
\chapter{Maxwell perturbations on Kerr-AdS black holes}
\label{ch:KerrAdS}


\section{Introduction}
\label{Maxse:intro}

The global structure of asymptotically AdS spacetimes allows interesting novel features, as compared to asymptotically flat spacetimes. For instance, when a rotating BH exists in the bulk of an asymptotically AdS spacetime, superradiant instabilities can be triggered by a massless field, in contrast to the asymptotically flat case wherein such instabilities only arise for massive fields. Recently scalar~\cite{Uchikata:2009zz} and gravitational perturbations~\cite{Cardoso:2013pza} on Kerr-AdS BHs have been addressed, and have been shown to exhibit a rich physics.

In this chapter we complete the analysis of bosonic fields on this background by studying Maxwell perturbations. To do so, one has to assign physically relevant boundary conditions which depend on the specific problem. In the context of quasinormal modes in asymptotically AdS BHs, the most studied perturbations are those of scalar fields, for which \textit{field vanishing} boundary conditions are usually imposed, see e.g.~\cite{Uchikata:2009zz}. For other spin fields, the problem has only been partly addressed. The quasinormal modes for the Maxwell field and gravitational field on Schwarzschild-AdS BHs have been obtained using the Regge-Wheeler method~\cite{Regge:1957td}, instead of the Teukolsky equation, in~\cite{Cardoso:2001bb,Cardoso:2003cj}, exploring the spherical symmetry of the background. Furthermore, these works impose field vanishing boundary conditions. For non-spherically symmetric backgrounds, like in Kerr-AdS BHs, one must, however, use the Teukolsky formalism  and, since this formalism uses a different set of variables, it is not obvious how to impose boundary conditions for non-zero spin fields. Recently, superradiant instabilities of the gravitational field on Kerr-AdS BHs have been studied~\cite{Cardoso:2013pza} with boundary conditions chosen as to preserve the asymptotic global AdS structure of the background~\cite{Dias:2013sdc}. The general principle to impose boundary conditions for arbitrary spin fields on Kerr-AdS BHs, however, is still missing.

The AdS boundary may be regarded as a perfectly reflecting mirror, in the sense that no energy flux can cross it. We will take this viewpoint as our basic principle to impose boundary conditions for linear perturbations of asymptotically AdS spacetimes. It suggests taking \textit{vanishing energy flux} (VEF) boundary conditions, which should be contrasted to the \textit{field vanishing} boundary conditions we mentioned before. We will first illustrate how this simple physical principle, can lead to two different Robin boundary conditions, using the Maxwell field as an example.

These boundary conditions are then used to study quasinormal modes, superradiantly unstable modes, and vector clouds. To address superradiant instabilities triggered by the Maxwell field on Kerr-AdS BHs, we use both an analytical matching scheme as well as a numerical method to explore the problem. The former method provides an intuitive way to understand how these two boundary conditions produce different instabilities, in the small BH and slow rotation regime. The latter method provides a technique to understand the problem in a larger region of the parameter space. We find that Maxwell fields can trigger stronger instabilities than scalar fields~\cite{Uchikata:2009zz} in the superradiant regime, for both boundary conditions.

Stationary clouds~\cite{Hod:2012px} are bound \mbox{state} solutions of test fields on a rotating background, at the linear level. They exist at the threshold of the superradiant instabilities triggered by that test field. Recently, a considerable number of studies of such clouds has \mbox{appeared} in the literature, mostly in asymptotically flat spacetimes~\cite{Hod:2012px,Hod:2013zza,Shlapentokh-Rothman:2013ysa,Herdeiro:2014goa,Hod:2014baa,Benone:2014ssa,Hod:2014sha,Hod:2014pza,Herdeiro:2014pka,Benone:2014nla,Hod:2015ota,Wilson-Gerow:2015esa,Herdeiro:2015gia,Hod:2015ynd}, but also in asymptotically AdS \mbox{spacetimes}~\cite{Dias:2011at,Cardoso:2013pza}. Most of these studies have addressed scalar field clouds\footnote{Superradiance onset curves for gravitational perturbations on Kerr-AdS, which --  as the existence lines for stationary clouds -- identify the backgrounds supporting the zero-mode of the perturbation,  have been studied in~\cite{Cardoso:2013pza}, and the corresponding ``hairy'' BH solutions have been constructed in~\cite{Dias:2015rxy}.} (even though \textit{marginal} clouds have been considered for a charged Proca field on a charged BH background in Chapter~\ref{ch:ChargedClouds}). Here we perform a study of Maxwell clouds, which can exist around rotating BHs in asymptotically AdS spacetimes. As a comparison with the Maxwell clouds on Kerr-AdS BHs we also consider scalar clouds in the same background.

It was proposed in~\cite{Herdeiro:2014goa,Herdeiro:2014ima} that the existence of stationary clouds of a given test field, as a \textit{zero-mode} of the superradiant instability, indicates the existence of new families of ``hairy'' BH solutions, at fully nonlinear level, such as the Kerr BHs with scalar hair found (numerically) in ~\cite{Herdeiro:2014goa}, whose existence was recently formally proved~\cite{Chodosh:2015oma}. It is an open issue if these hairy BHs may be formed dynamically, as the end point of the instability. Interesting evidence in this direction was reported recently~\cite{Dolan:2015dha} for the case of Reissner-Nordstr\"om BHs in a cavity, following the earlier discussion of superradiant instabilities in this setup~\cite{Herdeiro:2013pia,Hod:2013fvl,Degollado:2013bha}. The existence of Maxwell clouds on Kerr-AdS BHs has, therefore, the interesting implication that new families of solutions of charged rotating BHs exist, within the Einstein-Maxwell-AdS system, besides the well known Kerr-Newman-AdS family, and branching off from the latter.




The structure of this chapter is as follows. In Section~\ref{Maxsec:eqs} we present the Teukolsky equations for the Maxwell field on Kerr-AdS BHs. Then the corresponding boundary conditions are studied in Section~\ref{Maxsec:BC}, by requiring VEF. In Section~\ref{Maxsec:AM}, a standard analytical matching method is developed to study quasinormal modes in the small BH and slow rotation regime. In Section~\ref{Maxsec:NM}, we introduce two numerical methods, to solve the radial and angular Teukolsky equations, respectively, in a larger parameter space. In Section~\ref{Maxsec:SAdS}, we take the Schwarzschild-AdS BH as an example to show, even in this simpler case, that there is a new branch of quasinormal modes which has not been explored yet. In Section~\ref{Maxsec:instability}, numerical results for quasinormal modes, superradiant unstable modes and vector clouds are presented. As a comparison, scalar clouds are studied as well. We summarize our results in the last section.

\section{Teukolsky equations of the Maxwell field}
\label{Maxsec:eqs}
%
In the Newman-Penrose formalism, the Maxwell fields are described in terms of three complex scalars, two of which are independent. These two scalars are denoted by $\phi_0$ and $\phi_2$, and can be expanded as
\begin{equation}
\phi_0=R_{+1}(r)S_{+1}(\theta)\;,\;\;\;\;\;\;\phi_2=\dfrac{1}{2(\bar{\rho}^\ast)^2}R_{-1}(r)S_{-1}(\theta)\;,\label{fielddeccom}
\end{equation}
where $\bar{\rho}^\ast=r-ia\cos\theta$, and the radial functions $R_{\pm1}$ obey Eqs.~\eqref{Rpluseq} and~\eqref{Rminuseq} while the angular functions $S_{\pm1}$ obey Eqs.~\eqref{Spluseq} and~\eqref{Sminuseq}.

In the following we present these equations (both the radial and angular equations) explicitly, governing a spin $s (s=\pm1)$ perturbation.
\\
The radial equation is
\begin{equation}
\Delta_r^{-s}\dfrac{d}{dr}\left(\Delta_r^{s+1}\dfrac{d R_{s}(r)}{dr}\right)+H(r)R_{s}(r)=0\;,\label{radialeq}
\end{equation}
with
\begin{eqnarray}
H(r)=\dfrac{\Xi^2K^2-is\Xi K\Delta_r^\prime}{\Delta_r}+2is\Xi K^\prime+\dfrac{s+|s|}{2}\Delta_r^{\prime\prime}
+\dfrac{a^2}{L^2}-\lambda\;,\nonumber
\end{eqnarray}
where $K$ is given in Eq.~\eqref{KHeq}.
The angular equation is
\begin{equation}
\dfrac{d}{du}\left(\Delta_u\dfrac{dS_{lm}}{du}\right)+A(u)S_{lm}=0\;,\label{angulareq}
\end{equation}
with $u=\cos\theta$, and
\begin{equation}
A(u)=-\dfrac{K_u^2}{\Delta_u}-4smu\dfrac{\Xi}{1-u^2}+\lambda-|s|-2(1-u^2)\dfrac{a^2}{L^2}\;,\nonumber
\end{equation}
where
\begin{align}
K_u&=\Big(\omega a (1-u^2)+(s u-m)\Big)\Xi\;,\nonumber\\
\Delta_u&=(1-u^2)\left(1-\dfrac{a^2}{L^2}u^2\right)\;.\nonumber
\end{align}

\section{Boundary conditions}
\label{Maxsec:BC}

To solve a differential equation, such as the radial equation~\eqref{radialeq} or the angular equation~\eqref{angulareq}, one has to impose physically relevant boundary conditions. For the angular equation~\eqref{angulareq}, one usually requires its solutions to be regular at the singular points $\theta=0$ and $\theta=\pi$. This determines uniquely the set of angular functions labeled by $\ell$ and $m$. For the radial equation~\eqref{radialeq}, we have to impose conditions both at the horizon and at infinity. At the horizon, ingoing boundary conditions are imposed. At infinity, however, the boundary conditions are more subtle. For the often studied case of a scalar field on Kerr-AdS BHs, the boundary conditions typically imposed require the field itself to vanish~\cite{Uchikata:2009zz,Cardoso:2003cj}, when looking for quasinormal modes. For the Maxwell field and in the Teukolsky formalism, the asymptotic boundary conditions have not been explored yet. In this section, we are going to discuss them for the general Kerr-AdS background.

We propose that in the Teukolsky formalism, when looking for quasinormal modes of the Maxwell field on Kerr-AdS BHs, VEF boundary conditions should be imposed, following the spirit that the AdS boundary is a perfectly reflecting mirror so that no energy flux can cross it. For the particular case of the electromagnetic field, these boundary conditions create an analogy between the AdS boundary and a perfect conductor. In fact the conductor condition for the Maxwell field has been considered in the Kerr-mirror system~\cite{Brito:2015oca}. But the VEF boundary conditions, which for a scalar field can yield both standard Dirichlet and Neumann boundary conditions and for a Maxwell field can yield perfectly conducting boundary conditions, are a  general principle for any spin field, based on a sound physical rationale.

The energy-momentum tensor for the Maxwell field is\footnote{Note that the energy-momentum tensor for the Maxwell field used in Eq.~\eqref{EMTensor} has a minus sign difference with the usual definition.}
\begin{equation}
T_{\mu \nu}=F_{\mu\sigma}F^\sigma_{\;\;\;\nu}+\dfrac{1}{4}g_{\mu\nu}F^2\;,\label{EMTensor}
\end{equation}
with the Maxwell tensor $F_{\mu\nu}$~\cite{Teukolsky:1973ha}
\begin{equation}
F_{\mu\nu}=2\left(\phi_1(n_{[\mu} l_{\nu]}+m_{[\mu} \bar{m}_{\nu]})+\phi_2 l_{[\mu} m_{\nu]}+\phi_0 \bar{m}_{[\mu} n_{\nu]}\right)+{\rm c.c}\;,\nonumber
\end{equation}
where square brackets on subscripts stand for anti-symmetrization, the tetrad $\{l_\mu, n_\mu, m_\mu, \bar{m}_\mu\}$ is given in Eq.~\eqref{pre:tetrad}, and c.c stands for complex conjugate of the preceding terms.

With the Maxwell scalars defined in Eq.~\eqref{Maxwellscalars}, we are now able to calculate the radial energy flux $T^r_{\;\;t}$, by substituting all of the above ingredients into Eq.~\eqref{EMTensor}, which gives
\begin{equation}
T^r_{\;\;t}=T^r_{\;\;t,\;\uppercase\expandafter{\romannumeral1}}+T^r_{\;\;t,\;\uppercase\expandafter{\romannumeral2}}\;,\nonumber
\end{equation}
where
\begin{equation}
T^r_{\;\;t,\;\uppercase\expandafter{\romannumeral1}}=\dfrac{1}{2\Xi}\left(4|\phi_2|^2-\dfrac{\Delta_r^2}{\rho^4}|\phi_0|^2\right)\;,\label{rtcom}
\end{equation}
while $T^r_{\;\;t,\;\uppercase\expandafter{\romannumeral2}}$ becomes irrelevant at infinity, so we do not show its expression here.\\
%
With the fields decomposition in Eq.~\eqref{fielddeccom}, integrating $T^r_{\;\;t,\;\uppercase\expandafter{\romannumeral1}}$ over a sphere, we obtain the energy flux
\begin{equation}
\mathcal{F}|_r=\int_{S^2} \sin\theta d\theta d\varphi\; r^2 T^r_{\;\;t,\;\uppercase\expandafter{\romannumeral1}}
=\dfrac{r^2}{2\;\Xi\;\rho^4}(|R_{-1}|^2-\Delta_r^2|R_{+1}|^2)\;,\label{flux}
\end{equation}
up to an irrelevant normalization, and where the angular functions $S_{\pm1}(\theta)$ are normalized to unity.
To get the asymptotic boundary condition for $R_{-1}$, we expand Eq.~\eqref{radialeq} with $s=-1$ asymptotically as
\begin{equation}
R_{-1} \sim \;\alpha^{-} r+\beta^{-}+\mathcal{O}(r^{-1})\;,\label{asysol}
\end{equation}
where $\alpha^{-}$ and $\beta^{-}$ are two integration constants. In order to get the explicit form of the boundary conditions, the Starobinsky-Teukolsky identities are required. As we proved in Chapter~\ref{ch:prelim}, these identities can be written explicitly as
\begin{align}
BR_{+1}&=\left(\dfrac{d}{dr}-\dfrac{i\Xi K}{\Delta_r}\right)\left(\dfrac{d}{dr}-\dfrac{i\Xi K}{\Delta_r}\right)R_{-1}\;,\label{identity1}\\
BR_{-1}&=\Delta_r\left(\dfrac{d}{dr}+\dfrac{i\Xi K}{\Delta_r}\right)\left(\dfrac{d}{dr}+\dfrac{i\Xi K}{\Delta_r}\right)P_{+1}\;.\label{identity2}
\end{align}

Keeping in mind the Starobinsky-Teukolsky identity~\eqref{identity1}, making use of the radial equation~\eqref{radialeq} with $s=-1$ and the asymptotic expansion in Eq.~\eqref{asysol}, at the asymptotic boundary, the energy flux in Eq.~\eqref{flux} becomes
\begin{eqnarray}
\mathcal{F}|_{r,\infty}=B^2|\alpha^{-}|^2-|(\lambda-2\omega^2\Xi^2L^2)\alpha^{-}+2i\beta^{-}\omega\Xi|^2\;,\label{fluxinf}
\end{eqnarray}
where an overall constant of proportionality has been ignored.
In order to impose the VEF boundary conditions, i.e. $\mathcal{F}|_{r,\infty}=0$, we have
\begin{equation}
B^2|\alpha^{-}|^2-|(\lambda-2\omega^2\Xi^2L^2)\alpha^{-}+2i\beta^{-}\omega\Xi|^2=0\;.
\end{equation}
Note that $\alpha^{-}$ and $\beta^{-}$ are two independent integration constants; we can re-scale them so that the modulus in the above equation can be dropped\footnote{Indeed there still might be a phase factor between these two constants, but this phase factor can be fixed by calculating the normal modes.}. Then it is easy to solve this quadratic equation and obtain the two solutions
\begin{equation}
\dfrac{\alpha^{-}}{\beta^{-}}=\dfrac{2i\omega\Xi}{B-\lambda+2\omega^2\Xi^2L^2}\;,\;\;\;
\dfrac{\alpha^{-}}{\beta^{-}}=\dfrac{2i\omega\Xi}{-B-\lambda+2\omega^2\Xi^2L^2}\;.\label{bc}
\end{equation}
We have also proved in Appendix~\ref{app:angmomflux} that, the angular momentum flux of the Maxwell field vanishes asymptotically if the above boundary conditions are satisfied, similarly to the gravitational case~\cite{Cardoso:2013pza}.

For Schwarzschild-AdS BHs, Eq.~\eqref{bc} simplifies to
\begin{equation}
\dfrac{\alpha^{-}}{\beta^{-}}=\dfrac{i}{\omega L^2}\;,\;\;\;\dfrac{\alpha^{-}}{\beta^{-}}=\dfrac{i\omega}{-\ell(\ell+1)+\omega^2L^2}\; .\label{bcSAdS}
\end{equation}
These are, apparently, two distinct Robin boundary conditions, but at this moment it is unclear if they lead to physically different modes or if they are isospectral.

We can also follow the same procedure to calculate boundary conditions for the Teukolsky equation with $s=+1$. Instead of using $R_{+1}(r)$, we use $P_{+1}(r)$ for convenience, which relates to $R_{+1}(r)$ through $P_{+1}(r)=\Delta_r R_{+1}(r)$. As before, we expand $P_{+1}(r)$ from Eq.~\eqref{radialeq} with $s=+1$ asymptotically
\begin{equation}
P_{+1} \sim \;\alpha^{+} r+\beta^{+}+\mathcal{O}(r^{-1})\;,\label{asysol2}
\end{equation}
where $\alpha^{+}$ and $\beta^{+}$ are two integration constants. Using the Starobinsky-Teukolsky identity in Eq.~\eqref{identity2}, the asymptotic expansion in Eq.~\eqref{asysol2}, the Teukolsky equation with $s=+1$ in Eq.~\eqref{radialeq} and the transformation $P_{+1}(r)=\Delta_r R_{+1}(r)$, then Eq.~\eqref{flux} gives the  conditions
\begin{equation}
\dfrac{\alpha^{+}}{\beta^{+}}=-\dfrac{2i\omega\Xi}{B-\lambda+2\omega^2\Xi^2L^2}\;,\;\;\;
\dfrac{\alpha^{+}}{\beta^{+}}=-\dfrac{2i\omega\Xi}{-B-\lambda+2\omega^2\Xi^2L^2}\;,\label{bc2}
\end{equation}
after imposing the VEF boundary conditions. Comparing the two boundary conditions in Eq.~\eqref{bc} and in Eq.~\eqref{bc2}, we find that there is only a sign difference, or in other words, they are complex conjugate of each other. This is a consequence that $P_{+1}(r)$ and $R_{-1}(r)$ are proportional to complex conjugate functions of each other. We have checked that solving the radial equation~\eqref{radialeq} for $s=-1$ and $s=+1$ with the corresponding boundary conditions~\eqref{bc} and~\eqref{bc2}, for Schwarzschild-AdS BHs, the same quasinormal frequencies are obtained, which is consistent with the argument that these two Teukolsky equations encode the same information. Thus, for concreteness and without loss of generality, in the following we specify $s=-1$, and consider the corresponding boundary conditions.

\section{Analytical matching}
\label{Maxsec:AM}
%
In this section, we present an analytical calculation of quasinormal frequencies for a Maxwell field on a Kerr-AdS BH, with the two Robin boundary conditions discussed in Section~\ref{Maxsec:BC}. Such calculations can be used to illustrate how these Robin boundary conditions generate unstable modes.

Making use of the standard matching procedure, we shall first divide the space outside the event horizon into two regions: the \textit{near region}, defined by the condition $r-r_+\ll1/\omega$, and the \textit{far region}, defined by the condition $r_+\ll r-r_+$. Then we further require the condition $r_+\ll1/\omega$ so that an overlapping region exists where the solutions obtained in the near region and in the far region are both valid. In the following analysis we focus on small AdS BHs $(r_+\ll L)$ with slow rotation $(a\ll r_+)$. The former condition allows treating the frequencies for the Kerr-AdS BH as a perturbation of the AdS normal frequencies; the latter condition together with $\omega r_+\ll1$, implying $\omega a\ll 1$ and $a\ll L$, allow approximating the angular equation for the spin-weighted AdS-spheroidal harmonics by the spin-weighted spherical harmonics, so that the separation constant becomes
\begin{equation}
 \lambda\simeq\ell(\ell+1)\;,\;\;\;\text{with}\;\;\ell=1,2,3\;,\;\cdot\cdot\cdot\;,\label{eigenvalues}
\end{equation}
where $\ell$ is the angular momentum quantum number.

\subsection{Near region solution}
In the near region, under the small BH, $r_+\ll L$, and the slow rotation, $a\ll r_+$, approximations, Eq.~\eqref{radialeq} becomes
\begin{eqnarray}
\Delta_r R_{-1}''+\left(\dfrac{(r_+-r_-)^2\hat{\omega}}{\Delta_r}-\lambda\right)R_{-1}=0 \;,\label{neareq1}
\end{eqnarray}
with
\begin{equation}
\hat{\omega}=\left(\bar{\omega}+\dfrac{i}{2}\right)^2+\dfrac{1}{4}\;,\;\;\;\bar{\omega}=\left(\omega-m\Omega_H\right)\Xi\dfrac{r_+^2+a^2}{r_+-r_-}\;,\nonumber
\end{equation}
where $\Omega_H$ is the angular velocity of the event horizon, given by Eq.~\eqref{RHorizonV}.
It is convenient to define a new dimensionless variable
\begin{equation}
z\equiv \dfrac{r-r_+}{r-r_-}\;,\nonumber
\end{equation}
to transform Eq.~\eqref{neareq1} into
\begin{equation}
z(1-z)\frac{d^2R_{-1}}{dz^2}-2z\frac{dR_{-1}}{dz}+\left(\dfrac{\hat{\omega}(1-z)}{z}-\dfrac{\lambda}{1-z}\right)R_{-1}=0\;.\label{neareq2}
\end{equation}
The above equation can be solved in terms of the hypergeometric function
\begin{equation}
R_{-1}\sim z^{1-i\bar{\omega}}(1-z)^\ell\;F(\ell+1,\ell+2-2i\bar{\omega},2-2i\bar{\omega};z)\;,\label{nearsol}
\end{equation}
where an ingoing boundary condition has been imposed.

The near region solution, Eq.~\eqref{nearsol}, must be expanded for large $r$, in order to perform the matching with the far region solution below. To achieve this we take the $z\rightarrow1$ limit, and obtain
\begin{equation}
R_{-1} \sim \Gamma(2-2i\bar{\omega})\left[\dfrac{R^{\rm near}_{-1,1/r}}{r^{\ell}} +R^{\rm near}_{-1,r} r^{\ell+1}\right]
\;,\label{nearsolfar}
\end{equation}
by using the properties of the hypergeometric function~\cite{abramowitz+stegun}, and where
\begin{align}
R^{\rm near}_{-1,1/r} & \equiv  \dfrac{\Gamma(-2\ell-1) (r_+ - r_-)^\ell}{\Gamma(-\ell)\Gamma(1-\ell-2i\bar{\omega})} \ ,\nonumber \\
R^{\rm near}_{-1,r} &\equiv \dfrac{\Gamma(2\ell+1)(r_+-r_-)^{-\ell-1}}{\Gamma(\ell+1)\Gamma(\ell+2-2i\bar{\omega})} \ .
\end{align}

\subsection{Far region solution}
In the far region, $r-r_+\gg r_+$, the BH effects can be neglected ($M\rightarrow0, a\rightarrow0$) so that
\begin{equation}
\Delta_r\simeq r^2 \left(1+\dfrac{r^2}{L^2}\right)\; .\nonumber
\end{equation}
Then Eq.~\eqref{radialeq} becomes
\begin{equation}
\Delta_rR_{-1}''(r)+\left(\dfrac{K_r^2+i K_r \Delta_r^\prime}{\Delta_r}-2iK_r^\prime
-\ell(\ell+1)\right)R_{-1}(r)
=0\;,\label{fareq1}
\end{equation}
with $K_r=\omega r^2$.
\\The general solution for Eq.~\eqref{fareq1} is
\begin{align}
R_{-1}=&\;r^{\ell+1}(r-iL)^{\frac{\omega L}{2}}(r+iL)^{-\ell-\frac{\omega L}{2}}\Big[C_1F\Big(+\omega L,2\ell+2;\dfrac{2r}{r+iL}\Big)\Big.\nonumber \\ & \Big.\;-2^{-2\ell-1}C_2\left(1+\dfrac{iL}{r}\right)^{2\ell+1}F\Big(-\ell-1,-\ell+\omega L,-2\ell;\dfrac{2r}{r+iL}\Big)\Big]\;,\label{farsol}
\end{align}
where $C_1$, $C_2$ are two integration constants, and they will be constrained in the following to satisfy the boundary conditions.\\
The first boundary condition in Eq.~\eqref{bc}, in the far region, becomes
\begin{equation}
\dfrac{\alpha^{-}}{\beta^{-}}=\dfrac{i}{\omega L^2}\;.\nonumber
\end{equation}
In order to impose this boundary condition, we first expand Eq.~\eqref{farsol} at large $r$, in the form of Eq.~\eqref{asysol}; then one obtains the first relation between $C_1$ and $C_2$
\begin{equation}
\dfrac{C_2}{C_1}=-2^{2\ell+1}\dfrac{\ell}{\ell+1}\dfrac{F(\ell+1,\ell+1+\omega L,2\ell+2;2)}{F(-\ell,-\ell+\omega L,-2\ell;2)}\;.\label{c1c2rel1}
\end{equation}
The second boundary condition in Eq.~\eqref{bc}, in the far region, turns into
\begin{equation}
\dfrac{\alpha}{\beta}=\dfrac{i\omega}{-\ell(\ell+1)+\omega^2L^2}\;.\nonumber
\end{equation}
To impose the second boundary condition above, again expanding Eq.~\eqref{farsol} at large $r$, to extract $\alpha^{-}$ and $\beta^{-}$, then one gets the second relation between $C_1$ and $C_2$
\begin{equation}
\dfrac{C_2}{C_1}=2^{2\ell+1}\left(\dfrac{\ell}{\ell+1}\right)^2\dfrac{\ell+1+\omega L}{\ell-\omega L}\dfrac{\mathcal{A}_1}{\mathcal{A}_2}\;,
\end{equation}
where
\begin{align}
&\mathcal{A}_1=(\ell+1)F(\ell,\ell+1+\omega L,2\ell+2;2)+\omega L F(\ell+1,\ell+2+\omega L,2\ell+3;2)\;,\nonumber\\
&\mathcal{A}_2=\ell F(-\ell-1,-\ell+\omega L,-2\ell;2)-\omega L F(-\ell,-\ell+1+\omega L,1-2\ell;2)\;.
\end{align}
In order to match this solution to the near region solution, we expand Eq.~\eqref{farsol} for small $r$, to obtain
\begin{equation}
R_{-1}\;\sim\;\dfrac{R^{\rm far}_{-1,1/r}}{r^\ell}+R^{\rm far}_{-1,r}r^{\ell+1}\;,\label{farsolnear}
\end{equation}
with
\begin{align}
&R^{\rm far}_{-1,1/r}  \equiv   -i L C_2 \ ,\nonumber \\
&R^{\rm far}_{-1,r} \equiv (-1)^\ell2^{2\ell+1}L^{-2\ell} C_1\;.\nonumber
\end{align}

\subsection{Overlap region}
To match the near region solution Eq.~\eqref{nearsolfar} and the far region solution Eq.~\eqref{farsolnear} in the intermediate region, we impose the matching condition $R^{\rm near}_{-1,r}R^{\rm far}_{-1,1/r}=R^{\rm far}_{-1,r}R^{\rm near}_{-1,1/r}$, then we obtain
\begin{align}
&\dfrac{\Gamma(-2\ell-1)}{\Gamma(-\ell)}\dfrac{\Gamma(\ell+1)}{\Gamma(2\ell+1)}\dfrac{\Gamma(\ell+2-2i\bar{\omega})}{\Gamma(1-\ell-2i\bar{\omega})}\left(\dfrac{r_+-r_-}{L}\right)^{2\ell+1}
\nonumber\\&=\;i\;(-1)^\ell \dfrac{\ell}{\ell+1}\dfrac{F(\ell+1,\ell+1+\omega L,2\ell+2;2)}{F(-\ell,-\ell+\omega L,-2\ell;2)}\;,\label{match1}
\end{align}
with the first boundary condition given by Eq.~\eqref{bc}, and
\begin{align}
&\dfrac{\Gamma(-2\ell-1)}{\Gamma(-\ell)}\dfrac{\Gamma(\ell+1)}{\Gamma(2\ell+1)}\dfrac{\Gamma(\ell+2-2i\bar{\omega})}{\Gamma(1-\ell-2i\bar{\omega})}\left(\dfrac{r_+-r_-}{L}\right)^{2\ell+1}
\nonumber\\
&=\;i\;(-1)^{\ell+1}\left(\dfrac{\ell}{\ell+1}\right)^2\dfrac{\ell+1+\omega L}{\ell-\omega L}\;\dfrac{\mathcal{A}_1}{\mathcal{A}_2}\;,\label{match2}
\end{align}
with the second boundary condition given by Eq.~\eqref{bc}.
\\
Both Eqs.~\eqref{match1} and~\eqref{match2} can be solved perturbatively to look for the imaginary part of the quasinormal frequencies, in the small BH $(r_+\ll L)$ and slow rotation $(a\ll r_+)$ approximations.
In order to do so, we first look for normal modes. For a small BH, the left term in Eqs.~\eqref{match1} and~\eqref{match2} vanish at the leading order, then we have to require the right hand side of both equations to vanish as well. These conditions give the normal modes for pure AdS
\begin{align}
&F(\ell+1,\ell+1+\omega L,2\ell+2;2)=0\;\;\Rightarrow\;\;\omega_{1,N}L=2N+\ell+2\;,\label{normalmode1}\\
&\mathcal{A}_1=0\;\;\Rightarrow\;\;\omega_{2,N}L=2N+\ell+1\;,\label{normalmode2}
\end{align}
where $N=0,1,2,\cdot\cdot\cdot$, and $\ell=1,2,3,\cdot\cdot\cdot$. The two sets of modes are, in this case, isospectral up to one mode.
\\
When the BH effects are taken into account, a correction to the frequency will be introduced
\begin{equation}
\omega_j L=\omega_{j,N} L+i\delta_j\;,\label{normalmode}
\end{equation}
where $j=1,2$ for the two different boundary conditions, and $\delta$ is used to describe the damping (growth) of the quasinormal modes, and we replace $\omega L$ appearing in the second line of Eqs.~\eqref{match1} and~\eqref{match2} by $\omega_1 L$ and $\omega_2 L$ in Eq.~\eqref{normalmode}, respectively. Then, from each of these two equations, we obtain $\delta_j$ perturbatively, at leading order in $a$.

It turns out that the general expression for $\delta_j$ is quite complicated. As such, we only show here a few explicit examples. For $\ell=1$ and $N=0$, from Eq.~\eqref{match1}, we obtain
%
\begin{align}
\delta_1&=-\dfrac{16}{\pi}\dfrac{r_+^4}{L^4}+m\dfrac{16}{3\pi}\dfrac{ar_+^2}{L^3}+\mathcal{O}\left(\frac{a}{L},\frac{r_+^4}{L^4}\right)=-\dfrac{16}{3\pi}\dfrac{r_+^2}{L^2}\left(3\dfrac{r_+^2}{L^2}-m\dfrac{a}{L}\right)+\cdot\cdot\cdot\nonumber\\
&\simeq-\dfrac{16}{3\pi}\dfrac{r_+^4}{L^3}\left(\omega_{1,0}-m\Omega_H\right)+\cdot\cdot\cdot\;,\label{analy1}
\end{align}
where the angular velocity has been approximated by $\Omega_H\sim a/r_+^2$. It is manifest, from \eqref{analy1}, that $\delta_1<0$ when $\omega_{1,0}>m\Omega_H$, while $\delta_1>0$ when $\omega_{1,0}<m\Omega_H$. Thus we find growing modes within the superradiant regime, as expected.
\\
Keeping the same parameters as in the previous paragraph, i.e. $\ell=1$ and $N=0$, from Eq.~\eqref{match2}, we obtain
%
\begin{align}
\delta_2&=-\dfrac{8}{3\pi}\dfrac{r_+^4}{L^4}+m\dfrac{4}{3\pi}\dfrac{ar_+^2}{L^3}+\mathcal{O}\left(\frac{a}{L},\frac{r_+^4}{L^4}\right)=-\dfrac{4}{3\pi}\dfrac{r_+^2}{L^2}\left(2\dfrac{r_+^2}{L^2}-m\dfrac{a}{L}\right)+\cdot\cdot\cdot\nonumber\\
&\simeq-\dfrac{4}{3\pi}\dfrac{r_+^4}{L^3}\left(\omega_{2,0}-m\Omega_H\right)+\cdot\cdot\cdot\;,\label{analy2}
\end{align}
which also shows clearly that $\delta_2<0$ when $\omega_{2,0}>m\Omega_H$, but $\delta_2>0$ when $\omega_{2,0}<m\Omega_H$, signaling again superradiant instabilities.

Furthermore, for both cases, $\delta_1=0$ and $\delta_2=0$ give $\omega_{1,0}=m\Omega_H$ and $\omega_{2,0}=m\Omega_H$, which are the conditions to form clouds, with the two different boundary conditions.
%
%

\section{Numerical method}
\label{Maxsec:NM}
When the BH parameters lie beyond the small size and slow rotation approximations considered in the last section, the analytical method fails and we have to solve the problem numerically. Since the radial equation~\eqref{radialeq} and the angular equation~\eqref{angulareq} are coupled through their eigenvalues, we have to solve both equations simultaneously. In this section, we describe the numerical method used to solve both equations.

The radial equation~\eqref{radialeq}, will be solved with a direct integration method, which has been used in Chapters~\ref{ch:NeutralP} and~\ref{ch:ChargedP}. To be complete, here we briefly outline the procedure.
\\
We first use Frobenius' method to expand $R_{-1}$ close to the event horizon
\begin{equation}
R_{-1}=(r-r_+)^\rho \sum_{j=0}^\infty c_j\;(r-r_+)^j\;,\nonumber
\end{equation}
to initialize Eq.~\eqref{radialeq}. The series expansion coefficients $c_j$ can be derived directly after inserting these expansions into Eq.~\eqref{radialeq}.
The parameter $\rho$ is chosen as
\begin{equation}
\rho=1-\dfrac{i(\omega-m\Omega_H)}{4\pi T_H}\;,\nonumber
\end{equation}
so that an ingoing boundary condition is satisfied. The angular velocity $\Omega_H$ and the Hawking temperature $T_H$ are given in Eq.~\eqref{RHorizonV} and Eq.~\eqref{RTemp}.
At infinity, the asymptotic behavior of $R_{-1}$ is given by Eq.~\eqref{asysol}. The expansion coefficients, $\alpha^{-}$ and $\beta^{-}$, can be extracted from $R_{-1}$ and its first derivative. For that purpose, we define two new fields $\left\{\chi,\psi\right\}$, which asymptote respectively to $\left\{\alpha^{-},\beta^{-}\right\}$ at infinity. Such a transformation can be written in the matrix form
\begin{equation}
\mathbf{V}= \left(\begin{array}{cc} r & 1 \vspace{2mm}\\ 1 & 0 \end{array}\right) \mathbf{\Psi} \equiv \mathbf{T} \mathbf{\Psi} \;,\nonumber
\end{equation}
by defining the vector $\mathbf{\Psi}^T=(\chi,\psi)$ for the new fields, and another vector $\mathbf{V}^T=(R_{-1},\frac{d}{dr}R_{-1})$ for the original field and its derivative.
To obtain a first order system of ordinary differential equations for the new fields, we define another matrix $\mathbf{X}$, through
\begin{equation}
\dfrac{d\mathbf{V}}{dr}=\mathbf{X}\mathbf{V} \; ,
\end{equation}
which can be read off from the original radial equation~\eqref{radialeq} directly.
Then the radial equation~\eqref{radialeq} becomes
\begin{equation}
\dfrac{d\mathbf{\Psi}}{dr}=\mathbf{T}^{-1}\left(\mathbf{X}\mathbf{T}-\dfrac{d\mathbf{T}}{dr}\right) \mathbf{\Psi} \;.\label{radialmatrix}
\end{equation}
This is the final equation we are going to solve.

The angular equation~\eqref{angulareq}, will be solved using a spectral method, to look for the separation constant $\lambda$. By observing Eq.~\eqref{angulareq} and considering the constraint on rotation $a<L$, one finds two regular singularities at $u=\pm 1$. To impose regular boundary conditions at these regular singularities, we require
\begin{equation}
S\;\sim\;
\begin{cases}
(1-u)^{\frac{|m+s|}{2}}\;\;\;\;\;\text{when}\;\;\;u\rightarrow 1\;,\\(1+u)^{\frac{|m-s|}{2}}\;\;\;\;\;\text{when}\;\;\;u\rightarrow -1\;,
\end{cases}
\end{equation}
where as announced in Section~\ref{Maxsec:BC}, $s=-1$. These asymptotic behaviors can be factored out by defining a new function $\hat{S}$
\begin{equation}
S=(1-u)^{\frac{|m+s|}{2}}(1+u)^{\frac{|m-s|}{2}}\hat{S}\;.\label{trans}
\end{equation}
Then the angular equation~\eqref{angulareq} becomes
\begin{equation}
Y(u)\hat{S}=\lambda \hat{S}\;,\label{eigenfunc}
\end{equation}
where the operator $Y(u)$ can be obtained straightforwardly after inserting the transformation~\eqref{trans} into the angular equation~\eqref{angulareq}.
We choose a Chebyshev grid as the collocation points to discretize the operator $Y(u)$, which becomes a matrix. Then Eq~\eqref{eigenfunc} becomes a linear algebraic equation, and $\lambda$ is obtained by looking for the eigenvalues of the matrix $Y(u)$.

\section{Warm up for Schwarzschild-AdS black holes}
\label{Maxsec:SAdS}
We shall now apply the VEF boundary conditions to Maxwell perturbations on Schwarzschild-AdS BHs, in the Teukolsky formalism. We show that even in this simpler case, there is a new branch of quasinormal modes which has not been explored yet.

As we mentioned in Section~\ref{Maxse:intro}, the quasinormal modes for the Maxwell field on Schwarzschild-AdS BHs have been studied using the Regge-Wheeler formalism~\cite{Cardoso:2003cj}. Here we will tackle the same problem in the Teukolsky formalism, imposing the boundary conditions discussed in Section~\ref{Maxsec:BC}. We find that:
\begin{itemize}
\item[$\bullet$] when the first of the two boundary conditions in Eq.~\eqref{bcSAdS} is imposed, we recover the results given in the literature~\cite{Cardoso:2003cj,Berti:2003ud};
\item[$\bullet$] when the second of the two boundary conditions in Eq.~\eqref{bcSAdS} is imposed, there is one new branch of quasinormal modes.
\end{itemize}
To be complete and for comparison, we will show both results in the following. In the numerical calculations all physical quantities are normalized by the AdS radius $L$, so we set $L=1$. Also, observe that we use $\omega_1$ ($\omega_2$) to represent the quasinormal frequency corresponding to the first (second) boundary conditions.

The numerical method we are using here has already been illustrated in Section~\ref{Maxsec:NM}. Since the angular function in this case becomes the standard spherical harmonics, we do not need to solve the angular equation numerically. The results are as follows.

In Table~\ref{EM1}, we list a few fundamental $(N=0)$ quasinormal frequencies of $\omega_1$ (with $\ell=1$) and $\omega_2$ (with $\ell=2$), for different BH sizes. As observed in Section~\ref{Maxsec:AM}, the normal modes displayed in Eqs.~\eqref{normalmode1} and~\eqref{normalmode2}, are isospectral under the mapping
\begin{equation}
\ell_2\leftrightarrow \ell_1+1 \ ,
\end{equation}
except for one mode for the $\omega_2$ branch. Here $\ell_1$ and $\ell_2$ refer to the angular momentum quantum number in the spectrum of $\omega_1$ and $\omega_2$. The presence of a BH, however, breaks the isospectrality. To show this, we present in Table~\ref{EM1}, the two sets of quasinormal frequencies, with $\ell_1=1$ and $\ell_2=2$, respectively.
One observes that the degeneracy between $\omega_1$ and $\omega_2$ gets broken, especially in the small BH and intermediate BH regimes. For large BHs, these two modes are, again, almost isospectral, which seems to be a general feature for any type of perturbation~\cite{Berti:2003ud,Cardoso:2003cj}.
Furthermore, for large BHs, the real part of the frequency for either of the sets vanishes, while the imaginary part scales linearly with the BH size $r_+$. This scaling can be equally stated in terms of the Hawking temperature, which relates to the BH size through $T_H=3r_+/(4\pi L^2)$ for large BHs, supporting the arguments given in~\cite{Horowitz:1999jd}, where a similar linear relation was found for scalar fields.
We remark that the numerical data for $\omega_1$ displayed in Table~\ref{EM1} coincides with the numerical results presented in~\cite{Cardoso:2001bb,Cardoso:2003cj}, at least within $4$ significant digits, which can be used as a check for our numerical method.
\begin{table}
\caption{\label{EM1} Quasinormal frequencies of the Maxwell field on Schwarzschild-AdS. Some fundamental modes are shown, for different BH sizes $r_+$ and for the two sets of modes.}
\small
\begin{tabular*}{\textwidth}{@{\extracolsep{\fill}} l l l }
\hline
\hline
$r_+$ & $\omega_1 (\ell=1)$ & $\omega_2 (\ell=2)$ \\
\hline
0.2 & 2.6384 - 5.7947$\times 10^{-2}$ i & 2.9403 - 1.0466$\times 10^{-4}$ i\\
0.5 & 2.2591 - 0.6573 i & 2.7804 - 0.07549 i\\
0.8 & 2.1758 - 1.2870 i & 2.6923 - 0.2721 i\\
1.0 & 2.1630 - 1.6991 i & 2.6647 - 0.4061 i\\
5.0 & 0 - 8.7948 i & 0 - 5.0528 i\\
10 & 0 - 15.5058 i & 0 - 13.8198 i\\
50 & 0 - 75.0958 i & 0 - 74.7533 i\\
100 & 0 - 150.048 i & 0 - 149.876 i\\
\hline
\hline
\end{tabular*}
\end{table}

\begin{figure*}
\begin{center}
\begin{tabular}{c}
\hspace{-5mm}\includegraphics[width=0.95\textwidth]{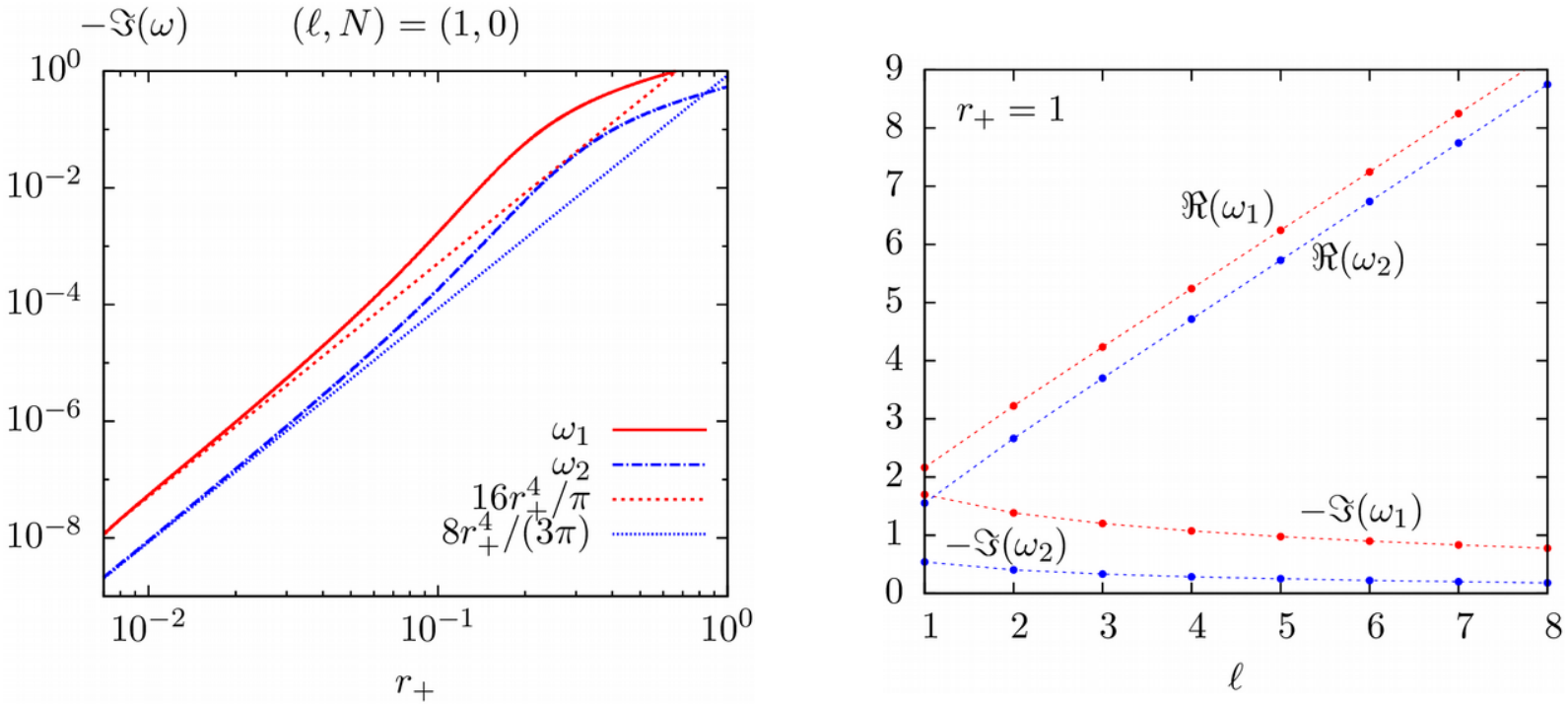}
\end{tabular}
\end{center}
\caption{\label{Mf} {\em Left:} Comparison of the imaginary part for quasinormal frequencies between the analytical approximation of small BHs (thin dashed lines) and the numerical data (thick lines) for the fundamental modes of each branch of solutions. {\em Right:} Effect of the angular momentum quantum number $\ell$ on the quasinormal frequencies for intermediate size BHs with $r_+=1$, and $N=0$. The red line is for $\omega_1$ and the blue line is for $\omega_2$.}
\end{figure*}

For small BHs, it can be shown using a perturbative analytical matching method given in Section~\ref{Maxsec:AM}, that the real part of the frequencies approaches the normal modes of empty AdS~\cite{Konoplya:2002zu}, given by Eqs~\eqref{normalmode1} and~\eqref{normalmode2} for the two different boundary conditions. On the other hand, the imaginary part for both modes approaches zero as
\begin{equation}
-\omega_{j,I} \propto r_+^{2\ell+2}\;,\nonumber
\end{equation}
which also seems to be a general feature for any type of perturbations~\cite{Berti:2009wx}. In Fig.~\ref{Mf}, left panel, we display the numerical data (thick lines) for the fundamental modes of each branch against the leading behavior obtained from the perturbative matching method in Section~\ref{Maxsec:AM}. We find a good agreement for small $r_+$, which is another check for our numerical method.

In Table~\ref{EM2}, we consider intermediate size BHs to exemplify the effect of the angular momentum quantum number $\ell$ on the frequencies\footnote{We have checked that the effect of varying $\ell$ is qualitatively similar for small BHs.}. As one can see, for both modes, the real (imaginary) part of the quasinormal frequencies increases (decreases) in magnitude as $\ell$ increases. These behaviors are more clearly shown in the right panel of Fig.~\ref{Mf}. Observe that the increase of the real part of the frequency with $\ell$ is qualitatively similar to the one observed for empty AdS.
\begin{table}
\caption{\label{EM2} Same as Table \ref{EM1}, but fixing now the BH size to be $r_+=1$. Some fundamental modes are shown, considering different angular momentum quantum number $\ell$.}
\small
\begin{tabular*}{\textwidth}{@{\extracolsep{\fill}} l l l }
\hline
\hline
$\ell$ & $\omega_1$ & $\omega_2$ \\
\hline
1 & 2.16302 - 1.69909 i & 1.55360 - 0.541785 i\\
2 & 3.22315 - 1.38415 i & 2.66469 - 0.406058 i\\
3 & 4.23555 - 1.20130 i & 3.69923 - 0.334088 i\\
4 & 5.23994 - 1.07445 i & 4.71659 - 0.286828 i\\
5 & 6.24294 - 0.97775 i & 5.72784 - 0.252025 i\\
6 & 7.24598 - 0.89976 i & 6.73632 - 0.224622 i\\
7 & 8.24941 - 0.83447 i & 7.74335 - 0.202110 i\\
8 & 9.25327 - 0.77838 i & 8.74952 - 0.183072 i\\
\hline
\hline
\end{tabular*}
\end{table}

Finally, let us remark that, in the above, we have focused on fundamental modes because, on the one hand, our main interest has been to explore the new set of modes which arises even for $N=0$ and, on the other hand, these low lying modes are expected to dominate the late time behavior of time evolutions.

\section{Quasinormal modes and superradiant instabilities}
\label{Maxsec:instability}
%
In this section, we now turn on rotation, to study superradiant instabilities for the Maxwell field on a Kerr-AdS BH. With the numerical strategies described in Section~\ref{Maxsec:NM}, and the boundary conditions given in Eq.~\eqref{bc}, the eigenvalues $\{\omega,\lambda\}$ of the coupled system in Eqs.~\eqref{radialmatrix} and~\eqref{eigenfunc} can be obtained iteratively. This is achieved by starting with an initial guess for $\omega$ or $\lambda$, which are roots of a characteristic equation, and then polishing the solution until $\{\omega,\lambda\}$ become stable. The initial values for $\omega$ or $\lambda$ can be chosen from the results in Schwarzschild-AdS BHs~\cite{Wang:2015goa} or $\ell(\ell+1)$, respectively.

Note that we use $\omega_1$ $(\omega_2)$ to represent the quasinormal frequency and $\lambda_1$ ($\lambda_2$) to stand for the separation constant, corresponding to the first (second) boundary conditions.

A few selected eigenvalues for $\omega$ and $\lambda$ are tabulated in Tables~\ref{Table1-1}$-$\ref{Table3}, with two boundary conditions for various BH sizes. Since it is a generic feature for superradiant instabilities that lower order modes exhibit a stronger instability, we focus on the lowest fundamental modes, characterized by $N=0,\;\ell=1$ and $m=0,\;\pm1$.

In Tables~\ref{Table1-1} and~\ref{Table1-2} we consider a small BH with size $r_+=0.1$. The first observation from these two tables is that superradiant instabilities exist for both boundary conditions, with positive $m$; this is because only positive $m$ modes can meet the superradiance condition, assuming positive frequencies.

\begin{table}
\caption{\label{Table1-1} Quasinormal frequencies and separation constants of the Maxwell field with the first boundary condition, for $\ell=1$ fundamental modes, on a Kerr-AdS BH with size $r_+=0.1$.}
\small
\begin{tabular*}{\textwidth}{@{\extracolsep{\fill}} l l l l }
\hline
\hline
$(\ell,m)$ & $a$ & $\omega_1$ & $\lambda_1$ \\
\hline
(1,\;0) & 0 & 2.8519 - 1.7050$\times 10^{-3}$ i & 2\\
  & 0.01 & 2.8505 - 1.7295$\times 10^{-3}$ i & 2.0005 - 5.9147$\times 10^{-7}$ i\\
  & 0.05 & 2.8151 - 2.5818$\times 10^{-3}$ i & 2.0133 - 2.1677$\times 10^{-5}$ i\\
  & 0.1 & 2.6740 - 2.2847$\times 10^{-2}$ i & 2.0480 - 7.1650$\times 10^{-4}$ i\\
\hline
(1,\;1) & 0.01 & 2.8436 - 9.5962$\times 10^{-4}$ i & 1.9149 + 2.8541$\times 10^{-5}$ i\\
        & 0.05 & 2.7837 + 5.5800$\times 10^{-4}$ i & 1.5879 - 8.0057$\times 10^{-5}$ i\\
        & 0.1 & 2.6493 + 2.0481$\times 10^{-3}$ i & 1.2278 - 5.6148$\times 10^{-4}$ i\\
\hline
(1,\;-1) & 0.01 & 2.8572 - 2.7984$\times 10^{-3}$ i & 2.0859 - 8.4666$\times 10^{-5}$ i\\
         & 0.05 & 2.8422 - 1.7571$\times 10^{-2}$ i & 2.4296 - 2.7419$\times 10^{-3}$ i\\
         & 0.1 & 2.7398 - 1.2611$\times 10^{-1}$ i & 2.8269 - 4.0538$\times 10^{-2}$ i\\
\hline
\hline
\end{tabular*}
\end{table}

\begin{table}
\caption{\label{Table1-2} Quasinormal frequencies and separation constants of the Maxwell field with the second boundary condition, for $\ell=1$ fundamental modes, on a Kerr-AdS BH with size $r_+=0.1$.
}
\small
\begin{tabular*}{\textwidth}{@{\extracolsep{\fill}} l l l l }
\hline
\hline
$(\ell,m)$ & $a$ & $\omega_2$ & $\lambda_2$ \\
\hline
(1,\;0) & 0 & 1.9533 - 1.8240$\times 10^{-4}$ i & 2\\
        & 0.01 & 1.9529 - 1.8329$\times 10^{-4}$ i & 2.0003 - 4.2946$\times 10^{-8}$ i\\
        & 0.05 & 1.9452 - 2.0829$\times 10^{-4}$ i & 2.0071 - 1.2092$\times 10^{-6}$ i\\
        & 0.1 & 1.9160 - 1.0036$\times 10^{-3}$ i & 2.0276 - 2.2602$\times 10^{-5}$ i\\
\hline
(1,\;1) & 0.01 & 1.9436 - 7.9809$\times 10^{-5}$ i & 1.9417 + 2.3800$\times 10^{-6}$ i\\
        & 0.05 & 1.8989 + 1.9474$\times 10^{-4}$ i & 1.7156 - 2.8302$\times 10^{-5}$ i\\
        & 0.1  & 1.8292 + 7.8282$\times 10^{-4}$ i & 1.4552 - 2.1958$\times 10^{-4}$ i\\
\hline
  (1,\;-1) & 0.01 & 1.9622 - 3.2541$\times 10^{-4}$ i & 2.0589 - 9.8188$\times 10^{-6}$ i\\
           & 0.05 & 1.9900 - 2.1380$\times 10^{-3}$ i & 2.2975 - 3.2933$\times 10^{-4}$ i\\
           & 0.1  & 1.9987 - 1.9186$\times 10^{-2}$ i & 2.5914 - 6.0286$\times 10^{-3}$ i\\
\hline
\hline
\end{tabular*}
\end{table}

\begin{figure*}
\begin{center}
\begin{tabular}{c}
\hspace{-4mm}\includegraphics[clip=true,width=0.41\textwidth]{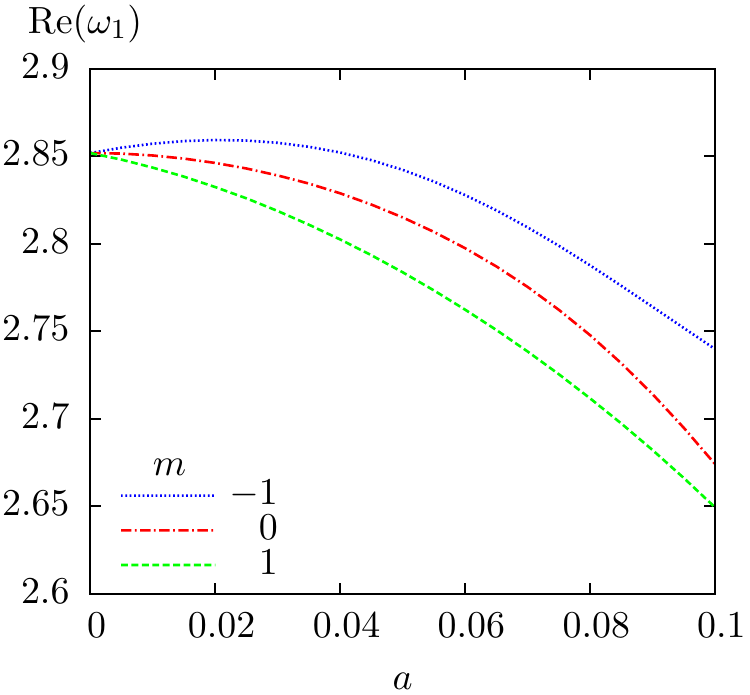}\hspace{15mm}\includegraphics[clip=true,width=0.41\textwidth]{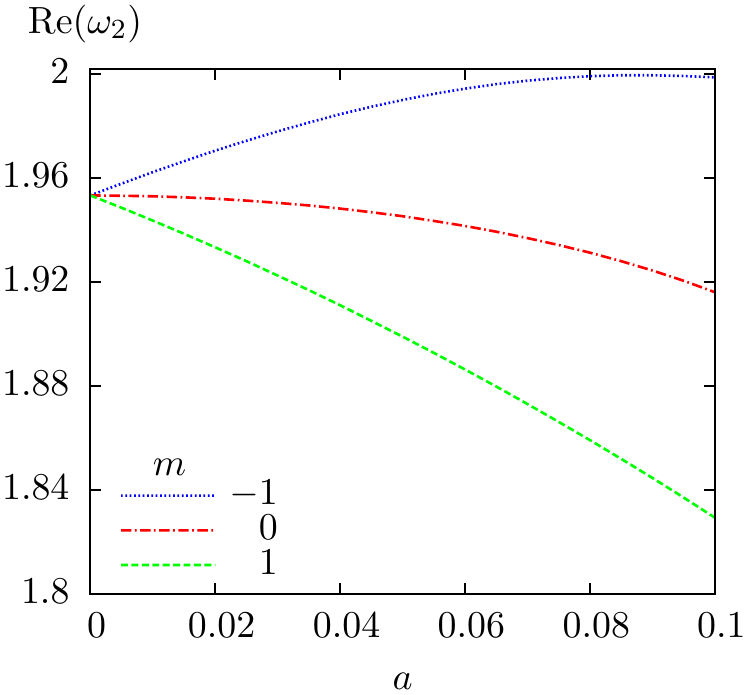}
\end{tabular}
\end{center}
\caption{\label{rp01refre} Variation of Re($\omega$) with varying rotation parameter, for fixed $r_+=0.1$ and $\ell=1$ but for different values of $m$. The left panel is for the first boundary condition while the right panel is for the second boundary condition.}
\end{figure*}

\begin{figure*}
\begin{center}
\begin{tabular}{c}
\hspace{-4mm}\includegraphics[clip=true,width=0.41\textwidth]{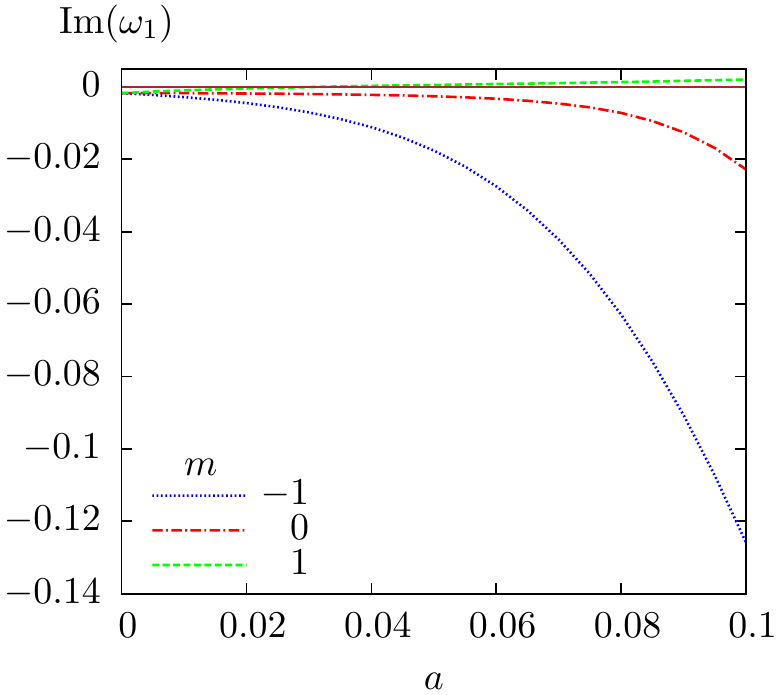}\hspace{15mm}\includegraphics[clip=true,width=0.41\textwidth]{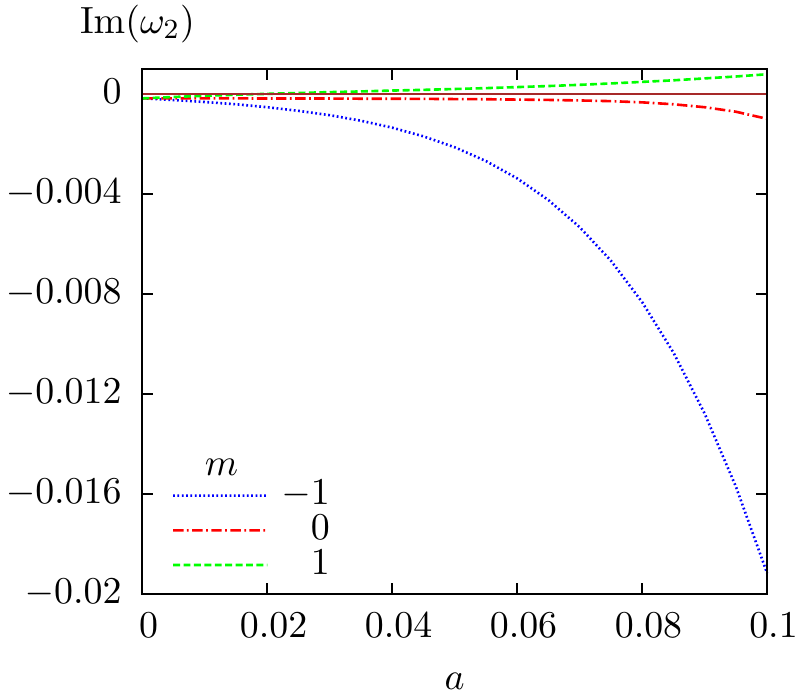}
\end{tabular}
\end{center}
\caption{\label{rp01imfre} Variation of Im($\omega$) with varying rotation parameter, for fixed $r_+=0.1$ and $\ell=1$ but for different values of $m$. The left panel is for the first boundary condition while the right panel is for the second boundary condition. The brown solid thin line corresponds to Im($\omega$)=0, to exhibit more clearly superradiant instabilities.}
\end{figure*}

\begin{figure*}
\begin{center}
\begin{tabular}{c}
\hspace{-4mm}\includegraphics[clip=true,width=0.41\textwidth]{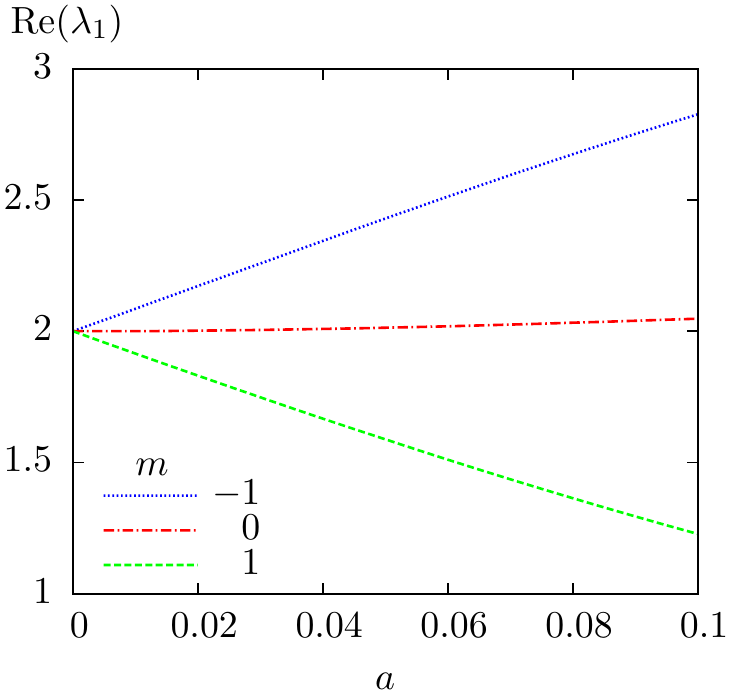}\hspace{15mm}\includegraphics[clip=true,width=0.41\textwidth]{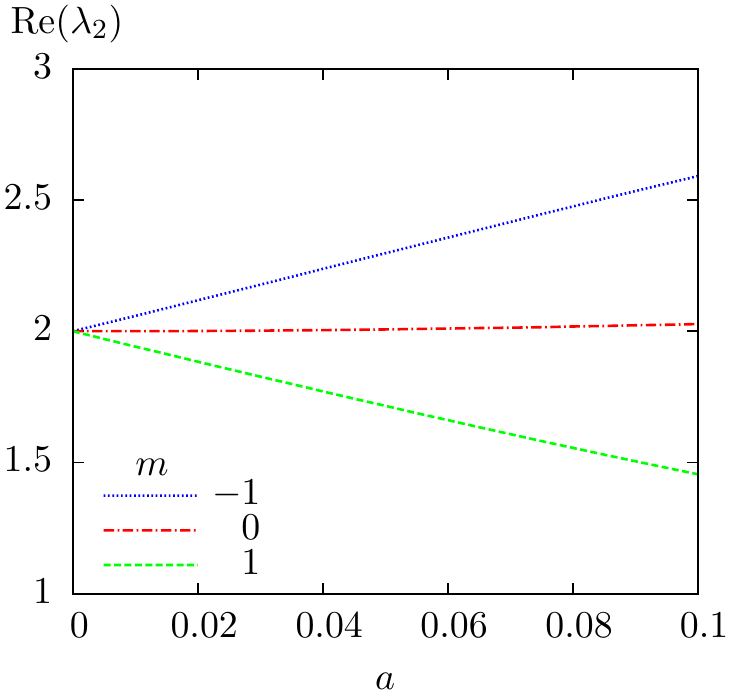}
\end{tabular}
\end{center}
\caption{\label{rp01relam} Variation of Re($\lambda$) with varying rotation parameter, for fixed $r_+=0.1$ and $\ell=1$ but for different values of $m$. The left panel is for the first boundary condition while the right panel is for the second boundary condition.}
\end{figure*}

\begin{figure*}
\begin{center}
\begin{tabular}{c}
\hspace{-4mm}\includegraphics[clip=true,width=0.41\textwidth]{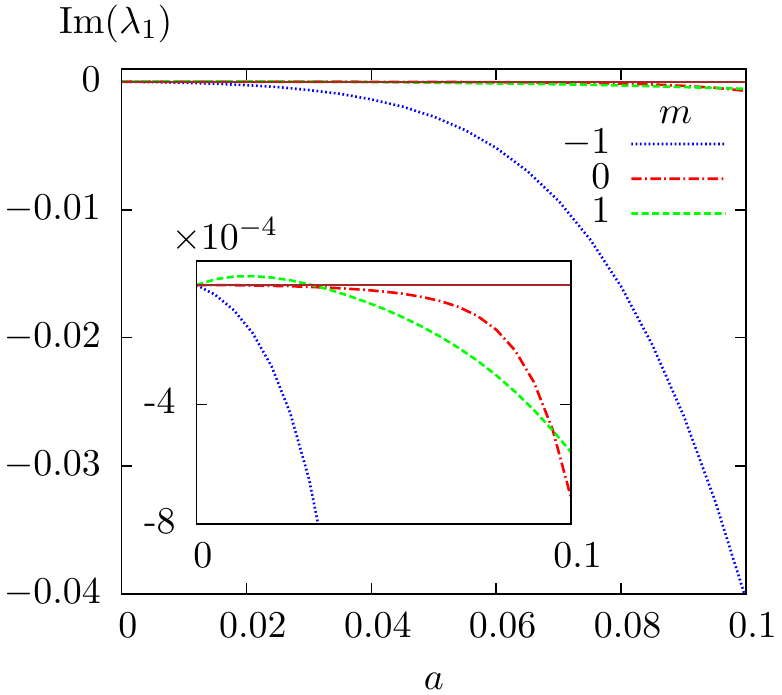}\hspace{15mm}\includegraphics[clip=true,width=0.41\textwidth]{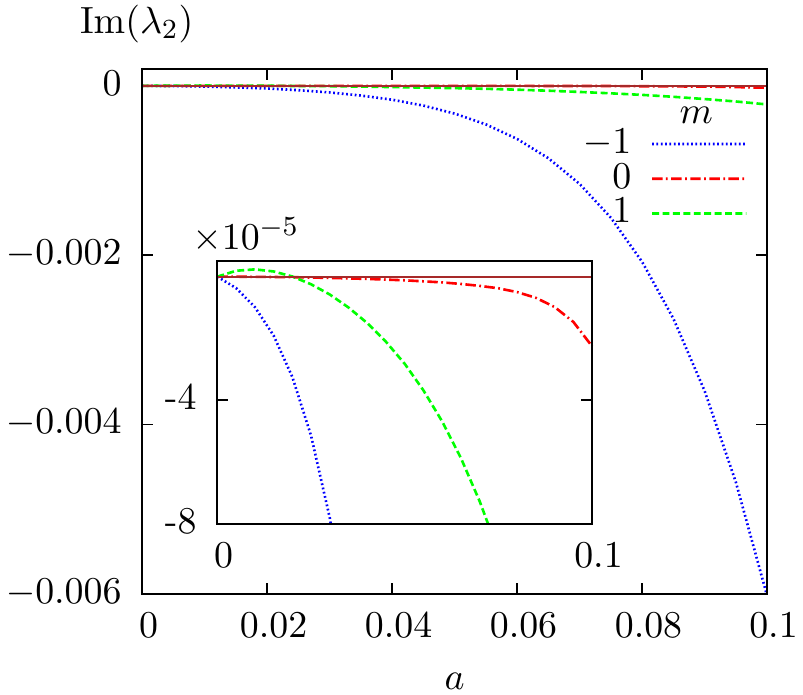}
\end{tabular}
\end{center}
\caption{\label{rp01imlam} Variation of Im($\lambda$) with varying rotation parameter, for fixed $r_+=0.1$ and $\ell=1$ but for different values of $m$. The left panel is for the first boundary condition while the right panel is for the second boundary condition. Again, the brown solid thin line corresponds to Im($\lambda$)=0, to exhibit that the sign of Im($\lambda$) changes when the superradiant instability occurs (seen in the insets).}
\end{figure*}

The effect of varying the rotation parameter on both eigenvalues, for different values of $m$ with fixed $\ell=1$, are shown in Figs.~\ref{rp01refre}$-$\ref{rp01imlam}. In Fig.~\ref{rp01refre}, the real part of the frequency is shown. An immediate first impression from Fig.~\ref{rp01refre}, is that it seems the rotation impacts differently on the $m=-1$ modes, for the two boundary conditions. Checking carefully the numerical data for Re($\omega_2$), however, we find that its value decreases sightly when $a$ is approaching $0.1$. Thus,  for  $m=-1$ and for both boundary conditions, Re($\omega$) starts by increasing with increasing rotation but then decreases. For the other two values of $m$, Re($\omega$) always decreases with increasing rotation.

%

In Fig.~\ref{rp01imfre}, the imaginary part of the frequency is shown, for both boundary conditions. Im($\omega$) increases with increasing rotation when $m=1$, eventually becoming positive, signaling the presence of superradiant unstable modes.


In Fig.~\ref{rp01relam} and Fig.~\ref{rp01imlam}, we show the real and imaginary part of the separation constant, respectively, for both boundary conditions. The real part increases with increasing rotation both when $m=-1$ and $m=0$, albeit only slightly in the latter case, and decreases with the rotation when $m=1$. As for the imaginary part, for $m=1$, it increases with the rotation initially in a small range; but it starts decreasing afterwards. For the other two values of $m$, the imaginary part of the separation constants always decrease with the rotation. We also note that Im($\lambda$), for $m=0$, with the first boundary condition decays faster than its counterpart with the second boundary condition. From these first four figures, we conclude that the effect of varying the rotation on the eigenvalues is qualitatively similar for the two boundary conditions. In the following, therefore, we only show the rotation effect on the eigenvalues with the first boundary condition, when considering other BH sizes.

\begin{figure*}
\begin{center}
\begin{tabular}{c}
\hspace{-4mm}\includegraphics[clip=true,width=0.41\textwidth]{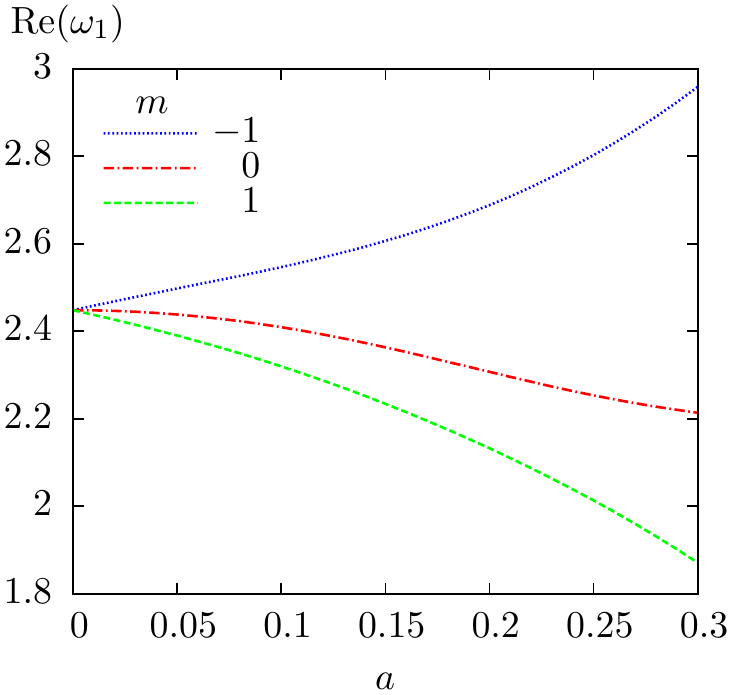}\hspace{15mm}\includegraphics[clip=true,width=0.41\textwidth]{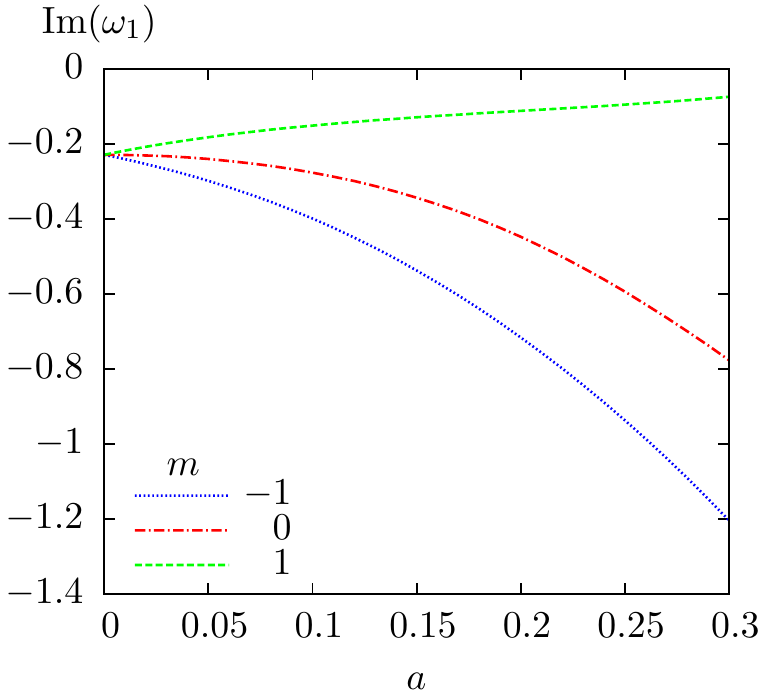}
\end{tabular}
\end{center}
\caption{\label{rp03fre} Variation of $\omega$ with varying rotation parameter, for different values of $m$. The BH size is fixed as $r_+=0.3$ and the first boundary condition has been imposed. The left panel is for Re($\omega$) while the right panel is for Im($\omega$).}
\end{figure*}

\begin{figure*}
\begin{center}
\begin{tabular}{c}
\hspace{-4mm}\includegraphics[clip=true,width=0.41\textwidth]{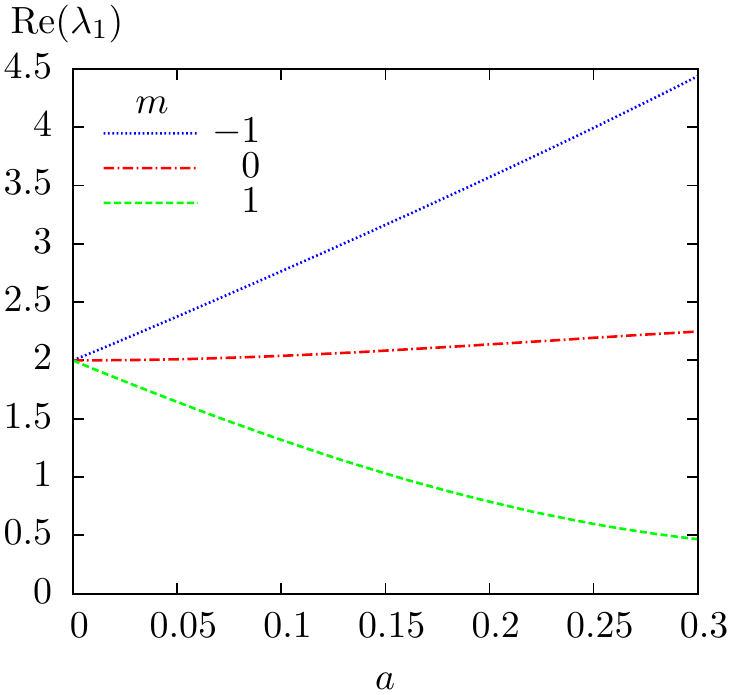}\hspace{15mm}\includegraphics[clip=true,width=0.41\textwidth]{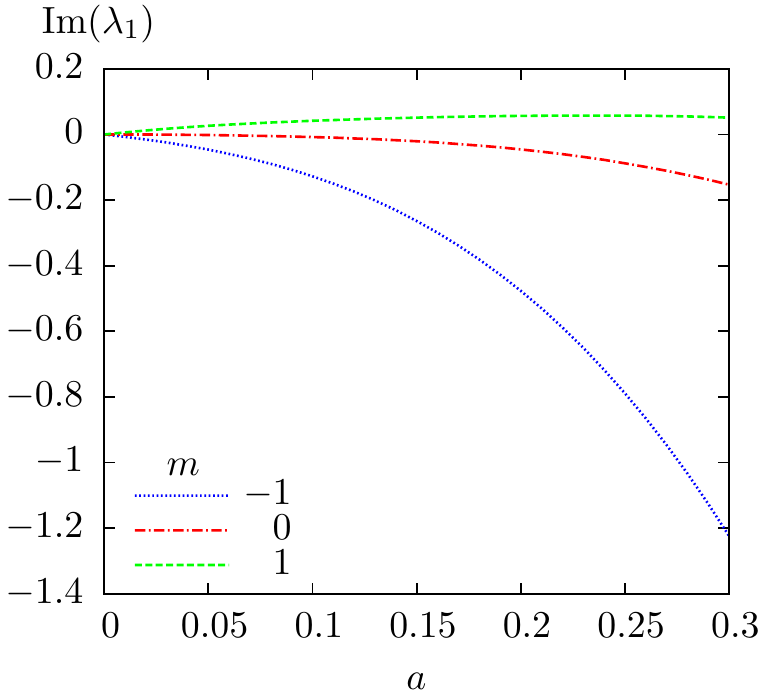}
\end{tabular}
\end{center}
\caption{\label{rp03lam} Variation of $\lambda$ with varying rotation parameter, for different values of $m$. The BH size is fixed as $r_+=0.3$ and the first boundary condition has been imposed. The left panel is for Re($\lambda$) while the right panel is for Im($\lambda$).}
\end{figure*}

We continue our study by varying the BH size. In Tables~\ref{Table2-1} and~\ref{Table2-2} we list a few selected eigenvalues for $r_+=0.3$. The interesting feature that now emerges is that superradiant instabilities only occur for the second boundary condition. This implies that the second boundary condition may produce unstable modes in a larger parameter space. This feature will be shown more clearly in the parameter space for the vector clouds. The effect of varying the rotation on the eigenvalues is shown in Figs.~\ref{rp03fre}$-$\ref{rp03lam}, with the first boundary condition. In Fig.~\ref{rp03fre}, it displays that, Re($\omega$) increases with increasing rotation for the $m=-1$ mode but decreases for both $m=0$ and $m=1$ modes, while Im($\omega$) increases with increasing rotation for the $m=1$ mode but decreases for both $m=0$ and $m=-1$ modes. The behaviors of the separation constant, shown in Fig.~\ref{rp03lam}, are similar to the counterparts in the $r_+=0.1$ case\footnote{Notice that for the $m=1$ mode, Im($\lambda$) starts decreasing with rotation around $a\simeq0.25$.}.

\begin{table}
\centering
\caption{\label{Table2-1} Quasinormal frequencies and separation constants of the Maxwell field with the first boundary condition, for $\ell=1$ fundamental modes, on a Kerr-AdS BH with size $r_+=0.3$.}
\small
\begin{tabular*}{\textwidth}{@{\extracolsep{\fill}} l l l l }
\hline
\hline
$(\ell,m)$ & $a$ & $\omega_1$ & $\lambda_1$ \\
\hline
(1,\;0) & 0 & 2.4481 - 0.2291 i & 2 \\
  & 0.1 & 2.4093 - 0.2768 i & 2.0397 - 7.8286$\times 10^{-3}$ i\\
  & 0.2 & 2.3071 - 0.4480 i & 2.1373 - 4.5479$\times 10^{-2}$ i \\
  & 0.3 & 2.2136 - 0.7769 i & 2.2476 - 1.5300$\times 10^{-1}$ i \\
  \hline
(1,\;1) & 0.1 & 2.3197 - 0.1512 i & 1.3185 + 4.1838$\times 10^{-2}$ i \\
  & 0.2 & 2.1325 - 0.1120 i & 0.7889 + 5.6598$\times 10^{-2}$ i \\
& 0.3 & 1.8707 - 7.4377$\times 10^{-2}$ i & 0.4681 + 5.1126$\times 10^{-2}$ i \\
\hline
  (1,\;-1) & 0.1 & 2.5468 - 0.3994 i & 2.7643 - 0.1276 i \\
           & 0.2 & 2.6884 - 0.7162 i & 3.5705 - 0.4764 i \\
           & 0.3 & 2.9588 - 1.2044 i & 4.4377 - 1.2239 i \\
\hline
\hline
\end{tabular*}
\end{table}

\begin{table}
\centering
\caption{\label{Table2-2} Quasinormal frequencies and separation constants of the Maxwell field with the second boundary condition, for $\ell=1$ fundamental modes, on a Kerr-AdS BH with size $r_+=0.3$.
}
\small
\begin{tabular*}{\textwidth}{@{\extracolsep{\fill}} l l l l }
\hline
\hline
$(\ell,m)$ & $a$ & $\omega_2$ & $\lambda_2$ \\
\hline
(1,\;0) & 0 & 1.8152 - 3.8034$\times 10^{-2}$ i & 2\\
        & 0.1 & 1.8092 - 4.7641$\times 10^{-2}$ i & 2.0252 - 1.0134$\times 10^{-3}$ i\\
        & 0.2 & 1.7921 - 8.8195$\times 10^{-2}$ i & 2.0949 - 6.9871$\times 10^{-3}$ i\\
        & 0.3 & 1.7875 - 1.8395$\times 10^{-1}$ i & 2.1957 - 2.9422$\times 10^{-2}$ i\\
\hline
(1,\;1) & 0.1 & 1.7265 - 1.5809$\times 10^{-2}$ i & 1.4841 + 4.4473$\times 10^{-3}$ i\\
        & 0.2 & 1.6328 - 2.8939$\times 10^{-3}$ i & 1.0447 + 1.4994$\times 10^{-3}$ i\\
        & 0.3 & 1.5304 + 5.2612$\times 10^{-3}$ i & 0.7048 - 3.6993$\times 10^{-3}$ i\\
\hline
(1,\;-1) & 0.1 & 1.9063 - 8.8702$\times 10^{-2}$ i & 2.5624 - 2.7793$\times 10^{-2}$ i\\
         & 0.2 & 2.0270 - 1.9387$\times 10^{-1}$ i & 3.1487 - 1.2392$\times 10^{-1}$ i\\
         & 0.3 & 2.2192 - 3.6493$\times 10^{-1}$ i & 3.7662 - 3.4830$\times 10^{-1}$ i\\
\hline
\hline
\end{tabular*}
\end{table}


\begin{figure*}
\begin{center}
\begin{tabular}{c}
\hspace{-4mm}\includegraphics[clip=true,width=0.41\textwidth]{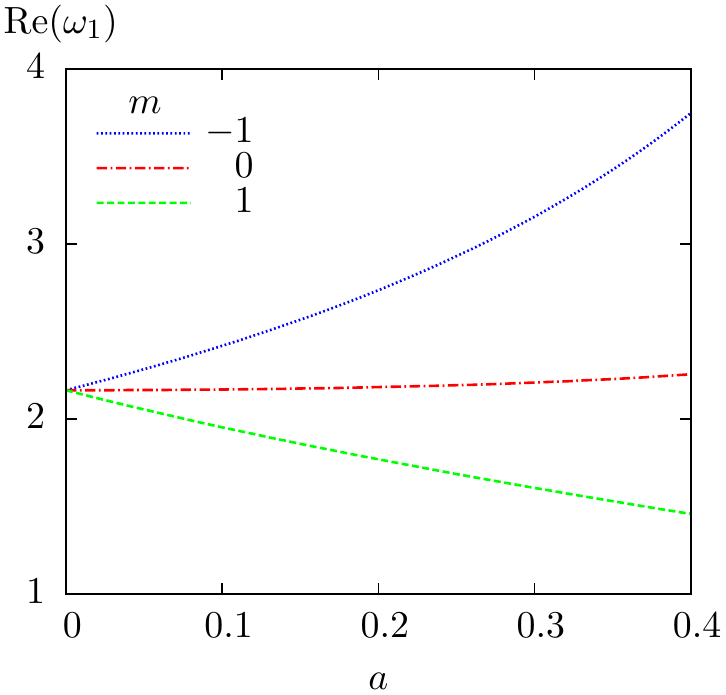}\hspace{15mm}\includegraphics[clip=true,width=0.41\textwidth]{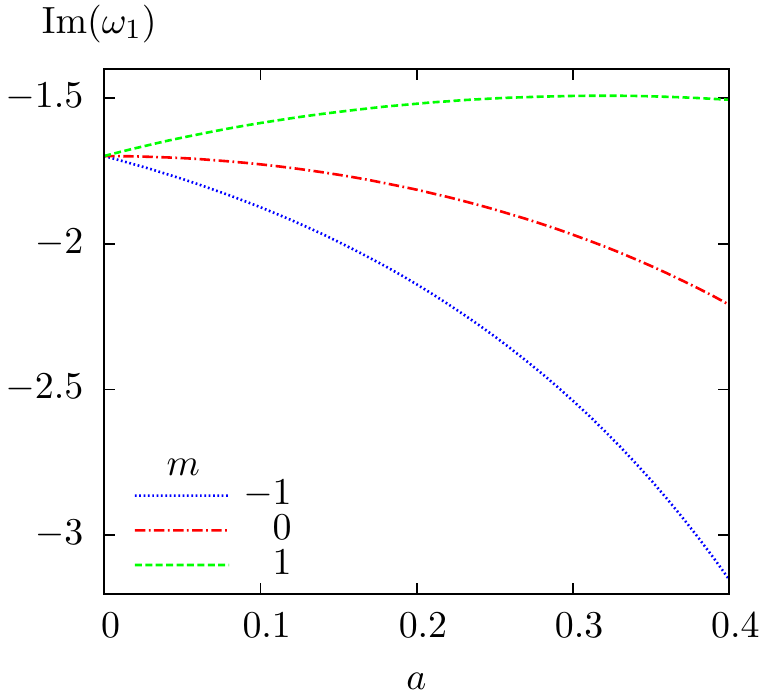}
\end{tabular}
\end{center}
\caption{\label{rp1fre} Variation of $\omega$ with varying rotation parameter, for different values of $m$. The BH size is fixed as $r_+=1$ and the first boundary condition has been imposed. The left panel is for Re($\omega$) while the right panel is for Im($\omega$).}
\end{figure*}

\begin{figure*}
\begin{center}
\begin{tabular}{c}
\hspace{-4mm}\includegraphics[clip=true,width=0.41\textwidth]{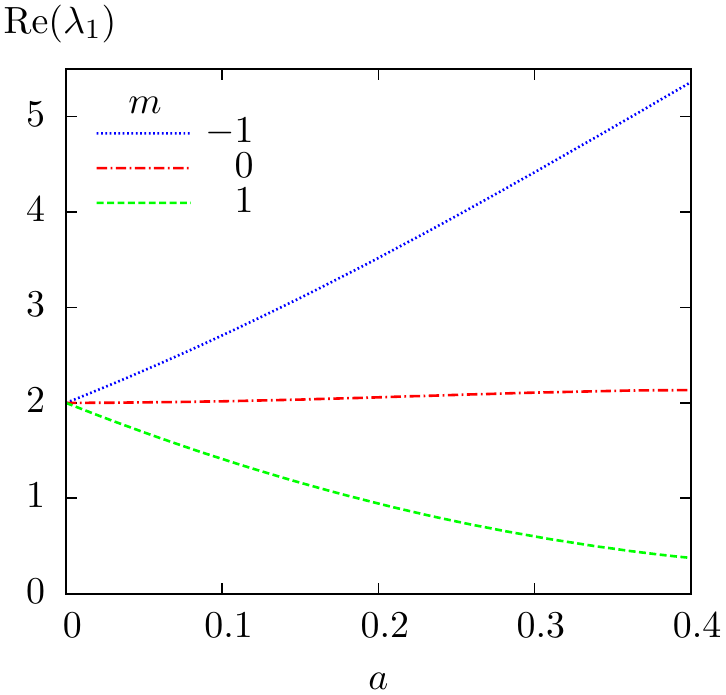}\hspace{15mm}\includegraphics[clip=true,width=0.41\textwidth]{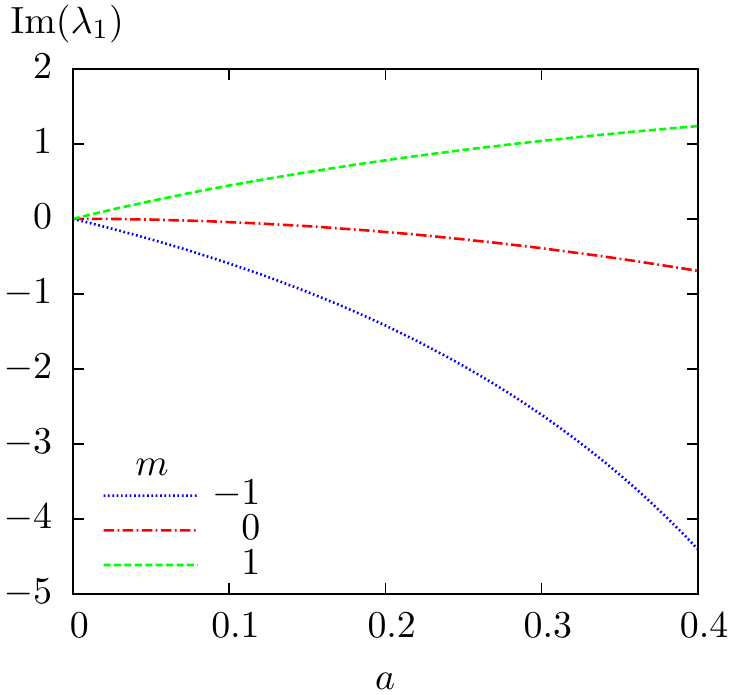}
\end{tabular}
\end{center}
\caption{\label{rp1lam} Variation of $\lambda$ with varying rotation parameter, for different values of $m$. The BH size is fixed as $r_+=1$ and the first boundary condition has been imposed. The left panel is for Re($\lambda$) while the right panel is for Im($\lambda$).}
\end{figure*}

The results for $r_+=1$ are presented in Table~\ref{Table3} and Figs.~\ref{rp1fre}$-$\ref{rp1lam}. From Table~\ref{Table3} one observes that there is no superradiant instability for any of the boundary conditions.

\begin{table}
\centering
\caption{\label{Table3} Quasinormal frequencies and separation constants of the Maxwell field with the two different Robin boundary conditions, for $\ell=1$ fundamental modes, on a Kerr-AdS BH with size $r_+=1$.}
\small
\begin{tabular*}{\textwidth}{@{\extracolsep{\fill}} l l l l l l }
\hline
\hline
$(\ell,m)$ & $a$ & $\omega_1$ & $\lambda_1$ & $\omega_2$ & $\lambda_2$ \\
\hline
(1,\;0) & 0 & 2.1630 - 1.6991 i & 2 & 1.5536 - 0.5418 i & 2\\
  & 0.1 & 2.1672 - 1.7274 i & 2.0162 - 4.4059$\times 10^{-2}$ i & 1.5627 - 0.5510 i & 2.0186 - 1.0131$\times 10^{-2}$ i\\
  & 0.2 & 2.1805 - 1.8146 i & 2.0580 - 0.1757 i & 1.5909 - 0.5795 i & 2.0727 - 4.0857$\times 10^{-2}$ i\\
  & 0.3 & 2.2067 - 1.9686 i & 2.1067 - 0.3931 i &  1.6416 - 0.6297 i & 2.1574 - 9.2910$\times 10^{-2}$ i\\
  \hline
(1,\;1) 
  & 0.1 & 1.9512 - 1.5858 i & 1.4111 + 0.4430 i & 1.4277 - 0.4948 i & 1.5675 + 0.1404 i\\
  & 0.2 & 1.7679 - 1.5195 i & 0.9445 + 0.7797 i & 1.3209 - 0.4643 i & 1.2046 + 0.2445 i\\
  & 0.3 & 1.6048 - 1.4929 i & 0.5998 + 1.0384 i & 1.2273 - 0.4463 i & 0.9151 + 0.3204 i\\
  \hline
  (1,\;-1) 
  & 0.1 & 2.4169 - 1.8746 i & 2.7065 - 0.5962 i & 1.7067 - 0.6116 i & 2.4983 - 0.1905 i\\
  & 0.2 & 2.7349 - 2.1390 i & 3.5193 - 1.4235 i & 1.8996 - 0.7147 i & 3.0586 - 0.4532 i\\
  & 0.3 & 3.1548 - 2.5377 i & 4.4164 - 2.6136 i & 2.1522 - 0.8684 i & 3.6770 - 0.8235 i\\
\hline
\hline
\end{tabular*}
\end{table}

In Fig.~\ref{rp1fre}, we present the real and imaginary parts of the frequency for the first boundary condition and $r_+=1$. Re($\omega$) increases with rotation for the $m=-1$ mode, decreases with rotation for the $m=1$ mode, and increases sightly with rotation for the $m=0$ mode. The behavior of Im($\omega$) is almost the opposite, since Im($\omega$) decreases with rotation for both $m=0$ and $m=-1$ modes, but for the $m=1$ mode, it increases first and then starts to decrease around $a\simeq0.33$. Comparing Fig.~\ref{rp1lam} with Fig.~\ref{rp1fre}, shows that the effect of rotation on $\lambda$ (both real part and imaginary part) mimics closely that on $\omega$, except for Im($\lambda$) of the $m=1$ mode, which always increases with the rotation parameter.

\section{Stationary vector and scalar clouds}
\label{Maxsec:clouds}
Stationary clouds are bound state solutions with real frequency of test fields around a rotating background, computed at the linear level. The existence of clouds indicates nonlinear hairy BH solutions~\cite{Herdeiro:2014goa,Herdeiro:2014ima}, but the converse needs not be true~\cite{Brihaye:2014nba,Herdeiro:2015kha}. In order to find such solutions, we demand $\omega=m\Omega_H$; in other words, stationary clouds are the zero modes of superradiance. Imposing this condition leads to a constraint on the BH parameters: BHs are quantized in the sense that only BHs with specific parameters can support a cloud with a given set of ``quantum'' numbers. This quantization defines \textit{existence lines} in the BH parameter space. In the practical implementation of our numerical calculations, we use the same method as before, with the condition $\omega=m\Omega_H$, to look for the rotation parameter. Note that all the results presented in this subsection are for fundamental modes, characterized by $N=0$.

The vector clouds we have obtained shall be presented in a parameter space spanned by $R_+$ and $\hat{\Omega}_H$, where 
\begin{eqnarray}
R_+=\sqrt{\dfrac{r_+^2+a^2}{\Xi}}\;,\;\;\;\;\;\;\hat{\Omega}_H=\Omega_H\Xi+a\;,\label{NewDef}
\end{eqnarray}
in which $\hat{\Omega}_H$ has been defined in Eq.~\eqref{NRHAV}.
The reason to use this pair of parameters, instead of $r_+$ and $\Omega_H$, is as follows. $\Omega_H$, as defined in Eq.~\eqref{RHorizonV}, is the horizon angular velocity measured relatively to a rotating frame at infinity, while $\hat{\Omega}_H$, defined in Eq.~\eqref{NRHAV} (also in Eq.~\eqref{NewDef}), is the horizon angular velocity measured with respect to a non-rotating observer at infinity. The latter one is more relevant in BH thermodynamics~\cite{Gibbons:2004ai}. In the practical calculations, one can use either of them since they are simply related by Eq.~\eqref{NewDef}. As one may check, $\hat{\Omega}_H$ is a monotonic function of $a$, in terms of $R_+$, but not of $r_+$. Also there is an intuitive geometric meaning for $R_+$, which is the areal horizon radius.

\begin{figure*}
\begin{center}
\begin{tabular}{c}
\hspace{-4mm}\includegraphics[clip=true,width=0.45\textwidth]{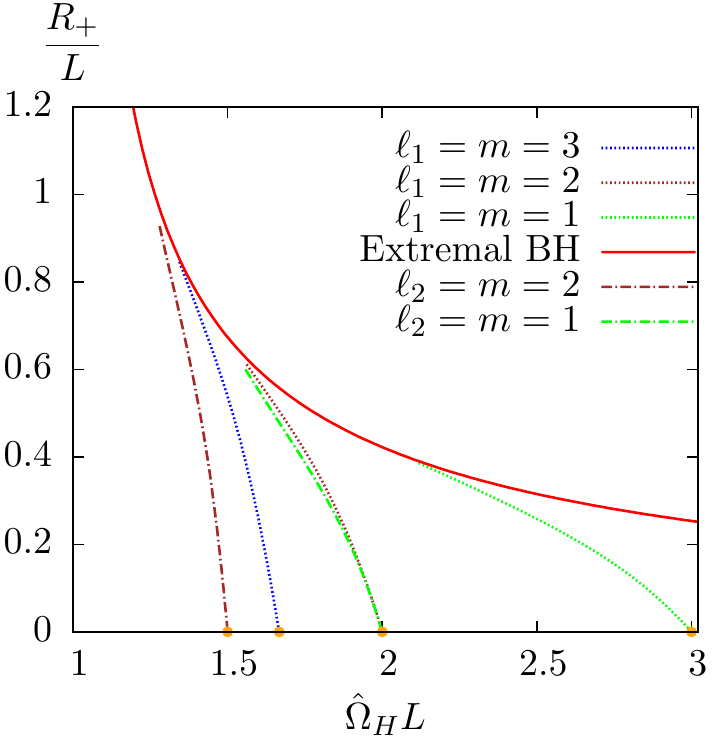}\hspace{15mm}\includegraphics[clip=true,width=0.45\textwidth]{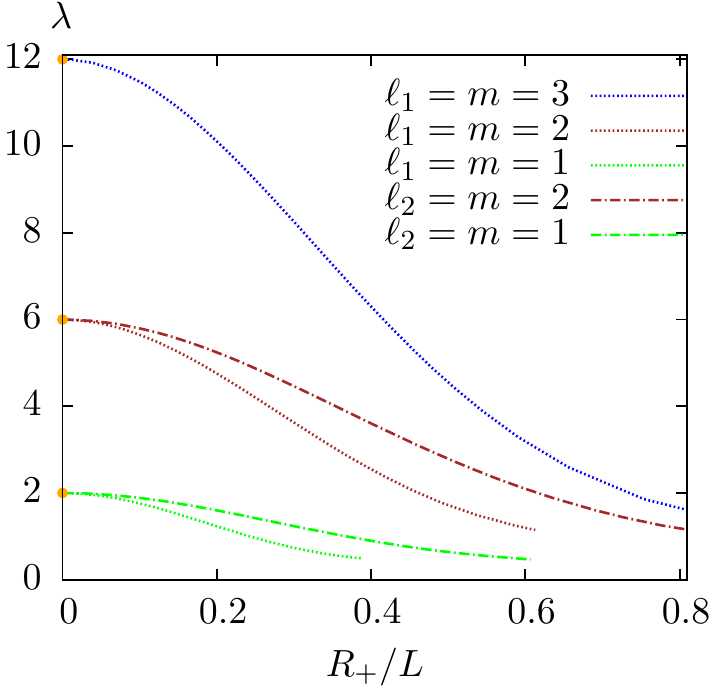}
\end{tabular}
\end{center}
\caption{\label{vec}Vector clouds (left panel) and the corresponding separation constants (right panel) in $R_+$ versus $\hat{\Omega}_H$ and $\lambda$ versus $R_+$ plots, respectively. $\ell_1 (\ell_2)$ refer to the results obtained by imposing the first (second) boundary condition.}
\end{figure*}

In Fig.~\ref{vec}, the existence lines for some examples of vector clouds are displayed (left panel) together with the corresponding separation constants (right panel). In the left panel, the red solid line corresponds to extremal BHs, and regular BHs only exist below this extremality line. The first three existence lines (with $\ell=m=1,2,3$) for the first boundary condition, and the first two existence lines (with $\ell=m=1,2$) for the second boundary condition, are presented by dotted and dot dashed lines, respectively. These lines start from bound state solutions (normal modes), denoted by orange dots in Fig.~\ref{vec}, of the Maxwell field on empty AdS, i.e.
\begin{eqnarray}
\hat{\Omega}_{H,1\;|R_+=0}=1+\dfrac{2}{\ell}\;,\;\;\;\;\;\;\hat{\Omega}_{H,2\;|R_+=0}=1+\dfrac{1}{\ell}\;.\label{bsAdS}
\end{eqnarray}
These are obtained by equating the superradiance condition\footnote{$\Omega_H$ is the same as $\hat{\Omega}_H$ in pure AdS.}, $\omega=m\Omega_H$, to the normal mode conditions in Eqs.~\eqref{normalmode1} and~\eqref{normalmode2}, together with $m=\ell$, and $N=0$. Observe, in particular, that although the two sets of normal modes in AdS are isospectral, the existence lines for the two boundary conditions only coincide as $R_+\rightarrow 0$, when taking $\ell=1$ for the first boundary condition and $\ell=2$ for the second.

An existence line with a particular $\ell=m$, separates the superradiantly stable Kerr-AdS BHs (to the left side of the existence line) and the superradiantly unstable ones (to the right side of the existence line), against that particular mode. Therefore, as one may observe from the left panel of Fig.~\ref{vec}, the stable region in the parameter space against the mode, say $\ell=m=n$, where $n$ is some integer, with the second boundary condition is also stable against all the modes with the first boundary condition from $\ell=m=1$ up to $\ell=m=n+1$. From the data in this figure, together with the relation between $R_+$ and $r_+$, Eq.~(\ref{NewDef}), it can be concluded that: with the first boundary condition, BHs with $r_+\leq0.25$ are superradiantly unstable against the $\ell=m=1$ fundamental mode; while with the second boundary condition, BHs with $r_+\leq0.34$ are superradiantly unstable against the $\ell=m=1$ fundamental mode. These observations also explain the fact that the superradiant instability only appears for modes with the second boundary condition in Table~\ref{Table2-2}.

\begin{figure*}
\begin{center}
\begin{tabular}{c}
\hspace{-4mm}\includegraphics[clip=true,width=0.45\textwidth]{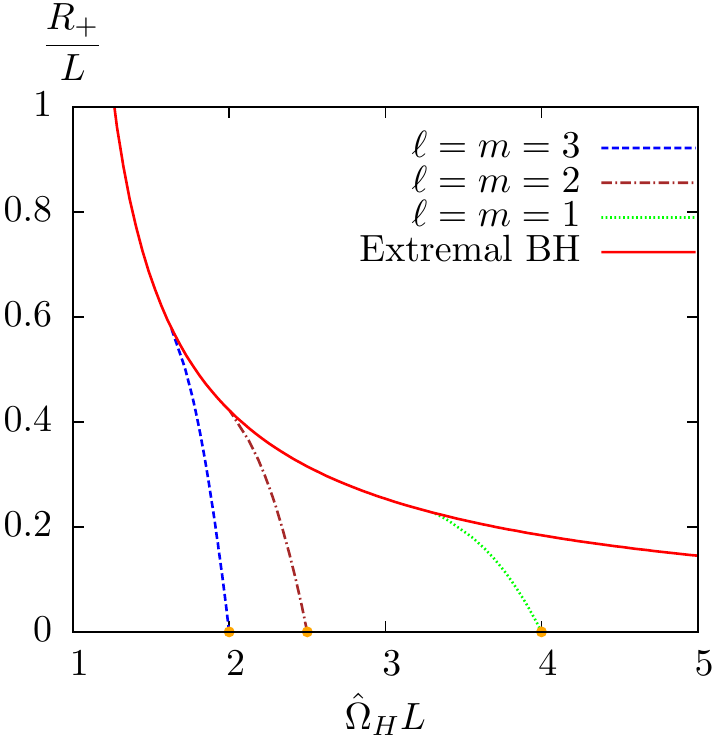}\hspace{15mm}\includegraphics[clip=true,width=0.45\textwidth]{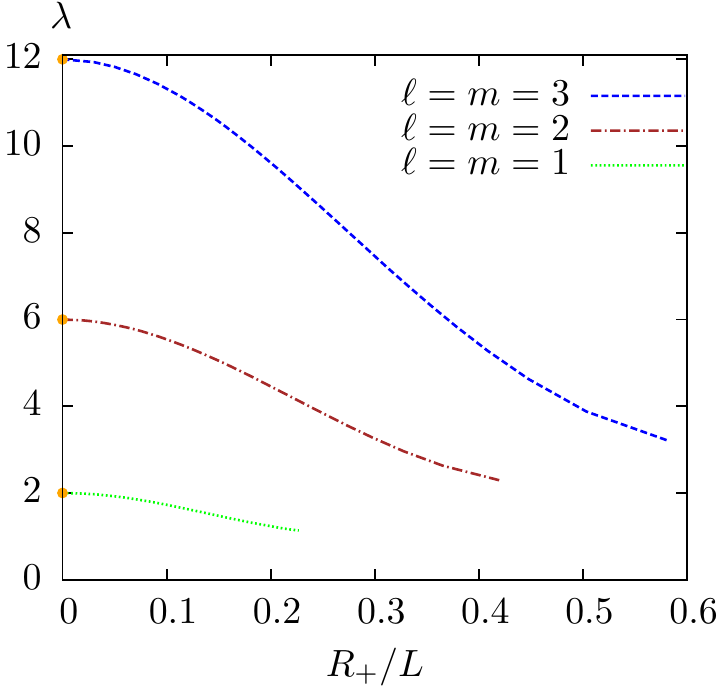}
\end{tabular}
\end{center}
\caption{\label{SCs}Scalar clouds (left panel) and the corresponding separation constants (right panel) in $R_+$ versus $\hat{\Omega}_H$ and $\lambda$ versus $R_+$ plots, respectively.}
\end{figure*}

As a comparison with the Maxwell stationary clouds reported above, we have also computed stationary scalar clouds, by solving the massless Klein-Gordon equation on Kerr-AdS BHs, with VEF boundary condition which in this case is the same as the usual field vanishing boundary condition. For this case there is a single set of modes. The results for the scalar clouds are exhibited in Fig.~\ref{SCs}, in terms of the same parameters $R_+$ and $\hat{\Omega}_H$.


In the left panel of Fig.~\ref{SCs}, the red solid line stands, as before, for extremal BHs, so that regular BHs only exist below this line. The first three existence lines, corresponding to the modes with $\ell=m=1,2,3$, are described by dotted, dot dashed and dashed lines, respectively. The corresponding separation constants are also shown in the right panel of Fig.~\ref{SCs}. The orange dots in both panels indicate again the normal modes and eigenvalues of the angular function in pure AdS, which are
\begin{eqnarray}
\hat{\Omega}_{H,{\rm scalar}}=1+\dfrac{3}{\ell}\;,\;\;\;\;\;\lambda=\ell(\ell+1)\;.
\end{eqnarray}
Again, the existence line with a particular $\ell=m$, divides the parameter space into two regions: BHs in the left region are superradiantly stable while BHs in the right region are superradiantly unstable, against that particular mode.

Comparing Figs.~\ref{vec} and~\ref{SCs} it becomes clear that the existence lines for stationary vector clouds appear to the \textit{left} of the existence line of a stationary scalar cloud with the same quantum numbers. Thus, there are BHs that are stable against the scalar mode but become unstable against the vector mode. In a sense, vector superradiance is stronger. Qualitatively, this conclusion agrees with the computation of the amplification factors for scalar and vector modes in superradiant scattering, in asymptotically flat spacetimes.

\section{Summary}
\label{Maxsec:con}
The behavior of test fields on an asymptotically AdS spacetime depends sensitively on the boundary conditions, since such spacetimes contain a timelike boundary. The AdS boundary is often regarded as a perfectly reflecting mirror in the sense that no flux (both energy flux and angular momentum flux) can cross it. Various types of boundary conditions in asymptotically AdS spacetimes have been explored, in particular consistent with this simple requirement. We took this requirement as the \textit{guiding principle} to impose boundary conditions on test fields. Following this principle, two boundary conditions were found for Maxwell fields on asymptotically AdS spacetimes, of which only one had been previously discussed on the Schwarzschild-AdS background. These two boundary conditions were then used to study quasinormal modes, superradiant unstable modes and vector clouds for the Maxwell field on Kerr-AdS BHs.


To find quasinormal modes and superradiant modes, we have solved the Teukolsky equations both analytically and numerically. In the small BH and slow rotation regime, an analytical matching method was applied to exhibit how these two boundary conditions work and how they produce superradiant instabilities. A numerical method was then used to explore the parameter space away from the small BH and slow rotation approximations. We found that for small BHs characterized by $r_+=0.1$, unstable superradiant modes appear with both boundary conditions. Increasing the BH size, as exemplified for $r_+=0.3$, superradiant instabilities only appear with the second boundary condition, and eventually disappear for both boundary conditions, as exemplified when $r_+=1$. Our analysis also shows that superradiant instabilities for the Maxwell field may exist for (moderately) larger BH sizes, when comparing with the scalar case, for which superradiant instabilities appear in the regime $r_+\leq0.16$~\cite{Uchikata:2009zz}.

To study stationary vector clouds, which can occur for massless fields in AdS due to the box-like global structure, we have solved the Teukolsky equations at the onset of the superradiant instability, i.e. for $\omega=m\Omega_H$. We found that both boundary conditions can yield vector clouds, and that these clouds are bounded by the extremal BHs, as for the scalar clouds on asymptotically flat Kerr BHs~~\cite{Hod:2012px,Hod:2013zza,Herdeiro:2014goa}. This behaviour differs from that observed for gravitational perturbations, for which only one of the sets of clouds are bounded by the extremal BHs~\cite{Cardoso:2013pza}. The existence of clouds at the linear level indicates nonlinear hairy BH solutions~\cite{Herdeiro:2014goa,Herdeiro:2014ima}, so it would be interesting to find the nonlinear realization of these vector clouds. There is already a well-known exact BH family within the Einstein-Maxwell-AdS system: the Kerr-Newman-AdS family. It will then be interesting to understand the interplay between this well known family and the new family of ``hairy'' BHs.


%% file: conclusion.tex
\chapter{Conclusions and outlook}
\label{ch:conclusion}
%
%
In this final chapter we draw our conclusions, and address some open questions.

This thesis covers studies of Hawking radiation and superradiance, for scalar and vector fields in the probe limit, using perturbative methods.

In Chapter~\ref{ch:GeneralInFlat}, we have studied the wave equations of a Proca field on spherically symmetric higher dimensional spacetimes. Such background was chosen to avoid the non-separation of variables for the Proca equations in a rotating spacetime; while still being an interesting setup in TeV gravity models. Using the Kodama-Ishibashi formalism, we achieved separation of variables, and obtained a set of coupled equations, as well as some decoupled ones.

These equations were used to study Hawking radiation, for a neutral Proca field on a $D$-dimensional Schwarzschild black hole in Chapter~\ref{ch:NeutralP} and a charged Proca field on a brane charged black hole in Chapter~\ref{ch:ChargedP}. We have designed a numerical strategy to solve the coupled equations and showed that the coupled systems may be treated with an {\bf S}-matrix type formalism which allows decoupling in the asymptotic regions. This {\bf S}-matrix was used to define a transmission matrix the eigenvalues of which give us the transmission factors. Then the Hawking fluxes were calculated using the standard formulas. For a neutral Proca field, we found distinctive features by introducing the mass term, such as the lifting of the degeneracy of the two transverse modes in four dimensions, the appearance of longitudinal modes and in particular the $s$-wave. When both background and Proca field charges were included, we observed the existence of superradiant modes, and a charge splitting effect for small energies and for two or more extra dimensions. We also compared the Proca bulk-to-brane ratio of energy emission, showing that most of the energy is emitted on the brane.

In Chapter~\ref{ch:ChargedClouds}, we have studied quasi-bound states for both the charged massive scalar field and the charged Proca field in Reissner-Nordstr\"om black holes. We established that no such states exist in the superradiant regime for the Proca field, a similar behavior to that known for the scalar field. For both fields, however, decaying quasi-bound states with an arbitrary small imaginary part of the frequency exist and thus which are arbitrarily long lived. In the limit of vanishing imaginary part of the frequency, the fields did not trivialize and we dubbed the corresponding configurations as marginal scalar or Proca clouds, since they were only marginal bound.

A problem which is still open is the study of the Proca field in the Kerr black hole without any approximations. Indeed, this problem has been partially addressed for slow rotation~\cite{paniPRL,paniPRD}. This is still interesting because of the following open questions: (1) is it possible to construct a modified Newman-Penrose formalism to deal with linear perturbations for massive fields in general? (2) Is it possible to establish a numerical method to solve the type of two dimensional partial differential equation arising in this problem?

In Chapter~\ref{ch:scalarHD}, we have studied superradiant instabilities for a charged scalar field in a $D$-dimensional Reissner-Nordstr\"om-AdS black hole. By employing an analytic matching method and a numerical method, we proved that superradiant instabilities do exist for all $\ell$ modes in higher dimensions. Inspired by the large $D$ general relativity~\cite{Emparan:2013moa}, it would also be interesting to develop another analytic treatment to look for superradiant modes.

In Chapter~\ref{ch:KerrAdS}, we have studied Maxwell perturbations on Kerr-AdS black holes. From the viewpoint that the AdS boundary may be regarded as a perfectly reflecting mirror, we proposed \textit{vanishing energy flux} boundary conditions which are physically motivated. Imposing such conditions, we obtained a set of two Robin boundary conditions even for a Schwarzschild-AdS black hole, where only one of them has been reported in the literature. Applying these two boundary conditions to Kerr-AdS black holes, we have studied superradiant instabilities, and observed that the new branch of quasinormal modes may be unstable in a larger parameter space. Our results also showed that superradiant instabilities for the Maxwell field may exist for (moderately) larger black hole size, when comparing with the scalar case.

To study stationary vector clouds, we have solved the Teukolsky equations at the onset of the superradiant instability. We found that both boundary conditions can yield vector clouds, which are characterized by existence lines in the parameter space. These lines are bounded by pure AdS spaces and the extremal black holes, which differs from that observed for gravitational perturbations, for which only one of the sets of clouds are bounded by the extremal black holes~\cite{Cardoso:2013pza}. The existence of clouds at the linear level indicates nonlinear hairy black hole solutions, so the open question is to find the nonlinear realization of these vector clouds. There is already a well-known exact black hole family within the Einstein-Maxwell-AdS system: the Kerr-Newman-AdS family. It will then be very interesting to study properties of the new type of ``hairy'' black hole solutions, and understand the interplay between the Kerr-Newman-AdS black holes and the new family of ``hairy'' black holes.

%% file: app_neutralP.tex
\chapter{Functions and matrices for the neutral Proca case}
\label{app:neutralP}

This appendix contains details on the functions, recurrence relations and matrices, used in Chapter~\ref{ch:NeutralP}.

The functions appeared in Eqs.~\eqref{NPsysterm1} and~\eqref{NPsysterm2} are $(\kappa_s^2=\ell(\ell+n-1))$
\begin{align}
M(r)&\equiv \sum^{2n-1}_{m=0}{\alpha_m y^m}= r\left(r^{n-1}-1\right)^2 \; ,\nonumber\\
N(r)&\equiv \sum^{2n-2}_{m=0}{\beta_m y^m}=n\left(r^{n-1}-1\right)^2  \; ,\nonumber\\
P(r)&\equiv \sum^{2n-1}_{m=0}{\gamma_m y^m}=-(\kappa_s^2+\mu_p^2r^2)\left(r^{2n-3}-r^{n-2}\right) +\omega^2r^{2n-1}  \; ,\nonumber\\
Q(r)&\equiv \sum^{2n-2}_{m=0}{\sigma_m y^m}= i\omega r^{n-1}\left(2r^{n-1}-n-1\right) \; ,\nonumber\\
\tilde M(r)&\equiv \sum^{2n}_{m=0}{\tilde \alpha_m y^m}=r^2\left(r^{n-1}-1\right)^2  \; ,\nonumber\\
\tilde N(r)&\equiv \sum^{2n-1}_{m=0}{\tilde \beta_m y^m}=(n-2)r\left(r^{n-1}-1\right)^2\;,\nonumber\\
\tilde P(r)&\equiv \sum^{2n}_{m=0}{\tilde \gamma_m y^m}=-\left(\kappa_s^2+\mu_p^2r^2\right)\left(r^{2n-2}-r^{n-1}\right)
+\omega^2r^{2n}-(n-2)\left(r^{n-1}-1\right)^2\;,\nonumber\\
\tilde Q(r)&\equiv \sum^{n}_{m=0}{\tilde \sigma_m y^m}=-i\omega(n-1)r^n\;.\nonumber
\end{align}
The recurrence relations are
\begin{align}
\mu_0&=\nu_0\;,\nonumber\\
\mu_1&=-\frac{\left(\rho(\rho-1)\alpha_3+\rho\beta_2+\gamma_1+\sigma_1\right)\nu_0+\sigma_0\nu_1}{\rho(\rho+1)\alpha_2+\gamma_0}\;,\nonumber\\
\mu_j&=\frac{\omega^2+(n-1)^2(\rho+j)(\rho+j-1)}{D_j}f_j+\frac{i\omega(n-1)}{D_j}\tilde
f_j\;,\nonumber\\
\nu_j&=\frac{\omega^2+(n-1)^2(\rho+j)(\rho+j-1)}{D_j}\tilde
f_j+\frac{i\omega(n-1)}{D_j}f_j\;,
\label{NPrecurone}
\end{align}
with
\begin{align}
D_j&=(n-1)^2\omega^2+\left(\omega^2+(n-1)^2(\rho+j)(\rho+j-1)\right)^2\;,\nonumber\\
f_j&=-\sum^j_{m=1}{\Big[\big(\alpha_{m+2}(\rho+j-m)(\rho+j-m-1)+\beta_{m+1}(\rho+j-m)+\gamma_m\big)\mu_{j-m}+\sigma_m\nu_{j-m}\Big]}\;,\nonumber\\
\tilde
f_j&=-\sum^j_{m=1}{\left[\big(\tilde\alpha_{m+2}(\rho+j-m)(\rho+j-m-1)+\tilde\beta_{m+1}(\rho+j-m)+\tilde\gamma_m\big)\nu_{j-m}+\tilde\sigma_m\mu_{j-m}\right]}\;.\nonumber
\end{align}
The coefficients used in the asymptotic expansion in the text are
\begin{equation}\label{eq:NPcpm}
c^\pm=\dfrac{i}{2\omega}\left[-\kappa_s^2+\dfrac{n(2-n)}{4}+(2\omega^2-\mu_p^2)\left(\delta_{2,n}+\delta_{3,n}\right)+\left(\pm i\dfrac{\mu_p^2}{2k}-\left(\dfrac{\mu_p^2}{2k}\right)^2\right)\delta_{2,n}\right] \; .
\end{equation}
The matrices used in the text are as follows
\begin{equation}\label{eq:NPT}
\mathbf{T}=\frac{1}{r^{\frac{n-2}{2}}}\left(\begin{array}{cccc}\frac{e^{i\Phi}}{r} & \frac{e^{-i\Phi}}{r} & e^{i\Phi} & e^{-i\Phi} \vspace{2mm}\\ \frac{ike^{i\Phi}}{r} & -\frac{ike^{-i\Phi}}{r} & \left[ik+\frac{i\varphi-\frac{n-2}{2}}{r}\right]e^{i\Phi} & -\left[ik+\frac{i\varphi+\frac{n-2}{2}}{r}\right]e^{-i\Phi} \vspace{2mm}\\ 0 & 0 & \left(-\frac{k}{\omega}+\frac{c^+}{r}\right)e^{i\Phi}  & \left(\frac{k}{\omega}+\frac{c^-}{r}\right)e^{-i\Phi} \vspace{2mm}\\ 0 & 0 & \left[-\frac{ik^2}{\omega}+\frac{ikc^+-\frac{k}{\omega}(i\varphi-\frac{n-2}{2})}{r}\right]e^{i\Phi}& -\left[\frac{ik^2}{\omega}+\frac{ikc^-+\frac{k}{\omega}(i\varphi+\frac{n-2}{2})}{r}\right]e^{-i\Phi} \end{array}\right) \ ,
\end{equation}
\begin{equation}\label{eq:X}
\mathbf{X}=\left(\begin{array}{cccc}0 & 1 & 0 & 0 \vspace{2mm} \\ -\dfrac{P}{M} & -\dfrac{N}{M}& -\dfrac{Q}{M}  & 0 \vspace{2mm}\\0 & 0 & 0 & 1 \vspace{2mm} \\ -\dfrac{\tilde Q}{\tilde M}  & 0& -\dfrac{\tilde P}{\tilde M}&-\dfrac{\tilde N}{\tilde M} \end{array}\right) \ ,
\end{equation}
and the 2-vectors
\begin{equation}\label{eq:NPyplus}
\mathbf{y}^{\pm}=\left(\begin{array}{c}\sqrt{\dfrac{\kappa_s^2 k}{\omega}} a_0^{\pm} \\ \sqrt{\dfrac{\omega k}{\mu_p^2}}\left[(\pm i\varphi-\frac{n-2}{2}+i\omega c^{\pm}\mp ik\delta_{n,2})a_0^{\pm}\pm ika_1^{\pm}\right] \end{array}\right) \ ,
\end{equation}
\begin{equation}\label{eq:NPhminus}
\mathbf{h}^-=\left(\begin{array}{c}\sqrt{\kappa_s^2} \nu_0 \\ \dfrac{i\omega(\mu_1-\rho(\frac{n}{2}\nu_0+\nu_1-\mu_1))+\kappa_s^2\nu_0}{\mu_p} \end{array}\right) \ .
\end{equation} 

%% file: app_chargedP.tex
\chapter{Functions and matrices for the charged Proca case}
\label{app:chargedP}

This appendix contains details on the functions and matrices used to rewrite the second order radial differential equations into a first order form, and recurrence relations used to initialize these radial differential equations close to the event horizon. This was used in Chapter~\ref{ch:ChargedP}.

The functions that used in the text are (where $\kappa_s^2=\ell(\ell+1)$):
\begin{align}
A(r)&\equiv \sum^{2n+1}_{m=0}{a_m y^m}=r \left[r^n-(1+Q^2)r+Q^2r^{n-2}\right]^2 \; ,\nonumber\\
B(r)&\equiv \sum^{2n}_{m=0}{b_m y^m}=2\left[r^n-(1+Q^2)r+Q^2r^{n-2}\right]^2\; ,\nonumber\\
C(r)&\equiv \sum^{2n+1}_{m=0}{c_m y^m}= (\omega r-qQ)^2r^{2n-1}
-(\kappa_s^2+\mu_p^2r^2)r^{n-1}\left[r^n-(1+Q^2)r+Q^2r^{n-2}\right]  \; ,\nonumber\\
E(r)&\equiv \sum^{2n}_{m=0}{e_m y^m}=iqQr^{n-1}\left[(1+Q^2)(n-1)r-2Q^2r^{n-2}\right]\nonumber\\& \;\;\;\;\;\;\;\;\;\;\;\;\;\;\;\;\;\;\;\;\;\;\;\;+i\omega r^n\left[2r^{n}-(1+Q^2)(n+1)r+4Q^2r^{n-2}\right] \, ,\nonumber\\
\tilde A(r)&\equiv \sum^{2n}_{m=0}{\tilde a_m y^m}=\left[r^n-(1+Q^2)r+Q^2r^{n-2}\right]^2  \; ,\nonumber\\
\tilde B(r)&\equiv 0 \;,\nonumber
\end{align}
\begin{align}
\tilde C(r)\equiv \sum^{2n}_{m=0}{\tilde c_m y^m}=& (\omega r-qQ)^2r^{2n-2}
 -\left(\kappa_s^2+\mu_p^2r^2\right)r^{n-2}\left[r^n-(1+Q^2)r+Q^2r^{n-2}\right] \;,\nonumber\\
\tilde E(r)\equiv \sum^{2n-2}_{m=0}{\tilde e_m y^m}=&iqQ r^{n-1}\left[(n-3)(1+Q^2)+2r^{n-1}\right]\nonumber\\&-i\omega r^{n-1}\left[(1+Q^2)(n-1)r-2Q^2r^{n-2}\right]\;.\nonumber
\end{align}
The recurrence relations are
\begin{align}
\mu_0&=\nu_0\;,\nonumber\\
\mu_1&=-\frac{\left[a_3\rho(\rho-1)+b_2\rho +c_1+e_1\right]\nu_0+ e_0 \nu_1}{a_2\rho(\rho+1)+c_0}\;,\nonumber\\
\mu_j&=\frac{\tilde a_2(\rho+j)(\rho+j-1)+\tilde c_0}{D_j}f_j-\frac{e_0}{D_j}\tilde
f_j\;,\nonumber\\
\nu_j&=\frac{a_2(\rho+j)(\rho+j-1)+c_0}{D_j}\tilde
f_j-\frac{\tilde e_0}{D_j}f_j\;,
\label{CPrecurone}
\end{align}
with
\begin{align}
D_j&=\left[a_2(\rho+j)(\rho+j-1)+c_0\right]\left[\tilde a_2(\rho+j)(\rho+j-1)+\tilde c_0\right]-\tilde e_0 e_0\;,\nonumber\\
f_j&=-\sum^j_{m=1}\left[(a_{m+2}(\rho+j-m)(\rho+j-m-1)+b_{m+1}(\rho+j-m)+c_m\big)\mu_{j-m}+e_m\nu_{j-m}\right]\;,\nonumber\\
\tilde f_j&=-\sum^j_{m=1}\left[\big(\tilde a_{m+2}(\rho+j-m)(\rho+j-m-1)+\tilde c_m\big)\nu_{j-m}+\tilde e_m\mu_{j-m}\right]\;.\nonumber
\end{align}
\begin{text}
The coefficients used in the asymptotic expansion in the text are
\begin{align}
c^\pm=&\dfrac{i}{2\omega}\Big[-\kappa_s^2+Q^2(\mu_p^2-2\omega^2)-\dfrac{q^2Q^2\mu_p^2}{k^2}\mp i\dfrac{qQ\omega}{k}+(2\omega^2-\mu_p^2)(1+Q^2)\delta_{3,n}\nonumber\\&+\Big(\pm i\dfrac{\mu_p^2(1+Q^2)}{2k}-\Big(\dfrac{\mu_p^2(1+Q^2)}{2k}\Big)^2+(1+Q^2)^2(2\omega^2-\mu_p^2)\nonumber\\&+\dfrac{qQ\omega(1+Q^2)(\mu_p^2-2k^2)}{k^2}\Big)\delta_{2,n}\Big] \; .\nonumber\label{CPcpm}
\end{align}
The relation between the new first order radial functions $\mathbf{\Psi}$ used in~\eqref{eq:ODEcoupled}, and the 4-vector $\mathbf{V}^T=(\psi,d_r\psi,\chi,d_r\chi)$ for the original fields and derivatives is found from Eqs.~\eqref{asymptoticpsi},~\eqref{asymptoticpchi} and their derivatives. The corresponding $r$-dependent matrix transformation $\mathbf{T}$ is defined
\begin{equation}
\mathbf{V}= \mathbf{T} \mathbf{\Psi} \; ,
\end{equation}
and its form is
\begin{equation}\label{eq:CPT}
\mathbf{T}=\left(\begin{array}{cccc}\frac{e^{i\Phi}}{r} & \frac{e^{-i\Phi}}{r} & e^{i\Phi} & e^{-i\Phi} \vspace{2mm}\\ \frac{ike^{i\Phi}}{r} & -\frac{ike^{-i\Phi}}{r} & \left[ik+\frac{i\varphi}{r}\right]e^{i\Phi} & -\left[ik+\frac{i\varphi}{r}\right]e^{-i\Phi} \vspace{2mm}\\ 0 & 0 & \left(-\frac{k}{\omega}+\frac{c^+}{r}\right)e^{i\Phi}  & \left(\frac{k}{\omega}+\frac{c^-}{r}\right)e^{-i\Phi} \vspace{2mm}\\ 0 & 0 & \left[-\frac{ik^2}{\omega}+\frac{ikc^+-\frac{ik\varphi}{\omega}}{r}\right]e^{i\Phi}& -\left[\frac{ik^2}{\omega}+\frac{ikc^-+\frac{ik\varphi}{\omega}}{r}\right]e^{-i\Phi} \end{array}\right) \ ,
\end{equation}
On another hand, the original system~\eqref{originalsys1} and
~\eqref{originalsys2}, can be written in a first order form
\begin{equation}
\dfrac{d\mathbf{V}}{dr}=\mathbf{X}\mathbf{V} \; ,
\end{equation}
where the matrix $\mathbf{X}$ is
\begin{equation}\label{eq:CPX}
\mathbf{X}=\left(\begin{array}{cccc}0 & 1 & 0 & 0 \vspace{2mm} \\ -\dfrac{C}{A} & -\dfrac{B}{A}& -\dfrac{E}{A}  & 0 \vspace{2mm}\\0 & 0 & 0 & 1 \vspace{2mm} \\ -\dfrac{\tilde E}{\tilde A}  & 0& -\dfrac{\tilde C}{\tilde A}& 0 \end{array}\right) \ ,
\end{equation}
and the 2-vectors
\begin{equation}\label{eq:CPyplus}
\mathbf{y}^{\pm}=\left(\begin{array}{c}\sqrt{\dfrac{\kappa_s^2 k}{\omega^2}} a_0^{\pm} \\ i\sqrt{\dfrac{k}{\mu_p^2}}\left[\left(\pm \varphi+\omega c^{\pm}\mp k(1+Q^2)\delta_{n,2}\pm\dfrac{kqQ}{\omega}\right)a_0^{\pm}\pm ka_1^{\pm}\right] \end{array}\right) \ ,
\end{equation}
\begin{equation}\label{eq:hminus}
\mathbf{h}^-=\left(\begin{array}{c}\sqrt{t} \nu_0 \\ \sqrt{\dfrac{\mu_p^2}{\mu_p^4+q^2Q^2}}\left(-i\kappa_s^2\nu_0+(\omega-qQ)d \right) \end{array}\right) \; ,
\end{equation}
with
\begin{align}
t&=\dfrac{\kappa_s^2}{\mu_p^4+q^2Q^2}\Big[(\omega-qQ)\left(qQa+b\mu_p^2+2i\rho \mu_p^2\alpha+\dfrac{2qQ\mu_p^2}{\beta}\right)-\kappa_s^2\mu_p^2\Big]\;,\nonumber\\
d&=\left[\left(1+\rho-\dfrac{iqQ}{\mu_p^2}\rho\right)(a+ib)+\left(1+\dfrac{iqQ}{\mu_p^2}\right)\left(\dfrac{iqQ}{\beta}-\alpha\rho\right)\right]\nu_0+(1+2\rho)\left(\dfrac{iqQ}{\mu_p^2}-1\right)\nu_1 \;,\nonumber\\
a&=\dfrac{qQ}{\omega-qQ}\;,\;\;\;\;\;\;\;b=\dfrac{qQ(2+n(n-3)(1+Q^2))}{\beta^2}-\dfrac{2\omega\alpha}{\beta}+\dfrac{\kappa_s^2+\mu_p^2}{\omega-qQ}\;,\nonumber\\
\alpha&=\dfrac{n(n-1)+(n-3)(n+2)Q^2}{2\beta}\;,\;\;\;\;\;\;\;\beta=(n-1)+(n-3)Q^2\;.\nonumber
\end{align}
\end{text}


%% file: app_ST1.tex
\chapter{Proof of Theorem 1}
\label{app:ST1}
In this appendix, we prove Theorem~\ref{thm1} given in Chapter~\ref{ch:prelim}, by taking $\bar{\mathscr{L}_0}\bar{\mathscr{L}_1}S_{+1}=BS_{-1}$, as an example. To do so, we start from Eq.~\eqref{Spluseq}, by applying the operator $\bar{\mathscr{L}_0}\bar{\mathscr{L}_1}$ on both sides,
\begin{align}
-\lambda\bar{\mathscr{L}_0}\bar{\mathscr{L}_1}S_{+1}&=\bar{\mathscr{L}_0}\bar{\mathscr{L}_1}(\bar{\mathscr{L}_0}^\dag\bar{\mathscr{L}_1}-2a\omega\cos\theta\Xi)S_{+1}\nonumber\\
&=\bar{\mathscr{L}_0}\bar{\mathscr{L}_1}\bar{\mathscr{L}_0}^\dag\bar{\mathscr{L}_1}S_{+1}-2a\omega\Xi\bar{\mathscr{L}_0}(\cos\theta\bar{\mathscr{L}_1}-\sin\theta\sqrt{\Delta_\theta})S_{+1}\nonumber\\
&=\bar{\mathscr{L}_0}\bar{\mathscr{L}_1}\bar{\mathscr{L}_0}^\dag\bar{\mathscr{L}_1}S_{+1}-2a\omega\Xi\cos\theta\bar{\mathscr{L}_0}\bar{\mathscr{L}_1}S_{+1}+4a\omega\Xi\sqrt{\Delta_\theta}\sin\theta \bar{\mathscr{L}_1}S_{+1}\;,\label{angularTeukol1}
\end{align}
where relations 
\begin{equation}
\bar{\mathscr{L}_n}^\dag\cos\theta=\cos\theta\bar{\mathscr{L}_n}^\dag-\sin\theta\sqrt{\Delta_\theta}\;,\;\;\;\;\;\;
\sqrt{\Delta_\theta}\sin\theta\bar{\mathscr{L}}_{n+1}=\bar{\mathscr{L}_n}\sqrt{\Delta_\theta}\sin\theta\;,
\end{equation}
have been used. The first term in Eq.~\eqref{angularTeukol1} can be further simplified
\begin{align}
\bar{\mathscr{L}_0}\bar{\mathscr{L}_1}\bar{\mathscr{L}_0}^\dag\bar{\mathscr{L}_1}S_{+1}&=\bar{\mathscr{L}_0}\bar{\mathscr{L}_1}(\bar{\mathscr{L}_0}+2\sqrt{\Delta_\theta}\mathcal{Q})\bar{\mathscr{L}_1}S_{+1}\nonumber\\
&=\bar{\mathscr{L}_0}(\bar{\mathscr{L}_1}^\dag-2\sqrt{\Delta_\theta}\mathcal{Q})\bar{\mathscr{L}_0}\bar{\mathscr{L}_1}S_{+1}+2\bar{\mathscr{L}_0}\bar{\mathscr{L}_1}\sqrt{\Delta_\theta}\mathcal{Q}\bar{\mathscr{L}_1}S_{+1}\;,
\end{align}
in which
\begin{align}
\bar{\mathscr{L}_1}\sqrt{\Delta_\theta}\mathcal{Q}&=\left(\bar{\mathscr{L}_0}+\dfrac{1}{\sin\theta}\dfrac{d}{d\theta}(\sqrt{\Delta_\theta}\sin\theta)\right)\sqrt{\Delta_\theta}\mathcal{Q}\nonumber\\
&=\sqrt{\Delta_\theta}\mathcal{Q}\bar{\mathscr{L}_0}+\sqrt{\Delta_\theta}\dfrac{d}{d\theta}(\sqrt{\Delta_\theta}\mathcal{Q})+\dfrac{\sqrt{\Delta_\theta}\mathcal{Q}}{\sin\theta}\dfrac{d}{d\theta}(\sqrt{\Delta_\theta}\sin\theta)\nonumber\\
&=\sqrt{\Delta_\theta}\mathcal{Q}\bar{\mathscr{L}_0}+\dfrac{d}{d\theta}(\Xi H)+\Xi H\cot\theta=\sqrt{\Delta_\theta}\mathcal{Q}\bar{\mathscr{L}_0}+2a\omega\Xi\cos\theta\;,
\end{align}
where $\mathcal{Q}=\tfrac{\Xi H}{\Delta_\theta}$, as defined in Eq.~\eqref{Qdef},
and where the following relation is used
\begin{equation}
\dfrac{dH}{d\theta}+H\cot\theta=2a\omega\cos\theta\;.
\end{equation}
Then Eq.~\eqref{angularTeukol1} becomes
\begin{align}
-\lambda\bar{\mathscr{L}_0}\bar{\mathscr{L}_1}S_{+1}&=\bar{\mathscr{L}_0}(\bar{\mathscr{L}_1}^\dag-2\sqrt{\Delta_\theta}\mathcal{Q})\bar{\mathscr{L}_0}\bar{\mathscr{L}_1}S_{+1}+2\bar{\mathscr{L}_0}(\sqrt{\Delta_\theta}\mathcal{Q}\bar{\mathscr{L}_0}+2a\omega\Xi\cos\theta)\bar{\mathscr{L}_1}S_{+1}\nonumber\\
&\;\;\;\;-2a\omega\Xi\cos\theta\bar{\mathscr{L}_0}\bar{\mathscr{L}_1}S_{+1}+4a\omega\Xi\sqrt{\Delta_\theta}\sin\theta \bar{\mathscr{L}_1}S_{+1}\nonumber\\
&=\bar{\mathscr{L}_0}\bar{\mathscr{L}_1}^\dag\bar{\mathscr{L}_0}\bar{\mathscr{L}_1}S_{+1}+4a\omega\Xi\bar{\mathscr{L}_0}\cos\theta\bar{\mathscr{L}_1}S_{+1}-2a\omega\Xi\cos\theta\bar{\mathscr{L}_0}\bar{\mathscr{L}_1}S_{+1}\nonumber\\&\;\;\;\;+4a\omega\Xi\sqrt{\Delta_\theta}\sin\theta \bar{\mathscr{L}_1}S_{+1}\nonumber\\
&=\bar{\mathscr{L}_0}\bar{\mathscr{L}_1}^\dag\bar{\mathscr{L}_0}\bar{\mathscr{L}_1}S_{+1}+4a\omega\Xi(\cos\theta\bar{\mathscr{L}_0}-\sin\theta\sqrt{\Delta_\theta})\bar{\mathscr{L}_1}S_{+1}\nonumber\\&\;\;\;\;-2a\omega\Xi\cos\theta\bar{\mathscr{L}_0}\bar{\mathscr{L}_1}S_{+1}
+4a\omega\Xi\sqrt{\Delta_\theta}\sin\theta \bar{\mathscr{L}_1}S_{+1}\nonumber\\
&=(\bar{\mathscr{L}_0}\bar{\mathscr{L}_1}^\dag+2a\omega\Xi\cos\theta)\bar{\mathscr{L}_0}\bar{\mathscr{L}_1}S_{+1}\;.\label{angularTeukol2}
\end{align}
Comparing Eq.~\eqref{angularTeukol2} with Eq.~\eqref{Sminuseq} and remembering the definition in Eq.~\eqref{newangoperator}, one concludes that $\bar{\mathscr{L}_0}\bar{\mathscr{L}_1}S_{+1}$ is proportional to $S_{-1}$. We identify the proportionality constant as $B$, such that
\begin{equation}
\bar{\mathscr{L}_0}\bar{\mathscr{L}_1}S_{+1}=BS_{-1}\;.\label{eqB1}
\end{equation}
Following the same procedures as above, one can also prove that
\begin{equation}
\bar{\mathscr{L}_0}^\dag\bar{\mathscr{L}_1}^\dag S_{-1}=BS_{+1}\;,\label{eqB2}
\end{equation}
where the same proportionality constant $B$ is used. This property is guaranteed by the normalization conditions given in Eq.~\eqref{angnorm}.
\\
Therefore, Theorem~\ref{thm1} is proved. 

%% file: app_angmomflux.tex
\chapter{Angular momentum flux}
\label{app:angmomflux}

This appendix contains details on derivation of the angular momentum flux for the Maxwell field on the Kerr-AdS background. Our purpose is to show that the vanishing of the energy flux, a physical principle used in Chapter~\ref{ch:KerrAdS} to impose boundary conditions for the Maxwell field on Kerr-AdS, leads to a vanishing angular momentum flux.

From the definition of the energy-momentum tensor for the Maxwell field
\begin{equation}
T_{\mu \nu}=F_{\mu\sigma}F^\sigma_{\;\;\;\nu}+\dfrac{1}{4}g_{\mu\nu}F^2\;,
\end{equation}
we can calculate the angular momentum flux
\begin{equation}
\mathcal{J}=\int_{S^2} \sin\theta d\theta d\varphi\; r^2 \left(T^r_{\;\;\varphi,\;\uppercase\expandafter{\romannumeral1}}+T^r_{\;\;\varphi,\;\uppercase\expandafter{\romannumeral2}}\right)\;,\label{angmomf}
\end{equation}
with
\begin{align}
&T^r_{\;\;\varphi,\;\uppercase\expandafter{\romannumeral1}}=-a\sin^2\theta\;T^r_{\;\;t,\;\uppercase\expandafter{\romannumeral1}}\;,\label{relation1}\\
&T^r_{\;\;\varphi,\;\uppercase\expandafter{\romannumeral2}}=-\dfrac{i\sin\theta\sqrt{\Delta_\theta}(r^2+a^2)}{2\Xi\rho^4}\Phi_1(\Phi_2^\ast+\Delta_r\Phi_0^\ast)+c.c\;,\label{relation2}
\end{align}
where $T^r_{\;\;t,\;\uppercase\expandafter{\romannumeral1}}$ is given in Eq.~\eqref{rtcom}, $c.c$ stands for the complex conjugate of the preceding terms, and
\begin{equation}
\Phi_0=\phi_0\;,\;\;\;\;\;\Phi_2=2\bar{\rho}^\ast\phi_2\;,\;\;\;\;\;\bar{\rho}=r+ia\cos\theta\;.
\end{equation}

From Eq.~\eqref{relation1}, and considering the vanishing energy flux boundary conditions, i.e.
\begin{equation}
\int_{S^2} \sin\theta d\theta d\varphi\; r^2 T^r_{\;\;t,\;\uppercase\expandafter{\romannumeral1}}\rightarrow0\;,
\end{equation}
asymptotically, one concludes that there is no contributions for the angular momentum flux from the first term $T^r_{\;\;\varphi,\;\uppercase\expandafter{\romannumeral1}}$.

For the second term, from Eq.~\eqref{relation2}, we notice that $\Phi_1$ is involved, which has been derived in Eq.~\eqref{Phi1eq}. For convenience, we list this solution and related expressions below. The solution for $\Phi_1$ is
\begin{equation}
\bar{\rho}^\ast\Phi_1=g_{+1}\bar{\mathscr{L}}_1S_{+1}-ia f_{-1}\mathscr{D}_0P_{-1}\;,
\end{equation}
with
\begin{align}
&g_{+1}=\frac{1}{B}(r\mathscr{D}_0P_{-1}-P_{-1})\;,\\
&f_{-1}=\frac{1}{B}(\cos\theta\bar{\mathscr{L}}_1S_{+1}+\sin\theta\sqrt{\Delta_\theta}S_{+1})\;,\\
&\bar{\mathscr{L}}_1S_{+1}=\dfrac{(2a\omega\Xi\cos\theta-\lambda)S_{+1}-BS_{-1}}{2\mathcal{Q}\sqrt{\Delta_\theta}}\;,
\end{align}
where
\begin{equation}
\mathscr{D}_0=\dfrac{\partial}{\partial r}-\dfrac{iK_r}{\Delta_r}\;,\;\;\;\mathcal{Q}=\dfrac{\Xi(a\omega\sin^2\theta-m)}{\sin\theta\Delta_\theta}\;,\;\;\;P_{-1}=BR_{-1}\;,
\end{equation}
and the constant $B$ is given by Eq.~\eqref{Bvalue}, $S_{+1}$ and $S_{-1}$ are spin weighted AdS spheroidal harmonics. From the properties of these spheroidal harmonic functions
\begin{equation}
S_{+1}(\pi-\theta)=S_{-1}(\theta)\;,\;\;\;\;\;\;S_{-1}(\pi-\theta)=S_{+1}(\theta)\;,
\end{equation}
which are guaranteed by the angular equations~\eqref{Spluseq} and~\eqref{Sminuseq}, then we have the following properties
\begin{align}
\int_0^\pi d\theta\sin\theta f_{odd}(\theta)S_{+1}(\theta)S^\ast_{+1}(\theta)=-\int_0^\pi d\theta\sin\theta f_{odd}(\theta)S_{-1}(\theta)S^\ast_{-1}(\theta)\;,\nonumber\\
\int_0^\pi d\theta\sin\theta f_{even}(\theta)S_{+1}(\theta)S^\ast_{+1}(\theta)=\int_0^\pi d\theta\sin\theta f_{even}(\theta)S_{-1}(\theta)S^\ast_{-1}(\theta)\;,\nonumber\\
\int_0^\pi d\theta\sin\theta f_{odd}(\theta)S_{-1}(\theta)S^\ast_{+1}(\theta)=-\int_0^\pi d\theta\sin\theta f_{odd}(\theta)S_{+1}(\theta)S^\ast_{-1}(\theta)\;,\nonumber\\
\int_0^\pi d\theta\sin\theta f_{even}(\theta)S_{-1}(\theta)S^\ast_{+1}(\theta)=\int_0^\pi d\theta\sin\theta f_{even}(\theta)S_{+1}(\theta)S^\ast_{-1}(\theta)\;,\label{intepro}
\end{align}
where
\begin{equation}
f_{odd}(\pi-\theta)=-f_{odd}(\theta)\;,\;\;\;\;\;\;f_{even}(\pi-\theta)=f_{even}(\theta)\;.\nonumber
\end{equation}
With all of these expressions at hand, and making use of the integration properties of the spin weighted AdS spheroidal harmonics given in~\eqref{intepro}, Eq.~\eqref{relation2} becomes
\begin{eqnarray}
&&T^r_{\;\;\varphi,\;\uppercase\expandafter{\romannumeral2}}=-\dfrac{i\sin\theta\sqrt{\Delta_\theta}(r^2+a^2)}{2\Xi\rho^4}(\mathcal{C}_1S_{+1}S^\ast_{-1}+\mathcal{C}_2S_{-1}S^\ast_{-1})
+c.c\;,\label{Trphi2rep}
\end{eqnarray}
where terms that vanish under the angular integration, due to the properties listed above, have been discarded. The expressions for $\mathcal{C}_1$ and $\mathcal{C}_2$ are messy in general, but they can be simplified asymptotically. The asymptotic expression for $\mathcal{C}_1$ goes as
\begin{equation}
\mathcal{C}_1\sim c_0+\mathcal{O}(1/r)\;,\label{c1expasy1}
\end{equation}
where $c_0$ is proportional to $T^r_{\;\;t,\;\uppercase\expandafter{\romannumeral1}}$ asymptotically, so that finally $\mathcal{C}_1\sim \mathcal{O}(1/r)$. Similar analysis can be done for $\mathcal{C}_2$ as well. The asymptotic expression for $\mathcal{C}_2$ is
\begin{equation}
\mathcal{C}_2\sim \hat{c}_0+\mathcal{O}(1/r)\;,\label{c1expasy2}
\end{equation}
and, as in the former case, $\hat{c}_0$ vanishes after the vanishing energy flux boundary conditions in Eq.~\eqref{bc} are imposed. Then from Eq.~\eqref{Trphi2rep}, together with Eqs~\eqref{c1expasy1} and~\eqref{c1expasy2}, we conclude that
\begin{eqnarray}
r^2T^r_{\;\;\varphi,\;\uppercase\expandafter{\romannumeral2}}\sim \mathcal{O}(1/r)\;,
\end{eqnarray}
asymptotically, which leads to the vanishing of the angular momentum flux of Eq.~\eqref{angmomf}.

%% file: pub-list.tex
\chapter{List of publications}
This thesis is based on the following published papers by the author:
\begin{enumerate}
\item
{\textit{Hawking radiation for a Proca field in D-dimensions},\\
Carlos~Herdeiro, Marco~O.~P.~Sampaio, Mengjie~Wang,\\
{}\href{http://journals.aps.org/prd/abstract/10.1103/PhysRevD.85.024005}{Phys.\ Rev.\ D {\bf 85} (2012) 2, 024005}\;$(${}\href{http://arxiv.org/abs/1110.2485}{\tt arXiv:1110.2485}$)$.
}

\item
{\textit{Hawking radiation for a Proca field in D dimensions. II. charged field in a brane charged black hole},\\
Mengjie~Wang, Marco~O.~P.~Sampaio, Carlos~Herdeiro,\\
{}\href{http://journals.aps.org/prd/abstract/10.1103/PhysRevD.87.044011}{Phys.\ Rev.\ D {\bf 87} (2013) 4, 044011}\;$(${}\href{http://arxiv.org/abs/1212.2197}{\tt arXiv:1212.2197}$)$.
}

\item
{\textit{Superradiant instabilities in a D-dimensional small Reissner-Nordstr\"om-anti-de Sitter black hole},\\
Mengjie~Wang, Carlos~Herdeiro,\\
{}\href{http://journals.aps.org/prd/abstract/10.1103/PhysRevD.89.084062}{Phys.\ Rev.\ D {\bf 89} (2014) 8, 084062}\;$(${}\href{http://arxiv.org/abs/1403.5160}{\tt arXiv:1403.5160}$)$.
}

\item
{\textit{Marginal scalar and Proca clouds around Reissner-Nordstr\"om black holes},\\
Marco~O.~P.~Sampaio, Carlos~Herdeiro, Mengjie~Wang,\\
{}\href{http://journals.aps.org/prd/abstract/10.1103/PhysRevD.90.064004}{Phys.\ Rev.\ D {\bf 90} (2014) 6, 064004}\;$(${}\href{http://arxiv.org/abs/1406.3536}{\tt arXiv:1406.3536}$)$.
}

\item
{\textit{Maxwell perturbations on asymptotically anti-de Sitter spacetimes: generic boundary conditions and a new branch of quasinormal modes},\\
Mengjie~Wang, Carlos~Herdeiro, Marco O. P. Sampaio,\\
{}\href{http://journals.aps.org/prd/abstract/10.1103/PhysRevD.92.124006}{Phys.\ Rev.\ D {\bf 92} (2015) 12, 124006}\;$(${}\href{http://arxiv.org/abs/1510.04713}{\tt arXiv:1510.04713}$)$.
}

\item
{\textit{Maxwell perturbations on Kerr-anti-de Sitter: quasinormal modes, superradiant instabilities and vector clouds},\\
Mengjie~Wang, Carlos~Herdeiro,\\
{}\href{https://journals.aps.org/prd/abstract/10.1103/PhysRevD.93.064066}
{Phys.\ Rev.\ D {\bf 93} (2016) 6, 064066}\;$(${}\href{http://arxiv.org/abs/1512.02262}{\tt arXiv:1512.02262}$)$.
}
\end{enumerate}




